\documentclass[3p,preprint,hyperref]{elsarticle}
\input{FIFF_aux}

\begin{document}

\begin{frontmatter}
\title{Feynman integrals and Fox functions}

\author[torino]{Giampiero Passarino}
\ead{giampiero@to.infn.it}

\address[torino]{\csuma}


\begin{abstract}
 \noindent
In this work we discuss the connection between Feynman integrals and Fox functions. Illustrative examples are
given.
\end{abstract}

\end{frontmatter}

\tableofcontents

\newpage
%
%
%
\section{Introduction \label{intro}}
The connection between higher transcendental functions (HTF) and Feynman integrals is a well{-}known 
subject, see \Brefs{Kalmykov:2008ofy,Abreu:2019xep,Blumlein:2021hbq,Klausen:2023gui}; for instance, a large
class of Feynman integrals can be expressed in terms of Horn functions~\cite{HTF}.
As a matter of fact \Brefs{TR,TRNS} suggested to consider Feynman integrals as a kind of generalized hypergeometric 
functions (see also \Brefs{Ponzano:1970ch,Barucchi:1973zm}).

In this work we will discuss techniques for the evaluation of one{-}loop and two{-}loop Feynman integrals with 
emphasis on the representation of the integrals through the generalized, multivariate, 
Fox function~\cite{oFox,compH,HS,Hus}. 

Following the work of \Bref{TR} it was suggested in \Bref{Kershaw:1973km} that, by studying Feynman integrals
as power series we can \textit{develop the idea} of Feynman integrals as generalized hypergeometric functions and
we want to present the idea that the necessay tools are the Fox functions.

The Fox $\mrH$ functions are alternative representations for Mellin{-}Barnes (hereafter MB) contour 
integrals~\cite{Boos:1990rg}; they provide
a general framework and a concise notation. 
Furthermore, Mellin–Barnes integrals and various generalizations of them play an important
role in the theory of hypergeometric and other special functions~\cite{Nauka}. 
However, they have to be solved numerically. Quoting \Bref{FFR}
``the most up-to-date symbolic softwares still have problems $\dots$ the Fox function is usually not even implemented''.
Stated differently, we can always ``invent'' some ``special'' function but the strict requirement is that we must
know all the properties of the function: series representation, integral representation, analytic continuation,
recurrence relations, contiguity relations \etc Finally we must have a computable form for the function.

It is not the purpose of this work to discuss renormalization in quantum field theory; renormalization in the MB
representation of Feynman integrals has been discussed in \Bref{deCalan:1979ii}. 

The outline of the paper is as follows: in \sect{HTF} we discuss properties of the most common HTFs. We will
present comparisons among different results, taking into account that different authors adopt different conventions.
After discussing generalized hypergeometric functions we will introduce the Meijer $\mrG$ function in \sect{MGfun}
and Fox $\mrH$ function in \sect{Ffun}, including its representations in \sect{repHfun}, special cases
in \sect{SCFfun}. The relation between $\mrG$ and $\mrH$ is presented in \sect{fFtM}, multivariate
$\mrH$ functions are introduced in \sect{mHfun}, analytic continuation (including the behavior on
the unit disk) in \sect{hmff}. Generalized, multivariate $\mrH$ functions are introduced in \sect{GMFF}, partial
differential equations in \sect{PDE}.

From \sect{FD} we introduce Feynman integrals;
in \sect{OLLP} we present examples for one{-}loop Feynman integrals (in arbitrary space{-}time dimensions)
using alternative representations.
In \sect{TLFI} we start discussing two{-}loop Feynman integrals, presenting a procedure for partial quadratization 
of the Symanzik polynomials~\cite{Bogner:2010kv,Weinzierl:2013yn}; after introducing the basic aspects of the
two{-}loop strategy in \sect{str} we proceed by discussing few explicit examples.
Connection with Landau equations are discussed in \sect{LEQS}.
The alternative approach of a numerical integration (ab initio) of the Feynman integrals is discussed in \sect{numi}.
We draw our conclusions in \sect{Conc}.
\section{Higher transcendental functions \label{HTF}}
In this Section, we describe methods for the evaluation of the most relevant HTFs, especially
when they are connected to Feynman integrals in the physical region.
A large fraction of this Section has been written for the reader's convenience. 
A significant progress has been made in the theory of hypergeometric functions of several variables,
most of the results are well{-}known in 
the literature and we evaluated how the new results fit in the pre{-}existing literature by including the needed 
in{-}text citations. In spite of that this Section should not be taken as a review of known results; the
collection of results will prove to be useful for the evaluation of Feynman integrals.

The most common HTF is the Gauss hypergeometric funcion~\cite{HTF,abramowitz+stegun}:
\bq
\hyp{\mra}{\mrb}{\mrc}{\mrz} =
\frac{\eG{\mrc}}{\eG{\mrb}\,\eG{\mrc - \mrb}}\,
\int_0^1 \mrd \mrx\,\mrx^{\mrb-1}\,\lpar 1 - \mrx \rpar^{\mrc - \mrb - 1}\,
\lpar 1 - \mrz\,\mrx \rpar^{-\mra} \spc
\eq
valid for $\mid \marg(1 - \mrz) \mid < \pi$ and $\Re \mrc > \Re \mrb > 0$; to avoid the second condition requires
replacing the path of integration $[0\,,\,1]$ with the Pochhammer contour. 
A second type of integral representation is given by a MB integral~\cite{HTF}
\bq
\hyp{\mra}{\mrb}{\mrc}{\mrz} =
\frac{\eG{\mrc}}{\eG{\mra}\,\eG{\mrb}}\,\int_{\mrL}\,\frac{\mrd \mrs}{2\,i\,\pi}\,
\frac{\eG{\mra + \mrs}\,\eG{\mrb + \mrs}\,\eG{ - \mrs}}{\eG{\mrc + \mrs}}\,
\lpar - \mrz \rpar^{\mrs}\,
\label{F21MB}
\eq
where $\mid \marg( - \mrz ) \mid < \pi$ and the contour $\mrL$ separates the poles of $\eG{ - \mrs}$ from the
poles of $\eG{\mra + \mrs}$ and of $\eG{\mrb + \mrs}$.
\subsection{Generalized hypergeometric functions \label{GHF}}
The relevant generalized hypergeometric functions are the
${}_{\scriptstyle{\mrp+1}}\,\mrF_{\scriptstyle{\mrp}}$.
In computing them we will use the following result~\cite{TIF}:
\bq
\int_0^1 \mrd \mrx \,\mrx^{\mrs-1}\,\lpar 1 - \mrx \rpar^{\nu-1}\,
{}_{\scriptstyle{\mrp}}\,\mrF_{\scriptstyle{\mrq}}\lpar \mra_1\,\dots\,\mra_{\mrp}\,;\,\mrb_1\,\dots\,
\mrb_{\mrq}\,;\,\alpha\,\mrx \rpar =
\eB{\nu}{\mrs}\,
{}_{\scriptstyle{\mrp+1}}\,\mrF_{\scriptstyle{\mrq+1}}\lpar \mra_1\,\dots\,\mra_{\mrp}\,,\mrs\,;\,
\mrb_1\,\dots\,\mrb_{\mrq}\,,\mrs + \nu\,;\,\alpha \rpar \spc
\label{pFq}
\eq
where $\Re s > 0$ and $\Re \nu > 0$. $\mrB$ is the Euler Beta function. When $\mrp = \mrq + 1$ we can use
\bq
{}_{\scriptstyle{\mrp+1}}\,\mrF_{\scriptstyle{\mrp}}
\lpar \mra_1\,\dots\,\mra_{\mrp+1}\,;\,\mrb_1\,\dots\,\mrb_{\mrp}\,;\,\mrz \rpar =
\prod_{\mrn=1}^{\mrp}\,\frac{\eG{\mrb_{\mrn}}}{\eG{\mra_{\mrn}}\,\eG{\mrb_{\mrn} - \mra_{\mrn}}}\,
\int_0^1\,\Bigl[ \prod_{\mrn=1}^{\mrp}\,\mrd \mrx_{\mrn}\,\mrx_{\mrn}^{\mra_{\mrn}-1}\,
\bigl( 1 - \mrx_{\mrn} \bigr)^{\mrb_{\mrn}-\mra_{\mrn}-1} \Bigr]\,
\Bigl(1 - \mrz\,\prod_{\mrn=1}^{\mrp}\,\mrx_{\mrn} \Bigr)^{-\mra_{\mrp+1}} \spc
\label{pp1Fp}
\eq
where $\Re \mrb_{\mrn} > \Re \mra_{\mrn} > 0$ and $\mid \marg (1 - \mrz) \mid < \pi$. 
Furthermore we have a MB representation:
\bq
{}_{\scriptstyle{\mrp+1}}\,\mrF_{\scriptstyle{\mrp}}
\lpar \mra_1\,\dots\,\mra_{\mrp+1}\,;\,\mrb_1\,\dots\,\mrb_{\mrp}\,;\,\mrz \rpar =
\frac{\prod_{\mrj=1}^{\mrp}\,\eG{\mrb_{\mrj}}}
     {\prod_{\mrj=1}^{\mrp+1}\,\eG{\mra_{\mrj}}}\,
\int_{\mrL}\,\frac{\mrd \mrs}{2\,i\,\pi}\,
\eG{ - \mrs}\,
\frac{\prod_{\mrj=1}^{\mrp+1}\,\eG{\mrs + \mra_{\mrj}}}
     {\prod_{\mrj=1}^{\mrp}\,\eG{\mrs +\mrb_{\mrj}}}\,
     ( - \mrz)^{\mrs} \spc
\label{GHFMB}
\eq    
where $\mrL$ is a standard contour such that only the poles of $\eG{ - \mrs}$ lie to the right of $\mrL$
and $\mid \marg( - \mrz ) \mid < \pi$.
When \eqns{pp1Fp}{GHFMB} are used in evaluating Feynman integrals the Feynman prescription is instrumental in
ensuring that we are working in
\bqa
\Lf^{\mrN} &:=& \{{\mathbf{z}} \in \Cf^{\mrN}\;:\; \mid \marg(1 - \mrz_{\mrj}) \mid < \pi\,,\, \mrj = 1\dots \mrN\} \spc
\nl
\Sf^{\mrN} &:=& \{{\mathbf{z}} \in \Cf^{\mrN}\;:\; \mid \marg( - \mrz_{\mrj}) \mid < \pi\,,\, \mrj = 1\dots \mrN\} \spp
\eqa
Another relevant relation is given by B\"uhring's recurrence~\cite{doi:10.1137/0519088,Bunit}:
\bq
\ghyp{\mrp}{\mrp-1}\lpar {\mathbf a}\,;\,{\mathbf b}\,;\,\mrz \rpar =
\frac{\prod_{\mrj=1}^{\mrp-1}\,\eG{\mrb_{\mrj}}}
     {\prod_{\mrj=1}^{\mrp}\,\eG{\mra_{\mrj}}}\,\int_{\mrL}\,\frac{\mrd \mrs}{2\,i\,\pi}\,
{\tilde{\mrA}}^{(\mrp)}\,
\frac{\eG{\mra_1}\,\eG{\mra_2}}
     {\eG{\mrc + \mra_1 + \mra_2 +\mrs}}\,
\hyp{\mra_1}{\mra_2}{\mrc + \mra_1 + \mra_2 + \mrs}{\mrz} \spc
\eq
where $\mrc = \sum_{\mrj}\,\mrb_{\mrj} - \sum_{\mrj}\,\mra_j$ and the coefficients ${\tilde{\mrA}}^{(\mrp)}$ are given in
Proposition~(4.8) of \Bref{Scheidegger:2016ysn}. 
\subsection{Lauricella functions \label{Lfun}}
A general description can be found in \Brefs{Ext,FDMB}. As we will show in \sect{FD}, the Lauricella functions have a 
fundamental role in evaluating Feynman integrals.
Consider for example the Lauricella function $\mrF^{(3)}_{\sPD}$ for wich we can write
\bq
\mrF^{(3)}_{\sPD}\lpar \mra\,;\,\mrb_1\,,\,\mrb_2\,,\,\mrb_3\,;\,\mrc\,;\,\mrz_1\,,\,\mrz_2\,,\,\mrz_3 \rpar =
\Gamma\lpar \mrc\,,\,\mra \rpar\,\int_0^1 \mrd \mrx\,\mrx^{\mra-1}\,\lpar 1 - \mrx \rpar^{\mrc-\mra-1}\,
\prod_{n=1}^{3}\,\lpar 1 - \mrz_{\mrn}\,\mrx \rpar^{-\mrb_{\mrn}} \spc
\eq
with $\Re \mrc > \Re \mra > 0$ and $\Gamma\lpar \mrc\,,\,\mra \rpar = 1/\eB{\mra}{\mrc}$. 

The Lauricella $\mrF_{\sPD}$ is connected to the hypergeometric function $\mrR$ introduced in \Bref{CarlsonR} and
representable as follows:
\bq
\eB{\mra}{\mrap}\,\mrR\lpar \mra\,;\,\mrb_1,\,\dots\,,\mrb_\mrn\,;\,\mrz_1,\,\dots\,,\mrz_\mrn \rpar =
\int_0^{\infty} \mrd \mrx\,\mrx^{\mrap - 1}\,\prod_{i=1}^{\mrn} \lpar \mrx + \mrz_i \rpar^{-\mrb_i} \spc
\eq
where $\mra + \mrap = \sum_i\,\mrb_i$. The following identity holds:
\bq
\mrF_{\sPD}\lpar \mra\,;\,\mrb_1,\,\dots\,,\mrb_{\mrn-1}\,;\,\mrc\,;\,1 - \frac{\mrz_1}{\mrz_{\mrn}}\,,\,
1 - \frac{\mrz_{\mrn-1}}{\mrz_{\mrn}} \rpar = \mrz_{\mrn}^{\mra}\,
\mrR\lpar \mra\,;\,\mrb_1,\,\dots\,,\mrb_\mrn\,;\,\mrz_1,\,\dots\,,\mrz_\mrn \rpar \spc
\quad
\mrb_{\mrn} = \mrc - \sum_{i=1}^{\mrn-1}\,\mrb_i \spp
\eq
The $\mrR$ function can also be represented by a multiple MB integral,
\bqa
\mrB\lpar \mrb_1,\,\dots\,,\mrb_\mrn \rpar\,
 \mrR\lpar \mra\,;\,\mrb_1,\,\dots\,,\mrb_\mrn\,;\,\mrz_1,\,\dots\,,\mrz_\mrn \rpar &=&
\Bigl[ \prod_{i=1}^{\mrn}\,\int_{\mrL_i}\,\frac{\mrd \mrs_i}{2\,\pi\,i} \Bigr]\,
\mrB\lpar \mra + \sum_i \mrs_i\,,\, - \mrs_1,\,\dots\,,\ - \mrs_\mrn \rpar
\nl
{}&\times&
\mrB\lpar \mrb_1 + \mrs_1,\,\dots\,,\mrb_\mrn + \mrs_{\mrn} \rpar\,
\prod_{i=1}^{\mrn}\,\lpar \mrz_i - 1 \rpar^{\mrs_i} \spc
\eqa
\bq
\mrB\lpar \mra_1,\,\dots\,,\mra_\mrn \rpar = \frac{\prod_{i=1}^{\mrn}\,\eG{\mra_i}}{\eG{\sum_{i=1}^{\mrn}\,\mra_i}} \spc
\eq
where it is assumed that none of $\mra$, $\mrb_1,\,\dots\,,\mrb_\mrn$ is zero or a negative integer.
In the general case we have
\bq
\mrF^{(\mrN)}_{\sPD}\lpar \mra\,;\,\mrb_1\,,\dots\,,\mrb_{\mrN}\,;\,\mrc\,;\,
\mrz_1\,,\dots\,,\mrz_{\mrN}\rpar =
\Gamma\lpar \mrc\,,\,\mra \rpar\,\int_0^1 \mrd \mrx\,\mrx^{\mra-1}\,\lpar 1 - \mrx \rpar^{\mrc-\mra-1}\,
\prod_{n=1}^{\mrN}\,\lpar 1 - \mrz_{\mrn}\,\mrx \rpar^{-\mrb_{\mrn}} \spc
\label{FDEM}
\eq
with $\Re \mrc > \Re \mra > 0$ and $\mid \marg(1 - \mrz_i) \mid < \pi$. There is an integral representation
of MB type, Eq.~(1.7) of \Bref{FDMB},
\bqa
\mrF^{(\mrN)}_{\sPD}\lpar \mra\,;\,\mrb_1\,,\dots\,,\mrb_{\mrN}\,;\,\mrc\,;\,
\mrz_1\,,\dots\,,\mrz_{\mrN}\rpar &=&
\frac{\eG{\mrc}}{\eG{\mra}\,\prod_{\mrj}\,\eG{\mrb_{\mrj}}}\,
\Bigl[ \prod_{\mrj=1}^{\mrN}\,\int_{\mrL_{\mrj}}\,\frac{\mrd \mrs_{\mrj}}{2\,i\,\pi} \Bigr]\,
\frac{\eG{\mra + \sum_{\mrj}\,\mrs_{\mrj}}}{\eG{\mrc + \sum_{\mrj}\,\mrs_{\mrj}}}
\nl
{}&\times&
\prod_{\mrj=1}^{\mrN}\,\eG{\mrb_{\mrj} + \mrs_{\mrj}}\,\eG{ - \mrs_{\mrj}}\,\lpar - \mrz_{\mrj} \rpar^{\mrs_{\mrj}} \spc
\label{FDMB}
\eqa
where $\mrL_{\mrj}$ is a deformed imaginary axis curved so that only the poles of $\eG{ - \mrs_{\mrj}}$ lie
to the right of $\mrL_{\mrj}$. 

When dealing with Feynman integrals in $\mrd$ dimensions we need an expansion~\cite{Bera:2022ecn}
in $\mrd - 4$ which, in our approach, involves $\uppsi{-}$deformed Lauricella functions, \eg
\bqa
\eG{\mrd - 4 + \mrb_1 + \mrs_1} &=& \eG{\mrb_1 + \mrs_1}\,\Bigl[ 1 + (\mrd - 4)\,\uppsi(\mrb_1 + \mrs_1) +
\ord{(\mrd - 4)^2} \Bigr] \spc
\nl
\mrF^{(3)}_{\sPD\,,\,\uppsi}\lpar \mra\,;\,\mrb_1\,,\dots\,,\mrb_{3}\,;\,\mrc\,;\,
\mrz_1\,,\dots\,,\mrz_{3}\rpar &=&
\frac{\eG{\mrc}}{\eG{\mra}\,\prod_{\mrj}\,\eG{\mrb_{\mrj}}}\,
\Bigl[ \prod_{\mrj=1}^{3}\,\int_{\mrL_{\mrj}}\,\frac{\mrd \mrs_{\mrj}}{2\,i\,\pi} \Bigr]\,
\frac{\eG{\mra + \sum_{\mrj}\,\mrs_{\mrj}}}{\eG{\mrc + \sum_{\mrj}\,\mrs_{\mrj}}}
\nl
{}&\times&
\prod_{\mrj=1}^{3}\,\eG{\mrb_{\mrj} + \mrs_{\mrj}}\,\eG{ - \mrs_{\mrj}}\,\lpar - \mrz_{\mrj} \rpar^{\mrs_{\mrj}}\,
\uppsi(\mrb_1 + \mrs_1) \spp
\eqa
Using the integral representation for the $\uppsi$ function~\cite{HTF} we obtain
\bqa
\mrF^{(3)}_{\sPD\,,\,\uppsi}\lpar \mra\,;\,\mrb_1\,,\dots\,,\mrb_{3}\,;\,\mrc\,;\,\mrz_1\,,\dots\,,\mrz_{3}\rpar &=&
\uppsi(1)\,\mrF^{(3)}_{\sPD}\lpar \mra\,;\,\mrb_1\,,\dots\,,\mrb_{3}\,;\,\mrc\,;\,\mrz_1\,,\dots\,,\mrz_{3}\rpar
\nl
{}&+& \int_0^1\,\frac{\mrd \mrx}{1 - \mrx}\,\Bigl[
\mrF^{(3)}_{\sPD}\lpar \mra\,,\,\mrb_1\,;\dots\,,\mrb_{3}\,;\,\mrc\,;\,\mrz_1\,,\dots\,,\mrz_{3}\rpar 
\nl
{}&-&
\mrx^{\mrb_1 - 1}\,
\mrF^{(3)}_{\sPD}\lpar \mra\,;\,\mrb_1\,,\dots\,,\mrb_{3}\,;\,\mrc\,;\,\mrx\,\mrz_1\,,\dots\,,\mrx\,\mrz_{3}\rpar 
\Bigr] \spp
\eqa
The result can be simplified if $\mrb_1$ is an integer, \eg
\bqa
\mrF^{(3)}_{\sPD\,,\,\uppsi}\lpar \mra\,;\,1\,,\dots\,,\mrb_{3}\,;\,\mrc\,;\,\mrz_1\,,\dots\,,\mrz_{3}\rpar &=&
\frac{\eG{\mrc}}{\eG{\mra}\,\eG{\mrc - \mra}}\,\int_0^1 \mrd \mrx\,
\mrx^{\mra - 1}\,(1 - \mrx)^{\mrc - \mra - 1}\,\prod_{i=2}^{3}\,(1 - \mrz_i\,\mrx)^{ - \mrb_1}
\nl
{}&\times& (1 - \mrz_1\,\mrx)^{-1}\,\Bigl[ \uppsi(1) - \ln(1 - \mrz_1\,\mrx) \Bigr] \spp
\eqa
 
\subsection{Meijer G functions \label{MGfun}}
There are various generalizations of the Gauss hypergeometric function; here we will discuss the Meijer $\mrG\,${-}functions.
They are a particular case of Fox functions~\cite{oFox,compH,HS} and can be used in their numerical evaluation.
There are different notations in the literature for the $\mrG\,${-}function. Here we follow \Bref{HTF} and write 
\[
\mrG^{\mrm\,,\,\mrn}_{\mrp\,,\,\mrq} \, \left(
\mrz\,,\; \Biggl [
\begin{array}{ccc}
\mra_1 & \dots & \mra_{\mrp} \\
\mrb_1 & \dots & \mrb_{\mrq} \\
\end{array}
\Biggr ]
\right)
= \frac{1}{2\,\pi\,i}\,\int_{\mrL} \mrd \mrs\, \theta(\mrs)\,\mrz^{\mrs} 
\]
\bq
\theta(\mrs) = \frac{\prod_{\mrj=1}^{\mrm} \eG{\mrb_{\mrj} - \mrs}\,\prod_{\mrj=1}^{\mrn} \eG{1 - \mra_{\mrj} + \mrs}}
                    {\prod_{\mrj=\mrm+1}^{\mrq} \eG{1 - \mrb_{\mrj} + \mrs}\,\prod_{\mrj=\mrn+1}^{\mrp} 
                     \eG{\mra_{\mrj} - \mrs}} \spc
\label{Gtheta}
\eq
where details on $\mrL$ can be found in Sect.($5.3$) of \Bref{HTF}. Let us define the following quantities:
\bq
\alpha = 2\,(\mrm + \mrn) - \mrp - \mrq \spc \qquad \mu = \mrq - \mrp \spp
\eq
To summarize, given the MB representation there are $3$ choices for the contour $\mrL$, defined in Sect.~$5.3$ of
\Bref{HTF}:
\begin{enumerate}

\item[] $\mrL_{-\,\infty}$, the integral converges if $\mu > 0$ and $\mrz \not= 0$ or $\mu = 0$ and $\mid \mrz \mid < 1$ \spc

\item[] $\mrL_{+\,\infty}$, the integral converges if $\mu < 0$ and $\mrz \not= 0$ or $\mu = 0$ and $\mid \mrz \mid > 1$ \spc

\item[] $\mrL_{i \gamma\,\infty}$, the integral converges if $\alpha > 0$ and $\mid \marg\,\mrz \mid < \pi\alpha/2$.
As explained in Sect.~$1.19$ of \Bref{HTF} we should always understand that the integral coverges absolutely for
$\mid \marg\,\mrz \mid < \pi\alpha/2$ and defines a function analytic in the sector
$\mid \marg\, \mrz \mid < \mathrm{min}(\pi \alpha/2\,,\,\pi)$, the point $\mrz = 0$ excluded.

\end{enumerate}

It is worth noting that when the integral converges for more than one of these paths, the result of integration can be 
shown to agree, as stated in \Bref{HTF}; the general proof of this statement can be found in Appendix~A of \Bref{KPc}.

If no two $\mrb_{\mrj}$ differ by an integer and
$\mrp \le \mrq$ and $\mid \mrz \mid < 1$ or no two $\mra_{\mrj}$ differ by an integer and $\mrq \le \mrp$
and $\mid \mrz \mid > 1$ the Meijer $\mrG$ function can be computed by expressing it as a combination of
generalized hypergeometric functions, see Sect.($5.3$) of \Bref{HTF}.
To summarize: consider
\bq
\mrG^{\mrm\,,\,\mrn}_{\mrp\,,\,\mrq}\bigl[ \mrz \bigr] \spc \quad \alpha > 0 \spc \quad
\mid \marg\, \mrz \mid < \pi\alpha/2 \spp
\eq
If the integral converges along $\mrL_{+\,\infty}$ then the Meijer function can be expressed through $\shyp{\mrp}{\mrq - 1}$
functions with $\mrp \le \mrq$ and $\mid \mrz \mid < 1$. If the integral converges along $\mrL_{-\,\infty}$
the Meiger function can be expressed through $\shyp{\mrq}{\mrp - 1}$ functions with $\mrp \ge \mrq$ and 
$\mid \mrz \mid > 1$.

Alternatively we can use a Gauss{-}Jacobi quadrature based on the following relation~\cite{HTF}:
\[
\mrG^{\mrm\,,\,\mrn+1}_{\mrp+1\,,\,\mrq+1} \, \left(
\mrz\,,\; \Biggl [
\begin{array}{cccc}
\alpha &\mra_1 & \dots       & \mra_{\mrp} \\
\mrb_1 & \dots & \mrb_{\mrq} & \beta\\
\end{array}
\Biggr ]
\right)
= \frac{1}{\eG{\alpha - \beta}}\,\int_0^1 \mrd \mrx\,\mrx^{-\alpha}\,\lpar 1 - \mrx \rpar^{\alpha - \beta - 1}\,
\mrG^{\mrm\,,\,\mrn}_{\mrp\,,\,\mrq} \, \left(
\mrz\,\mrx\,,\; \Biggl [
\begin{array}{ccc}
\mra_1 & \dots & \mra_{\mrp} \\
\mrb_1 & \dots & \mrb_{\mrq} \\
\end{array}
\Biggr ]
\right)
\]
\bq
{}\vspace{-0.3cm}
\label{EMG}
\eq
subject to the conditions
\bqa
{}&{}& \mrp + \mrq < 2\,(\mrm + \mrn) \spc \qquad
\mid \marg\,\mrz \mid < (\mrm + \mrn - \frac{1}{2}\,\mrp - \frac{1}{2}\,\mrq) \spc
\nl
{}&{}& \Re \beta < \Re \alpha < \Re \mrb_{\mrj} + 1 \spc \quad \mrj = 1,\,\dots\,,\mrm \spp
\eqa
For instance we obtain $\mrG^{1,1}_{3,3}$ from $\mrG^{1,2}_{2,2}$ which is proportional to a Gauss hypergeometric function.  

In this work we will use the Gauss quadrature rules for real exponents. Gauss{-}Christoffel quadrature rules with 
complex Jacobi weight functions $\alpha, \beta$, satisfying $\Re \alpha > - 1$, $\Re \beta > - 1$ have been introduced 
in \Bref{GJc}.
Gauss quadrature rule for PV integrals have been discussed in \Bref{GCPV}. 

Some of the results in the following Sections will be based on \Bref{compH};
it is worth noting that in \Bref{compH} the function $\theta$ of \eqn{Gtheta} is defined as follows:
\bq
\theta(\mrs) = \frac{\prod_{\mrj=1}^{\mrn} \eG{1 - \mra_{\mrj} - \mrs}\,
                     \prod_{\mrj=1}^{\mrm} \eG{\mrb_{\mrj} + \mrs}}
                    {\prod_{\mrj=\mrn+1}^{\mrp} \eG{\mra_{\mrj} + \mrs}\,
                     \prod_{\mrj=\mrm+1}^{\mrq} \eG{1 - \mrb_{\mrj} - \mrs}} \spp
\eq
\subsection{Fox H functions \label{Ffun}}
The $\mrH\,${-}function has uses in fluids, special functions, Mellin transforms,
fractional calculus, \etc
Using the notations of \Bref{compH} the Fox $\mrH\,${-}function is defined as follows:
\[
\mrH^{\mrm\,,\,\mrn}_{\mrp\,,\,\mrq} \, \left(
\mrz\,,\; \Biggl [
\begin{array}{cc}
 & \lpar \mra_{\mrj}\,,\,\mrA_{\mrj} \rpar_{\mrp} \\
 & \lpar \mrb_{\mrj}\,,\,\mrB_{\mrj} \rpar_{\mrq} \\
\end{array}
\Biggr ]
\right)
= \frac{1}{2\,\pi\,i}\,\int_{\mrL} \mrd \mrs\,\theta(\mrs)\,\mrz^{\mrs}
\]
\bq
\theta(\mrs) =
\frac{\prod_{\mrj=1}^{\mrm}\,\eG{\mrb_{\mrj} + \mrB_{\mrj}\,\mrs}\,
      \prod_{\mrj=1}^{\mrn}\,\eG{1 - \mra_{\mrj} - \mrA_{\mrj}\,\mrs}}
      {\prod_{\mrj=\mrm+1}^{\mrq}\,\eG{1 - \mrb_{\mrj} - \mrB_{\mrj}\,\mrs}\,
      \prod_{\mrj=\mrn+1}^{\mrp}\,\eG{\mra_{\mrj} + \mrA_{\mrj}\,\mrs}} \spc
\label{Hdef}
\eq
and where an empty product is interpreted as unity; furthermore,
\bq
\lpar \mra_{\mrp}\,,\,\mrA_{\mrp} \rpar =
\lpar \mra_1\,,\,\mrA_1 \rpar\;;\; \dots \;;\;\lpar \mra_\mrp\,,\,\mrA_\mrp \rpar \spc
\eq
\etc Following \Bref{compH} we define auxiliary quantities as follows:
\bq
\mu = \sum_{\mrj=1}^{\mrq}\,\mrB_{\mrj} - \sum_{\mrj=1}^{\mrp}\,\mrA_{\mrj} \spc 
\qquad
\beta = \prod_{\mrj=1}^{\mrp}\,\lpar \mrA_{\mrj} \rpar^{-\mrA_{\mrj}}\,\prod_{\mrj=1}^{\mrq}\,\mrB_{\mrj}^{\mrB_{\mrj}} \spc
\label{murho}
\eq
\bq
\alpha = \sum_{\mrj=1}^{\mrn}\,\mrA_\mrj - 
         \sum_{\mrj=\mrn+1}^{\mrp}\,\mrA_\mrj +
         \sum_{\mrj=1}^{\mrm}\,\mrB_\mrj - 
         \sum_{\mrj=\mrm+1}^{\mrq}\,\mrB_\mrj \spp
\eq
Under the assumption that the poles of the integrand in \eqn{Hdef} are simple, the integral converges
under the following conditions (taken from \Bref{HTF}):
\begin{enumerate}

\item[] $\mrL = \mrL_{-\,\infty}$; the integral converges for all $\mrz$ if $\mu > 0$ and $\mrz \not= 0$ or $\mu = 0$
and $\mid \mrz \mid < \beta$.

\item[] $\mrL = \mrL_{+\,\infty}$; the integral converges for all $\mrz$ if $\mu < 0$ and $\mrz \not= 0$ or $\mu = 0$
and $\mid \mrz \mid > \beta$.

\item[] $\mrL= \mrL_{i \gamma\,\infty}$; the integral converges if $\alpha > 0$ and 
$\mid \marg\, \mrz \mid < \alpha \pi/2$.

\end{enumerate}
Furthermore for $\mid \mrz \mid \to 0$ (when the poles are simple) we have
$\mrH(\mrz) = \mrO\lpar \mid \mrz \mid^\mre \rpar$ and
\begin{enumerate}

\item[] For $\mu \ge 0$ or $\mu = 0$, $\alpha > 0$, $\mid \marg\, \mrz \mid < \alpha \pi/2$
$\quad \mapsto \quad$
$\mre = \mathrm{min}\,\Bigl( \frac{\Re \mrb_{\mrj}}{\mrB_{\mrj}} \bigr) \spc \quad 1 \le \mrj \le \mrm$. 
\item[] For $\mu \le 0$ or $\mu = 0$, $\alpha > 0$, 
$\quad \mapsto \quad$
$\mre = \mathrm{min}\,\Bigl( \frac{\Re \mra_{\mrj} - 1}{\mrA_{\mrj}} \bigr) \spc \quad 1 \le \mrj \le \mrn$.
\end{enumerate}
\paragraph{Mellin transforms} \hspace{0pt} \\

Finally we consider the following Mellin transforms~\cite{GHfun}:
\bq
\mrM^{(1)}(\mrz) = \int_0^1 \mrd \mrx\,\mrx^{\alpha - 1}\,\lpar 1 - \mrx \rpar^{\beta - 1}\,
\mrH^{\mrm\,,\,\mrn}_{\mrp\,,\,\mrq}\,\lpar \mrz\,\mrx \rpar \spc
\quad
\mrH^{\mrm\,,\,\mrn}_{\mrp\,,\,\mrq}\,\lpar \mrz \rpar = \frac{1}{2\,\pi\,i}\,\int_{\mrL} \mrd \mrs\,
\theta(\mrs)\,\mrz^{\mrs} \spp 
\eq
Assuming the conditions for absolute convergence and $\Re \alpha + \mre > 0\;$, $\Re \beta > 0$
we obtain
\[
\mrM^{(1)}(\mrz) = \eG{\beta}\,
\mrH^{\mrm\,,\,\mrn+1}_{\mrp+1\,,\,\mrq+1} \, \left(
\mrz\,,\; \Biggl [
\begin{array}{cc}
\dots \quad&\quad (1 - \alpha\,,\,1) \\
\dots \quad&\quad (1 - \alpha - \beta\,,\,1) \\
\end{array}
\Biggr ]
\right)
\]
\bq
\label{MHfuna}
\eq
Next, with $\Re \gamma > 0$ we have
\[
\mrM^{(2)}(\mrz) = \int_0^1 \mrd \mrx\,\mrx^{\alpha - 1}\,\lpar 1 - \mrx^{1/\gamma} \rpar^{\beta - 1}\,
\mrH^{\mrm\,,\,\mrn}_{\mrp\,,\,\mrq}\,\lpar \mrz\,\mrx \rpar =
\gamma\,\eG{\beta}
\mrH^{\mrm\,,\,\mrn+1}_{\mrp+1\,,\,\mrq+1} \, \left(
\mrz\,,\; \Biggl [
\begin{array}{cc}
\dots \quad&\quad (1 - \gamma\,\alpha\,,\,\gamma) \\
\dots \quad&\quad (1 - \gamma\,\alpha - \beta\,,\,\gamma) \\
\end{array}
\Biggr ]
\right)
\]
\bq
\label{MHfunb}
\eq
Consider the following integral:
\bq
\mrM^{(3)}(\mrz) = \int_0^1 \mrd \mrx\,\mrx^{\alpha - 1}\,\lpar 1 - \mrx \rpar^{\beta - 1}\,
\mrH^{\mrm\,,\,\mrn}_{\mrp\,,\,\mrq}\,\lpar \frac{1}{\mrz\,\mrx} \rpar \spp
\eq
The existence of the integral depends on the behavior of $\mrH(\mrz)$ when $\mid \mrz \mid \to \infty$; 
the detailed classification can be found in \Bref{BRaa,HS} or in Sect ($5.4.1$) of \Bref{HTF} for those cases where 
the $\mrH$ function reduces to a $\mrG$ function. Here we will assume that the $\mrH$ function does not become
exponentially infinite in the limit, without discussing the range of parameters where this occurs. We obtain
\[
\mrM^{(3)}(\mrz) = \eG{\beta}\,
\mrH^{\mrm+1\,,\,\mrn}_{\mrp+1\,,\,\mrq+1} \, \left(
\mrz^{-1}\,,\; \Biggl [
\begin{array}{cc}
\dots \quad&\quad (\alpha + \beta\,,\,1) \\
\dots \quad&\quad (\alpha\,,\,1) \\
\end{array}
\Biggr ]
\right)
\]
\bq
\label{MHfunaa}
\eq
These relations can be used to evaluate H functions recursively.
\subsubsection{Representation of the Fox \texorpdfstring{$\mrH$}{H} function \label{repHfun} }
In order to discuss representations of the Fox function we summarize a few (well{-}known) 
facts~\cite{HTF,compH,KPa,KPb,spring}. 
For the sake of simplicity let us
work with the Meijer $\mrG$ function of \eqn{Gtheta} where $\mrL = \mrL_{i\,\infty}$ separating the poles at
$\mrs = \mrb_{\mrj} - \mrk$ from the poles at $\mrs = \mra_{\mrj} - 1 - \mrl$. The $\mrG$ and $\mrH$
functions needed in computing Feynman integrals always derive from MB representations
of $\mrF^{(\mrN)}_{\sPD}$ functions where $\mrL = \mrL_{i\,\infty}$. 
We recall the definition of $\alpha$ in \Bref{HTF}, $\alpha= 2\,(\mrm + \mrn) - \mrp - \mrq$, and obtain the
following results:
\begin{enumerate}

\item If $\mrq > \mrp$ and $0 \not= \mrz \in \Cf$ then the $\mrG$ function converges for $\mrL = \mrL_{-\,\infty}$
and can be computed as a sum of residues. It is important to observe that the contour can be deformed into
$\mrL_{i\,\infty}$ if $\alpha > 0$ and $\mid \marg(\mrz) \mid < \frac{1}{2}\,\alpha\,\pi$.

\item If $\mrq < \mrp$ and $0 \not= \mrz \in \Cf$ then the $\mrG$ function converges for $\mrL = \mrL_{+\,\infty}$
and can be computed as a sum of residues. It is important to observe that the contour can be deformed into
$\mrL_{i\,\infty}$ if $\alpha > 0$ and $\mid \marg(\mrz) \mid < \frac{1}{2}\,\alpha\,\pi$.

\item When, however, $\mrq = \mrp$ the function converges  

\bei

\item[$3\mra)$] for $\mrL = \mrL_{+\,\infty}$ if $\mid \mrz \mid < 1$,

\item[$3\mrb)$] for $\mrL = \mrL_{-\,\infty}$ if $\mid \mrz \mid > 1$.

\eei

\end{enumerate}
The generalization with explicit results is given in \Bref{compH} (see also \Bref{nam}) and will not be reported here; 
specific examples will be given in \sect{hmff}. 
When needed we can use the following relation ($\sigma > 0$):
\[
\mrH^{\mrm\,,\,\mrn}_{\mrp\,,\,\mrq} \, \left(
\mrz\,,\; \Biggl [
\begin{array}{cc}
 & \lpar \mra_{\mrp}\,,\,\mrA_{\mrp} \rpar \\
 & \lpar \mrb_{\mrq}\,,\,\mrB_{\mrq} \rpar \\
\end{array}
\Biggr ]
\right)
=
\mrH^{\mrn\,,\,\mrm}_{\mrq\,,\,\mrp} \, \left(
\mrz^{-1}\,,\; \Biggl [
\begin{array}{cc}
 & \lpar 1 - \mrb_{\mrq}\,,\,\mrB_{\mrq} \rpar \\
 & \lpar 1 - \mra_{\mrp}\,,\,\mrA_{\mrp} \rpar \\
\end{array}
\Biggr ]
\right)
\]
\bq
\label{Hinv}
\eq
\[
\mrH^{\mrm\,,\,\mrn}_{\mrp\,,\,\mrq} \, \left(
\mrz\,,\; \Biggl [
\begin{array}{cc}
 & \lpar \mra_{\mrp}\,,\,\mrA_{\mrp} \rpar \\
 & \lpar \mrb_{\mrq}\,,\,\mrB_{\mrq} \rpar \\
\end{array}
\Biggr ]
\right)
= \sigma\,
\mrH^{\mrn\,,\,\mrm}_{\mrq\,,\,\mrp} \, \left(
\mrz^{\sigma}\,,\; \Biggl [
\begin{array}{cc}
 & \lpar \mra_{\mrp}\,,\,\sigma\,\mrA_{\mrp} \rpar \\
 & \lpar \mrb_{\mrq}\,,\,\sigma\mrB_{\mrq} \rpar \\

\end{array}
\Biggr ]
\right)
\]
\bq
\label{Hscal}
\eq
The relation between $\mrG$ and generalized hypergeometric function is given in \Bref{HTF}.

Let us define the following quantities:
\bq
\mra_{\mrh,\mrj} = \mra_{\mrh} - \mra_{\mrj} \spc \quad
\mrb_{\mrj,\mrh} = \mrb_{\mrj} - \mrb_{\mrh} \spc \quad
\mrc_{\mrj,\mrh} = \mrb_{\mrj} - \mrh_{\mrj} \spc 
\eq
\bqa
\{ \mrA_{\mrh,\mrp} \} = \{1 - \mra_{1,\mrh}\,,\,\dots\,,*\,,\,\dots\,,\,1 - \mra_{\mrh,\mrp} \} \spc
&\quad&
\{ \mrB_{\mrh,\mrq} \} = \{1 - \mrb_{1,\mrh}\,,\,\dots\,,*\,,\,\dots\,,\,1 - \mrb_{\mrh,\mrq} \} \spc
\nl
\{ \mrC_{\mrh,\mrp} \} = \{1 + \mrc_{\mrh,1}\,,\,\dots\,,\,1 + \mrc_{\mrh,\mrp} \} \spc
&\quad&
\{ \mrC_{\mrq,\mrh} \} = \{1 + \mrc_{1,\mrh}\,,\,\dots\,,\,1 + \mrc_{\mrq,\mrh} \} \spp
\label{GtoHypo}
\eqa
If the integral coverges along $\mrL_{i \gamma\,\infty}$ and along $\mrL_{+\,\infty}$ we can use
\bqa
\mrG^{\mrm\,,\,\mrn}_{\mrp\,,\,\mrq} &=& \sum_{\mrh=1}^{\mrm}\,\frac{\mrN_{\mrm,\mrn}}{\mrD_{\mrm,\mrn}}\,
\mrz^{\mrb_{\mrh}}\,
{}_{\scriptstyle{\mrp}}\,\mrF_{\scriptstyle{\mrq - 1}} \lpar
\{ \mrC_{\mrh\,,\,\mrp} \}\,;\,\{ \mrB_{\mrh\,,\,\mrq} \}\,;\,\mrz_{\mrp} \rpar \spc
\nl
\mrN_{\mrm\,,\,\mrn} &=& \prod_{\mrj=1}^{\mrm} \eG{\mrb_{\mrj,\mrh}}\,
                         \prod_{\mrj=1}^{\mrn} \eG{1 + \mrc_{\mrh,\mrj}} \spc
\nl
\mrD_{\mrm\,,\,\mrn} &=& \prod_{\mrj=\mrm+1}^{\mrq} \eG{1 - \mrb_{\mrj,\mrh}}\,
                         \prod_{\mrj=\mrn+1}^{\mrp} \eG{1 - \mrc_{\mrh,\mrj}} \spp
\label{GtoHypt}
\eqa
where $\mid \mrz \mid < 1$ for $\mrp = \mrq$.
If the integral coverges along $\mrL_{i \gamma\,\infty}$ and along $\mrL_{-\,\infty}$ we can use
\bqa
\mrG^{\mrm\,,\,\mrn}_{\mrp\,,\,\mrq} &=& \sum_{\mrh=1}^{\mrn}\,\frac{\mrN_{\mrn,\mrm}}{\mrD_{\mrn,\mrm}}\,
\mrz^{\mra_{\mrh} - 1}\,
{}_{\scriptstyle{\mrq}}\,\mrF_{\scriptstyle{\mrp - 1}} \lpar
\{ \mrC_{\mrq\,,\,\mrh} \}\,;\,\{ \mrA_{\mrh\,,\,\mrp} \}\,;\,\mrz^{-1}_{\mrq} \rpar \spc
\nl
\mrN_{\mrn\,,\,\mrm} &=& \prod_{\mrj=1}^{\mrn} \eG{\mra_{\mrh,\mrj}}\,
                         \prod_{\mrj=1}^{\mrm} \eG{1 - \mrc_{\mrj,\mrh}} \spc
\nl
\mrD_{\mrn\,,\,\mrm} &=& \prod_{\mrj=\mrn+1}^{\mrp} \eG{1 - \mra_{\mrh,\mrj}}\,
                         \prod_{\mrj=\mrm+1}^{\mrq} \eG{ - \mrc_{\mrj,\mrh}} \spp
\eqa
where $\mid \mrz \mid > 1$ for $\mrp = \mrq$ and where we have defined
\bq
\mrz_{\mrp} = ( - 1 )^{\mrp - \mrm - \mrn}\,\mrz \spc \qquad
\mrz_{\mrq} = ( - 1 )^{\mrq - \mrm - \mrn}\,\mrz \spp
\eq
\subsubsection{Special cases of the Fox H function\label{SCFfun}}
When all the parameters $\mrA_\mrj$ and $\mrB_\mrj$ are integers the Fox function can be transformed into a Meijer
function by using the multiplication formula,
\bq
\eG{1 - \mrc_\mrj + \mrk_\mrj\,s} = \eG{\mrk_\mrj\,\lpar s + \frac{1 - \mrc_\mrj}{\mrk_\mrj} \rpar} \spc
\quad
\eG{\mrk\,\mrz} = (2\,\pi)^{1/2 - 1/2\,\mrk}\,\mrk^{\mrk\,\mrz - 1/2}\,
\prod_{i=0}^{\mrk-1}\,\eG{\mrz + \frac{i}{\mrk}} \spp
\label{mth}
\eq
We give a simple example:
\[
\mrH^{1,1}_{1,1}
\left(
\mrz \;,\; \Biggl [
\begin{array}{cc}
\mrc & 2 \\
\mrd & 1 \\
\end{array}
\Biggr ]
\right)
=
2^{-\mrc}\,\pi^{-1/2}\,\mrG^{1,2}_{2,1}
\left(
4\,\mrz \;,\; \Biggl[
\begin{array}{ccc}
\frac{1}{2}\,\mrc &\qquad& \frac{1}{2} + \frac{1}{2}\,\mrc \\
                 &\mrd&                                   \\
\end{array}
\Biggr ]
\right)
\]
A general differential{-}reduction algorithm has been developed in \Bref{Bytev:2009kb}.
\subsubsection{From Fox to Meijer \label{fFtM}}
The Fox $\mrH\,${-}function is defined by a Mellin{-Barnes} type integral in \eqn{Hdef}. The integrand contains 
Euler Gamma functions of the form $\eG{\mrd + \mrD\,\mrs}$. The corresponding Meijer $\mrG\,${-}function has 
$\mrD = 1$. If one or more parameters $\mrc_{\mrp}$ or $\mrd_{\mrq}$ are small we can use the following 
identity~\cite{compH}:
\[
\mrH^{\mrm\,,\,\mrn}_{\mrp\,,\,\mrq} \, \left(
\mrz\,,\; \Biggl [
\begin{array}{cc}
 & \lpar \mrc_{\mrj}\,,\,\mrC_{\mrj} \rpar_{1\,,\mrp} \\
 & \lpar \mrd_{\mrj}\,,\,\mrD_{\mrj} \rpar_{1\,,\mrq} \\
\end{array}
\Biggr ]
\right)
=
\mrz^{- \sigma}\,
\mrH^{\mrm\,,\,\mrn}_{\mrp\,,\,\mrq} \, \left(
\mrz\,,\; \Biggl [
\begin{array}{cc}
 & \lpar \mrc_{\mrj} + \sigma\,\mrC_{\mrj}\,,\,\mrC_{\mrj} \rpar_{1\,,\mrp} \\
 & \lpar \mrd_{\mrj} + \sigma\,\mrD_{\mrj}\,,\,\mrD_{\mrj} \rpar_{1\,,\mrq} \\
\end{array}
\Biggr ]
\right)
\]
\bq
\label{malab}
\eq
to make sure that $
\mid \mrc_{\mrj} + \sigma\,\mrC_{\mrj} \mid > 1 \spc \; \mid \mrd_{\mrj} + \sigma\,\mrd_{\mrj} \mid > 1 
\; \forall \mrj$.
Therefore, we will be in the situation where the function contains $\eG{\mrd + \mrD\,\mrs}$ with
$\mid \mrd \mid > 1$,
and we can use~\cite{HTF,HTgam}
\bq
\eG{\mrd + \mrD\,\mrs} \sim \eG{\mrd + \mrs}\,\sum_{\mrn=0}^{\infty}\,\mrc_{\mrn}\,\mrd^{\mrD_{-}\,\mrs - \mrn} \spc
\eq
where $\mrD_{\pm} = \mrD \pm 1$ and
\bq
\mrc_0 = 1 \spc \quad
\mrc_1 = \frac{1}{2}\,\mrD_{-}\,\lpar \mrD_{+}\,\mrs - 1 \rpar\,\mrs \spc 
%
\label{Gratio}
\eq
For the determination of higher order coefficients in \eqn{Gratio} ($\mrc_2 , \dots$) see Eq.~(12) of \Bref{HTgam}.
Approximations for more general quotients of Gamma functions can be found in \Bref{BRaa}.

To see how it works we will consider one particular example:
\[
\mrI =
\mrH^{1\,,\,1}_{1\,,\,2}
\, \left(
\mrz\,,\; \Biggl [
\begin{array}{cc}
& \lpar 1 - \gamma\,,\,1 \rpar \\
  \lpar 0\,,\,1 \rpar & \lpar 1 - \beta\,,\,\alpha \rpar \\
\end{array}
\Biggr ]
\right)
\]
\bq
\label{pexa}
\eq
The integral in \eqn{pexa} has the following representation:
\bq
\mrI = \frac{1}{2\,i\,\pi}\,\int_{\mrL}\,\mrd \mrs\,
\frac{\eG{\mrs}\,\eG{1- \gamma - \mrs}}{\eG{\beta - \alpha\,\mrs}}\,\mrz^{- \mrs} \spp
\eq
After performing the $\sigma\,${-}shift of \eqn{malab} we will have $1 - \beta = \Delta$ with $\mid \Delta \mid > 1$. 
We can write
\bq
\eG{\Delta - \alpha\,\mrs} = \eG{\Delta - \mrs}\,\Delta^{(1 - \alpha)\mrs}\,\Bigl\{
1 - \frac{1}{2\,\Delta}\,\Bigl[ (1 - \alpha^2)\,\mrs^2 + (1 - \alpha)\,\mrs \Bigr] + \ord{\Delta^{-2}} \Bigr\} \spp
\eq
Therefore, the result is
\bq
\mrI = \frac{1}{2\,i\,\pi}\,\int_{\mrL} \mrd \mrs\,
\frac{\eG{\mrs}\,\eG{1 - \gamma - \mrs}}{\eG{\Delta - \mrs}}\,\bigl[ \mrz\,\Delta^{1 - \alpha}\,\Bigr]^{- \mrs}\,
\Bigl[ 1 + \frac{1 - \alpha}{2\,\Delta}\,\mrs + \frac{1 - \alpha^2}{2\,\Delta}\,\mrs^2 +
\ord{\Delta^{-2}} \Bigr] \spp
\eq
Next we can use
\bq
\mrs\,\lpar \mra\,\mrz \rpar^{- \mrs} = - \mrz\,\frac{\mrd}{\mrd \mrz}\,\lpar \mra \mrz\rpar^{- \mrs} \spc
\eq
To obtain
\[
\mrI =
\mrH^{1\,,\,1}_{1\,,\,2}
\, \left(
\mrz\,\Delta^{1 - \alpha}\,,\; \Biggl [
\begin{array}{cc}
& \lpar 1 - \gamma\,,\,1 \rpar \\
  \lpar 0\,,\,1 \rpar & \lpar \Delta\,,\,\alpha \rpar \\
\end{array}
\Biggr ]
\right) 
- \frac{1 - \alpha}{2\,\Delta}\,\mrz \frac{\mrd}{\mrd \mrz}\,
\mrH^{1\,,\,1}_{1\,,\,2}
\, \left(
\mrz\,\Delta^{1 - \alpha}\,,\; \Biggl [
\begin{array}{cc}
& \lpar 1 - \gamma\,,\,1 \rpar \\
  \lpar 0\,,\,1 \rpar & \lpar \Delta\,,\,\alpha \rpar \\
\end{array}
\Biggr ]
\right) 
+ \dots
\]
As far as the derivatives are concerned we can use
\[
\mrz\,\frac{\mrd}{\mrd \mrz}
\mrH^{\mrm\,,\,\mrn}_{\mrp\,,\,\mrq} \, \left(
\mrz\,,\; \Biggl [
\begin{array}{cc}
 & \lpar \mra_1\,,\,\mrA_1 \rpar\,\dots\,\lpar \mra_{\mrp}\,\,\mrA_{\mrp} \rpar \\
 & \lpar \mrb_1\,,\,\mrB_1 \rpar\,\dots\,\lpar \mrb_{\mrq}\,\,\mrB_{\mrq} \rpar \\
\end{array}
\Biggr ]
\right)
=
\]
\[
\frac{\mra_1 - 1}{\mrA_1}\,
\mrH^{\mrm\,,\,\mrn}_{\mrp\,,\,\mrq} \, \left(
\mrz\,,\; \Biggl [
\begin{array}{cc}
 & \lpar \mra_1\,,\,\mrA_1 \rpar\,\dots\,\lpar \mra_{\mrp}\,\,\mrA_{\mrp} \rpar \\
 & \lpar \mrb_1\,,\,\mrB_1 \rpar\,\dots\,\lpar \mrb_{\mrq}\,\,\mrB_{\mrq} \rpar \\
\end{array}
\Biggr ]
\right)
\; + \frac{1}{\mrA_1}\,
\mrH^{\mrm\,,\,\mrn}_{\mrp\,,\,\mrq} \, \left(
\mrz\,,\; \Biggl [
\begin{array}{cc}
 & \lpar \mra_1 - 1\,,\,\mrA_1 \rpar\,\dots\,\lpar \mra_{\mrp}\,\,\mrA_{\mrp} \rpar \\
 & \lpar \mrb_1\,,\,\mrB_1 \rpar\,\dots\,\lpar \mrb_{\mrq}\,\,\mrB_{\mrq} \rpar \\
\end{array}
\Biggr ]
\right)
\]
which is valid for $\mrn > 1$. The corresponding expressions for $\mrn < \mrp - 1$ or $\mrm \le \mrq - 1$ can
be found in Eq.(1.76) or Eq.(1.77) of \Bref{compH}. 
As a result we can replace $\mrH\,${-}functions with $\mrG\,${-}functions in the $1/\Delta$ expansion.

A simpler result can be obtained when we have $\eG{\alpha + \mri/\mrk\,\mrs}$ where $\mri, \mrk$ are integers. For
instance we start from
\bqa
\mrH^{\mrm\,,\,\mrn}_{\mrp\,,\,\mrq}( \mrz ) &=&
\int_{\mrL}\,\frac{\mrd \mrs}{2\,i\,\pi}\,\frac{\mrN}{\mrD}\,\mrz^{ - \mrs} \spc
\nl
\mrN = \prod_{\mrj=1}^{\mrm}\,\eG{\mrb_{\mrj} - \mrs}\,
         \prod_{\mrj=1}^{\mrn - 1}\,\eG{1 - \mra_{\mrj} + \mrs}\,\eG{1 - \mra_{\mrn} + \frac{\mri}{\mrk}\,\mrs} \spc
&\qquad&
\mrD = \prod_{\mrj=\mrm+1}^{\mrq}\,\eG{1 - \mrb_{\mrj} + \mrs}\,\prod_{\mrj=\mrn+1}^{\mrp}\,\eG{\mra_j - \mrs} \spc
\eqa
and, using a well{-}known propery of the Fox function, derive
\bqa
\mrH^{\mrm\,,\,\mrn}_{\mrp\,,\,\mrq}( \mrz ) &=&
\mrk\,\int_{\mrL}\,\frac{\mrd \mrs}{2\,i\,\pi}\,\frac{\mrN^{\prime}}{\mrD^{\prime}}\,\mrz^{ - \mrk\,\mrs} \spc
\nl
\mrN^{\prime} = \prod_{\mrj=1}^{\mrm}\,\eG{\mrb_{\mrj} - \mrk\,\mrs}\,
         \prod_{\mrj=1}^{\mrn - 1}\,\eG{1 - \mra_{\mrj} + \mrk\,\mrs}\,\eG{1 - \mra_{\mrn} + \mri\,\mrs} \spc
&\qquad&
\mrD^{\prime} = \prod_{\mrj=\mrm+1}^{\mrq}\,\eG{1 - \mrb_{\mrj} + \mrk\,\mrs}\,
         \prod_{\mrj=\mrn+1}^{\mrp}\,\eG{\mra_j - \mrk\,\mrs} \spp
\eqa
Next we use \eqn{mth} and obtain a Meijer function. A simple example is as follows: given 
\bq
\mrH^{1,2}_{2,2}(\mrz) = \int_{\mrL}\,\frac{\mrd \mrs}{2\,i\,\pi}\,
\frac{\eG{\mrb_1 - \mrs}\,\eG{1 - \mra_1 + \mrs}\,\eG{1 - \mra_2 + \frac{2}{3}\,\mrs}}
     {\eG{1 - \mrb_2 + \mrs}}\,\mrz^{ - \mrs} \spc
\eq
we derive a $\mrG^{3,5}_{5,6}$ Meijer G function,
\bqa
\mrH^{1,2}_{2,2}(\mrz) &=& 3\,\lpar 2\,\pi \rpar^{ - 3/2}\,2^{1/2 - \mra_2}\,3^{\mrb_1 + \mrb_2 - \mra_1 - 1/2}\,
\mrG^{3,5}_{5,6}\lpar \frac{9}{4}\,\mrz^3 \rpar \spc
\qquad
\mrG^{3,5}_{5,6} = \int_{\mrL}\,\frac{\mrd \mrs}{2\,i\,\pi}\,\frac{\mrN}{\mrD}\,
\lpar \frac{9}{4}\,\mrz^3 \rpar^{- \mrs} \spc
\nl
\mrN &=&
\eG{1 - \frac{1}{3}\,\mra_1 + \mrs}\,
\eG{1 - \frac{1}{2}\,\mra_2 + \mrs}\,
\eG{\frac{1}{2} - \frac{1}{2}\,\mra_2 + \mrs}\,
\eG{\frac{1}{3} - \frac{1}{3}\,\mra_1 + \mrs}\,
\eG{\frac{2}{3} - \frac{1}{3}\,\mra_1 + \mrs}
\nl
{}&\times&
\eG{\frac{1}{3}\,\mrb_1 - \mrs}\,
\eG{\frac{1}{3} + \frac{1}{3}\,\mrb_1 - \mrs}\,
\eG{\frac{2}{3} + \frac{1}{3}\,\mrb_1 - \mrs} \spc
\nl
\mrD &=&
\eG{1 - \frac{1}{3}\,\mrb_2 + \mrs}\,
\eG{\frac{1}{3} - \frac{1}{3}\,\mrb_2 + \mrs}\,
\eG{\frac{2}{3} - \frac{1}{3}\,\mrb_2 + \mrs} \spp
\eqa
\subsubsection{Multivariate Fox functions \label{mHfun}}
The next generalization of the hypergeometric function is given by the multivariate $\mrH$ Fox function.
Determining the region of convergence of a multivariate Fox function depends on the definition of the function.
Different authors use different definitions; that is why we will indulge in presenting more results than 
actually needed. 
There is a definition given in \Bref{compH} that we present in order to avoid confusion in the notations:
\[
\mrH\lpar \mrz_1\,\dots\,\mrz_\mrr \rpar =
\mrH^{0\,,\,n\,;\,(\mrm_\mrj\,,\,\mrm_\mrj)_{1\,,\,\mrr}}_{\mrp\,,\,\mrq\,;\,(\mrp_\mrj\,,\,\mrq_\mrj)_{1\,,\,\mrr}} 
\left(
\mrz_1\,\dots\,\mrz_\mrr\,,\; \Biggl [
\begin{array}{ccc}
(\mra_\mrj\,,\,\alpha^1_\mrj\,\dots\,\alpha^\mrr_\mrj)_{1\,,\,\mrp}
&\quad ; \quad&
(\mrc^1_\mrj\,,\,\gamma^1_\mrj)_{1\,,\,\mrp_1}\,\dots\,(\mrc^\mrr_\mrj\,,\,\gamma^\mrr_\mrj)_{1\,,\,\mrp_\mrr} \\
(\mrb_\mrj\,,\,\beta^1_\mrj\,\dots\,\beta^\mrr_\mrj)_{1\,,\,\mrp}
&\quad ; \quad&
(\mrd^1_\mrj\,,\,\delta^1_\mrj)_{1\,,\,\mrq_1}\,\dots\,(\mrd^\mrr_\mrj\,,\,\delta^\mrr_\mrj)_{1\,,\,\mrq_\mrr} \\
\end{array}
\Biggr ]
\right)
\]
\bq
\label{mHdef}
\eq
\bqa
\mrH\lpar \mrz_1\,\dots\,\mrz_\mrr \rpar &=& \Bigl [ \prod_{i=1}^{\mrr}\,\int_{\mrL_i}\,
\frac{\mrd \mrs_i}{2\,\pi\,i} \Bigr ]\,\Uppsi\lpar \mrz_1\,,\dots\,,\mrz_\mrr \rpar\,
\prod_{i=1}^{\mrr}\,\theta_i(\mrs_i)\,\mrz_i^{\mrs_i} \spc
\nl
\Uppsi &=&
 \frac{\prod_{\mrj=1}^{\mrn}\,\eG{1 - \mra_\mrj + \sum_{i}\,\alpha^i_\mrj\,\mrs_i}}
      {\prod_{\mrj=\mrn+1}^{\mrp}\,\eG{\mra_\mrj - \sum_{i}\,\alpha^i_\mrj\,\mrs_i} \,
       \prod_{\mrj=1}^{\mrq}\,\eG{1 - \mrb_\mrj + \sum_{i}\,\beta^i_\mrj\,\mrs_i}} \spc
\nl
\theta_i &=&
 \frac{\prod_{\mrj=1}^{\mrm_i}\,\eG{\mrd^i_\mrj - \delta^i_\mrj\,\mrs_i} \,
       \prod_{\mrj=1}^{\mrn_i}\,\eG{1 - \mrc^i_\mrj + \gamma^i_\mrj\,\mrs_i}}
      {\prod_{\mrj=\mrn_i+1}^{\mrp_i}\,\eG{\mrc^i_\mrj - \gamma^i_\mrj\,\mrs_i} \,
       \prod_{\mrj=\mrm_i+1}^{\mrq_i}\,\eG{1 - \mrd^i_\mrj + \delta^i_\mrj\,\mrs_i}} \spc
\eqa
with the conditions 
\bq
\Lambda_i = 
\sum_{\mrj=1}^{\mrq}\,\beta^i_\mrj +
\sum_{\mrj=1}^{\mrq_i}\,\delta^i_\mrj -
\sum_{\mrj=1}^{\mrp}\,\alpha^i_\mrj -
\sum_{\mrj=1}^{\mrp_i}\,\alpha^i_\mrj \ge 0 \spp
\label{Ldef}
\eq
Furthermore $\mid \marg (\mrz_i) \mid < \Omega_i\,\pi/2$ with
\bq
\Omega_i = 
\sum_{\mrj=1}^{\mrn}\,\alpha^i_{\mrj} -
\sum_{\mrj=\mrn+1}^{\mrp}\,\alpha^i_{\mrj} -
\sum_{\mrj=1}^{\mrq}\,\beta^i_{\mrj} +
\sum_{\mrj=1}^{\mrn_i}\,\gamma^i_{\mrj} -
\sum_{\mrj=\mrn_i+1}^{\mrp_i}\,\gamma^i_{\mrj} +
\sum_{\mrj=1}^{\mrm_i}\,\delta^i_{\mrj} -
\sum_{\mrj=\mrm_i+1}^{\mrq_i}\,\delta^i_{\mrj} > 0 \spc
\label{Hconv}
\eq
and $\mrz_i = 0$ excluded. If we define
\bq
\mre_i = \stackrel{\mathrm{min}}{\scriptscriptstyle{1 \le \mrj \le \mrm_i}}\,
\Bigl [ \frac{\Re \mrd^i_{\mrj}}{\delta^i_{\mrj}} \Bigr] \spc
\eq
we obtain
\bq
\mrH = \mrO\lpar \mid \mrz_1 \mid^{\mre_1}\,,\,\dots\,,\mid \mrz_\mrr \mid^{\mre_{\mrr}} \rpar
\quad \hbox{for} \quad \stackrel{\mathrm{max}}{\scriptscriptstyle{1 \le \mrj \le \mrr}}\,\mid \mrz_{\mrj} \mid \to 0 \spp
\label{zbeh}
\eq
Consider $\mrH^{\mrm_1\,,\,\mrn_1}_{\mrp_1\,,\,\mrq_1}(\mrz_1)$ where the corresponding MB contour is in the
complex $\mrs_1$ plane and runs from $\sigma_1 - i\,\infty$ to $\sigma_1 + i\,\infty$. Let 
$\Re \mrg > \Re \mrf + \sigma_1$,
the following relation holds:
\bq
\mrH^{0\,,\,\mrn\,;\,(\mrm_1+1\,,\,\mrn_1)\,,\,(1\,,\,1)}_{1\,,\,0\,;\,(\mrp_1\,,\,\mrq_1+1)\,,\,(2\,,\,1)}\lpar \mrz_1\,,\,- \mrz_2 \rpar =
\eG{\mre}\,\int_0^1 \mrd \mrx\,\mrx^{\mrf-1}\,\lpar 1 - \mrx \rpar^{\mrg - \mrf - 1}\,
\lpar 1 - \mrz_2\,\mrx \rpar^{-\mre}\,\mrH^{\mrm_1\,,\,\mrn_1}_{\mrp+1\,,\,\mrq_1}\lpar \mrz_1\,\mrx \rpar \spc
\eq
where the new parameters are
\bqa 
\mra_1 &=& 1 - \mrf \spc\quad 
\mrc^2_1 = 1 - \mre \spc \quad
\mrd^1_{\mrm_1+1} = \mrg - \mrf \spc\quad 
\mrd^2_1 = 0 \spc\quad 
\mrd^2_2 = 1 - \mrg \spc\quad 
\nl
\alpha_1 &=& \alpha_2 = \gamma^2_1 = \delta^1_{\mrm_1+1} = \beta^2_1 = 1 \spp
\eqa
When the various Gamma functions appear with a power which may take non{-}integer values the Fox $\mrH$ function is denoted by 
$\mrI$ and the list of arguments is modified as follows:
\bq
(\mra_\mrj\,,\,\alpha^1_\mrj\,\dots\,\alpha^\mrr_\mrj)_{1\,,\,\mrp} \quad \to \quad
(\mra_\mrj\,,\,\alpha^1_\mrj\,\dots\,\alpha^\mrr_\mrj\,,\,\xi_\mrj)_{1\,,\,\mrp} \spc
\eq
where $\xi_\mrj$ is the corresponding power. 
\subsubsection{Handling multivariate Fox functions \label{hmff}}
Our goal is to present multiple series representations of multivariate Fox functions.
Therefore, we will discuss series expansions for the $\mrH$ function, suitable for numerical computation under a certain
set of conditions~\cite{rathie2012new}.
Always considering the definition given in \Bref{compH} we introduce the following example:
\bqa
\mrI\lpar \mrz_1\,,\,\mrz_2\rpar &=& \Bigl [ \prod_{i=1}^{2}\,\int_{\mrL_i}\,
\frac{\mrd \mrs_i}{2\,\pi\,i} \Bigr ]\,\Uppsi\lpar \mrs_1\,,\,\mrs_2 \rpar\,
\prod_{i=1}^{2}\,\theta_i(\mrs_i)\,\mrz_i^{ - \mrs_i} \spc
\nl
\Uppsi &=& \frac{\eG{1 - \mra_1 + \mrs_1 + \mrs_2}\,\eG{1 - \mra_2 + \mrs_1 + \mrs_2}}
              {\eG{1 - \mrb_1 + 2\,\mrs_1 + \mrs_2}} \spc
\nl
\theta_1 &=& \frac{\eG{\mrd_1 - \mrs_1}\,\eG{\mrd_2 - \mrs_1}\,\eG{1 - \mrc_1 + \mrs_1}\,\eG{1 - \mrc_2 + \mrs_1}}
                  {\eG{\mrc_3 - \mrs_1}\,\eG{1 - \mrd_3 + \mrs_1}} \spc \qquad
\theta_2 = \eG{ - \mrs_2} \spc
\label{Itwo}
\eqa
corresponding to a bivariate Fox function of \sect{GMFF} with parameters $\mrr = 2, \mrm = 7$ and $\mrn = 3$. Following the
results of Sect.~6 in \Bref{rathie2012new} we define
\bq
\mrJ = \int_{\mrL_2}\,\frac{\mrd \mrs_2}{2\,i\,\pi}\,
\frac{\eG{1 - \mra_1 + \mrs_1 + \mrs_2}\,\eG{1 - \mra_2 + \mrs_1 + \mrs_2}\,\eG{ - \mrs_2}}
     {\eG{1 - \mrb_1 + 2\,\mrs_1 + \mrs_2}}\,\mrz_2^{ - \mrs_2} = \mrH^{2\,\,1}_{2\,,\,2}( \mrz_2 ) \spc
\eq
\[
\mrH^{2\,\,1}_{2\,,\,2}( \mrz_2 ) = 
\mrH^{2\,,\,1}_{2\,,\,2}
\left(
\mrz_2\;,\; \Biggl [
\begin{array}{cc}
1                   &  1 - \mrb_1 + 2\,\mrs_1 \\
1 - \mra_1 + \mrs_1  &  1 - \mra_2 + \mrs_1    \\
\end{array}
\Biggr ]
\right)
\]
The parameters are such that the two sets of poles,
\bq
\mrA_2)\;\; \mrs_2 = \mrn \ge 0 \spc \qquad \mid \qquad \mrB_2)\;\; \mrs_2 = \mra_{\mrj} - 1 - \mrs_1 - \mrk < 0 \; (j = 1,2) \spc
\label{s2poles}
\eq
where $\mrn, \mrk \in \Zf^*$, are separated by $\mrL_2$.
\bei

\item[\ovalbox{$\mrB_2\,,\,\mrA_1$}] In this case we obtain
\bqa
\mrI &=& \sum_{\mrn=0}^{\infty}\,\frac{( - 1)^{\mrn}}{\mrn\,!}\,
\int_{\mrL_1}\,\frac{\mrd \mrs_1}{2\,i\,\pi}\,
\Bigl[
\Gamma_1\,\mrz_1^{ - \mrs_1}\,\mrz_2^{1 - \mra_1 + \mrs_1 + \mrn} +
\Gamma_2\,\mrz_1^{ - \mrs_1}\,\mrz_2^{1 - \mra_2 + \mrs_1 + \mrn} \Bigr] \spc
\nl\nl
\Gamma_1 &=&
\frac{
\eG{\mrd_1 - \mrs_1}\,
\eG{\mrd_2 - \mrs_1}\,
\eG{1 - \mrc_1 + \mrs_1}\,
\eG{1 - \mrc_2 + \mrs_1}\,
\eG{ - \mra_2 + \mra_1 - \mrn}\,
\eG{1 - \mra_1 + \mrs_1 + \mrn}}
{\eG{\mrc_3 - \mrs_1}\,
\eG{1 - \mrd_3 + \mrs_1}\,
\eG{ - \mrb_1 + \mra_1 + \mrs_1 - \mrn}} \spc 
\nl\nl
\Gamma_2 &=& 
\frac{
\eG{\mrd_1 - \mrs_1}\,
\eG{\mrd_2 - \mrs_1}\,
\eG{1 - \mrc_1 + \mrs_1}\,
\eG{1 - \mrc_2 + \mrs_1}\,
\eG{\mra_2 - \mra_1 - \mrn}\,
\eG{1 - \mra_2 + \mrs_1 + \mrn}}
{\eG{\mrc_3 - \mrs_1}\,
\eG{1 - \mrd_3 + \mrs_1}\,
\eG{ - \mrb_1 + \mra_2 + \mrs_1 - \mrn}} \spp
\eqa 
We have poles at
\bq
\mrA_1)\;\; \mrs_1 = \mrd_{\mrj} + \mrm \qquad \mid \qquad \mrB_1)\;\; \mrs_1 = \mrc_{\mrj} - 1 - \mrm \spc \quad
\mrs_1 = \mra_{\mrj} - 1 - \mrn - \mrm \spc
\eq
where $\mrn, \mrm \in \Zf^*$.
We require that the two sets are separated by $\mrL_1$ and that the condition in \eqn{s2poles} is
satisfied; therefore, we select the set $\mrA_1$ and require that
\bq
\mrd_\mrj \ge 0 \spc \quad
\mra_\mrj < 1 \spc \quad \mrc_\mrj < 1 \spc \quad
\mra_\mrj - \mrd_\mrk < 1 \spp
\eq
The resulting expansion is
\bqa
\mrI &=& \sum_{\mrn,\mrm=0}^{\infty}\,\frac{( - 1)^{\mrn + \mrm}}{\mrn\,!\;\mrm\,!}\,
 \mrz_2^{\mrn}\,\lpar \frac{\mrz_2}{\mrz_1} \rpar^{\mrm}\,\Bigl[
\mrz_1^{ - \mrd_1}\,(\mrz_2^{1 + \mrd_1 - \mra_1}\,\Gamma_1 +
\mrz_2^{1 + \mrd_1 - \mra_2}\,\Gamma_2) +
\mrz_1^{ - \mrd_2}\,(\mrz_2^{1 + \mrd_2 - \mra_1}\,\Gamma_3 +
\mrz_2^{1 + \mrd_2 - \mra_2}\,\Gamma_4)  \Bigr] \spc
\nl\nl
\Gamma_1 &=&
\frac{ 
\eG{ - \mra_2 + \mra_1 - \mrn}\,
\eG{\mrd_2 - \mrd_1 - \mrm}\,
\eG{1 + \mrd_1 - \mrc_1 + \mrm}\,
\eG{1 + \mrd_1 - \mrc_2 + \mrm}\,
\eG{1 + \mrd_1 - \mra_1 + \mrn + \mrm}}
{\eG{\mrc_3 - \mrd_1 - \mrm}\,
\eG{1 - \mrd_3 + \mrd_1 + \mrm}\,
\eG{\mrd_1 - \mrb_1 + \mra_1 - \mrn + \mrm}} \spc
\nl
\Gamma_2 &=&
\frac{
\eG{\mra_2 - \mra_1 - \mrn}\,
\eG{\mrd_2 - \mrd_1 - \mrm}\,
\eG{1 + \mrd_1 - \mrc_1 + \mrm}\,
\eG{1 + \mrd_1 - \mrc_2 + \mrm}\,
\eG{1 + \mrd_1 - \mra_2 + \mrn + \mrm}}
{\eG{\mrc_3 - \mrd_1 - \mrm}\,
\eG{1 - \mrd_3 + \mrd_1 + \mrm}\,
\eG{\mrd_1 - \mrb_1 + \mra_2 - \mrn + \mrm}} \spc
\nl
\Gamma_3 &=&
\frac{
\eG{ - \mra_2 + \mra_1 - \mrn}\,
\eG{ - \mrd_2 + \mrd_1 - \mrm}\,
\eG{1 + \mrd_2 - \mrc_1 + \mrm}\,
\eG{1 + \mrd_2 - \mrc_2 + \mrm}\,
\eG{1 + \mrd_2 - \mra_1 + \mrn + \mrm}}
{\eG{\mrc_3 - \mrd_2 - \mrm}\,
\eG{1 - \mrd_3 + \mrd_2 + \mrm}\,
\eG{\mrd_2 - \mrb_1 + \mra_1 - \mrn + \mrm}} \spc
\nl          
\Gamma_4 &=&
\frac{
\eG{\mra_2 - \mra_1 - \mrn}\,
\eG{ - \mrd_2 + \mrd_1 - \mrm}\,
\eG{1 + \mrd_2 - \mrc_1 + \mrm}\,
\eG{1 + \mrd_2 - \mrc_2 + \mrm}\,
\eG{1 + \mrd_2 - \mra_2 + \mrn + \mrm}}
{\eG{\mrc_3 - \mrd_2 - \mrm}\,
\eG{1 - \mrd_3 + \mrd_2 + \mrm}\,
\eG{\mrd_2 - \mrb_1 + \mra_2 - \mrn + \mrm}} \spc
\eqa
corresponding to $\mid \mrz_2 \mid < 1$ and $\mid \mrz_1 \mid > \mid \mrz_2 \mid$.  

\item[\ovalbox{$\mrA_2\,,\,\mrA_1$}] is the other possible case. We obtain
\bqa
\mrI &=& \sum_{\mrn\,,\,\mrm=0}^{\infty}\,\frac{( - 1)^{\mrn + \mrm}}{\mrn\,!\;\mrm\,!}\,
\Bigl[ \mrz_1^{ - \mrd_1}\,\Gamma_1 + \mrz_1^{ - \mrd_2}\,\Gamma_2 \bigr]\,
\mrz_1^{ - \mrm}\,\mrz_2^{ - \mrn} \spc
\nl\nl
\Gamma_1 &=&
\frac{
\eG{\mrd_2 - \mrd_1 - \mrm}\,
\eG{1 + \mrd_1 - \mrc_1 + \mrm}\,
\eG{1 + \mrd_1 - \mrc_2 + \mrm}\,
\eG{1 + \mrd_1 - \mra_1 + \mrk}\,
\eG{1 + \mrd_1 - \mra_2 + \mrk}}
{\eG{\mrc_3 - \mrd_1 - \mrm}\,
\eG{1 - \mrd_3 + \mrd_1 + \mrm}\,
\eG{1 + 2\,\mrd_1 - \mrb_1 + \mrn + 2\,\mrm}} \spc
\nl
\Gamma_2 &=&
\frac{
\eG{ - \mrd_2 + \mrd_1 - \mrm}\,
\eG{1 + \mrd_2 - \mrc_1 + \mrm}\,
\eG{1 + \mrd_2 - \mrc_2 + \mrm}\,
\eG{1 + \mrd_2 - \mra_1 + \mrk}\,
\eG{1 + \mrd_2 - \mra_2 + \mrk}}
{\eG{\mrc_3 - \mrd_2 - \mrm}\,
\eG{1 - \mrd_3 + \mrd_2 + \mrm}\,
\eG{1 + 2\,\mrd_2 - \mrb_1 + \mrn + 2\,\mrm}} \spc
\label{dsum}
\eqa
where $\mrk= \mrn + \mrm$, corresponding to $\mid \mrz_i \mid > 1$.

\end{itemize}

Other cases can be obtained by using the inversion relation of \eqn{Hinv}.

\paragraph{A general algorithm  \label{Hpassare}} \hspace{0pt} \\
A more general algorithm can be found in \Bref{Passare:1996db}; we briefly summarize the procedure using the results
of Sect.~2 of \Bref{Passare:1996db}. The $\mrH$ function is more conveniently written as
\bq
\mrH = \Bigl[ \prod_{\mrj=1}^{\mrn}\,\int_{\mrL_{\mrj}}\,\frac{\mrd \mrs_{\mrj}}{2\,i\,\pi} \Bigr]\,
\frac{\prod_{\mrj=1}^{\mrp}\,
      \eG{\sprod{\mra_{\mrj}}{\mrs} + \mrb_{\mrj}}
     }
     {\prod_{\mrj=1}^{\mrq}\,
      \eG{\sprod{\mrc_{\mrj}}{\mrs} + \mrd_{\mrj}}
     }\,
     \prod_{\mrj=1}^{\mrn}\,( - \mrz_{\mrj} )^{ - \mrs_{\mrj}} \spc
\label{PH}
\eq
where ${\mathbf a} , {\mathbf c} \in \Rf^{\mrn}$ and $\mrb_{\mrj} , \mrd_{\mrj} \in \Rf$. The contours of integration are
$\mrL_{\mrj}$, connecting $- i\,\infty$ to $+ i\,\infty$ never intersect the hyperplanes 
\bq
\mrL^{\mrk}_{\mrj} = \Bigl\{ \sprod{\mra_{\mrj}}{\mrs} + \mrb_{\mrj} = - \mrk \Bigr\} \spp
\eq
\Bref{Passare:1996db} defines local Grothendiek residues at the point of intersection of the hyperplanes $\mrL^{\mrk}_{\mrj}$,
denoted by $\mrs^{\mrm}_{\mrJ}$,
\bq
\mathrm{Res}_{_{\mrs^{\mrm}_{\mrJ}}}\,\mrh = \int_{\gamma^{\mrm}_{\mrJ}}\,\mrh \spc
\eq
where $\mrh$ is the integrand in \eqn{PH} and $\gamma^{\mrm}_{\mrJ}$ is the cycle
\bq 
\gamma^{\mrm}_{\mrJ} = \Bigl\{ {\mathbf s}\,;\,\mid\,
    \sprod{\mra_{\mrj_1}}
          {\mrs} + \mrb_{\mrj_1} + \mrm\,\mid\,
=\,\dots\,=\;\mid\,
    \sprod{\mra_{\mrj_{\mrn}}}
          {\mrs} + \mrb_{\mrj_{\mrn}} + \mrm\,\mid\;=\,\ep \Bigr\} \spc
\eq
with $\ep \to 0_{+}$.  The result is
\bq
\mathrm{Res}_{_{\mrs^{\mrm}_{\mrJ}}}\,\mrh =
   \frac{( - 1)^{\mid \mrm \mid}
        }
        {\Delta_{\mrJ}\,\mrm\,!
        }
   \frac{
         \prod_{\mrj \not= \mrJ}\,
         \eG{\sprod{\mra_{\mrj}}
                   {\mrs^{\mrm}_{\mrj}} + \mrb_{\mrj}
            }
        }
        {
         \prod_{\mrk=1}^{\mrq}\,
         \eG{\sprod{\mrc_{\mrk}}{\mrs^{\mrm}_{\mrJ}} + \mrd_{\mrk}
            }
        }\,
      \prod_{\mrj=1}^{\mrn}\,( - \mrz_{\mrj} )^{(\mrs^{\mrm}_{\mrJ})_{\mrj}} \spc
\label{Pres}
\eq
where 
\bq
\Delta_{\mrJ} = \mathrm{det} (\mra_{\mrj_1}\,\dots\,\mra_{\mrj_{\mrn}}) \spc \quad
\mid \mrm \mid = \sum_{\mrj=1}^{\mrn}\,\mrm_{\mrj} \spc \quad
\mrm\,! = \prod_{\mrj=1}^{\mrn}\,\mrm_{\mrj}\,! \spp
\eq
The last part of Sect.~2 of \Bref{Passare:1996db} answers the question of expressing the integral in \eqn{PH}
through the residues of \eqn{Pres}. We will not present this last part of the algorithm but refer to
Claim~1 of \Bref{Passare:1996db}.  
Investigation of multiple MB integrals by means of multidimensional residues is also dicussed in \Bref{ZT}.

In a second approach MB integrals are computed by introducing conic hulls or triangulations of point configurations.  
Both the resonant and non{-}resonant cases can be handled by these methods which are automatized in the
{\tt{MBConicHulls.w1}}
\footnote{MBConicHulls webpage \url{https://github.com/SumitBanikGit/MBConicHulls}}
package~\cite{banik2024geometrical}.  
\paragraph{More on analytic continuation} \hspace{0pt} \\
Inspite of the progress made in the theory of hypergeometric functions,
many important questions which are fully understood for the Gauss function
remain unresolved in the multidimensional case.

A general treatment of analytic continuation is illustrated by the following example:
\bq
\mrI = \Bigl[ \prod_{i=1}^{2}\,\int_{\mrL_i}\,\frac{\mrd \mrs_i}{2\,i\,\pi} \Bigr]\,
\frac{
\eG{ - \mrs_1}\,
\eG{ - \mrs_2}\,
\eG{\mrs_1 - \alpha_2}\,
\eG{\mrs_2 + \alpha_2 - \alpha_1}\,
\eG{2 + \mrs_2 + \mrs_1 - \alpha_2 + \alpha_1}
}
{
\eG{3 + \mrs_2 + \mrs_1 - \alpha_2 + \alpha_1}
}\,
\mrz_1^{\mrs_1}\,\mrz_2^{\mrs_2} \spc
\eq
where $\mid \mrz_i \mid > 1$ and $\mid \mrz_1\mid > \mid \mrz_2\mid$. The $\mrs_2$ integral gives
\bqa
\mrI &=& \int_{\mrL_1}\,\frac{\mrd \mrs_1}{2\,i\,\pi}\,
 \eG{ - \mrs_1}\,\eG{\mrs_1 - \alpha_2}\,
 \mrG^{12}_{22}\,\lpar {\mathbf a}\,;\,{\mathbf b}\,;\,\mrz_2 \rpar\,\mrz_1^{\mrs_1} \spc
\nl
{\mathbf a} &=& \{1 - \alpha_2 + \alpha_1 \,,\, - 1 - \mrs_1 + \alpha_2 - \alpha_1\} \spc
\nl
{\mathbf b} &=& \{0 \,,\, - 2 - \mrs_1 + \alpha_2 - \alpha_1\} \spp 
\eqa
Using Eq.~(5.3.6) of \Bref{HTF} we can write the Meijer $\mrG\,${-}function in terms of 
\bq
\hyp
{\alpha_2 - \alpha_1}{ - 2 - \mrs_1 + 2\,\alpha_2 - 2\,\alpha_1}{ - 1 - \mrs_1 + 2\,\alpha_2 - 2\,\alpha_1}{ - \mrz_2^{-1}} 
\spp
\eq
Using the MB representation gives      
\bq
\mrI= \mrz_2^{\alpha_1 - \alpha_2}\,
\Bigl[ \prod_{i=1}^{2}\,\int_{\mrL_i}\,\frac{\mrd \mrs_i}{2\,i\,\pi} \Bigr]\,\Upphi_2\,
\mrz_1^{\mrs_1}\mrz_2^{ - \mrs2} +
\mrz_2^{\alpha_2 - \alpha_1 - 2}\,
\int_{\mrL_1}\,\frac{\mrd \mrs_1}{2\,i\,\pi}\,\Upphi_1\,\mrz_1^{\mrs_1}\,\mrz_2^{ - \mrs_1} \spc
\label{inres}
\eq
where $\Upphi_2 = \mrN/\mrD$,
\bq
\Upphi_1 =
\eG{ - \mrs_1}\,
\eG{\mrs_1 - \alpha_2}\,
\eG{ - 2 - \mrs_1 + 2\,\alpha_2 - 2\,\alpha_1}\,
\eG{2 + \mrs_1 - \alpha_2 + \alpha_1} \spc
\eq
\bqa 
\mrN &=&
\eG{ - \mrs_1}\,
\eG{ - \mrs_2}\,
\eG{\mrs_1 - \alpha_2}\,
\eG{\mrs_2 + \alpha_2 - \alpha_1}\,
\eG{ - 1 - \mrs_1 + 2\,\alpha_2 - 2\,\alpha_1}
\nl
{}&\times&
\eG{2 + \mrs_1 - 2\,\alpha_2 + 2\,\alpha_1}\,
\eG{ - 2 + \mrs_2 - \mrs_1 + 2\,\alpha_2 - 2\,\alpha_1} \spc
\nl
\mrD &=&
\eG{ - 2 - \mrs_1 + 2\,\alpha_2 - 2\,\alpha_1}\,
\eG{3 + \mrs_1 - 2\,\alpha_2 + 2\,\alpha_1}\,
\eG{ - 1 + \mrs_2 - \mrs_1 + 2\,\alpha_2 - 2\,\alpha_1} \spp
\eqa
The integral over $\mrs_1$ in \eqn{inres} give rise to
a combination of $\mrG^{22}_{22}$ and $\mrG^{32}_{44}$, transformed into a combination of $\ghyp{2}{1}$ and
$\ghyp{4}{3}$; for the generalized hypergeometric functions we use the MB representation of \eqn{GHFMB}, obtaining
\bq
\mrI = \mrB_1\,\mrz_1^{\alpha_2 - \alpha_1 - 2}\,\mrI_1 +
\mrB_2\,\mrz_1^{\alpha_2}\,\mrz_2^{ - \alpha_1 - 2}\,\mrI_2 +
\mrB_3\,\mrz_1^{\alpha_2}\,\mrz_2^{\alpha_1 - \alpha_2}\,\mrI_3 \spc
\eq
\bq
\mrB_1 = \eG{ - 2 - \alpha_1}\,\eG{3 + \alpha_1} \spc \quad
\mrB_2 = \eG{ - 1 - \alpha_1}\,\eG{2 + \alpha_1} \spc \quad
\mrB_3 =
\frac{\eG{ - 1 + \alpha_2 - 2\,\alpha_1}\,\eG{2 - \alpha_2 + 2\,\alpha_1}}
     {\eG{ - 2 + \alpha_2 - 2\,\alpha_1}\,\eG{3 - \alpha_2 + 2\,\alpha_1}} \spp
\eq
The three integrals are
\bqa
\mrI_1 &=& \int_{\mrL_1}\,\frac{\mrd \mrs_1}{2\,i\,\pi}\,
\frac{
\eG{ - \mrs_1}\,
\eG{\mrs_1 + \alpha_2 - \alpha_1}\,
\eG{2 + \mrs_1 - \alpha_2 + \alpha_1}
}
{
\eG{3 + \mrs_1 + \alpha_1}
}\,
\lpar \frac{\mrz_2}{\mrz_1}\rpar^{\mrs_1} \spc 
\nl
\mrI_2 &=&
\int_{\mrL_1}\,\frac{\mrd \mrs_1}{2\,i\,\pi}\,
\frac{
\eG{ - \mrs_1}\,
\eG{\mrs_1 - \alpha_2}\,
\eG{ - 2 + \mrs_1 + \alpha_2 - 2\,\alpha_1}
}
{
\eG{ - 1 + \mrs_1 - \alpha_1} 
}\,
\lpar \frac{\mrz_2}{\mrz_1}\rpar^{\mrs_1} \spc 
\nl
\mrI_3 &=&
\frac{
\eG{ - \mrs_1}\,
\eG{ - \mrs_2}\,
\eG{\mrs_1 - \alpha_2}\,
\eG{\mrs_2 + \alpha_2 - \alpha_1}\,
\eG{ - 2 + \mrs_2 + \mrs_1 + \alpha_2 - 2\,\alpha_1}
}
{
\eG{ - 1 + \mrs_2 + \mrs_1 + \alpha_2 - 2\,\alpha_1} 
}\,
\mrz_1^{ - \mrs_1}\,\mrz_2^{ - \mrs_2} \spp
\eqa
The integrals can now be computed by adding the residues of the poles at $\mrs_1 = \mrn_1$ and
$\mrs_2 = \mrn_2$.

These results assume that we are not in the resonant case. To illustrate the resonant case we present
a simple example:
\bq
\mrB = \int_0^1 \mrd \mrx\,\ln(1 - \lambda\,\mrx + \lambda\,\mrx^2 - i\,\delta) \spp
\eq
The integral will give
\bq
\mrB = - \frac{\sqrt{2\,\pi}}{16}\,\lambda\,\int_{\mrL}\,\frac{\mrd \mrs}{2\,i\,\pi}\,
\eG{ - \mrs}\,\frac{\eGs{1 + \mrs}}{\eG{5/2 + \mrs}}\,\lpar - \frac{1}{4}\,\lambda - i\,\delta \rpar^{\mrs} \spp
\eq
When $\mid \lambda \mid < 1$ we can compute the residues at $\mrs = \mrn$. To deal with the case
$\mid \lambda \mid > 1$ we write
\bq
\mrB = - \frac{1}{6\,\sqrt{2}}\,\lambda\,\hyp{1}{1}{\frac{5}{2}}{\frac{1}{4}\,\lambda + i\,\delta} \spc
\eq
where the hypergeometric function is in the resonant case; therefore, the correct expansion is
\bq
\mrB = - \frac{3}{2}\,\mrz^{ - 3/2}\,(\mrz - 1)^{1/2}\,\ln( - \mrz)
- \frac{3}{2}\,\mrz^{-1}\,\sum_{\mrn=0}^{\infty}\,\frac{( - 1/2)_{\mrn}}{\mrn\,!}\,\mrh_{\mrn}\,\mrz^{ - \mrn} \spc
\eq
where~\cite{HTF} $\mrh_{\mrn} = \uppsi(1 + \mrn) - \uppsi(3/2 - \mrn)$ and $\mrz = 1/4\,\lambda + i\,\delta$.
As soon as we have to deal with resonant Meijer $\mrG\,${-} functions or resonant generalized hypergeometric
functions the analytic continuation becomes more complicated but we can use the results of Sect.~(2.2) and of 
Sect.~(6.1) of \Bref{Scheidegger:2016ysn} which are based on \Bref{Ncoef}.
We give one example concerning the function
\bq
\ghyp{3}{2}\lpar {\mathbf a}\,;\,{\mathbf b}\,;\,\mrz \rpar \spc
\eq
using Proposition~(4.8) of \Bref{Scheidegger:2016ysn}. The result is 
\bqa
\ghyp{3}{2}\lpar {\mathbf a}\,;\,{\mathbf b}\,;\,\mrz \rpar &=&
\frac{\eG{\mrb_1}\,\eG{\mrb_2}}{\eG{\mra_3}}],
\int_{\mrL}\,\frac{\mrd \mrs}{2\,i\,\pi}\,
\frac{{\tilde{\mrA}}^{(3)}(\mrs)}{\eG{\mrc + \mra_1 + \mra_2 + \mrs}}\,
\hyp{\mra_1}{\mra_2}{\mrc + \mra_1 + \mra_2 + \mrs}{\mrz} \spc
\nl
{\tilde{\mrA}}^{(3)}(\mrs) &=& \exp\{i\,\pi\,\mrs\}\,\eG{ - \mrs}\, 
\frac{\eG{c + \mra_1 + \mra_2 - \mrb_1 + \mrs}\,\eG{\mrb_1 - \mra_3 + \mrs}}
     {\eG{\mrb_1 - \mra_3}\,\eG{\mrb_2 - \mra_3}} \spc
\label{Brec}
\eqa
where $\mrc = \sum\,\mrb - \sum\,\mra$. We now use Eq.~(2.10.2) of \Bref{HTF} and obtain
\bq
\ghyp{3}{2}\lpar {\mathbf a}\,;\,{\mathbf b}\,;\,\mrz \rpar =
\sum_{\mrn=0}^{\infty}\,\frac{1}{\mrn\,!}\,\lpar \mrI_1 + \mrI_2 \rpar \spc \quad
\mrI_{\mrj} = \lpar - \mrz\rpar^{ - \mra_{\mrj}}\,\Upphi_{\mrj}\,\Uppsi_{\mrj} \spc
\eq
\bqa
\Upphi_1 &= &
\frac{\eG{\mrb_1}\,\eG{\mrb_2}\,\eG{\mra_2 - \mra_1}}
     {\eG{\mrb_1 - \mra_3}\,\eG{\mrb_2 - \mra_3}} \spc
\qquad
\Upphi_2 = 
\frac{\eG{\mrb_1}\,\eG{\mrb_2}\,\eG{\mra_1 - \mra_2}}
     {\eG{\mrb_1 - \mra_3}\,\eG{\mrb_2 - \mra_3}} \spc
\nl\nl
\Uppsi_1 &=&
\frac{\eG{\mrb_1 - \mra_3 + \mrn}\,\eG{\mrb_2 - \mra_3 + \mrn}}
     {\eG{\mrb_1 + \mrb_2 - \mra_1 - \mra_3 + \mrn}}
\nl
{}&\times& \hyp{\mra_1}{1 - \mrb_1 - \mrb_2 + \mra_1 + \mra_3 - \mrn}{1 + \mra_1 - \mra_2}{\mrz^{-1}} \spc
\nl\nl
\Uppsi_2 &=&
\frac{\eG{\mrb_1 - \mra_3 + \mrn}\,\eG{\mrb_2 - \mra_3 + \mrn}}
     {\eG{\mrb_1 + \mrb_2 - \mra_2 - \mra_3 + \mrn}}
\nl
{}&\times& \hyp{\mra_2}{1 - \mrb_1 - \mrb_2 + \mra_2 + \mra_3 - \mrn}{1 + \mra_2 - \mra_1}{\mrz^{-1}} \spc
\eqa
with the following contraints: $\mra_1 - \mra_2 \not\in \Zf$, $\mra_3 - \mrb_1$ and $\mra_3 - \mrb_2$ are not
negative integers and where the arguments of $\hyp{\mra}{\mrb}{\mrc}{\mrz}$ in \eqn{Brec} are such
that $\mra - \mrb \not\in \Zf$ and $\mrc$ is not a negative integer.

Finally, the analytic continuation of the special Meijer $\mrG^{\mrm,\mrp}_{\mrp,\mrp}$ in the resonant case is
given in Proposition~(2.5) and in Corollary~(2.6) of \Bref{Scheidegger:2016ysn}.
\paragraph{The logarithmic (resonant) case} \hspace{0pt} \\
The presence of double or higher poles introduces logarithms; for instance, given
$\eG{\mrg_1}\,\eG{\mrg_2}$ where $\mrg_{\mrj}({\mathbf s})$ are bivariate polynomials in $\Rf^2$, the relevant quantity
is their zero set, $\mrZ({\mathbf g})$, which can be a non{-}empty subset of $\Rf^2$. 
For instance,
\bq
\eG{\mra + \mrs_1 + \mrs_2}\,\eG{\mrb + \mrs_1 + \mrs_2} \spc
\eq
gives simple poles as long as $\mra - \mrb$ in not an integer. The case
\bq
\eG{\mra + \mrs_1 + \mrs_2}\,\eG{\mrb + \mrs_1 - \mrs_2} \spc
\eq
is such that at the points
\bq
\mrs_1^{\mrk \mrj} = - \frac{1}{2}\,(\mra + \mrb + \mrk + \mrj) \spc \qquad
\mrs_2^{\mrk \mrj} = - \frac{1}{2}\,(\mra - \mrb + \mrk - \mrj) \spc
\eq
both Gamma functions have a pole. The best way to handle this configuration is to define 
(we will only discuss the case $\mrk = \mrj = 0$)
\bq
\mrg_1 = \mra + \mrs_1 + \mrs_2 \spc \qquad
\mrg_2 = \mrb + \mrs_1 - \mrs_2 \spc \qquad
\mrf_{\mri} = \mrA_{\mri \mrj}\,\mrg_{\mrj} \spp
\eq
We put $\mra = \sigma_1 + \sigma_3$, where $\sigma_{\mrj}$ are Pauli matrices and obtain
\bq
\mrf_1 = \mra + \mrb + 2\,\mra_1 \spc \qquad
\mrf_2 = \mrb - \mra - 2\,\mrs_2 \spc \qquad
\mathrm{det}\,\mrA = 2 \spp
\eq
Given
\bq
\mrI = \Bigl[ \prod_{\mrj=1}^{2}\,\int_{\mrL_{\mrj}}\,\frac{\mrd \mrs_{\mrj}}{2\,i\,\pi} \Bigr]\,
\eG{\mra + \mrs_1 + \mrs_2}\,\eG{\mrb + \mrs_1 - \mrs_2}\,\mrh(\mrs_1\,,\,\mrs_2) \spc
\eq
around $\mrs^{00}_1, \mrs^{00}_2$ we can write
\bq
\mrI_{00} = \Bigl[ \prod_{\mrj=1}^{2}\,\int_{\mrL_{\mrj}}\,\frac{\mrd \mrs_{\mrj}}{2\,i\,\pi} \Bigr]\,
\Bigl[ \frac{1}{\mrg_1\,\mrg_2} + \ord{1} \Bigr]\,\mrh(\mrs_1\,,\,\mrs_2) =
 \pm \frac{1}{2}\,\mrh\lpar - \frac{1}{2}\,(\mra + \mrb)\,,\, - \frac{1}{2}\,(\mra - \mrb) \rpar \spc
\eq
where we have used~\cite{Catt,Tsi} 
\bq
\mathrm{Res}\,\frac{\mrh}{\mrg_1\,\mrg_2} = 2\,\mathrm{Res}\,\frac{\mrh}{\mrf_1\,\mrf_2} \spp
\eq

Results in the resonant case are based on the so{-}called
converse mapping theorem~\cite{Friot:2011ic}; given
\bq
\mrM(\mrf) = \int_0^{\infty} \mrd \mrx\,\mrx^{\mrs - 1}\,\mrf(\mrx) \spc \quad
\mrf(x) = \int_{\mrL}\,\frac{\mrd \mrs}{2\,i\,\pi}\mrx^{ - \mrs}\,\mrM(\mrf) \spc
\eq
if $\mrf$ has a MB transform in the strip $[\,\alpha\,,\,\beta\,]$ and $\mrM(\mrf) = \ord{\mrs^{ - \mrn}}$
for $\mrn > 1$ we obtain
\bqa
\mrM(\mrf) \sim \sum_{\rho > \beta\,,\,\mrn}\,\frac{\mra_{\rho\,,\,\mrn}}{(\mrs - \rho)^{\mrn}}
\quad &\leftrightarrow& \quad
\mrf(x) \sim \sum_{\rho > \beta\,,\,\mrn}\,\frac{( - 1)^{\mrn}}{\eG{\mrn + 1}}\,\mra_{\rho\,,\,\mrn}\,
\mrx^{- \rho}\,\ln^{\mrn - 1} \mrx \spc
\nl
\mrM(\mrf) \sim \sum_{\rho < \alpha\,,\,\mrn}\,\frac{\mrb_{\rho\,,\,\mrn}}{(\mrs + \rho)^{\mrn}}
\quad &\leftrightarrow& \quad
\mrf(x) \sim \sum_{\rho < \alpha\,,\,\mrn}\,\frac{( - 1)^{\mrn - 1}}{\eG{\mrn + 1}}\,\mrb_{\rho\,,\,\mrn}\,
\mrx^{\rho}\,\ln^{\mrn - 1} \mrx \spp
\eqa
We emphasize that these are not always the only logarithms in the answer and we should consider the resummation
of the residues of the single poles. A trivial but illustrative example is the following:
\bq
\mrM = - \int_{\mrL}\,\frac{\mrd \mrs}{2\,i\,\pi}\,\eG{ - \mrs}\,\eG{\mrs}\,( - \mra + i\,\delta )^{\mrs} \spc
\eq
where $0 < \mra < 1$ and $ - 1 < \Re(\mrs) < 0$. There is a double pole at $\mrs = 0$ and simple poles
at $\mrs = \mrn$, $\mrn \ge 1$. We obtain
\bq
\mrM = - \ln( - \mra + i\,\delta)  - \sum_{\mrn=1}^{\infty}\,\frac{1}{n}\,(\mra - i\,\delta)^{\mrn} =
- \ln( - \mra + i\,\delta) + \ln(1 - \mra + i\,\delta) \spp
\eq
The extension of the theorem to the multidimensional case requires the Grothendieck residue 
theorem~\cite{tajima2020effective,Soares}.

We proceed with the following example:
\bqa
\mrI_2\lpar \mrz_1\,,\,\mrz_2\rpar &=& \Bigl [ \prod_{i=1}^{2}\,\int_{\mrL_i}\,
\frac{\mrd \mrs_i}{2\,\pi\,i} \Bigr ]\,\Uppsi\lpar \mrs_1\,,\,\mrs_2 \rpar\,
\prod_{i=1}^{2}\,\theta_i(\mrs_i)\,\mrz_i^{ - \mrs_i} \spc
\nl
\Uppsi &=& \frac{\eG{1 - \mra_1 + \mrs_1 + \mrs_2}\,\eG{1 - \mra_2 + \mrs_1 + \mrs_2}}
              {\eG{1 - \mrb_1 + 2\,\mrs_1 + \mrs_2}} \spc
\nl
\theta_1 &=& \frac{\eGs{\mrd_1 - \mrs_1}\,\eG{1 - \mrc_1 + \mrs_1}\,\eG{1 - \mrc_2 + \mrs_1}}
                  {\eG{\mrc_3 - \mrs_1}\,\eG{1 - \mrd_2 + \mrs_1}} \spc
\qquad
\theta_2 = \eG{ - \mrs_2} \spp
\label{Itwodp}
\eqa
The double pole gives
\bqa
\mrI_{\mathrm{dp}} &=&
\sum_{\mrn\,,\,\mrm=0}^{\infty}\,
\Bigl\{
2\,\uppsi(1)\,\frac{(-1)^{\mrn + \mrm}}{\mrn\,!\;\,\mrm\,!}\,
\mrz_1^{ - \mrd_1 - \mrm}\,\Bigl[
\mrH_1(\mrn\,,\,\mrm)\,\mrz_2^{1 + \mrd_1 - \mra_1 + \mrn + \mrm} +
\mrH_2(\mrn\,,\,\mrm)\,\mrz_2^{1 + \mrd_1 - \mra_2 + \mrn + \mrm} \Bigr]
\nl 
{}&+& \frac{( - 1)^{\mrn}}{\mrn\,!\,(\mrm\,!)^2}\,\mrz_1^{ - \mrd_1 - \mrm}\,\ln\frac{\mrz_2}{\mrz_1}\,\Bigl[
\mrH_1(\mrn\,,\,\mrm)\,\mrz_2^{1 + \mrd_1 - \mra_1 + \mrn + \mrm} +
\mrH_2(\mrn\,,\,\mrm)\,\mrz_2^{1 + \mrd_1 - \mra_2 + \mrn + \mrm} \Bigr]
\nl
{}&-& \frac{( - 1)^{\mrn}}{\mrn\,!\,(\mrm\,!)^2}\,
\mrz_1^{ - \mrd_1 - \mrm}\,\mrz_2^{1 + \mrd_1 - \mra_1 + \mrn + \mrm}\,\mrH_1(\mrn\,,\,\mrm)\,\Bigl[
             \uppsi(1 - \mrd_2 + \mrd_1 + \mrm) -
             \uppsi(1 + \mrd_1 - \mra_1 + \mrn + \mrm) 
\nl
{}&-&        \uppsi(1 + \mrd_1 - \mrc_2 + \mrm) 
           - \uppsi(1 + \mrd_1 - \mrc_1 + \mrm) -
             \uppsi(\mrc_3 - \mrd_1 - \mrm) +
             \uppsi(\mrd_1 - \mrb_1 + \mra_1 - \mrn + \mrm) \Bigr]
\nl
{}&-& \frac{( - 1)^{\mrn}}{\mrn\,!\,(\mrm\,!)^2}\,
\mrz_1^{ - \mrd_1 - \mrm}\,\mrz_2^{1 + \mrd_1 - \mra_1 + \mrn + \mrm}\,\mrH_2(\mrn\,,\,\mrm)\,\Bigl[
             \uppsi(1 - \mrd_2 + \mrd_1 + \mrm) -
             \uppsi(1 + \mrd_1 - \mra_2 + \mrn + \mrm) 
\nl
{}&-&        \uppsi(1 + \mrd_1 - \mrc_2 + \mrm) 
           - \uppsi(1 + \mrd_1 - \mrc_1 + \mrm) -
             \uppsi(\mrc_3 - \mrd_1 - \mrm) +
             \uppsi(\mrd_1 - \mrb_1 + \mra_2 - \mrn + \mrm) \Bigr]
         \Bigr\} \spc
\eqa
\bqa
\mrH_1 &=&
\frac{          
\eG{ - \mra_2 + \mra_1 - \mrn}\,
\eG{1 + \mrd_1 - \mrc_1 + \mrm}\,
\eG{1 + \mrd_1 - \mrc_2 + \mrm}\,
\eG{1 + \mrd_1 - \mra_1 + \mrn + \mrm}}
{\eG{\mrc_3 - \mrd_1 - \mrm}\,
\eG{1 - \mrd_2 + \mrd_1 + \mrm}\,
\eG{\mrd_1 - \mrb_1 + \mra_1 - \mrn + \mrm}} \spc
\nl
\mrH_2 &=&
\frac{
\eG{\mra_2 - \mra_1 - \mrn}\,
\eG{1 + \mrd_1 - \mrc_1 + \mrm}\,
\eG{1 + \mrd_1 - \mrc_2 + \mrm}\,
\eG{1 + \mrd_1 - \mra_2 + \mrn + \mrm}}
{\eG{\mrc_3 - \mrd_1 - \mrm}\,
\eG{1 - \mrd_2 + \mrd_1 + \mrm}\,
\eG{\mrd_1 - \mrb_1 + \mra_2 - \mrn + \mrm}} \spc
\eqa
where $\uppsi(\mrz)$ is the logarithmic derivative of the Gamma function.
There is the possibility to extend the previous results: consider 
\bq
\mrI_1 = \sum_{\mrn , \mrm = 0}^{\infty}\,\frac{\Gamma_1}{\mrn\,!\;\mrm\,!}\,
( - \mrz_1)^{ - \mrm}\,( - \mrz_2)^{ - \mrn} \spc
\eq
where $\Gamma_1$ is given in \eqn{dsum}. If $\Gamma_{1\,\mrn}$ is the part of $\Gamma_1$ depending on $\mrn$ 
we can write
\bqa
{}&{}& \sum_{\mrn = 0}^{\infty}\,\Gamma_{1\,\mrn}\,\frac{( - \mrz_2)^{ - \mrn}}{\mrn\,!} =  
       \frac{\eG{1 + \mrm + \mrd_1 - \mra_1}\,\eG{1 + \mrm + \mrd_1 - \mra_2}}
            {\eG{1 + 2\,\mrm + 2\,\mrd_1 - \mrb_1}}
\nl
{}&\times&
  \hyp{1 + \mrm + \mrd_1 - \mra_1}{1 + \mrm + \mrd_1 - \mra_2}{1 + 2\,\mrm + 2\,\mrd_1 - \mrb_1}{ - \frac{1}{\mrz_2}} \spc
\eqa
giving
\bqa
\mrI_1 &=& \sum_{\mrm = 0}^{\infty}\,\frac{\mrN_{\mrm}}{\mrD_{\mrm}}\,
 \hyp{1 + \mrm + \mrd_1 - \mra_1}{1 + \mrm + \mrd_1 - \mra_2}{1 + 2\,\mrm + 2\,\mrd_1 - \mrb_1}{ - \frac{1}{\mrz_2}}\,
\frac{( - \mrz_1)^{\mrm}}{\mrm\,!} \spc
\nl
\mrN_{\mrm} &=&
\eG{ - \mrm + \mrd_2 - \mrd_1}\,
\eG{1 + \mrm + \mrd_1 - \mrc_1}\,
\eG{1 + \mrm + \mrd_1 - \mrc_2}\,
\eG{1 + \mrm + \mrd_1 - \mra_1}\,
\eG{1 + \mrm + \mrd_1 - \mra_2} \spc
\nl
\mrD_{\mrm} &=&
\eG{ - \mrm - \mrd_1 + \mrc_3}\,
\eG{1 + \mrm - \mrd_3 + \mrd_1}\,
\eG{1 + 2\,\mrm + 2\,\mrd_1 - \mrb_1} \spp
\label{nac}
\eqa
The advantage of using \eqn{nac} is that the result gives an analytic continuation in $\mrz_2$.
Alternatively we can use
\bqa
\mrI_1 &=& \pi^{1/2}\,4^{\mrb_1/2 - \mrd_1}\,\sum_{\mrn = 0}^{\infty}\,\frac{\mrN_{\mrn}}{\mrD_{\mrm}}\,
\mrF_{\mrn}(\mrz_1)\,( 2\,\mrz_2)^{- \mrn} \spc
\nl
\mrD_{\mrn} &=&
\eG{\mrd_2 - \mrd_1}\,
\eG{1 + \mrd_1 - \mrc_1}\,
\eG{1 + \mrd_1 - \mrc_2}\,
\eG{1 + \mrd_1 - \mra_1 + \mrn}\,
\eG{1 + \mrd_1 - \mra_2 + \mrn} \spc
\nl
\mrD_{\mrn} &=&
\eG{ - \mrd_1 + \mrc_3}\,
\eG{1 - \mrd_3 + \mrd_1}\,
\eG{1 + \mrd_1 + \frac{1}{2}\,(\mrn - \mrb_1)}\,
\eG{\mrd_1 + \frac{1}{2}\,(1 + \mrn - \mrb_1)} \spc
\nl
\mrF_{\mrn}(\mrz_1) &=&
{}_{\scriptstyle{5}}\,\mrF_{\scriptstyle{4}}\lpar \mathbf{a}\,;\,\mathbf{b}\,;\,\frac{1}{4\,\mrz_1} \rpar \spc 
\eqa
with parameters
\bqa
\mathbf{a} &=& \Bigl(
1 + \mrd_1 - \mrc_1\,,\,
1 + \mrd_1 - \mrc_2\,,\,
1 + \mrd_1 - \mrc_3\,,\,
1 + \mrd_1 - \mra_1 + \mrn\,,\,
1 + \mrd_1 - \mra_2 + \mrn \Bigr) \spc
\nl
\mathbf{b} &=& \Bigl(
1 - \mrd_2 + \mrd_1\,,\,
1 - \mrd_3 + \mrd_1\,,\,
1 + \mrd_1 + \frac{1}{2}\,(\mrn - \mrb_1)\,,\,
\mrd_1 + \frac{1}{2}\,(1 + \mrn - \mrb_1) \Bigr) \spp
\eqa
The general startegy is as follows: we consider Fox functions where all the gamma functions are of the form
$\Gamma(\alpha \pm \mri/\mrk\,\mrs)$ and fractions are elminated by using properties of $\mrH$. Once we obtain $\mrH$ as 
a multiple sum we isolate the quotient of products of Gamma functions depending on index $\mrn$; next we use
the multiplication formula of Gauss and Legendre and
\bq
\Gamma(\alpha - \mrn) = ( - 1)^{\mrn}\,\frac{\eG{\alpha}\,\eG{1 - \alpha}}{\eG{1 - \alpha + \mrn}} \spp
\eq
As a result the sum over $\mrn$ can be written in terms of a generalized hypergeometric function $\pFq$ and we have the
following situation: the series for $\pFq$ converges for all finite $\mrz$ if $\mrp \le \mrq$
(for instance ${}_{\scriptstyle{2}}\,\mrF_{\scriptstyle{2}}$ may have an essential singular point at $\infty$), converges for
$\mid \mrz \mid < 1$ if $\mrp = \mrq + 1$. In the last case, for all parameters not being integer, the
function has a cut in the interval $(1\,,\,\infty)$ where it is continuous from below. 
\paragraph{Evaluation of generalized hypergeometric functions} \hspace{0pt} \\
The evaluation of ${}_{\scriptstyle{\mrq + 1}}\,\mrF_{\scriptstyle{\mrq}}$ can be performed by using the Gauss{-}Jacobi
quadrature; we will consider explicitly the case $\mrp = 3$ where we can write
\bq
\hyptt\lpar \mathbf{a}\,;\,\mathbf{b}\,;\,\mrz \rpar =
\frac{1}{\eB{\mrb_2 - \mra_3}{\mra_3}}\,
\int_0^1 \mrd \mrx\,\mrx^{\mra_3 - 1}\,(1 - \mrx)^{\mrb_2 - \mra_3 - 1}\,
\hyp{\mra_1}{\mra_2}{\mrb_1}{\mrz\,\mrx} \spc
\label{EM32}
\eq
which requires $\Re\,\mra_3 > 0$ and $\Re\,\mrb_2 - \Re\,\mra_3 > 0$. The case $\mrb_3 = \mra_2$ is trivial, so we consider
$\mrb_2 = \mra_3 - \Delta$ with $0 < \Delta < 1 $ and use
\bq
\mrb_2\;\hyptt(\mrb_1\,,\,\mrb_2) =
(\mrb_1 - 1)\;\hyptt(\mrb_1 - 1\,,\,\mrb_2 + 1) -
(\mrb_1 - \mrb_2 - 1)\;\hyptt(\mrb_1\,,\,\mrb_2 + 1) \spp
\eq
When $\mrz > 1$ (here we assume that $\mrz$ is real and positive) the point $\mrx = 1/\mrz \in [0\,,\,1]$ and we have
to examine the parameters of $\shyp{2}{1}$. When $\Re (\mrb_1 - \mra_1 - \mra_2) > 0$ we have that 
$\shyp{2}{1}(1)$ is finite; therefore, we will write 
\bq
\hyptt\lpar \mathbf{a}\,;\,\mathbf{b}\,;\,\mrz \rpar = \mrI_{<} + \mrI_{>} \spc
\eq
where
\bq
\mrI_{<} = \mrz^{ - \mra_3}\,\int_0^1 \mrd \mrx\,\mrx^{\mra_3 - 1}\,
\lpar 1 - \frac{\mrx}{\mrz} \rpar^{\mrb_2 - \mra_3 - 1}\,
\hyp{\mra_1}{\mra_2}{\mrb_1}{\mrx} \spc
\eq
and where (with $\mrz \equiv \mrz - i\,\delta$))
\bqa
\mrI_{>} &=& 
\mrz^{1 - \mrb_2}\,\int_1^{\mrz}\,\mrd \mrx\,\mrx^{\mra_3 - \mra_1 - 1}\,
(\mrz - \mrx)^{\mrb_2 - \mra_3 - 1}\,\Bigl[
\mrB_2\,\mrE( - \mra_2)\,\hyp{\mra_2}{1 - \mrb_1 + \mra_2}{1 + \mra_2 - \mra_1}{\mrx^{-1}} 
\nl
{}&+&
\mrB_1\,\mrE( - \mra_1)\,\hyp{\mra_1}{1 - \mrb_1 + \mra_1}{1 - \mra_2 + \mra_1}{\mrx^{-1}}
\Bigr] \spc
\eqa
where $\mrE(\alpha) = \exp\{i\,\alpha\,\pi\}$ and the coefficients $\mrB_{1,2}$ are defined by~\cite{HTF}
\bq
\mrB_1 = \frac{\eG{\mrc}\,\eG{\mrb - \mra}}
              {\eG{\mrb}\,\eG{\mrc - \mra}} \spc \quad
\mrB_2 = \frac{\eG{\mrc}\,\eG{\mra - \mrb}}
              {\eG{\mra}\,\eG{\mrc - \mrb}} \spp
\eq
When $\Re (\mrb_1 - \mra_1 - \mra_2) < 0$ we now proceed as follows: split the integral
\bq
\hyptt\lpar \mathbf{a}\,;\,\mathbf{b}\,;\,\mrz \rpar = \mrI_{<} + \mrI_{\mrc} +\mrI_{>} \spc
\eq
where the three integrals are in the intervals $[0\,,\mrx_{-}], [\mrx_{-}\,,\,\mrx_{+}]$ and $[\mrx_{+}\,,\,1]$, such
that $\mrx= 1/\mrz \in [\mrx_{-}\,,\,\mrx_{+}]$. It follows that
\bqa
\mrI_{\mrc} &=& \mrI_{\mrc}^{\reg} + \mrI_{\mrc}^{\mrH} \spc
\nl
\mrI_{\mrc}^{\reg} &= & \mrA_1\,\int_{\mrx_{-}}^{\mrx_{+}} \mrd \mrx\,\mrx^{\mra_3 - 1}\,(1 - x)^{\mrb_2 - \mra_3 - 1}\,
\hyp{\mra_1}{\mra_2}{\mra_1 + \mra_2 - \mrb_1 + 1}{1 - \mrz\,\mrx} \spc
\nl
\mrI_{\mrc}^{\mrH} &=& \mrA_2\,\int_{\mrx_{-}}^{\mrx_{+}} \mrd \mrx\,\mrx^{\mra_3 - 1}\,(1 - x)^{\mrb_2 - \mra_3 - 1}\,
 (1 - \mrz\,\mrx)^{\mrb_1 - \mra_1 - \mra_2}\,
\hyp{\mrb_1 - \mra_1}{\mrb_1 - \mra_2}{\mrb_1 - \mra_1 - \mra_2 + 1}{1 - \mrz\,\mrx} \spp
\eqa
\paragraph{CPV vs HFI} \hspace{0pt} \\
We have to consider the following integral:
\bq
\Upphi(\mrz) = \int_{\mra}^{\mrb} \mrd \mrx\,\frac{\mrf(\mrx)}{(\mrx - \mrz)^{\alpha}} \spc
\eq
where $\mrz$ does not lie along the contour of integration which is the straight line from
$\mra$ to $\mrb$. The value of $\Upphi(z)$ as $\mrz$ approaches any value $x_0 \in[a\,,\,b]$
depends on the approach, depending  on  whether the limit is from above or from 
below the real axis. Let $\alpha = \mrn + 1$, the limiting values are known as the boundary values of 
$\Upphi(\mrz)$ and they are given by the limiting values
\bq
\Upphi^{\pm}(\mrx_0) = \lim_{\delta \to 0}\,\int_{\mra}^{\mrb} \mrd \mrx\,
\frac{\mrf(\mrx)}{( \mrx - \mrx_0 \pm i\,\delta )^{\mrn + 1}} \spp
\eq
The Sokhotski-Plemelj-Fox Theorem is a statement on the relationship between
these values and the Cauchy Principal Value or the Hadamard Finite-Part Integrals~\cite{FPI},
in particular
in particular
\bq
\Upphi^{\pm}(\mrx_0) = {\mathcal H}\,\int_{\mra}^{\mrb} \mrd \mrx\,
\frac{\mrf(\mrx)}{( \mrx - \mrx_0 )^{\mrn + 1}} \pm
i\,\pi\,\frac{\mrf^{(\mrn)}(\mrx_0)}{\mrn\,!} \spp
\eq
where the integral takes on either the CPV ($\mrn = 0$) or the Hadamard FPI ($\mrn \ge 1$).
Let us consider the generalization where $\alpha$ is not an integer and define
\bq
\upphi^{\pm}_{\alpha}(\mrx_0) = \lim_{\delta \to 0}\,\int_{\mra}^{\mrb} \mrd \mrx\,
\lpar \mrx - \mrx_0 \pm i\,\delta \rpar^{-\alpha} \spc 
\label{mdiff}
\eq
with $\mrx_0 \in [\mra\,,\,\mrb]$. We obtain
\bq
\upphi^{\pm}_{\alpha}(\mrx_0) = \lim_{\delta \to 0}\,\lpar \pm i\,\delta \rpar^{- \alpha}\,\Bigl[
\mrX_\mrb\;{}_2\mrF_1\lpar \alpha\,,\,1\,;\,2\,;\, \pm i\,\frac{\mrX_\mrb}{\delta}\rpar -
\mrX_\mra\;{}_2\mrF_1\lpar \alpha\,,\,1\,;\,2\,;\, \mp i\,\frac{\mrX_\mra}{\delta}\rpar \Bigr] \spc
\eq
which is valid for $\mid \mathrm{arg} ( 1 \pm i\,\frac{\mrX_\mra}{\delta} ) \mid < \pi$ and
$\mid \mathrm{arg} ( 1 \mp i\,\frac{\mrX_\mrb}{\delta} ) \mid < \pi$, with $\mrX_\mra(\mrb)= \mra(\mrb) - \mrx_0$.
Using
\bq
{}_2\mrF_1\lpar \alpha\,,\,1\,;\,2\,;\,\mrz \rpar = \frac{1}{1 - \alpha}\,\Bigl[
( - \mrz )^{- \alpha}\,{}_2\mrF_1\lpar \alpha\,,\,\alpha - 1\,;\,2 - \alpha\,;\,\mrz^{-1}\rpar
 + \frac{1}{\mrz} \Bigr] \spc
\eq
which is valid for $\mid \mathrm{arg} ( - \mrz ) \mid < \pi$, we derive 
\bq
\upphi^{\pm}_{\alpha}(\mrx_0) = \frac{1}{1 - \alpha}\,\Bigl[ 
\lpar \mrb - \mrx_0 \pm i\,0 \rpar^{1 - \alpha} -
\lpar \mra - \mrx_0 \pm i\,0 \rpar^{1 - \alpha}\Bigr] \spc
\eq
for arbitrary values of $\alpha$. If $\mrf(\mrx)$ is analytic in $\mrx_0$ we obtain
\bqa
\Upphi^{\pm}(\mrx_0) &=&
\sum_{\mrn=0}^{\mrN - 1}\,\frac{\mrf^{(\mrn)}(\mrx_0)}{\mrn\,!}\,\upphi^{\pm}_{\alpha - \mrn}(\mrx_0) +
\int_{\mra}^{\mrb} \mrd \mrx\,{\overline{\mrf}}(\mrx)\,\lpar \mrx - \mrx_0 - i\,0\rpar^{- \alpha} \spc
\nl
{\overline{\mrf}}(\mrx) &=& \mrf(\mrx) - \sum_{\mrn=0}^{\mrN - 1}\,\mrf^{(\mrn)}(\mrx_0)\,(\mrx - \mrx_0)^\mrn \spc
\eqa
where $\mrN - \Re\,\alpha \ge 0$ and $\mrf^{(\mrn)}(\mrx_0)$ not singular for 
$\mrn \le \mrN - 1$.

It is worth noting that in \Bref{doi:10.1137/0519088}
the hypergeometric function ${}_{\mrp + 1}\mrF_{\mrp} (\mrz)$ is expressed in terms 
of power series in the variable $1 / (\mrz - \mrz_0 )$ that converge 
for $\mid (\mrz - \mrz_0 ) \mid > \max (\mid (\mrz_0 ) \mid\,,\,\mid (\mrz_0 - 1) \mid)$ where $\mrz_0 $ is any 
complex number.
\paragraph{Behavior on the unit disk} \hspace{0pt} \\
In the evaluation of Feynman integrals we will need Fox functions at $\mrz = 1$ or $\mrz= - 1 + i\,\delta$, so we need the
behavior on the unit disk. First let us consider the behavior of generalized hypergeometric functions. We start with
\bqa
\hyp{\mra}{\mrb}{\mrc}{\mrz} \sim \mrA_1 \spc &\qquad& \Re(\mrc - \mra - \mrb) > 0 \spc
\nl
\hyp{\mra}{\mrb}{\mrc}{\mrz} \sim \mrA_2\,(1 - \mrz)^{\mrc - \mra - \mrb} \spc &\qquad& \Re(\mrc - \mra - \mrb) < 0 \spc
\eqa
while for $\Re (\mrc - \mra - \mrb) = 0$ we are in the logarithmic case and~\cite{HTF}
\bq
\hyp{\mra}{\mrb}{\mra + \mrb}{\mrz} \sim - \frac{\ln(1 - \mrz)}{\eG{\mra}\,\eG{\mrb}} \spp
\eq
The coefficients are
\bq
\mrA_1 = \frac{\eG{\mrc}\,\eG{\mrc - \mra - \mrb}}
              {\eG{\mrc - \mra}\,\eG{\mrc - \mrb}} \spc
\qquad
\mrA_2= \frac{\eG{\mrc}\,\eG{\mra + \mrb - \mrc}}
             {\eG{\mra}\,\eG{\mrb}} \spp
\eq
Let $\rho = \mrc - \mra - \mrb$; the case $\Re \rho > 0$ does not exclude the presence of a branch point. Indeed, 
in the nonpolynomial cases the function has a branch cut along the interval $(1\,,\,\infty)$; the discontnuity around
the cut can be computed by using Eqs. ($25$ -- $44$) of Sect.~$2.9$ in \Bref{HTF}. 

For $\ppoFp$ we first define~\cite{GHFW}
\bq
\uppsi_{\mrp} = \sum_{\mrj=1}^{\mrp}\,\mrb_{\mrj} - \sum_{\mrj=1}^{\mrp+1}\,\mra_{\mrj} \spc \qquad
\Gamma_{\mrA} = \prod_{\mrj=1}^{\mrp+1}\,\eG{\mra_j} \spc \quad
\Gamma_{\mrB} = \prod_{\mrj=1}^{\mrp}\,\eG{\mrb_j} \spc 
\label{defpsi}
\eq
and derive the following behavior in the divergent case:
\bqa
\Re(\uppsi_{\mrp}) < 0 &\qquad& \ppoFp \sim \frac{\Gamma_{\mrB}}{\Gamma_{\mrA}}\,
\eG{ - \uppsi_{\mrp}}\,(1 - \mrz)^{\uppsi_{\mrp}} \spc
\nl
\Re(\uppsi_{\mrp}) = 0 &\qquad& \ppoFp \sim -\,\frac{\Gamma_{\mrB}}{\Gamma_{\mrA}}\,\ln(1 - \mrz) \spp
\label{GHFato}
\eqa
When $\uppsi_{\mrp} > 0$ we obtain
\bq
\ppoFp\lpar {\mathbf a}\,;\,{\mathbf b}\,;\,1 \rpar = \frac{\Gamma_{\mrB}}{\Gamma_{\mrA}}\,
\mrh_{\mrp}\lpar 0\,;\,{\mathbf a}\,;\,{\mathbf b} \rpar \spc
\eq
where the coefficients $\mrh_{\mrp}$ are given in Eq.~(3.2b) of \Bref{_etinkaya_2021}. 

Next we consider the class of Fox functions that can be transformed into Meijer functions. A Meijer $\mrG$
function has singular points at $\mrz = 0\,\,\infty$ if $\mrp \not= \mrq$ while for $\mrp = \mrq$
the singular points are at $\mrz = 0\,,\,\infty\,,\,(-1)^{\mrm + \mrn - \mrq}$.
When $\mrp = \mrq$ the relevant parameter is $\sigma = \mrm + \mrn - \mrp$. When needed we can always transform
\bq
\mrG^{\mrm\,,\,\mrn}_{\mrp\,,\,\mrp}(\mrz) \quad \to \quad \mrG^{\mrp\,,\,1}_{\mrp\,,\,\mrp}(\zeta) \spc
\qquad \zeta= ( - 1)^{\mrp - \mrm - \mrn + 1} \spc
\eq
which has $\sigma = 1$. The classification~\cite{MGW} depends also on
\bq
\uppsi = \sum_{\mrj=1}^{\mrp} \lpar \mra_{\mrj} - \mrb_{\mrj} \rpar \spc
\eq
and goes as follows:
\bei

\item[$\sigma = 1$] and $\mid \marg(\mrz) \mid < \pi$ we have a singular point at $\mrz = - 1$. $\mrG$ is continuous for
$\Re \uppsi > 0$ and has, in general, a logarithmic singularity for $\uppsi = 0$, while for $\Re \uppsi < 0$ has a power
singularity of order $ - \uppsi$.

\item[$\sigma = 0$] The singular point is at $\mrz = 1$. $\mrG$ is continuous for
$\Re \uppsi > 0$ and has, in general, a logarithmic singularity for $\uppsi = 0$, while for $\Re \uppsi < 0$ has a power
singularity of order $ - \uppsi$.

\item[$\sigma < 1$] but not $\sigma = 0$ the function is a piecewise analytic function of $\mrz$ with a discontinuity
on the circle $\mid \mrz \mid = 1$. If we need the point $\mrz = 1$ we meet a discontinuity, \ie the behavior
depends on the way $\mrz$ is approaching $1$. If we need the point $\mrz= - 1 + i\,\delta$ we are approaching the
unit disk from outside.

\eei 

These general rules can be illustrated with simple examples where we use \eqns{GtoHypo}{GtoHypt}.
\bei

\item[$\mrG^{1\,,\,2}_{2\,,\,1}$] Here $\mrp > \mrq$ and we can use \eqn{GtoHypt} and compute at $\mrz = 1$. Indeed
this function is given in terms of ${}_{\scriptstyle{1}}\,\mrF_{\scriptstyle{1}}(\mrz)$ (Kummer confluent function) which
is an entire function of $\mrz$.

\item[$\mrG^{2\,,\,1}_{2\,,\,2}$] Here $\mrp = \mrq$ and $\sigma = 1$, signalling a singular point at $\mrz = - 1$.
Indeed this function can be expressed in terms of
${}_{\scriptstyle{2}}\,\mrF_{\scriptstyle{1}}( -\,;\, - \mrz)$ or of
${}_{\scriptstyle{2}}\,\mrF_{\scriptstyle{1}}( -\,;\, - \mrz^{-1})$
which are singular at $\mrz = - 1$ when the value of the corresponding $\mrc - \mra - \mrb$ is not positive.

\item[$\mrG^{1\,,\,2}_{3\,,\,3}$] Here $\mrp = \mrq$ and $\sigma = 0$, signalling a singular point at $\mrz = 1$.
Indeed this function can be expressed in terms of
${}_{\scriptstyle{3}}\,\mrF_{\scriptstyle{2}}( -\,;\,\mrz)$ or of
${}_{\scriptstyle{3}}\,\mrF_{\scriptstyle{2}}( -\,;\,\mrz^{-1})$
which are singular at $\mrz = - 1$ when the value of the corresponding $\uppsi$ is not positive.

\eei
We also show an example illustrating the behavior on the unit disk. Given
\bq
\mrH(\mrz) = \int_{\mrL}\frac{\mrd \mrs}{2\,i\,\pi}\,
\frac{\eG{\mrb_1 - \mrs}\,\eG{\mrb_2 - \mrs}\,\eG{1 - \mra_1 + \mrs}\,\eG{1 - \mra_2 + \mrs}}
     {\eG{1 - \mrb_3 + \mrs}\,\eG{\mra_3 - \mrs}}\,\mrz^{\mrs} \spc
\label{udisk}
\eq
We consider the case $\mid \mrz \mid > 1$,
\bqa
\mrH_{>}(z) &=& 
\sum_{\mrn=0}^{\infty}\,\frac{( - 1)^{\mrn}}{\mrn\,!}\,\mrz^{ - \mrn - 1}\,\Bigl[
\mrz^{\mra_1}\,\Gamma_1 + \mrz^{\mra_2}\,\Gamma_2 \Bigr] \spc
\nl
\Gamma_1 &=&
\frac{
\eG{ - \mra_2 + \mra_1 - \mrn}\,
\eG{1 + \mrb_1 - \mra_1 + \mrn}\,
\eG{1 + \mrb_2 - \mra_1 + \mrn}}
{\eG{ - \mrb_3 + \mra_1 - \mrn}\,
\eG{1 + \mra_3 - \mra_1 + \mrn}} \spc
\nl 
\Gamma_2 &=&
\frac{        
\eG{\mra_2 - \mra_1 - \mrn}\,
\eG{1 + \mrb_1 - \mra_2 + \mrn}\,
\eG{1 + \mrb_2 - \mra_2 + \mrn}}
{\eG{ - \mrb_3 + \mra_2 - \mrn}\,
\eG{1 + \mra_3 - \mra_2 + \mrn}} \spc
\eqa
and the case $\mid \mrz \mid < 1$,
\bqa
\mrH_{<}(z) &=& -
\sum_{\mrn=0}^{\infty}\,\frac{( - 1)^{\mrn}}{\mrn\,!}\,\mrz^{\mrn}\,\Bigl[
\mrz^{\mrb_1}\,\Gamma_3 + \mrz^{\mrb_2}\,\Gamma_4 \Bigr] \spc
\nl
\Gamma_3 &=&
\frac{
\eG{\mrb_2 - \mrb_1 - \mrn}\,
\eG{1 + \mrb_1 - \mra_1 + \mrn}\,
\eG{1 + \mrb_1 - \mra_2 + \mrn}}
{\eG{ - \mrb_1 + \mra_3 - \mrn}\,
\eG{1 - \mrb_3 + \mrb_1 + \mrn}} \spc
\nl
\Gamma_4 &=&
\frac{
\eG{ - \mrb_2 + \mrb_1 - \mrn}\,
\eG{1 + \mrb_2 - \mra_1 + \mrn}\,
\eG{1 + \mrb_2 - \mra_2 + \mrn}}
{\eG{ - \mrb_2 + \mra_3 - \mrn}\,
\eG{1 - \mrb_3 + \mrb_2 + \mrn}} \spp
\eqa
The function in \eqn{udisk} is a $\mrG^{2\,,\,2}_{3\,,\,3}$ and corresponds to $\sigma = 1$; using \eqn{EMG} we 
transform it into the Mellin transform of a $\mrG^{2\,,\,1}_{2\,,\,2}$ and evaluate it at $\mrz = 1$. The procedure requires
\bq
\mra_{1,2} < 1 \spc \quad
\mrb_{1,2} > 0 \spc \quad
\mra_1 > \mrb_3 \spc \quad
\mrb_{1,2} > \mra_1 - 1 \spp
\eq
Under these conditions we obtain
\bqa
\mrH_{1} &=&
- \sum_{\mrn=0}^{\infty}\,\frac{( - 1)^{\mrn}}{\mrn\,!}\,\Bigl[
\frac{
\eG{ - \mrb_2 + \mrb_1 - \mrn}\,
\eG{1 + \mrb_2 - \mra_1 + \mrn}\,
\eG{1 + \mrb_2 - \mra_2 + \mrn}}
{\eG{ - \mrb_2 + \mra_3 - \mrn}\,
\eG{1 - \mrb_3 + \mrb_2 + \mrn}}
\nl
{}&+&
\frac{       
\eG{\mrb_2 - \mrb_1 - \mrn}\,
\eG{1 + \mrb_1 - \mra_1 + \mrn}\,
\eG{1 + \mrb_1 - \mra_2 + \mrn}}
{\eG{ - \mrb_1 + \mra_3 - \mrn}\,
\eG{1 - \mrb_3 + \mrb_1 + \mrn}}
\Bigr]
\eqa
To give an example we obtain excellent numerical agreement for
$\mrH_{>}(1.01)$,
$\mrH_{<}(0.99)$ and $\mrH_1$.
For more general arguments on hypergeometric functions at unit argument see \Bref{_etinkaya_2021}. 
\paragraph{More on the logarithmic case} \hspace{0pt} \\
The logarithmic (resonant) case can be more complicated when we start considering Feynman integrals. In evaluating
the corresponding Fox functions a few properties of the Euler Gamma function are useful (see \Bref{HTF}):
\bqa
\eG{\mrz - \mrn} &=& ( - 1)^{\mrn}\,
\frac{\eG{\mrz}\,\eG{1 - \mrz}}
     {\eG{\mrn + 1 - \mrz}} \spc
\nl\eG{\mrz + \mrn} &=& ( - 1)^{\mrn}\,
\frac{\eG{\mrz}\,\eG{1 - \mrz}}
     {\eG{1 - \mrn - \mrz}} =
\prod_{\mrj=0}^{\mrn-1}\,(\mrz + \mrj)\,\eG{\mrz} \spp
\eqa
The behavior of $\eG{\mrz}$ near $\mrz = - \mrk$ can be obtained by using
\bqa
\eG{\mrz} &=& \frac{\eG{1 - \mrk - \mrz}}{\eG{1 - \mrz}}\,\eG{\mrz + \mrk} \spc
\nl
\eG{\mrz} &=& \mrz^{-1}\,\exp\{ - \gamma\,\mrz + \sum_{\mrn=2}^{\infty}\,\frac{( - 1)^{\mrn}}{\mrn}\,
\zeta(\mrn)\,\mrz^{\mrn}\} \spc
\eqa
\bq
\eG{\mrz} = \frac{( - 1)^{\mrk}}{\mrk\,!}\,\Bigl[
\frac{1}{\mrz + \mrk} + \uppsi(\mrk + 1) + \sum_{\mrn=1}^{\infty}\,\mrc_{\mrn}\,(\mrz + \mrk)^{\mrn} \Bigr] \spp
\eq
The first $2$ coefficients are
\bqa
\mrc_1 &=& \zeta(2) + \frac{1}{2}\,\Bigl[ \uppsi^2(\mrk + 1 - \uppsi^{(1)}(\mrk + 1) \Bigr] \spc
\nl
\mrc_2 &=& \uppsi(\mrk + 1)\,\Bigl[
\zeta(2) + \frac{1}{6}\,\uppsi^2(\mrk + 1) - \frac{1}{2}\,\uppsi^{(1)}(\mrk + 1) \Bigr] +
\frac{1}{6}\,\uppsi^{(2)}(\mrk + 1) \spc
\eqa
where $\zeta(\mrz)$ is the Riemann zeta function and $\uppsi^{(n)}(\mrz)$ is the polygamma 
function~\cite{HTF,abramowitz+stegun}, with
$\uppsi(1) = - \gamma$, the Euler{-}Mascheroni constant.

The Taylor expansion of $\Gamma(\mrz)$ around $\mrz_0$ is given by
\bq
\eG{\mrz} = \eG{\mrz_0}\,\Bigl\{ 1 + (\mrz - \mrz_0)\,\uppsi(\mrz_0) +
\frac{1}{2}\,(\mrz - \mrz_0)^2\,\Bigl[ \uppsi^2(\mrz_0) + \uppsi^{(1)}(\mrz_0) +
\ord{(\mrz - \mrz_0)^3} \Bigr\} \spp
\eq
For instance $\eG{\mrn - \mrz}$ ($\mrn \in \Zf^{+}$) around $\mrz = - \mrk$ is given by
\bqa
\eG{\mrn - \mrz} &=& \eG{\mrn + \mrk}\,\Bigl[
1 + \sum_{\mrj=1}^{\infty}\,\mrd_{\mrj}\,(\mrn + \mrk)^{\mrj} \Bigr] \spc
\nl
\mrd_1 &=& - \uppsi(\mrn + \mrk) \spc \quad
\mrd_2 = \frac{1}{2}\,\Bigl[ \uppsi^2(\mrn + \mrk) + \uppsi^{(1)}(\mrn + \mrk) \Bigr] \spp
\eqa
The polygamma functions can be computed by using the following results~\cite{abramowitz+stegun}:
\bqa
\uppsi(\mrz + 1) &=& \uppsi(\mrz) + 1, \spc \qquad \uppsi(1)= - \gamma \spc
\nl
\uppsi^{(\mrn + 1)}(\mrz + 1) &=& \uppsi^{(\mrn)}(\mrz) + ( - 1)^{\mrn}\,\frac{\mrn\,!}{\mrz^{\mrn + 1}} \spc \qquad
\uppsi^{(\mrn)}(1) = ( - 1)^{\mrn + 1}\,\mrn\,!\;\zeta(\mrn + 1) \spp
\eqa
Consider now the following integral:
\bqa
\mrI &=& \int_{\mrL}\,\frac{\mrd \mrs}{2\,i\,\pi}\,
(2 + \mrn + \mrs)\,(\mrs + 1)\,(\mrs + 2)\,\frac{\mrN}{\mrD}\,\mrz^{\mrs} \spc
\nl
\mrN &=&
\eG{\mrs}\,
\eG{ - \mrs}\,
\eG{1 - \mrs}\,
\eG{1 + \mrs}^4 \,
\eG{\mrn - \mrs} \spc
\nl
\mrD &=&
\eG{2 + \mrs}\,
\eG{3 + \mrs}\,
\eG{1 + \mrn - \mrs} \spc
\eqa
where $\mrn \ge 0$ and $\mid \mrz \mid > 1$. We compute the following residues:
\bei

\item[\ovalbox{$\mrs = - 1$}]  First we use $\eG{\mrs} = \mrs^{-1}\,\eG{\mrs + 1}$ and use the Laurant expansion
of $\eG{\mrs + 1}$. Next we use the Taylor expansion of $\eG{ - \mrs}$, $\eG{1 - \mrs}$, of
$\eG{2 + \mrs}, \eG{3 + \mrs}$ and of $\eG{\mrn - \mrs}$. The result is
\bqa
\mathrm{Res}_{\mrs = - 1} &=& - \mrn_1^{-1}\,\mrz^{-1} \, \Bigl\{
\Bigl[
            \frac{1}{2}\,((5\,\zeta(2) + \gamma^2)\,\mrn_2 - 2\,\gamma)
          + (1 - \gamma\,\mrn_2)\,\mrn_1^{-1}
          + \mrn_2\,\mrn_1^{-2}
            \Bigr]\,\ln(\mrz)
\nl
{}&+&
            \frac{1}{2}\,\Bigl[
            1 - \gamma\,\mrn_2
          + \mrn_2\,\mrn_1^{-1}
            \Bigr]\,\ln^2(\mrz)
          + \frac{1}{6}\,\mrn_2\,\ln^3(\mrz)
\nl
{}&-&    
          \frac{1}{6}\,\Bigl[ (2\,\zeta(3) + 15\,\gamma\,\zeta(2) + \gamma^3)\,\mrn_2 - 3\,(5\,\zeta(2) + \gamma^2) \Bigr]
\nl
{}&+&     \frac{1}{2}\,\Bigl[ (5\,\zeta(2) + \gamma^2)\,\mrn_2 - 2\,\gamma \Bigr]\,\mrn_1^{-1}
          + (1 - \gamma\,\mrn_2)\,\mrn_1^{-2}
          + \mrn_2\,\mrn_1^{-3}
\Bigr\} \spc
\eqa
where $\mrn_{\mrj} = \mrn + \mrj$.
\item[\ovalbox{$\mrs = - 2$}] In this case we simplify again all the Gamma functions and use the Laurent 
expansion of $\eG{\mrs}$ while Taylor expanding the remaing ones. The result is
\bqa
\mathrm{Res}_{\mrs = - 2} &=& - \mrz^{-2} \, \Bigl\{
\Bigl[
            2 - \gamma
          + \gamma\,\mrn_2^{-1}
          - \mrn_2^{-2}
          \Bigr]\,\ln(\mrz)
\nl
{}&+&     \frac{1}{2}\,\Bigl(
          - \mrn_2^{-1}
            \Bigr)\,\ln^2(\mrz)
          + \frac{1}{2}\,(6 + 5\,\zeta(2) - 4\,\gamma + \gamma^2)
\nl
{}&+& \frac{1}{2}\,(2 - 5\,\zeta(2) - \gamma^2)\,\mrn_2^{-1}
          + \gamma\,\mrn_2^{-2}
          - \mrn_2^{-3} 
\Bigr\} \spp
\eqa
\item[\ovalbox{$\mrs = - \mrk$}] With $\mrk > 2$ we obtain
\bq
\mathrm{Res}_{\mrs = - \mrk}= \frac{\mrz^{ - \mrk}}{\eG{\mrk}\,\mrk_{\mrn}}\,\Bigl[
\mrL_0 + \mrL_1\,\ln(\mrz) + \mrL_2\,\ln^2(\mrz) \Bigr] \spc
\eq
\bqa
\mrL_1 &=&
(1 + \mrk_\mrn)\,\Bigl[ 3 + 2\,\uppsi(\mrk) - 3\,\uppsi(\mrk_+) \Bigr]\,\mrk_-^{-1}
- 2\,\Bigl[ 2\,\uppsi(\mrk) - 3\,\uppsi(\mrk_+) \Bigr]
\nl
{}&-& \Bigl[ (\mrk_- + \mrk_\mrn + \mrk_\mrn^2)\,\mrk + 
          (9\,\mrk_\mrn + 3\,\mrk_\mrn^2 - 2\,\mrk)\,\mrk_-^2 \Bigr]\,\mrk^{-1}\,\mrk_\mrn^{-1}\,\mrk_-^{-2} \spc
\nl\nl          
\mrL_2 &=& - \frac{1}{2}\,(1 - 2\,\mrk_- + \mrk_\mrn)\,\mrk_-^{-1} \spc
\nl\nl          
\mrL_0 &=&
3\,\Bigl[ 2 + 2\,\uppsi(\mrk) - 3\,\uppsi(\mrk_+) \Bigr]\,\mrk_\mrn\,\mrk^{-1}
+ 3\,\Bigl[ 2 + 6\,\uppsi(\mrk) - 9\,\uppsi(\mrk_+) \Bigr]\,\mrk^{-1}
\nl
{}&+& \Bigl[ 3 + 2\,\uppsi(\mrk) - 3\,\uppsi(\mrk_+))\,
       (\mrk_- + \mrk_\mrn + \mrk_\mrn^2 \Bigr]\,\mrk_\mrn^{-1}\,\mrk_-^{-2}
\nl
{}&+& \Bigl[ 6\,\zeta(2) + 4\,\uppsi(\mrk)\,
            \Bigl(\uppsi(\mrk) - 3\,\uppsi(\mrk_+)\Bigr) + 9\,\uppsi(\mrk_+)^2 + 
            2\,\uppsi^{(1)}(\mrk) - 3\,\uppsi^{(1)}(\mrk_+) \Bigr]
\nl
{}&-& 2\,\Bigl[ 2\,\uppsi(\mrk) - 3\,\uppsi(\mrk_+) \Bigr]\,\mrk_\mrn^{-1}
- \Bigl[ (\mrk_-^2 - 2\,\mrk_-^3 + \mrk_\mrn\,\mrk_- + \mrk_\mrn^2 + \mrk_\mrn^3)\,\mrk^2 - 
            3\,(3\,\mrk_\mrn^2 + \mrk_\mrn^3 - 3\,\mrk\,\mrk_\mrn)\,\mrk_-^3 \Bigr]\,
            \mrk^{-2}\,\mrk_\mrn^{-2}\,\mrk_-^{-3}
\nl
{}&-& \frac{1}{2}\,(1 + \mrk_\mrn)\,
            \Bigl[ 12 + 6\,\zeta(2) + 3\,\uppsi(\mrk)\,\Bigl(3 + \uppsi(\mrk) - 3\,\uppsi(\mrk_+)\Bigr) 
\nl
   {}&-& 9\,\uppsi(\mrk_+)\,\Bigl(2 - \uppsi(\mrk_+)\Bigr) + 2\,\uppsi^{(1)}(\mrk) 
            - 3\,\uppsi^{(1)}(\mrk_+) \Bigr]\,\mrk_-^{-1} \spc
\eqa
where $\mrk_{\pm} = \mrk \pm 1$ and $\mrk_{\mrn} = \mrk + \mrn$.           

\eei
One{-}fold MB integrals are simple, techniques for two{-}fold MB integrals have been developed
in \Bref{Friot:2011ic}.
It is worth noting that a novel technique for the analytic evaluation of multifold MB
integrals has been introduced in \Brefs{Banik:2023rrz,banik2024geometrical} which is based on triangulating a set of
points which can be assigned to a given MB integral. Computation of MB integrals
using the method of brackets is described in \Bref{Gonzalez:2021vqh}.
Analytic continuation of Mellin-Barnes integrals is described in \Bref{Czakon:2005rk},
multiple series representations in \Bref{Ananthanarayan:2020fhl}. Expansions of the $\mrG\,${-}function in
\Bref{MRG}. Numerical evaluation of MB integrals and related problems are described in \Bref{Freitas:2010nx}.
\subsection{Generalized, multivariate, Fox function \label{GMFF}}
The most general, multivariate Fox function has been introduced in \Bref{HS} (see also \Brefs{HY,BS}):
here ${\mathbf s} = [\mrs_1\,\dots\,\mrs_{\mrr} ]$,
${\mathbf \alpha} = [\alpha_1\,\dots\,\alpha_{\mrr} ]$,
${\mathbf \beta} = [\beta_1\,\dots\,\beta_{\mrr} ]$,
$\marg({\mathbf z}) = [\marg(\mrz_1)\,\dots\,\mrg(\mrz_{\mrr} ]$,
\bq
{\mathbf A} = \lpar \mra_{\mrj , \mrk} \rpar_{\mrm \times \mrr} \spc \qquad
{\mathbf B} = \lpar \mrb_{\mrj , \mrk} \rpar_{\mrn \times \mrr} \spc
\eq
\bq
\mrH\Bigl[ {\mathbf z}\,;\,({\bf \alpha}\,,\,{\mathbf A} )\,;\,({\bf \beta}\,,\,{\mathbf B}) \Bigr] =
\Bigl[ \prod_{i=1}^{\mrr}\,\int_{\mrL_i}\,\frac{\mrd \mrs_i}{2\,i\,\pi} \Bigr]\,
\Uppsi\,\,\prod_{i=1}^{\mrr}\,( \mrz_i)^{- \mrs_i} \spc \quad
\Uppsi = \frac{ \prod_{\mrj=1}^{\mrm}\,\eG{\alpha_{\mrj} + \sum_{\mrk}\,\mra_{\mrj , \mrk}\,\mrs_{\mrk}}}
            { \prod_{\mrj=1}^{\mrn}\,\eG{\beta_{\mrj} + \sum_{\mrk}\,\mrb_{\mrj , \mrk}\,\mrs_{\mrk}}} \spc
\label{gmff}
\eq
where $\mra$ and $\mrb$ are arbitrary real numbers. It should be clear that \Bref{HS} gives the most general 
criterion to determine the regions of convergence of a multivariate Fox function, adding more
to the $\Lambda_i, \Omega_i$ criterion of \eqns{Ldef}{Hconv}. From now on we will always use the definition of \Bref{HS}.  
The function in \eqn{gmff} is defined by the following matrix:
\[
{\mathcal A} =
\left(
\begin{array}{c}
{\mathbf A}_1 \\
\dots        \\
{\mathbf A}_{\mrm + \mrn} \\
\end{array}
\right)
\]
\bqa
{\mathbf A}_{\mrj} &=& \lpar \mra_{\mrj , 1}\,\,\dots\,,\mra_{\mrj , \mrr} \rpar \quad \hbox{for} \quad
\mrj = 1\,,\,\dots\,,\mrm \spc
\nl
{\mathbf A}_{\mrm + \mrj} &=& \lpar \mrb_{\mrj , 1}\,\,\dots\,,\mrb_{\mrj , \mrr} \rpar \quad \hbox{for} \quad
\mrj = 1\,,\,\dots\,,\mrn \spp
\eqa
We also define the following matrices
\[
{\mathbf R}_{\mrj , \mrj_1 , \dots , \mrj_{\mrr - 1}} =
\left(
\begin{array}{c}
{\mathbf A}_{\mrj} \\
{\mathbf A}_{\mrj_1} \\
\dots \\
{\mathbf A}_{\mrj_{\mrr - 1}} \\
\end{array}
\right)
\qquad\qquad
{\mathbf P}_{\mrj_1 , \dots , \mrj_{\mrr - 1}} =
\left(
\begin{array}{c}
\marg({\mathbf z}) \\
{\mathbf A}_{\mrj_1} \\
\dots \\
{\mathbf A}_{\mrj_{\mrr - 1}} \\
\end{array}
\right)
\]
It is important to realize that the multiple integral in \eqn{gmff} may be overall divergent although the iterate 
integrals converge~\cite{HS}. The convergenge problem has been discussed in \Bref{HS} whith the following theorem:
\begin{itemize}

\item[$\bullet$]
Define a sequence
\bq
1 \le \mrj_1 < \,\dots\, < \mrj_{\mrr-1} \le \mrm + \mrn \spc \quad
\mathrm{rank}\,{\mathbf R}_{\mrj_1 , \dots , \mrj_{\mrr-1}} = \mrr - 1
\label{Hseq}
\eq
then the generalized, multivariate, Fox function is given by a convergent integral if for any sequence
given in \eqn{Hseq} we have
\bqa
{}&1)& \quad  \rho\lpar \mrj_1\,,\,\dots\,,\,\mrj_{\mrr-1} \rpar =
\sum_{\mrj=1}^{\mrm + \mrn}\,\mathrm{sgn}\lpar \mrm + \frac{1}{2} - \mrj \rpar\,
\bmid\, \mathrm{det}\,{\mathbf R}_{\mrj , \mrj_1 , \dots , \mrj_{\mrr-1}} \bmid > 0 \spc
\nl
{}&2)& \quad
\frac{\pi}{2}\,\rho\lpar \mrj_1\,,\,\dots\,,\,\mrj_{\mrr-1} \rpar >
\bmid\,\mathrm{det}\,{\mathbf P}_{\mrj_1 , \dots , \mrj_{\mrr-1}} \bmid \spp
\label{ccond}            
\eqa

\end{itemize}
From \Bref{HS} we also obtain the following result: if there exists at least one sequence such that
the corrseponding $\rho$ is zero then the integral in \eqn{gmff} diverges for all 
${\mathbf z} \in \Cf^{\mrr}\,/\,\Rf^{\mrr}_{+}$.

Clearly,the theorem of \Bref{HS} is not providing any information about convergene{/}divergence
of the integral for ${\mathbf z} \in \Rf^{\mrr}_{+}$ for arbitrary values of $\mrr$. This follows from
the fact that when all variables $\mrz_{\mrj}$ are real and positive the determinant of ${\mathbf P}$ is trivially
zero. Actually the integral will converge with some additional conditions.
For $\mrr = 1$ a detailed description of these conditions can be found in Section $(1.19)$ of \Bref{HTF}.
For $\mrr = 2$ we refer to the work of \Bref{HY} where the Fox function of two real positive variables is
discussed in \S 4. Little is known for $\mrr > 2$, as stated in \Bref{HS}.

When $\mrr, \mrm$ and $\mrn$ are large the number of sequences in \eqn{Hseq} is huge. Our procedure is as follows:
\begin{enumerate}

\item for each sequence use a SVD for the matrix $\mathbf{R}^{\mrt}$, \ie
\bq
\mathbf{R}^{\mrt} = \mathbf{U}\,\mathbf{S}\,\mathbf{V}^{\mrt} \spc
\eq
where the diagonal entries of $\mathbf{S}$ are uniquely determined and are known as the singular values. 
The number of non-zero singular values is equal to the rank of the matrix.

\item Keep only the sequences satisfying \eqn{Hseq} and compute $\mathrm{det}\,\mathbf{R}$ and
the cofactors of $\mathbf{P}$. Assuming that $\rho > 0$ for all sequences compare
$\bmid\,\mathrm{det}\,{\mathbf{P}}_{\mrj_1 , \dots , \mrj_{\mrr-1}} \bmid$, sequence by sequence, and keep those that are different.
 
\end{enumerate}
We will give a few examples. Consider the following integral:
\bqa
\mrI_0\lpar \mra\,;\,\mrb_1\,,\,\mrb_2\,;\,\mrc_1\,,\,\mrc_2\,;\,\mrz_1\,,\,\mrz_2 \rpar &=&
\Bigl[ \prod_{i=1}^{2}\,\int_{\mrL_i}\,\frac{\mrd \mrs_i}{2\,\pi\,i} \Bigr]\,
 \frac{\eG{\mra + \mrs_1 + \mrs_2}}
      {\eG{\mrc_1 + \mrs_1 + \mrs_2}\,\eG{\mrc_2 + \mrs_1 + \mrs_2}}
\nl
{}&\times&
 \eG{ - \mrs_1}\,\eG{\mrb_1 + \mrs_1}\,
 \eG{ - \mrs_2}\,\eG{\mrb_2 + \mrs_2}\,\mrz_1^{ - \mrs_1}\,\mrz_2^{- \mrs_2} \spc
\label{F1MBp}
\eqa
where we assume that the contours $\mrL_i$ run from $\sigma_i - \infty$ to $\sigma_i + \infty$.
The integral in \eqn{F1MBp} is represented by the matrix
\[
{\mathcal{A}}^{\mrt} =
\left(
\begin{array}{ccccc}
1   \quad &\quad - 1 \quad &\quad 0 \quad &\quad 1 \quad &\quad 1 \\
&&&& \\
- 1 \quad &\quad 0   \quad &\quad 1 \quad &\quad 0 \quad & \quad1 \\
\end{array}
\right)
\]
Following the analysis of \Bref{compH}, see \eqns{Ldef}{Hconv}, we derive 
$\Lambda_1 = \Lambda_2 = - 1$ and $\Omega_1 = \Omega_2 = 1$ with the conclusion that the integral is convergent for 
$\mid \marg(\mrz_i) \mid < \pi/2$, see \eqns{Ldef}{Hconv}. 
According to the analysis of \Bref{HS} the integral $\mrI_0$ corresponds to $ \mrr = 2$ and $\mrm = 5\,,\,
\mrn = 2$. 
If $\phi_{\mrj}$ denotes the argument of $\mrz_{\mrj}$ we obtain the conditions
\bq
\mid \phi_{1,2} \mid < \frac{\pi}{2} \spc \qquad
\mid \phi_1 - \phi_2 \mid < 2\,\pi \spp
\eq
Consider now a second integral
\bqa
\mrI_1\lpar \mra\,;\,\mrb_1\,,\,\mrb_2\,;\,\mrc_1\,,\,\mrc_2\,;\,\mrz_1\,,\,\mrz_2 \rpar &=&
\Bigl[ \prod_{i=1}^{2}\,\int_{\mrL_i}\,\frac{\mrd \mrs_i}{2\,\pi\,i} \Bigr]\,
 \frac{\eG{\mra - \mrs_1 + \mrs_2}}
      {\eG{\mrc_1 + \mrs_1 + \mrs_2}\,\eG{\mrc_2 + \mrs_1 + \mrs_2}}
\nl
{}&\times&
 \eG{ - \mrs_1}\,\eG{\mrb_1 + \mrs_1}\,
 \eG{ - \mrs_2}\,\eG{\mrb_2 + \mrs_2}\,\mrz_1^{ - \mrs_1}\,\mrz_2^{ - \mrs_2} \spc
\eqa
where the parameters do not satisfy the definition to convert the integral into a standard Fox function (according
to the definition of \Bref{compH}). For
each integral we follow the prescription of Sect. $(5.3)$ of \Bref{HTF}. compute the parameter
$\alpha = 2\,( \mrm + \mrn) - \mrp - \mrq$
and set $\Omega_i = \alpha$. Consider the integral along $\mrL_1$ where we obtain
$\Omega_1 = \alpha = 1$. However, in this case we have $\rho(5) = 0$.
Another illustrative example is the following one:
\bq
\mrJ =
\Bigl[ \prod_{i=1}^{2}\,\int_{\mrL_i}\,\frac{\mrd \mrs_i}{2\,\pi\,i} \Bigr]\,
\frac{\eG{\mra_1 + 2\,\mrs_1 + 5\,\mrs_2}\,\eG{\mra_2 + \mrs_1 + 3\,\mrs_2}\,
      \eG{\mra_3 - 0.1\,\mrs_1}\,\eG{\mra_4 - 0.1\,\mrs_2}}
     {\eG{\mrb + 0.5\,\mrs_1 + 0.4\,\mrs_2}}\,\mrz_1^{ - \mrs_1}\,\mrz_2^{ - \mrs_2} \spp
\eq
If we consider the single integral over $\mrs_1$ or over $\mrs_2$ and follow \Bref{HTF} we find
$\alpha > 0$ in both cases. However it is easy to derive $\rho(1) = 0$

Consider another integral:
\bqa
\mrI_2\lpar \mra\,;\,\mrb_1\,,\,\mrb_2\,;\,\mrc_1\,,\,\mrc_2\,;\,\mrz_1\,,\,\mrz_2 \rpar &=&
\Bigl[ \prod_{i=1}^{2}\,\int_{\mrL_i}\,\frac{\mrd \mrs_i}{2\,\pi\,i} \Bigr]\,
 \frac{\eG{\mra + \mrs_1 + \mrs_2}}
      {\eG{\mrc_1 - \mrs_1 + \mrs_2}\,\eG{\mrc_2 + \mrs_1 + \mrs_2}}
\nl
{}&\times&
 \eG{ - \mrs_1}\,\eG{\mrb_1 + \mrs_1}\,
 \eG{ - \mrs_2}\,\eG{\mrb_2 + \mrs_2}\,\mrz_1^{ - \mrs_1}\,\mrz_2^{ - \mrs_2} \spc
\eqa
we obtain $\Omega_1 = 1$. In this case all values of $\rho$ are positive and
\bq
\mid \phi_{1,2} \mid < \frac{\pi}{2} \spc \quad
\mid \phi_1 - \phi_2 \mid < \pi \spc \quad
\mid \phi_1 + \phi_2 \mid < 2\,\pi \spp
\eq
A third example is as follows:
\bqa
\mrI_3\lpar \mra\,;\,\mrb_1\,,\,\mrb_2\,;\,\mrc_1\,,\,\mrc_2\,;\,\mrz_1\,,\,\mrz_2 \rpar &=&
\Bigl[ \prod_{i=1}^{2}\,\int_{\mrL_i}\,\frac{\mrd \mrs_i}{2\,\pi\,i} \Bigr]\,
 \frac{\eG{\mra - \mrs_1 + \mrs_2}}
      {\eG{\mrc_1 - \mrs_1 + \mrs_2}\,\eG{\mrc_2 + \mrs_1 + \mrs_2}}
\nl
{}&\times&
 \eG{ - \mrs_1}\,\eG{\mrb_1 + \mrs_1}\,
 \eG{ - \mrs_2}\,\eG{\mrb_2 + \mrs_2}\,\mrz_1^{ - \mrs_1}\,\mrz_2^{ - \mrs_2} \spc
\eqa
where we obtain $\Omega_1 = 1$. All values of $\rho$ are positive and
\bq
\mid \phi_{1,2} \mid < \frac{\pi}{2} \spc \quad
\mid \phi_1 - \phi_2 \mid < 2\,\pi \spc \quad
\mid \phi_1 + \phi_2 \mid < \pi \spp
\eq
It is worth noting that many times the additional conditions on the $\phi$s, \ie those on linear
combinations of the $\phi_i$, are less restrictive than the ones on single arguments. In these cases
the convergence of single, iterated integrals is enough to guarantee overall convergence.

Finally, we provide an example with $\mrr = 3$;
\bqa
\mrJ_3 &=&
\Bigl[ \prod_{i=1}^{3}\,\int_{\mrL_i}\,\frac{\mrd \mrs_i}{2\,\pi\,i} \Bigr]\,\frac{\mrN}{\mrD}\,
\mrz_1^{ - \mrs_1}\,\mrz_2^{ - \mrs_2}\,\mrz_3^{ - \mrs_3} \spc
\nl
\mrN &=& \eG{\mra_1 + \mrs_1 + \mrs_2}\,
         \eG{\mra_2 + \mrs_2 + \mrs_3}\,
         \eG{\mra_3 + \mrs_1 + \mrs_3}\,
         \prod_{i=1}^{3}\,\eG{\mra_{3+i} - \mrs_i} \spc \qquad
\mrD = \prod_{i=1}^{3}\,\eG{\mrb_i + \mrs_i} \spp
\eqa
The convergence region for $\mrJ_3$ is given in Example~$3$ of \Bref{HS}. Here we make the following observation:
there are $3$ sequences that should not be included in determining the region of convergence. Indeed,
from \eqn{Hseq} the sequences $(4\,,\,7)$, $(5\,,\,8)$ and $(6\,,\,9)$ correspond to matrices of rank one.

To summarize: the condition given in \eqn{ccond} can be rewritten as
\bq
\bmid \sum_{\mrj=1}^{\mrr}\,\mru_{\mrj}\,\marg( \mrz_{\mrj} ) \bmid \,<\, \frac{\pi}{2}\,\rho(j_1,\,\dots\,,i_{\mrr-1}) \spc
\quad \mru_{\mrj} = ( - 1)^{j - 1}\,\mathrm{det}\,\mrU_{\mrj} \spc
\label{nccond}
\eq
where $\mrU_{\mrj}$ is the matrix $\lpar \mathbf{A}_{\mrj_1},\,\dots\,,\mathbf{A}_{\mrj_{\mrr-1}}\rpar^{\mrt}$ where 
column $j$ has been deleted.

In the $\mrr\,${-}dimensional space $[ \marg(\mrz_1),\,\dots\,,\marg(\mrz_{\mrr}) ]$ the convergence region is the 
intersection of all the strips in \eqn{nccond}. From \Bref{HS} we have that there exists a sufficiently small
$\xi > 0$ such that convergence is given by $\mid \marg(\mrz_{\mrj}) \mid < \xi \pi/2$.
Therefore, after generating all possible sequences, we have to select those corresponding to matrices with rank 
equal to $\mrr - 1$, 
find the set of relevant strips and identify the convex polygon (polytope) intersection of the strips.
Many strips are irrelevant: for instance, if we have $\mid \phi_i \mid < \pi$ for $i=1,2,3$ then an example of
irrelevant strips is given by
\bq
\mid \phi_1 + 2\,\phi_2 - \phi_3 \mid < 4\,\pi \spc \qquad
\mid \phi_1 + 3\,\phi_2 - \phi_3 \mid < 5\,\pi \spp
\eq 
Finally there is another generalized function, defined by
\bq
\mrI\Bigl[ {\mathbf z}\,;\,({\bf \alpha}\,,\,{\mathbf A}\,,\,{\mathbf c} )\,;\,
                           ({\bf \beta}\,,\,{\mathbf B}\,,\,{\mathbf d}) \Bigr] =
\Bigl[ \prod_{i=1}^{\mrr}\,\int_{\mrL_i}\,\frac{\mrd \mrs_i}{2\,i\,\pi} \Bigr]\,
\Uppsi\,\,\prod_{i=1}^{\mrr}\,( \mrz_i)^{- \mrs_i} \spc \qquad
\Uppsi = \frac{ \prod_{\mrj=1}^{\mrm}\,\Gamma^{\mrc_{\mrj}}\lpar \alpha_{\mrj} + \sum_{\mrk}\,\mra_{\mrj , \mrk}\,\mrs_{\mrk}
\rpar}
            { \prod_{\mrj=1}^{\mrn}\,\Gamma^{\mrd_{\mrj}}\lpar \beta_{\mrj} + \sum_{\mrk}\,\mrb_{\mrj , \mrk}\,\mrs_{\mrk}
\rpar}
 \spp
\label{gmIff}
\eq
Most of the Fox functions arising in the calculation of Feynman integrals can be written as
MB integrals of a Lauricella $\mrF^{(\mrN)}_{\sPD}$ function, \eqn{FDEM}, an explicit example will be given 
in \eqn{FDexa}. Here we consider the simple example given by a Fox function corresponding to parameters 
$\mrr = 4, \mrm = 10$ and $\mrn = 5$ and given by
\bqa
\mrH &=& \Bigl[ \prod_{i=1}^{4}\,\int_{\mrL_i}\,\frac{\mrd \mrs_i}{2\,i\,\pi} 
\lpar - \mrz_i \rpar^{\mrs_i} \Bigr]\,\frac{\mrN}{\mrD} \spc
\nl\nl
\mrN &=&
\eG{ - \mrs_1}\,
\eG{\mrd + \mrs_1}\,
\eG{\mre + \mrs_1}\,
\eG{ - \mrs_2}\,
\eG{ - \mrs_3}\,
\eG{ - \mrs_4}
\nl {}&\times&
\eG{\beta_{10} + \mrs_2 + \mrs_1\,\beta_{11}}\,
\eG{\beta_{20} + \mrs_3 + \mrs_1\,\beta_{21}}\,
\eG{\beta_{30} + \mrs_4 + \mrs_1\,\beta_{31}}\,
\eG{\mra + \mrs_4 + \mrs_3 + \mrs_2} \spc
\nl\nl
\mrD &=&
\eG{\mrf + \mrs_1}\,
\eG{\beta_{10} + \mrs_1\,\beta_{11}}\,
\eG{\beta_{20} + \mrs_1\,\beta_{21}}\,
\eG{\beta_{30} + \mrs_1\,\beta_{31}}\,
\eG{\mrc + \mrs_4 + \mrs_3 + \mrs_2} \spp
\eqa
It is simple to obtain the following relation:
\bq
\mrH = \frac{\eG{\mra}}{\eG{\mrc}}\,\int_{\mrL_1}\,\frac{\mrd \mrs_1}{2\,i\,\pi}\,
\lpar - \mrz_1 \rpar^{\mrs_1}\,
\mrF^{(3)}_{\sPD}\lpar \mra\,;\,\beta_1\,,\,\beta_2\,,\,\beta_3\,;\,\mrc\,;\,\mrz_2\,,\,\mrz_3\,,\,\mrz_4 \rpar \spc 
\label{fLtF}
\eq
where $\beta_i = \beta_{i0} + \beta_{i1}\,\mrs_1$. The advantage of using \eqn{fLtF} is that, following \Bref{FDMB},
we have results for different domains of the arguments and of the parameters of the Lauricella function, for
instance for the analytic continuation into different domains of $\mathbf{z} \in \Cf^{\mrN}$; as a consequence,
\eqn{fLtF} and its generalizations can be used to define analytic continuations of a Fox function.

The interest in $\mrF^{(\mrN)}_{\sPD}$ is also due to previous applications in the calculation of 
Feynman integrals~\cite{Fleischer:2003rm,Bytev:2013gva}.
Furthermore, a Lauricella function of $\mrn$ variables can always be expressed in terms of the hypergeometric 
function $\mrR$ introduced in \Bref{CarlsonR}; this allows us to obtain the Euler transformations of $\mrF^{\mrN)}_{\sPD}$,
\eg $\mrz_i \to 1/\mrz_i$.
If needed it is also possible to write $\mrR$ as the integral of another $\mrR$ function with one less variable.
Another advantage of using \eqn{fLtF} is that approximations of $\mrF^{(\mrN)}_{\sPD}$ are known, going from 
one{-}dimensional numerical integration to one{-}dimensional series expansion~\cite{Ls} to Laplace approximation~\cite{LL}.

Another example is
\bqa
\mrH &=& \Bigl[ \prod_{i=1}^{2}\,\int_{\mrL_i}\,\frac{\mrd \mrs_i}{2\,i\,\pi}\,\frac{\mrN}{\mrD}\,
         (\mrz_1)^{\mrs_1}\,(\mrz_2)^{\mrs_2} \spc
\nl
\mrN &=& \eG{ - \mrs_1}\,\eG{ - \mrs_2}\,\eG{1 + \mrs_1}\,\eG{1 + \mrs_2}\,
         \eG{ - \mra_3 + \mrs_1}\,\eG{ - \mra_4 + \mrs_2}\,\eG{1 + \mra_1 + \mrs_2 + \mrs_1} \spc
 \nl
\mrD &=& \eG{2 + \mrs_2 + \mrs_1}\,\eG{2 + \mra_2 + \mra_1 + \mrs_2 + \mrs_1} \spc
\label{nexao}
\eqa
which corresponds to a Fox function of \sect{GMFF} with parameters $\mrr = 2, \mrm = 7$ and $\mrn = 2$. 
The same function can be rewitten as follows:
\bqa
\mrH &=& \frac{\eG{ - \mra_3}\,\eG{ - \mra_4}}{\eG{1 + \mra_2}}\,\eB{1 + \mra_1}{1 + \mra_2}\,\int_0^1\,\mrd \mrx
\nl {}&\times&
\mrF^{(2)}_{\sPD}\lpar 1 + \mra_1\,;\, - \mra_3\,,\, - \mra_4\,;\,2 + \mra_1 + \mra_2\,;\,
                       - \mrz_1\,\mrx\,,\, - \mrz_2\,(1 - \mrx) \rpar \spc
\label{nexat}
\eqa
giving the original Fox fuction as a one{-}dimensional integral of a Lauricella function, see \eqn{FDEM}.

The results of \sect{hmff} can be summarized with the help of the following examples where two additional ingredients 
are needed.
All the results in this Section assume that the parameters are such to make series converge.

The Gauss hypergeometric function of unit argument (when convergent) is given by a single term but for 
$\ghyp{\mrp + 1}{\mrp}$, with the exception of a few special cases, a series is needed.  
\bq
\ghyp{3}{2}\lpar {\mathbf a}\,;\,{\mathbf b}\,;\,\omega\,\mrz \rpar =
(1 - \mrz)^{ - \mra_1}\,\sum_{\mrk=0}^{\infty}\,\frac{(\mra_1)_{\mrk}}{\mrk\,!}\,
\ghyp{3}{2}\lpar - \mrk\,,\,\mra_2\,,\,\mra_3\,;\,{\mathbf b}\,;\,\omega \rpar\,
\Bigl( \frac{\mrz}{\mrz - 1} \Bigr)^{\mrk} \spc
\eq
\bq
\ghyp{3}{2}\lpar {\mathbf a}\,;\,{\mathbf b}\,;\,1\rpar = 
 \sum_{\mrk=1}^{2}\,\Upphi^{\mrk}_0\,\sum_{\mrn=0}^{\infty}\,\frac{\Upphi^{\mrk}(\mrn)}{\mrn\,!} \spc
\eq
\bq
\Upphi^{\mrk}_0 = \frac{\mrN^{\mrk}_0}{\mrD^{\mrk}_0} \spc \qquad 
\Upphi^{\mrk} = \frac{\mrN^{\mrk}}{\mrD^{\mrk}} \spp 
\eq
Given
\bq
\uppsi_2= \mrb_1 + \mrb_2 - \mra_1 - \mra_2 - \mra_3 \spc
\quad
\mrb_{\mri\mrj}= \mrb_{\mri} - \mrb_{\mrj} \spc \quad
\mra_{\mri\mrj}= \mra_{\mri} - \mra_{\mrj} \spc \quad
\mrc_{\mri\mrj}= \mrb_{\mri} - \mra_{\mrj} \spc 
\eq
we obtain
\bqa
\mrN^1_0 &=&       
\eG{\mrb_1}\,
\eG{\mrb_2}\,
\eG{\mrb_{12}}\,
\eG{1 - \mrb_{12}}\,
\eG{\uppsi_2}\,
\eG{1 - \uppsi_2} \spc
\nl
\mrD^1_0 &=&
\eG{\mra_3}\,
\eG{2 - \mrb_{11}}\,
\eG{\mrc_{11}}\,
\eG{\mrc_{12}}\,
\eG{\mrc_{13}}\,
\eG{\mrc_{23}}\,
\eG{1 - \mrc_{12}} \spc
\nl
\mrN^1 &=&
\eG{2 - \mrb_1 + \mrn}\,
\eG{\mrc_{23} + \mrn}\,
\eG{1 - \mrc_{12} + \mrn}\spc
\nl
\mrD^1 &=&
\eG{1 + \mrb_{12} + \mrn}\,
\eG{1 - \mrc_{12} + \mra_1 + \mrn} \spc 
\nl\nl
\mrN^2_0 &=&
\eG{\mrb_1}\,
\eG{\mrb_2}\,
\eG{ - \mrb_{12}}\,
\eG{1 + \mrb_{12}}\,
\eG{\uppsi_2}\,
\eG{1 - \uppsi_2} \spc
\nl
\mrD^2_0 &=&
\eG{\mra_3}\,
\eG{\mrc_{13}}\,
\eG{\mrc_{21}}\,
\eG{\mrc_{22}}\,
\eG{\mrc_{23}}\,
\eG{1 - \mrc_{21}}\,
\eG{1 - \mrc_{22}} \spc
\nl
\mrN^2 &=&
\eG{\mrc_{13} + \mrn}\,
\eG{1 - \mrc_{21} + \mrn}\,
\eG{1 - \mrc_{22} + \mrn} \spc
\nl
\mrD^2 &=&
\eG{1 - \mrb_{12} + \mrn}\,
\eG{1 - \mrc_{22} + \mra_1 + \mrn} \spp
\eqa
The series converges for $\Re(\mra_3) > 0$. If this condition is violated we can always use contiguity relations, 
see remark $3.3$ of \Bref{_etinkaya_2021}.
A similar result can be obtained for arbitrary $\mrp$, see \Bref{_etinkaya_2021}.
As already stated we assume that $\Re(\uppsi_2) > 0$ and $\mrb_{\mri \mrj} \not\in \Zf, \forall \mri, \mrj$ etc

In the following examples the strategy is:
\begin{enumerate}

\item select the poles at $\mrs_2 = \mrn$, assuming that $\mrL_2$ separate the poles and that the poles are simple;

\item sum the residues, eventually using
\bq \eG{\mrz - \mrn} = ( - 1)^{\mrn}\,
\frac{\eG{\mrz}\,\eG{1 - \mrz}}{\eG{1 - \mrz + \mrn}} \spc
\eq

\item use $\ghyp{\mrp + 1}{\mrp}(\mrz)$ at $\mrz = \pm 1 \pm i\,\delta$;

\item recombine in order to obtain a univariate Fox (Meijer) function or a function without $\pm 1 \pm i\,\delta$
as argument.

\end{enumerate}
\bei
\item[\ovalbox{Example $1$}] Consider the following function,
\bq
\mrE_1 = \Bigl[ \prod_{\mri=1}^{2}\,\int_{\mrL_{\mri}}\,\frac{\mrd \mrs_{\mri}}{2\,i\,\pi} \Bigr]\,
\frac{
 \eG{ - \mrs_1}\,
 \eG{\mra_1 + \mrs_1}\,
 \eG{ - \mrs_2}\,
 \eG{\mra_2 + \mrs_2}\, 
 \eG{\mrb + \mrs_1 + \mrs_2}}
 {\eG{\mrc + \mrs_1 + \mrs_2}}\,\mrz^{\mrs_1}\,(1 + i\,\delta)^{\mrs_2} \spc
\eq
which is an $\mrH$ function in two variables. We obtain
\bqa
\mrE_1 &=& \Bigl[ \prod_{\mri=1}^{2}\,\int_{\mrL_{\mri}}\,\frac{\mrd \mrs_{\mri}}{2\,i\,\pi} \Bigr]\,
\eG{ - \mrs_1}\,
\eG{ - \mrs_2}\,
\eG{\mra_1 + \mrs_1}
\nl
{}&\times&
\Bigl[ \frac{2^{ - \mra_2}}{\eG{\mrc - \mrb}}\,\Upphi_1 + \frac{2^{ - \mrb}}{\eG{\mrc - \mrb}}\,\Upphi_2 \Bigr]\,
(\frac{\mrz}{2})^{\mrs_1}\,(\frac{1}{2} - i\,\delta)^{\mrs2} \spc
\nl\nl
\Upphi_1 &=&
\frac{
\eG{\mra_1 + \mrs_1}\,
\eG{ - \mra_2 + \mrb + \mrs_1}\,
\eG{1 + \mra_2 - \mrb - \mrs_1}\,
\eG{\mra_2 + \mrs_2}\,
\eG{\mrc - \mrb + \mrs_2}
}
{\eG{ - \mra_2 + \mrc + \mrs_1}\,
\eG{1 + \mra_2 - \mrb + \mrs_2 - \mrs_1}
} \spc
\nl
\Upphi_2 &=&           
\frac{
\eG{\mra_1 + \mrs_1}\,
\eG{1 - \mra_2 + \mrb + \mrs_1}\,
\eG{ - \mra_2 + \mrc + \mrs_2 + \mrs_1}\,
\eG{\mrb + \mrs_2 + \mrs_1}\,
\eG{\mra_2 - \mrb - \mrs_1}
}
{
\eG{ - \mra_2 + \mrc + \mrs_1}\,
\eG{1 - \mra_2 + \mrb + \mrs_2 + \mrs_1}
} \spc
\eqa
which is again a Fox function of arguments $\mrz/2$ and $1/2 - i\,\delta$.
\item[\ovalbox{Example $2$}] Consider the following function,
\bq
\mrE_2 = \Bigl[ \prod_{\mri=1}^{2}\,\int_{\mrL_{\mri}}\,\frac{\mrd \mrs_{\mri}}{2\,i\,\pi} \Bigr]\,
\frac{
\eG{ - \mrs_1}\,
\eG{\mra_1 + \mrs_1}\,
\eG{ - \mrs_2}\,
\eG{\mra_2 + \mrs_2}\,
\eG{\mrb + \mrs_1 + \mrs_2}
}
{
\eG{\mrc + \mrs_1 + \mrs_2}
}\,
\mrz^{\mrs_1}\,( - 1 - i\,\delta)^{\mrs_2} \spp
\eq
We obtain
\bq
\mrE_2 =
\frac{
\eG{\mrb}\,
\eG{\mra_1}\,
\eG{\mra_2}\,
\eG{\mrc - \mrb - \mra_2}
}
{
\eG{\mrc - \mrb}\,
\eG{\mrc - \mra_2}
}\,
\ghyp{2}{1}\lpar \mra_1\,,\,\mrb\,;\,\mrc - \mra_2\,;\, - \mrz \rpar \spp
\eq
\item[\ovalbox{Example $3$}] Consider the following function,
\bq
\mrE_3 = \Bigl[ \prod_{\mri=1}^{2}\,\int_{\mrL_{\mri}}\,\frac{\mrd \mrs_{\mri}}{2\,i\,\pi} \Bigr]\,
\frac{          
\eG{ - \mrs_1}\,
\eG{\mra_1 + \mrs_1}\,
\eG{ - \mrs_2}\,
\eG{\mra_2 + \mrs_2}\,
\eG{\mrb + \mrs_1 + \mrs_2}
}
{
\eG{\mra_3 + \mrs_2}\,
\eG{\mrd - \mrs_2}\,
\eG{\mrc + \mrs_1 + \mrs_2}
}\,
\mrz^{\mrs_1}\,(1 + i\,\delta)^{\mrs_2} \spp
\eq
We obtain
\bq
\mrE_3 = \int_{\mrL}\,\frac{\mrd \mrs}{2\,i\,\pi}\,\mrz^{\mrs}\,
\frac{\eG{\mra_2}}{\eG{\mrd}\,\eG{\mrc - \mrb}}\,
\sum_{\mrk=1}^{2}\sum_{\mrn=0}^{\infty}\,\Upphi^{\mrk}_0\,\frac{\Upphi^{\mrk}(\mrn)}{\mrn\,!} \spc
\eq
\bqa
\mrN^1_0 &=&
\eG{ - 1 + \mra_3 - \mra_2 + \mrd + \mrc - \mrb}\,
\eG{2 - \mra_3 + \mra_2 - \mrd - \mrc + \mrb}\,
\eG{2 - \mra_3 + \mrn}\,
\eG{\mrc - \mrb + \mrn}\,
\eG{2 - \mra_3 - \mrd + \mrn} \spc
\nl
\mrD^1_0 &=&
\eG{2 - \mra_3}\,
\eG{\mra_3 - \mra_2}\,
\eG{ - 1 + \mra_3 + \mrd}\,
\eG{2 - \mra_3 - \mrd}\,
\eG{2 - \mra_3 + \mra_2 - \mrd + \mrn} \spc
\nl
\mrN^1 &=&
\eG{ - \mrs}\,
\eG{\mra_1 + \mrs}\,
\eG{\mra_3 - \mrc - \mrs}\,
\eG{1 - \mra_3 + \mrc + \mrs} \spc
\nl
\mrD^1 &=&
\eG{\mra_3 - \mrb - \mrs}\,
\eG{1 - \mra_3 + \mrc + \mrs + \mrn} \spc
\nl\nl
\mrN^2_0 &=&
\eG{ - 1 + \mra_3 - \mra_2 + \mrd + \mrc - \mrb}\,
\eG{2 - \mra_3 + \mra_2 - \mrd - \mrc + \mrb}\,
\eG{1 + \mra_2 - \mrb + \mrn} \spc
\nl
\mrD^2_0 &=&
\eG{1 + \mra_2 - \mrb} \spc
\nl
\mrN^2 &=&
\eG{ - \mrs}\,
\eG{\mra_1 + \mrs}\,
\eG{ - \mra_3 + \mrc + \mrs}\,
\eG{1 + \mra_3 - \mrc - \mrs}\,
\eG{\mra_3 - \mrb - \mrs + \mrn}\,
\eG{2 - \mrd - \mrc - \mrs + \mrn} \spc
\nl
\mrD^2 &=&
\eG{ - \mra_2 + \mrc + \mrs}\,
\eG{\mra_3 - \mrb - \mrs}\,
\eG{ - 1 + \mrd + \mrc + \mrs}\,
\eG{2 - \mrd - \mrc - \mrs}
\nl
{}&\times&
\eG{1 + \mra_3 - \mrc - \mrs + \mrn}\,
\eG{2 + \mra_2 - \mrd - \mrc - \mrs + \mrn} \spp
\eqa
\item[\ovalbox{Example $4$}] Consider the following function,
\bq
\mrE_3 = \Bigl[ \prod_{\mri=1}^{2}\,\int_{\mrL_{\mri}}\,\frac{\mrd \mrs_{\mri}}{2\,i\,\pi} \Bigr]\,
\frac{
\eG{ - \mrs_1}\,
\eG{\mra_1 + \mrs_1}\,
\eG{ - \mrs_2}\,
\eG{\mra_2 + \mrs_2}\,
\eG{\mra_3 + \mrs_2}\,
\eG{\mrb + \mrs_1 + \mrs_2}
}
{
\eG{\mrc + \mrs_2}\,
\eG{\mrd + \mrs_1 + \mrs_2}
}\,
\mrz^{\mrs_1}\,(1 + i\,\delta)^{\mrs_2}
\eq
We obtain
\bqa
\mrE_4 &=& \frac{\eG{\mra_3}}{\eG{\mrd - \mrb}}\,
\int_{\mrL}\,\frac{\mrd \mrs}{2\,i\,\pi}\,\mrz^{\mrs}\,
\sum_{\mrk=1}^{2}\,\sum_{\mrm=0}^{\infty}\,\sum_{\mrn=0}^{\infty}\,\lpar \frac{1}{2} \rpar^{\mrm}\,\Upphi^{\mrk}_0\,
\frac{\Upphi^{\mrk}}{\mrm\,!\;\mrn\,!} \spc
\nl\nl
\mrN^1_0 &=&
\eG{\mrm + \mra_2}\,
\eG{\mrd - \mrb + \mrn}\,
\eG{2 - \mrc + \mrn}\,
\eG{1 + \mra_3 - \mrc + \mrn}
\nl
{}&\times&
\eG{\mrm - \mra_3 + \mrd + \mrc - \mrb}
\eG{1 - \mrm + \mra_3 - \mrd - \mrc + \mrb} \spc
\nl
\mrD^1_0 &=&
\eG{2 - \mrc}\,
\eG{\mrm + \mrc}\,
\eG{ - \mra_3 + \mrc}\,
\eG{1 + \mra_3 - \mrc}\,
\eG{1 - \mrm + \mra_3 - \mrc + \mrn} \spc
\nl
\mrN^1 &=&
\eG{ - \mrs}\,
\eG{\mra_1 + \mrs}\,
\eG{ - \mrd + \mrc - \mrs}\,
\eG{1 + \mrd - \mrc + \mrs} \spc
\nl
\mrD^1 &=&
\eG{\mrc - \mrb - \mrs}\,
\eG{1 + \mrd - \mrc + \mrs + \mrn} \spc
\nl\nl
\mrN^2_0 &=&
\eG{\mrm + \mra_2}\,
\eG{1 - \mrm - \mrb + \mrn}\,
\eG{\mrm - \mra_3 + \mrd + \mrc - \mrb}\,
\eG{1 - \mrm + \mra_3 - \mrd - \mrc + \mrb} \spc
\nl
\mrD^2_0 &=&
\eG{1 - \mrm - \mrb} \spc
\nl
\mrN^2 &=&
\eG{ - \mrs}\,
\eG{\mra_1 + \mrs}\,
\eG{\mrd - \mrc + \mrs}\,
\eG{1 - \mrd + \mrc - \mrs}\,
\eG{\mrc - \mrb - \mrs + \mrn}\,
\eG{1 + \mra_3 - \mrd - \mrs + \mrn} \spc
\nl
\mrD^2 &=&
\eG{\mrc - \mrb - \mrs}\,
\eG{\mrm + \mrd + \mrs}\,
\eG{ - \mra_3 + \mrd + \mrs}\,
\eG{1 + \mra_3 - \mrd - \mrs}
\nl
{}&\times&
\eG{1 - \mrd + \mrc - \mrs + \mrn}\,
\eG{1 - \mrm + \mra_3 - \mrd - \mrs + \mrn} \spp
\eqa
\eei
\subsubsection{From multivariate Fox to multivariate Meijer \label{mFtomM}}
The results of \sect{fFtM} can be generalized to the case of multivariate functions, although the generalization
requires the introduction of series. The method is based on the inverse Gamma series,
\bq
\frac{\prod_{\mrj=1}^{\mrp}\,\eG{\mra + \mrx_{\mrj}}}
     {\prod_{\mrj=1}^{\mrp}\,\eG{\mra + \mry_{\mrj}}} =
\sum_{\mrn=0}^{\infty}\,\frac{\mrg^{\mrp}_{\mrn}({\mathbf x}\,;\,{\mathbf y})}
                             {\eG{\mra + \uppsi({\mathbf x}\,;\,{\mathbf y}) + \mrn}} \spc
\eq
where $\mrg^{\mrp}_{\mrn}$ are the N\o{}rdlund's coefficients~\cite{Ncoef,lopez2018uniformly}. A simple example, corresponding
to $\mrp = 2$, is based on~\cite{lopez2018uniformly,_etinkaya_2021} 
\bq
\mrg^2_{\mrn} = \frac{1}{\mrn\,!}\,(\mry_1 - \mrx_1)_{\mrn}\,(\mry_2 - \mrx_1)_{\mrn} \spc \qquad
\uppsi({\mathbf x}\,;\,{\mathbf y}) = \sum_{\mrj=1}^{\mrp} (\mry_{\mrj} - \mrx_{\mrj}) \spc
\eq
where $(\mra)_{\mrn}$ is the Pochhammer symbol. We obtain 
\bq
\eG{\mra + \mrs_1 + 2\,\mrs_2} = 2^{\mra - 1/2 + 2\,\mrs_2}\,
\frac{
\eG{\mra + \mrs_1}\,
\eG{\frac{1}{2}\,\mra + \mrs_2}\,
\eG{\frac{1}{2} + \frac{1}{2}\,\mra + \mrs_2}\,
}
{
\eG{\mra + \mrs_2 + \mrs_1}
}\,
\sum_{\mrn=0}^{\infty}\,\frac{( - \mrs_2)_{\mrn}\,(\mrs_2 - \mrs_1)_{\mrn}}
                             {\eG{\mra + \mrs_1 + \mrs_2 + \mrn}\,\mrn\,!} \spc
\eq
\bq
\frac{1}{\eG{\mra + \mrs_1 + 2\,\mrs_2}} =
\frac{
\eG{\mra + \mrs_2 + \mrs_1}
}
{
\eG{\mra + \mrs_2}\,
\eG{\mra + \mrs_2 + \mrs_1}
}\,
\sum_{\mrn=0}^{\infty}\,
\frac{(\mrs_1)_{\mrn}\,(\mrs_1 + \mrs_2)_{\mrn}}{\eG{\mra - \mrs_1 - \mrs_2 + \mrn}\,\mrn\,!} \spp
\eq
The general case proceeds by iterations and leads to multiple series, \eg
\bq
\eG{\mra + \mrj\,\mrs_1 + \mrk\,\mrs_2} =
\frac{
\eG{\mra + \mrs_1\,\mrj}\,
\eG{\mra + \mrs_2\,\mrk}
}
{
\eG{\mra + \mrs_2 + \mrs_1}
}\,
\sum_{\mrn=0}^{\infty}\,\frac{1}{\mrn\,!}\,
\frac{
\eG{\mrn - \mrs_1 + \mrs_2\,(\mrk-1)}\,
\eG{\mrn - \mrs_2 + \mrs_1\,(\mrj-1)}
}
{
\eG{ - \mrs_1 + \mrs_2\,(\mrk-1)}\,
\eG{ - \mrs_2 + \mrs_1\,(\mrj-1)}\,
\eG{\mra + \mrn + \mrs_2 + \mrs_1} \spc
}
\eq
where we iterate, \eg
\bq
\eG{\mrn - \mrs_1 + \mrs_2\,(\mrk-1)} =
\frac{
\eG{\mrn + \mrs_2}\,
\eG{\mrn + (\mrk - 1)\,\mrs_2}\,
}
{
\eG{\mrn + \mrs_2 - \mrs_1}\,
}\,
\sum_{\mrm=0}^{\infty}\,
\frac{(\mrs_1)_{\mrm}\,(\mrs_1 + (\mrk - 2)\,\mrs_2)_{\mrm}}{\eG{\mrn + \mrm}\,\mrm\,!} \spp
\eq
\subsubsection{Euler{-}Mellin integrals \label{EMI}}
Evaluation of the Meijer $\mrG$ function can be simplified by using \eqn{EMG}. We give an example
of the derivation in a simple case and discuss the generalization to Fox $\mrH$ functions.
We write~\cite{Dubovyk:2022obc,Matsubara-Heo:2023ylc}
\bq
\eG{\mra + \mrs}\,\eG{\mrb + \mrs} = \eB{\mra + \mrs}{\mrb + \mrs}\,\eG{\mra + \mrb + 2\,\mrs} \spc
\eq
and use the Euler{-}Mellin (hereafter EM) representation for the Euler beta function to obtain
\bqa
\hyp{\mra}{\mrb}{\mrc}{\mrz} &= & \frac{\eG{\mrc}}{\eG{\mra}\,\eG{\mrb}}\,
\int_{\mrL}\,\frac{\mrd \mrs}{2\,i\,\pi}\,
\frac{\eG{ - \mrs}\,\eG{\mra + \mrs}\,\eG{\mrb + \mrs}}{\eG{\mrc + \mrs}}\,\lpar - \mrz \rpar^{\mrs}
\nl
{}&=&
\frac{\eG{\mrc}}{\eG{\mra}\,\eG{\mrb}}\,\int_0^1\,\mrd \mrx\,\mrx^{\mra - 1}\,(1 - \mrx)^{\mrb - 1}\,
\int_{\mrL}\,\frac{\mrd \mrs}{2\,i\,\pi}\,
\frac{\eG{ - \mrs}\,\eG{\mra + \mrb + 2\,\mrs}}{\eG{\mrc + \mrs}}\,\lpar - \mrX\,\mrz \rpar^{\mrs} \spc
\eqa
where $\mrX = \mrx\,(1 - \mrx)$ and where the $\mrs\,${-} integral gives $\mrH^{1\,,\,1}_{2\,,\,1}$ with
$\mid \marg ( - \mrz ) \mid < \pi$ ($\mrX \in \Rf_{+}$).

Consider now the following integral
\bq
\mrI = \Bigl[ \prod_{i=1}^{2}\,\int_{\mrL_i}\,\frac{\mrd \mrs_i}{2\,i\,\pi} \Bigr]\,
\frac{\eG{1 + \ep + \mrs_1 + \mrs_2}\,\eG{\ep + \mrs_1 + \mrs_2}\,\eG{ - \mrs_2}}{\eG{\lambda + \mrs_2}}\,
\mrz_1^{ - \mrs_1}\,\mrz_2^{- \mrs_2} \spp
\eq
The integral converges for $\mid \phi_1 \mid < \pi$ and $\mid \phi_1 - \phi_2 \mid < \pi$. Use
\bq
 \eG{1 + \ep + \mrs_1 + \mrs_2}\,\eG{\ep + \mrs_1 + \mrs_2} = 
 \eB{1 + \ep + \mrs_1 + \mrs_2}{\ep + \mrs_1 + \mrs_2}\,\eG{1 + 2\,\ep + 2\,\mrs_1 + 2\,\mrs_2} \spc
\label{pairing}
\eq
use the EM representation for the Euler Beta function and obtain
\bq
\mrI = \int_0^1\,\mrd \mrx\,\mrx^{\ep}\,\lpar 1 - \mrx \rpar^{\ep - 1}\,\mrJ \spc
\quad
\mrJ = \Bigl[ \prod_{i=1}^{2}\,\int_{\mrL_i}\,\frac{\mrd \mrs_i}{2\,i\,\pi} \Bigr]\,
\frac{\eG{ - \mrs_2}\,\eG{1 + 2\,\ep + 2\,\mrs_1 + 2\,\mrs_2}}{\eG{\lambda + \mrs_2}}\,
\Bigl( \frac{\mrX}{\mrz_1} \Bigr)^{\mrs_1}\,\Bigl( \frac{\mrX}{\mrz_2} \Bigr)^{\mrs_2} \spp
\eq
Therefore, $\mrJ$ is the generalized Fox function of \eqn{gmff} corresponding to $\mrm = 2$ and $\mrn = 1$
while the original integral corresponds to $\mrm = 3$ and $\mrn = 1$. The regions of convergence of the
MB integrals are the same, $\mrJ$ is regular at the origin, the EM integral requires that 
$\Re\,\ep > 0$. It is worth noting that limits on $\marg(\mrz_i)$ follow from \Bref{HS} where
the integrals are written in terms of $\mrz_i^{-\mrs_i}$ while the behavior at the origin follows from
\Bref{compH} where the integrals are written in terms of $\mrz_i^{\mrs_i}$.

We call this procedure ``reducibility of a generalized Fox function'' which amounts to write a generalized Fox function
characterized by two integers $\mrm, \mrn$ (see \eqn{gmff}) as a EM integral of a generalized
Fox function where $\mrm + \mrn \to \mrm + \mrn - 1$.
The general rule is the following: pair two Gamma functions in the numerator and write the corresponding beta 
functions using the EM representation; check the convergence of the resulting MB
an EM integrals and repeat the operation as long as the integrals converge. 
However, in order to apply the theorem of \Bref{HS} for a Fox function we should stop the procedure if 
$\mrr > \mrm + \mrn + 1$.

If convenient it is also
possible to pair one Gamma function in the numerator with a Gamma function in the denominator.

Another example is the following:
\bq
\mrI_2 = \Bigl[ \prod_{i=1}^{2}\,\int_{\mrL_i}\,\frac{\mrd \mrs_i}{2\,i\,\pi} \Bigr]\,
\frac{\eG{\lambda_1 - \mrs_1}\,\eG{\lambda_2 - \mrs_2}\,\eG{\ep + \mrs_1 + \mrs_2}\,
      \eG{1 + \ep + \mrs_1 + \mrs_2}}{\eG{\mrs_1}\,\eG{\mrs_2}}\,
     \prod_{i=1}^{2}\,\mrz_i^{- \mrs_i} \spp
\eq
Using again \eqn{pairing} we arrive at
\bqa
\mrI_2 &=& \int_0^1\,\mrd \mrx\,\mrx^{\ep}\,\lpar 1 - \mrx \rpar^{\ep - 1}\,\mrJ_2 \spc
\nl
\mrJ_2 &=& \Bigl[ \prod_{i=1}^{2}\,\int_{\mrL_i}\,\frac{\mrd \mrs_i}{2\,i\,\pi} \Bigr]\,
 \frac{\eG{\lambda_1 - \mrs_1}\,\eG{\lambda_2 - \mrs_2}\,\eG{1 + 2\,\ep + 2\,\mrs_1 + 2\,\mrs_2}}
      {\eG{\mrs_1}\,\eG{\mrs_2}}\,
  \Bigl( \frac{\mrX}{\mrz_1} \Bigr)^{\mrs_1}\,\Bigl( \frac{\mrX}{\mrz_2} \Bigr)^{\mrs_2} \spp
\eqa
For $\mrX = \mrx\,(1 - \mrx) \to 0$ the function $\mrJ_2$ behaves like $\mrX^{\lambda_1 + \lambda_2}$,
therefore, the EM integral requires $\Re\,(\ep + \lambda_1 + \lambda_2) > 0$.
\section{Partial differential equations \label{PDE}}
In this Section, we present a brief summary of the Horn system of partial differential equations. 
The motivations can be illustrated by considering the Lauricella function $\mrF^{(\mrN)}_{\sPD}$ which can be
represented by an $\mrN\,${-}variate hypergeometric series converging in the unit polydisk. The function has
an integral representation of the MB type which gives an analytic continuation of the series; however, as
pointed out in \Bref{FDMB} the most adequate tools for the calculation of the function outside 
$\Uf^{\mrN}$ are based on solving a system of $\mrN$ partial differential equations of second order with
respect to the variables $\mrz_{\mrj}$. We want to analyze the same problem when dealing with Fox functions. 

Following \Bref{HTF} we start by considering a bivariate Fox function,
\bq
\mrH\lpar \mrx\,,\,\mry \rpar = \sum_{\mrm,\mrn = 0}^{\infty}\,\mrA_{\mrm\,,\,\mrn}\,\mrx^{\mrm}\,\mry^{\mrn} \spp
\label{H2series}
\eq
If we have
\bq
\mrA_{\mrm + 1\,,\,\mrn}/\mrA_{\mrm\,,\,\mrn} = \mrF(\mrm\,,\,\mrn)/\mrF^{\prime}(\mrm\,,\,\mrn) \spc \quad
\mrA_{\mrm\,,\,\mrn + 1}/\mrA_{\mrm\,,\,\mrn} = \mrG(\mrm\,,\,\mrn)/\mrG^{\prime}(\mrm\,,\,\mrn) \spc 
\eq
and $\mrF\,\,\mrF^{\prime}\,,\,\mrG\,,\,\mrG^{\prime}$ are polynomials as specified in Sect.~5.7.1 of
\Bref{HTF}, then $\mrH$ satisfies the following system of partial differential equations:
\bq
\Bigl[ \mrF^{\prime}(\theta\,,\,\phi)\,\mrx^{-1} - \mrF(\theta\,,\,\phi) \Bigr]\,\mrH = 0 \spc
\qquad
\Bigl[ \mrG^{\prime}(\theta\,,\,\phi)\,\mry^{-1} - \mrG(\theta\,,\,\phi) \Bigr]\,\mrH = 0 \spc
\eq
where $\theta$ and $\phi$ are
\bq
\theta = \mrx\,\frac{\partial}{\partial\,\mrx} \spc \qquad
\phi = \mry\,\frac{\partial}{\partial\,\mry} \spp
\eq
If we restrict to the case of partial differential equations of second order the solution is given in
Sect.~5.9 of \Bref{HTF} (Horn's list). We now consider \eqn{H2series} with
\bq
\mrA_{\mrn\,\,\mrm} = \frac{\eG{\mrb + 2\,\mrn + 2\,\mrm}\,\eG{\mra_1 + \mrn}\,\eG{\mra_2 + \mrm}}
                           {\mrm\,!\;\mrn\,!\;\eG{\mrc + \mrn + \mrm}} \spc
\eq
and obtain
\bqa
\mrF &=&
       (1 + \mrb)\,\mrb\,\mra_1 
       + 2\,(1 + 2\,\mrb)\,\mra_1 \, \mrm
       + (2\,\mra_1 + \mrb + 4\,\mrb\,\mra_1 + \mrb^2) \, \mrn
       + 4\,\mra_1 \, \mrm^2
       + 2\,(1 + 2\,\mra_1 + 2\,\mrb) \, \mrn^2
\nl
{}&+& 2\,(1 + 4\,\mra_1 + 2\,\mrb) \, \mrm\,\mrn
       + 4\,(\mrm + \mrn)^2\,\mrn \spc
\nl
\mrF^{\prime} &=& \mrc + (1 + \mrc)\,\mrn + (\mrm + \mrn)^2 \spc
\nl
\mrG &=&
       (1 + \mrb)\,\mrb\,\mra_2 
       + (2\,\mra_2 + \mrb + 4\,\mrb\,\mra_2 + \mrb^2) \, \mrm
       + 2\,(1 + 2\,\mrb)\,\mra_2 \, \mrn
       + 2\,(1 + 2\,\mra_2 + 2\,\mrb) \, \mrm^2
       + 4\,\mra_2 \, \mrn^2
\nl
{}&+& 2\,(1 + 4\,\mra_2 + 2\,\mrb) \, \mrm\,\mrn
       + 4\,(\mrm + \mrn)^2\,\mrm \spc
\nl
\mrG^{\prime} &=& \mrc + (1 + \mrc)\,\mrm + (\mrm + \mrn)^2 \spp
\eqa
Therefore, following \Bref{Ghs}, $\mrH$ in \eqn{H2series} is a Horn hypergeometric function and satisfies a 
system of $2$ partial differential equations of third order. The first is
\bqa
{}&{}&   4\,\mrx^3 \, \mrH_{\mrx\mrx\mrx} 
       + 4\,\mrx\,\mry^2 \, \mrH_{\mrx\mry\mry} 
       + 8\,\mrx^2\,\mry \, \mrH_{\mrx\mrx\mry} 
\nl
{}&-& \mrx\,\Bigl[ 1 - 2\,(7 + 2\,\mra_1 + 2\,\mrb)\,\mrx \Bigr] \,  \mrH_{\mrx\mrx} 
       + 4\,\mra_1\,\mry^2 \, \mrH_{\mry\mry} 
       - \mry\,\Bigl[ 1 - 2\,(7 + 4\,\mra_1 + 2\,\mrb)\,\mrx \Bigr] \, \mrH_{\mrx\mry} 
\nl
{}&-& \Bigl[ \mrc - (6 + 6\,\mra_1 + 5\,\mrb + 4\,\mrb\,\mra_1 + \mrb^2)\,\mrx \Bigr] \, \mrH_{\mrx} 
       + 2\,(3 + 2\,\mrb)\,\mra_1\,\mry \, \mrH_{\mry} 
       + (1 + \mrb)\,\mrb\,\mra_1 \, \mrH = 0 \spc
\eqa
where $\mrH_{\mrx} = \partial \mrH/\partial \mrx$ \etc
The second equation follows from the $\mrm \leftrightarrow \mrn$ and $\mra_1 \leftrightarrow \mra_2$ symmetry of
the coefficient $\mrA$.

The Horn's approach can be made more general as follows~\cite{FDMB}: given a power series
\bq
\mrH \lpar \mrz_1\,,\,\dots\,,\,\mrz_{\mrn} \rpar =
\sum_{\mathbf{k} \in \Zf^{\mrn}}\,\mrA\lpar \mrk_1\,,\,\dots\,,\,\mrk_{\mrn} \rpar\,
\prod_{i=1}^{\mrn}\,\mrz_i^{\mrk_i} \spc
\eq
consider the ratios
\bq
\frac{\mrA(\mathbf{k} + \mathbf{e}_j)}{\mrA(\mathbf{k})} = 
\frac{\mrP_j(\mathbf{k})}{\mrQ_j(\mathbf{k)}} \spp
\eq
If $\mrP_j$ and $\mrQ_j$ are polynomilas in $\mrn$ variables and 
$\mathbf{e}_j = (0,\dots,1,\dots\,0)$ ($1$ is in the jth position) then $\mrH$ satisfies a system of partial 
differenial equations,
\bq
\Bigl[ \mrQ_j(\theta)\,\mrz_j^{-1} - \mrP_j(\theta) \Bigr]\,\mrH = 0 \spc \quad
\theta_i = \mrz_i\,\frac{\partial}{\partial \mrz_i} \spp
\label{Hsystem}
\eq
This approach has been used in \Brefs{Kershaw:1973km,PhysRevD.9.370,PhysRevD.11.452}, where the system corresponding to 
the Feynman integral for a 
one{-}loop diagram has been constructed, and in \Bref{FDMB} where the Lauricella function $\mrF^{(\mrN)}_{\sPD}$ has
bee analyzed. See also \Bref{Ghs,Moriello:2019yhu,Armadillo:2022ugh}.
The form of $\mrA$ is determined by the Ore{-}Sato theorem~\cite{Ore,Sato,OSth}, giving $\mrA$ as a product of $\Gamma$
functions (and some multiplier having no crucial importance for the properties of the series),
\bq
\prod_i\,\eG{\alpha_i + \mra_i\,\mrk_1 + \mrb_i\,\mrk_2 + \dots} \spc
\eq
where the $\alpha_i \in \Cf$ are arbitrary constants but $\mra_i, \mrb_i, \dots$ are arbitrary integers, positive, 
negative or zero. The multiplier is of the form
\bq
\prod_i\,\lambda_i^{\mrk_i}\,\mrR(\mathbf{k}) \spc \qquad
\lambda_i \in \Cf \spc
\eq
and $\mrR$ is a rational function.

Clearly this poses a problem for arbitrary Fox functions; an extension can be provided for
coefficients belonging to $\mathrm{Frac}(\Rf)$. Let us consider a simple example,
\bq
\mrH = \frac{\eG{\mrc}}{\eG{\mrb}}\,\Bigl[ \prod_{i=1}^{2}\,\int_{\mrL_i}\,\frac{\mrd \mrs_i}{2\,i\,\pi} \Bigr]\,
\Bigl[ \prod_{i=1}^{2}\,\frac{\eG{ - \mrs_i}\,\eG{\mra_i + \mrs_i}}{\eG{\mra_i}} \Bigr]\,
\frac{\eG{\mrb + \frac{1}{2}\,\mrs_1 + \mrs_2}}{\eG{\mrc + \mrs_1 + \mrs_2}}\,
\prod_{i=1}^{2}\,(\mrz_i)^{ - \mrs_i} \spp
\eq
The integrals over $\mrs_1$ is
\bq
\mrH^{2,1}_{2,2}\Bigl[ \mrz_1\,,\{\mrp\} \Bigr] \spc
\eq
with parameters given by
\bqa
(1\,,\,1) &\quad& (\mrc + \mrs_2\,,\,1) \nl
(\mra_1\,,\,1) &\quad& (\mrb + \mrs_2\,,\,\frac{1}{2}) \spp
\eqa
Using a property of the $\mrH$ functions~\cite{compH} the same integral can be rewritten as
\bq
2\,\mrH^{2,1}_{2,2}\Bigl[ \mrz_1^2\,,\{\mrp^{\prime}\} \Bigr] \spc
\eq
where now the parameters are
\bqa
(1\,,\,2) &\quad& (\mrc + \mrs_2\,,\,2) \nl
(\mra_1\,,\,2) &\quad& (\mrb + \mrs_2\,,\,1) \spc
\eqa
producing 
\bq
\pi^{-1/2}\,\int_{\mrL_1}\,\frac{\mrd \mrs_1}{2\,i\,\pi}\,
\eG{ - \mrs_1}\,
\frac{\eG{\frac{1}{2} - \mrs_1}\,\eG{\mra_1 + 2\,\mrs_1}\,\eG{\mrb + \mrs_1 + \mrs_2}}
     {\eG{\mrc + 2\,\mrs_1 + \mrs_2}}\,\lpar 4\,\mrz_1^2 \rpar^{ - \mrs_1} \spp
\eq
The full integral can be written
as
\bq
\sum_{\mrm\,,\,\mrn =0}^{\infty}\,\mrA_{\mrm\,,\,\mrn}\,\lpar 4\,\mrz_1^2 \rpar^{\mrm}\,\mrz_2^{\mrn} \spc
\eq
where now $\mrA$ satisfies the Ore{-}Sato theorem.

A relevant question is the following one~\cite{Kalmykov:2012rr,Kalmykov:2016lxx}: can we derive the system of partial 
differential equations by working directly with the MB representation? In other words we start with
\bq
\mrH = \Bigl[ \prod_{i=1}^{\mrN}\,\int_{\mrL_i}\,\frac{\mrd \mrs_i}{2\,i\,\pi} \Bigr]\,\Uppsi(\mathbf{s})\,
\prod_{i=1}^{\mrN}\,\mrz_i^{\mrs_i} \spc
\qquad
\Uppsi = \frac{ \prod_{\mrj=1}^{\mrm}\,\eG{\alpha_{\mrj} + \mra^{\mrk}_{\mrj}\,\mrs_{\mrk}}}
              { \prod_{\mrj=1}^{\mrn}\,\eG{\beta_{\mrj} + \mrb^{\mrk}_{\mrj}\,\mrs_{\mrk}}} \spc
\eq
and, given the unit vector $\mathbf{e}_i$, we define
\bq
\mrP_i = \Uppsi\lpar \mathbf{s} + \mathbf{e}_i \rpar \spc \qquad
\mrQ_i = \Uppsi\lpar \mathbf{s} \rpar \spp
\label{MBrep}
\eq
We would like to understand under which conditions we can say that $\mrH$ satisfies
\bq
\Bigl[ \mrQ_{\mrj}(\mathbf{\theta})\,\mrz_{\mrj}^{- 1} - \mrP_{\mrj}(\mathbf{\theta}) \Bigr]\,\mrH = 0 \spc \quad 
\theta_i = \mrz_i\,\frac{\partial}{\partial \mrz_i} \spp
\eq
We will proceed by considering the following example:
\bq
\mrH = \Bigl[ \prod_{i=1}^{2}\,\int_{\mrL_i}\,\frac{\mrd \mrs_i}{2\,i\,\pi} \Bigr]\,
 \frac{\eG{\mrb + \mrs_1 + \mrs_2}}{\eG{\mrc + \mrs_1 + \mrs_2}}\,
 \eG{ - \mrs_1}\,\eG{\mra_1 + \mrs_1}\,\eG{\mra_2 + \mrs_2}\,\eG{\mra_3 - \mrs_2}\,
\mrz_1^{ - \mrs_1}\,\mrz_2^{ - \mrs_2} \spc
\eq
where we assume $\mra_{1,2,3} > 0$, $\mrb > 0$ and $\mra_3 + \mrb > \mra_1$. The contours are such that
$ - \mra_1 < \Re\,\mrs_1 < 0$ and $0 < \Re\,\mrs_2 < \mra_3$. We derive
\bq
\mrH = \mrz_2^{ - \mra_3}\,\sum_{\mrm\,,\,\mrn = 0}^{\infty}\,\mrA_{\mrm\,,\,\mrn}\,
\lpar - \mrz_1 \rpar^{ - \mrm}\,\lpar - \mrz_2 \rpar^{ - \mrn} \spc
\eq
with coefficients
\bq
\mrA_{\mrm\,,\,\mrn} = \frac{\eG{\mra_1 + \mrm}\,\eG{\mra_3 + \mrb + \mrm + \mrn}\,\eG{\mra_2 + \mra_3 + \mrn}}
                            {\mrm\,!\;\mrn\,!\;\eG{\mra_3 + \mrc + \mrm + \mrn}} \spp
\label{srep}
\eq
We derive the sets $\mrF\,,\,\mrF^{\prime}\,,\,\mrG\,,\,\mrG^{\prime}$ (according to the terminology of \Bref{HTF})
using \eqn{srep} or using \eqn{MBrep} and obtain the same result only for $\mra_3 = 0$. The
general conclusion is that the two procedures give the same result for a function $\mrH$ of the following form:
\bq
\mrH = \Bigl[ \prod_{i=1}^{\mrN}\,\int_{\mrL_i}\,\frac{\mrd \mrs_i}{2\,i\,\pi} \Bigr]\,
\Bigl[ \prod_{i=1}^{\mrN}\,\eG{ - \mrs_i} \Bigr]\,
\Uppsi(\mathbf{s})\,
\prod_{i=1}^{\mrN}\,\mrz_i^{\mrs_i} \spc
\quad
\Uppsi = \frac{ \prod_{\mrj=1}^{\mrm}\,\eG{\alpha_{\mrj} + \mra^{\mrk}_{\mrj}\,\mrs_{\mrk}}}
              { \prod_{\mrj=1}^{\mrn}\,\eG{\beta_{\mrj} + \mrb^{\mrk}_{\mrj}\,\mrs_{\mrk}}} \spc
\label{Hfact}
\eq
where $\mrL_{\mrj}$ is such that among all the poles of the integrand only the poles of $\eG{ - \mrs_{\mrj}}$
lie to the right of $\mrL_{\mrj}$. Stated differently, before using \eqn{MBrep} we must perform a change
of variables bringing $\mrH$ in the (factorized) form of \eqn{Hfact}.

It is beyond the scope of this work to discuss solution of \eqn{Hsystem}; for more details we refer to the
work of \Brefs{Ghs,SSib,Horns} where the necessary and sufficient conditions for a formal solution in the 
class of Horn series.
The use of partial differential equations is also important in order to understand singularities of
multivariate hypergeometric functions; in general the goal is to study the singularities of
hypergeometric functions which are defined by means of analytic continuation of hypergeometric series.  
We refer to the work of \Bref{passare2004singularities} where the
analysis is restricted to the case of nonconfluent hypergeometric functions, \ie those where
$\mathrm{deg}\,\mrP_i = \mathrm{deg}\,\mrQ_i$.
Theorems on the dimension of the space of hypergeometric functions and on
a basis in the space of hypergeometric functions can be found in \Bref{GZK} and in Sect.~5 of \Bref{Ghs}.

In the second part of this work we will discuss the connection between Feynman integrals and Fox functions. 
Here we have shown that a large class of Fox functions can be written in terms of Horn hypergeometric functions
(see Sect.~2 of \Bref{Ghs}). It is well{-}known that general Horn functions can also be defined as
solutions of the A{-}systems of \Bref{GZK}. The Horn series and the GKZ series are related, \ie
each GKZ series can be represented as a product of a monomial and a Horn series.
\section{Feynman diagrams \label{FD}}
Starting with this section we give a few examples of the treatment of Feynman integrals.
Numerical techniques for the evaluation of one{-}loop Feynman integrals are discussed in
\Bref{Ferroglia:2002mz}; for two{-} loop integrals in
\Brefs{Passarino:2001jd,Actis:2004bp,Passarino:2006gv,Passarino:2016zcd}.
Alternative methods have been introduced in \Brefs{Borowka:2019zhf,Smirnov:2021rhf}. 
Extensions to include the SMEFT can be found in \Brefs{Ghezzi:2015vva,Passarino:2019yjx,Camponovo:2022wwn}.
Series representations for generalized Feynman integrals with notes on numerical evaluation are discussed
in \Bref{Klausen:2023gui}. Canonical differential equations have been introduced in \Bref{Henn:2013pwa},
Yangian Bootstrap in \Bref{Loebbert:2020glj}.
Other results on MB expansion can be found in
\Brefs{Usyukina:1993ch,Anastasiou:2005cb,Gluza:2007bd,Smirnov:2009up}; some of the MB tools can be found
at {\url{http://projects.hepforge.org/mbtools/}}.

A generalized two-loop diagram~\footnote{
Note that in our metric spacelike $\mrp$ implies positive
$\mrp^2$. Further $\mrp_4 = i\,\mrp_0$, with $\mrp_0$ real for a physical four{-}momentum.} 
is defined with arbitrary, non-canonical, powers for its 
propagators and will be represented as
\bqa
{}&{}&
\mrG^{\sigma\delta\gamma}(\mu_1\,,\,\cdots\,,\,\mu_{\ssR}\,|\,
\nu_1\,,\,\cdots\,,\,\nu_{\ssS}\,;\,
\{\eta^1\,\mrp\}\,,\,\{\eta^{12}\,\mrp\}\,,\,\{\eta^2\,\mrp\}\,,\,
\{m\}_{a+b+c}) =
\nl
{}&{}&
\frac{\mu^{2 (4 - \mrd)}}{\pi^4}\,
\int\,\mrd^{\mrd} \mrq_1\,\mrd^{\mrd} \mrq_2\,\prod_{\mrr=1}^{\mrR}\,\mrq_{1\mu_r}\,
\prod_{\mrs=1}^{S}\,\mrq_{2\nu_{\mrs}}\,
\prod_{i=1}^{a}\,(\mrk^2_i + \mrm^2_i)^{-\sigma_i}\,
\!\!\!\!\prod_{j=a+1}^{a+c}\,(\mrk^2_{\mrj} + \mrm^2_{\mrj})^{-\gamma_{\mrj}}\,
\!\!\!\!\prod_{l=a+c+1}^{a+c+b} \!\!\!\!(\mrk^2_l + \mrm^2_l)^{-\delta_l},
\label{Gdiag}
\eqa
where $\mrd = 4 - \ep$, with $\mrd$ being the space-time dimension, and where
$a, b$ and $c$ give the number of lines (each with its multiplicity) in the 
$\mrq_1, \mrq_2$ and $\mrq_1-\mrq_2$ loops respectively. 

For standard functions (i.e.\ those where all the propagators have canonical power $-1$), 
we have $\sigma= a, \delta= b$ and $\gamma= c$; for generalized functions we have
$\sigma = \sum_{i=1}^a \sigma_i$ etc. Furthermore we have
\[
\ba{ll}
\mrk_i = \mrq_1+\sum_{j=1}^{\ssN}\,\eta^1_{ij}\,\mrp_{\mrj} \, , \;&\; i=1,\dots,a \,  , \\
\mrk_i = \mrq_1-\mrq_2+\sum_{j=1}^{\ssN}\,\eta^{12}_{ij}\,\mrp_{\mrj} \, , \;&\;
i=a+1,\dots,
a+c \,  , \\
\mrk_i = \mrq_2+\sum_{j=1}^{\ssN}\,\eta^2_{ij}\,\mrp_{\mrj} \, , \;&\;
i=a+c+1,\dots,
a+c+b \, ,
\ea
\]
$\mrN$ being the number of vertices, $\eta^a = \pm 1,$ or $0$ and
$\{\mrp\}$ the set of external momenta. Furthermore, $\tHs$ is the arbitrary
unit of mass. 
\subsection{Feynman parameters}
We will not consider the introduction of tensor integrals (for which we refer to \Bref{Actis:2004bp}), 
and introduce scalar integrals
\bq
\mrI = \int\,\mrd^{\mrd} \mrq_1\,\mrd^{\mrd} \mrq_2\,
\prod_{i=1}^{a}\,(\mrk^2_i + \mrm^2_i)^{-\sigma_i}\,
\!\!\!\!\prod_{j=a+1}^{a+c}\,(\mrk^2_{\mrj} + \mrm^2_{\mrj})^{-\gamma_{\mrj}}\,
\!\!\!\!\prod_{l=a+c+1}^{a+c+b} \!\!\!\!(\mrk^2_l + \mrm^2_l)^{-\delta_l} \spp
\eq
In the following $\mrL= a + b + c$ denotes the number of internal legs, $\mrN$ the number of 
external legs and
\[ 
\upnu_i = \left\{
\begin{array}{l}
\sigma_i, \quad i= 1\,\dots\,a \\
\gamma_i, \quad i= a+1\,\dots\,a+c \\
\delta_i, \quad i= a+c+1\,\dots\,a+c+b
\end{array}
\right.
\]
with $\upnu = \sum_{i=1}^{\mrL}\,\upnu_i$. Feynman parameters are introduced
\bq
\mrI = \frac{\Gamma(\upnu)}{\sum_{i=1,\mrL}\,\Gamma(\upnu_i)}\,
\int d^{\mrL}\alpha\,\delta\lpar 1 - \sum_{i=1,\mrL}\,\alpha_i \rpar\,
\prod_{i=1,\mrL}\,\alpha^{\upnu_i - 1}_i\,
\int\,\mrd^{\mrd} \mrq_1\,\mrd^{\mrd} \mrq_2\,\mathcal{I}^{-\upnu} \spc
\eq
where the function $\mathcal{I}$ is defined by
\bq
\mathcal{I} = \sum_{i,j=1,2}\,\mrq_i\,\mrM_{ij}\,\mrq_{\mrj} + 2\,\sum_{i=1,2}\,\spro{\mrQ_i}{\mrq_i} +
\mrJ
\eq
\bq
\mrM_{11} = \sum_{i=1,a+c}\,\mrx_i \spc \quad
\mrM_{22} = \sum_{i=a+c+1,a+c+b}\,\mrx_i \spc \quad
\mrM_{12} = - \frac{1}{2}\,\sum_{i=a+1,a+c}\,\mrx_i \spc 
\eq
\bq
\mrQ_1 = \sum_{j=1,\mrN}\,\Bigl[ \sum_{i=1}^{a}\,\eta^1_{ij} + 
                                 \sum_{i=a+1}^{a+c}\,\eta^{12}_{ij} \Bigr]\,\mrx_i\,\mrp_{\mrj} \spc
\quad
\mrQ_2 = \sum_{j=1,\mrN}\,\Bigl[ \sum_{i=a+c+1}^{a+c+b}\,\eta^2_{ij} - 
                                 \sum_{i=a+1}^{a+c}\,\eta^{12}_{ij} \Bigr]\,\mrx_i\,\mrp_{\mrj} \spc
\eq
\bq
\mrJ = \sum_{i=1,a}\,\mrx_i\,\Bigl[ \mrm^2_i + \sum_{j,k=1,\mrN}\,\eta^1_{ij}\,\eta^1_{ik}\,
\spro{P_{\mrj}}{\mrp_k} \Bigr] + \cdots
\eq
Integrating over momenta gives
\bq
\int\,\mrd^{\mrd} \mrq_1\,\mrd^{\mrd} \mrq_2\,\mathcal{I}^{-\upnu} = 
- \pi^{\mrd}\,\frac{\Gamma(\upnu - \mrd)}{\Gamma(\upnu)}\,
\mrS^{\upnu - 3/2\,\mrd}_1\,\mrS^{\mrd - \upnu}_2 \spc
\label{pfac}
\eq
where the two Symanzik polynomials~\cite{Weinzierl:2013yn} are
\bq
\mrS_1 = \mathrm{det}\,\mrM \spc
\qquad 
\mrS_2 = \mathrm{det}\,(\mrM)\,\Bigl( \mrJ - \mrQ\,\mrM^{-1}\,\mrQ \Bigr) \spp
\eq
In the rest of the paper we will omit the pre{-}factor of \eqn{pfac}.

A few remarks on the two Symanzik polynomials are in order: both polynomials
are homogeneous in the Feynman parameters, $\mrS_1$ is of degree $l$,
$\mrS_2$ is of degree $l+1$. The polynomial $\mrS_1$ is linear in each Feynman parameter. 
If all internal masses are zero, then also $\mrS_2$ is linear in each Feynman parameter. 
In expanded form each monomial of $\mrS_1$ has coefficient $+\,1$.

It is important to underline that the original MB{-}expansion of \Bref{Boos:1990rg} works in momentum space while
our approach starts in the Feynman parameter space.
\subsection{Polylogarithms, elliptic polylogarithms and harmonic polylogarithms\label{pleplhpl}}
Polylogarithms and their extensions are basic building blocks in quantum field theory. 
In this Section we present their representation in terms of Fox function.

\ovalbox{Nielsen polylogarithms}.
The role of generalized Nielsen polylogarithms~\cite{poll,Devoto:1983tc,Kolbig:1983qt,Goncharov:2001iea} in the 
evaluation of one{-}loop Feynman diagrams is well{-}know. To illustrate their connection with Fox functions we consider 
a simple example which underlines the various steps used in our procedure:
\bq
\mrJ_2 = \int_0^1 \mrd \mrx\,\mrd \mry\,\lpar 1 - \mrz\mrx \mry \rpar^{-1} \spp
\eq
Of course $\mrJ_2$ is the function $\li{2}{\mrz}$; however we can write
\bq
\mrJ_2 = - \int_0^1 \frac{\mrd \mrx}{x}\,\ln(1 - \mrz\,\mrx) = \mrz\,\int_0^1\, \mrd \mrx\,
\hyp{1}{1}{2}{\mrz\,\mrx} \spc
\eq
which requires $\mid \marg( 1 - \mrz\,\mrx ) \mid < \pi$. The Gauss hypergeometric function is rewitten as
\bq
\int_{\mrL}\,\frac{\mrd \mrs}{2\,i\,\pi}\,
\frac{\eG{ - \mrs}\,\eGs{1 + \mrs}}{\eG{2 + \mrs}}\,\lpar - \mrz\,\mrx \rpar^{\mrs} \spc
\eq
which requires $ - 1 < \Re \mrs < 0$ and the convergence relation $\mid \marg( - \mrz) \mid < \pi$. 
Therefore, the result is a bivariate Fox $\mrI$ function, see \eqn{gmIff}. The analytic properties can be analyzed 
by studying the case $\mid \mrz \mid > 1$. Note that for $\mrz \in \Rf$ we will use $\mrz \to \mrz - i\,\delta$ with 
$\delta \to 0_{+}$. The presence of the pole at $\mrs = - 1$ generates a residue given by
\bq
- \zeta(2) - \frac{1}{2}\,\ln^2( - \mrz) \spc
\eq
which gives the correct imaginary part to $\mrJ_2$. 
To understand the existence of branch cuts in Fox(Meijer) functions we consider the following simple example:
\bq
\mrJ_1 = \int_0^1 \mrd \mrx\,\ln(1 - \mrz\,\mrx) = 
\lpar1 - \frac{1}{\mrz} \rpar\,\ln(1 - \mrz) - 1 \spp
\eq
The integral can also be written as follows:
\bq
\mrJ_1 = \int_{\mrL}\,\frac{\mrd \mrs}{2\,i\,\pi}\,\frac{\eGs{1 + \mrs}\,\eG{ - \mrs}}{\eG{3 + \mrs}}\,
\lpar - \mrz \rpar^{\mrs + 1} \spc
\eq
which can be splitted into three contributions:
\begin{enumerate}

\item double pole at $\mrs = - 1\;$; $\mrJ_{11} = - 1 + \ln( - \mrz)$ ,

\item double pole at $\mrs = - 2\;$; $\mrJ_{12} = \frac{1}{\mrz} - \frac{1}{\mrz}\,\ln( - \mrz)$ ,

\item simple poles at $\mrs = - \mrn - 1$, giving 
$\mrJ_{13} = \sum_{\mrn=2}^{\infty}\,\frac{1}{\mrn\,(\mrn - 1)}\,\mrz^{ - \mrn}$ .

\end{enumerate}
The correct branch cut is obtained when we resum in $\mrJ_{13}$ obtaining
\bq
\mrJ_{13} = - \frac{1}{\mrz}\,\Bigl[ 1 + \ln(1 - \mrz) - \ln( - \mrz) \Bigr] - \ln( - \mrz) \spc
\eq
\ie the correct result with a cancellation of the spurious $\ln( - \mrz)$ terms. 
Analytic continuation of hypergeometric functions in the logarithmic (resonant) case requires special consideration; 
results for the Gauss hypergeometric function can be found in Sect.~$2.10$ of \Bref{HTF} and they have been extended 
to cover Lauricella functions in Sect.~$2.5$ of \Bref{FDMB} (see also \Bref{Bez}) and Meijer functions 
in \Bref{Scheidegger:2016ysn}.

The existence of a branch cut for $\mrG$ depends on the parameters of the function. The simplest example is 
as follows~\cite{TIF}:
\bq
\mrG^{1\,,\,1}_{1\,,\,1} = \int_{\mrL}\,\frac{\mrd \mrs}{2\,i\,\pi}\,\eG{\mrb + \mrs}\,\eG{1 - \mra - \mrs}\,
\mrz^{ - \mrs} = \eG{1 - \mra + \mrb}\,\mrz^{\mrb}\,\lpar 1 + \mrz \rpar^{\mra - \mrb - 1} \spc
\eq
where the cut is along $( - \infty\,,\,0 )$ if $\mrb \not\in \Zf$ and $\mrb - \mra \not\in \Zf$; however, when
$\mrb \in \Zf$ the cut is along $( - \infty\,,\, - 1 )$. When $\mrp = \mrq$ the branch cuts can appear from
more general hypergeometric functions. If we consider the linear differential equation satisfied by $\mrG$
(see Sect.~5.4 of \Bref{HTF}) in the case $\mrp = \mrq$ we find that $\mrz = ( - 1)^{\mrp - \mrm - \mrn}$ is
a regular singularity. Dealing with singularities of multivariate hypergeometric functions is a much more
complicated problem and we refer to the work of \Brefs{PST,matsubaraheo2023generalized}.

Other examples of Nielsen polylogarithms are
\bqa
\li{3}{\mrz} &=& - \int_{\mrL}\,\frac{\mrd \mrs}{2\,i\,\pi}\,
\frac{\eG{ - \mrs}\,\Gamma^4\lpar 1 + \mrs \rpar}{\Gamma^3\lpar 2 + \mrs \rpar}\,\lpar - \mrz \rpar^{1 + \mrs} \spc
\nl
\mrS_{1\,,\,2} &=& \frac{1}{2}\,\Bigl[ \prod_{i=1}^{2}\,\int_{\mrL_i}\,\frac{\mrd \mrs_i}{2\,i\,\pi} \Bigr]\,
\frac{\prod_{i=1}^{2}\,\eG{ - \mrs_i}\,\eGs{1 + \mrs_i}}{\prod_{i=1}^{2}\,\eG{2 + \mrs_i}}\,
\frac{\eG{2 + \mrs_1 + \mrs_2}}{\eG{3 + \mrs_1 + \mrs_2}}\,
\lpar - \mrz \rpar^{2 + \mrs_1 + \mrs_2} \spp
\eqa
\ovalbox{Elliptic polylogarithms}~\cite{Adams:2016xah,Adams:2016sob,Bloch:2016izu,Adams:2015ydq,Bloch:2013tra,Broedel:2017kkb} 
also play a role in computing Feynman diagrams. We need an expression for $\li{\mrn}{\mrz}$
which can be obtained recursively starting from $\li{2}{\mrz}$:
\bq
\li{\mrn}{\mrz} = \int_0^{\mrz}\,\frac{\mrd \mrx}{\mrx}\,\li{\mrn - 1}{\mrx} =
- \int_{\mrL}\,\frac{\mrd \mrs}{2\,i\,\pi}\,\eG{ - \mrs}\,\eG{1 + \mrs}\,
\frac{( - \mrz)^{\mrs+1}}{(\mrs + 1)^{\mrn}} \spp
\label{Lin}
\eq
The presence of $(\mrs + 1)^{ - \mrn}$ suggests the use of the following identity~\cite{BSid}:
\bq
\int_{\mrL}\,\frac{\mrd \mrs}{2\,i\,\pi}\,\phi(\mrs)\,\lpar \mrs + \eta \rpar^{ - \mrn - 1}\,\mrz^{ - \mrs} =
\frac{\mrz^{\eta}}{\eG{\mrn + 1}}\,\int_{\mrz}^{\infty}\,\mrd \mrx\,\ln^{\mrn}\lpar \frac{\mrx}{\mrz} \rpar\,
\mrx^{ - \eta - 1}\,\mrH(\mrx^{-1}) \spc
\quad
\mrH(\mrz) = \int_{\mrL}\,\frac{\mrd \mrs}{2\,i\,\pi}\,\phi(\mrs)\,\mrz^{\mrs} \spp
\eq
Following \Bref{Passarino:2016zcd} we consider the following function of depth two:
\bq
\mathrm{ELi}_{\mrn\,,\,\mrm} = \sum_{j=1}^{\infty}\,\frac{\mrx^j}{j^{\mrn}}\,\li{\mrm}{\mry\,\mrq^j} \spc
\eq
where $\mrq \in \Cf^x$, with $\mid \mrq \mid < 1$ and $\mry \in \Cf$ with $1 \not\in \mrq^{\Rf}\,\mry$. Therefore,,
for $\mrx < \mid \mrq^{-1} \mid$ the series converges absolutely. Using the result of \eqn{Lin} we obtain
\bq
\mathrm{ELi}_{\mrn\,,\,\mrm} = - \int_{\mrL}\,\frac{\mrd \mrs}{2\,i\,\pi}\,
\eG{ - \mrs}\,\eG{1 + \mrs}\,\frac{( - \mry )^{\mrs + 1}}{( \mrs + 1 )^{\mrm}}\,
\li{\mrn}{\mrx\,\mrq^{\mrs + 1}} \spp
\eq 
Alternatively, for $0 < q < 1$, we can write~\cite{Passarino:2016zcd} a Watson's contour integral~\cite{WCI}
\bq
\mathrm{ELi}_{1\,,\,0} = - \frac{\mrx\,\mry\,\mrq}{2\,\pi\,i}\int_{\mrL}\,\mrd \mrs
\frac{\pi}{\sin\,\pi \mrs}\,\frac{( - \mrx\,\mrq )^{\mrs}}{\mrs + 1}\,\frac{1}{1 - \mry\,\mrq^{\mrs+1}} \spc
\label{ELiMB}
\eq
where $\mid \mrx\,\mrq \mid < 1$ and $\mid \mathrm{arg}( - \mrx\,\mrq) \mid < \pi$. The contour of integration,
denoted by $\mrL$, runs from $-\,i\,\infty$ to $+\,i\,\infty$ so that the poles at
$\mrs \in \Zf^*$ lie to the right of the contour and the other poles, at $\mrs \in \Zf^{-}$ and
$\mrs = - 1 + (\ln \mry + 2\,\mrm\,\pi\,i)/\omega$ with $\omega= - \ln \mrq$ and $\mrm \in \Zf$,
lie to the left and the latter are at least some $\ep$ ($\ep \to 0_{+}$) distance away from the 
contour.
Note that \eqn{ELiMB} can be generalized~\cite{Passarino:2016zcd} to define the analytic
continuation of $\mathrm{ELi}_{\mrn\,,\,0}$ and can be extended to complex $\mrq$ inside the unit
disc.

A more compact set of results can be obtained by using
\bq
\ELi_{\mrn\,,\,\mrm}(\mrx\,,\,\mry\,;\,\mrq) =
\int_0^1 \frac{\mrd \mrz}{\mrz}\,\ELi_{\mrn - 1\,,\,\mrm}(\mrz \mrx\,,\,\mry\,;\,\mrq) =
\int_0^1 \frac{\mrd \mrz}{\mrz}\,\ELi_{\mrn\,,\,\mrm - 1}(\mrx\,,\,\mrz \mry\,;\,\mrq) \spc
\eq
\[
\frac{1}{1 - \mry\,\mrq^{\mrs + 1}} =
\mrG^{1,1}_{1,1}\,\left( - \mry\,\mrq^{\mrs + 1}\,,\; \Biggl[
\begin{array}{c}
0 \\
0 \\
\end{array}
\Biggr ]
\right)
\]
\[
\int_0^1 \mrd \mrz
\mrG^{\mrm,\mrn}_{\mrp,\mrq}\,\left( \mrz\,\mrx\,,\; \Biggl[
\begin{array}{ccc}
\mra_1 & \dots & \mra_{\mrp} \\
\mrb_1 & \dots & \mrb_{\mrq} \\
\end{array}
\Biggr ]
\right) =
\mrG^{\mrm,\mrn + 1}_{\mrp + 1,\mrq + 1}\,\left( \mrx\,,\; \Biggl[
\begin{array}{cccc}
0 & \mra_1 & \dots & \mra_{\mrp} \\
\mrb_1 & \dots & \mrb_{\mrq} & 1 \\
\end{array}
\Biggr ]
\right) 
\]
For instance we obtain
\[
\ELi_{1,1}(\mrx\,,\,\mry\,;\,\mrq) = - \mrx \mry \mrq\,\int_{\mrL}\,\frac{\mrd \mrs}{2\,i\,\pi}\,
\frac{\pi}{\sin \pi \mrs}\,\frac{( - \mrx\,\mrq^{\mrs})}{\mrs + 1}\,
\mrG^{1,2}_{2,2}\,\left( - \mry\,\mrq^{\mrs + 1}\,,\; \Biggl[
\begin{array}{cc}
0 & 0 \\
0 & 1 \\
\end{array}
\Biggr ]
\right) \spc
\]
\[
\ELi_{2,1}(\mrx\,,\,\mry\,;\,\mrq) = - \mrx \mry \mrq\,\int_{\mrL}\,\frac{\mrd \mrs}{2\,i\,\pi}\,
\frac{\pi}{\sin \pi \mrs}\,\frac{( - \mrx\,\mrq^{\mrs})}{(\mrs + 1)^2}\,
\mrG^{1,2}_{2,2}\,\left( - \mry\,\mrq^{\mrs + 1}\,,\; \Biggl[
\begin{array}{cc}
0 & 0 \\
0 & 1 \\
\end{array}
\Biggr ]
\right) \quad \mathrm{etc.}
\]

For the \ovalbox{harmonic polylogarithms} we use the notations of \Bref{Remiddi:1999ew}:
\bq
\vec{\mrm}_{\mrw} = \lpar \mra\,,\,\vec{\mrm}_{\mrw - 1} \rpar \spc \qquad
\mrH\lpar \vec{\mrm}_{\mrw}\,;\,\mrz \rpar = \int_0^{\mrz} \mrd \mrx\,\mrf( \mra\,;\,\mrx )\,
\mrH\lpar \vec{\mrm}_{\mrw - 1}\,;\,\mrx \rpar \spc
\eq
where we have
\bqa
{}&{}& \mrf( - 1\,;\,\mrx ) = \frac{1}{1 + \mrx} \spc \quad
\mrf(0\,;\,\mrx ) = \frac{1}{\mrx} \spc \quad
\mrf( 1\,;\,\mrx ) = \frac{1}{1 - \mrx} \spc 
\nl
{}&{}&\mrH( - 1\,;\,\mrx ) = \ln(1 + \mrx) \spc \quad
\mrH( 0\,;\,\mrx ) = \ln(\mrx) \spc \quad
\mrH( 1\,;\,\mrx ) = -\,\ln(1 - \mrx) \spp
\eqa
Using \eqn{Lin} we can transform harmonic polylogarithms into Fox function. We give two examples of weight four:
\bqa
\mrH\lpar - 1\,,\,0\,,\,0\,,\,1\,;\,\mrz \rpar &=&
\mrz^2\,\Bigl[ \prod_{i=1}^{2}\,\int_{\mrL_i}\,\frac{\mrd \mrs_i}{2\,i\,\pi} \Bigr]\,
\frac{\eG{ - \mrs_1}\,\eG{ - \mrs_2}\,\eG{3 + \mrs_1}\,\eG{1 + \mrs_2}\,\eG{2 + \mrs_1 + \mrs_2}}
{\eG{3 + \mrs_1 + \mrs_2}}
\nl
{}&\times&
\frac{( - \mrz)^{\mrs_1}\,\mrz^{\mrs_2}}{(\mrs_1 + 1)^4\,(\mrs_2 + 2)} \spc
\eqa
while a more complicated case is
\bqa
\mrH\lpar - 1\,,\,1\,,\,-1\,,\,0\,;\,\mrz \rpar &=& \sum_{\mrj=1}^{3}\,\mrH_{\mrj} \spc
\nl\nl
\mrH_1 &=& \mrz\,(\mrz - 1)\,\Bigl[ \prod_{i=1}^{3}\,\int_{\mrL_i}\,\frac{\mrd \mrs_i}{2\,i\,\pi} \Bigr]
\,\frac{\eG{ - \mrs_1}\,\eG{ - \mrs_2}\,\eG{ - \mrs_3}\,\eG{1 + \mrs_2}\,\eG{1 + \mrs_3}\,
          \eG{3 + \mrs_1}\,\eG{2+ \mrs_2 + \mrs_1}}
         {\eG{3 + \mrs_2 + \mrs_1}}
\nl
{}&\times&
          (\mrs_1 + 1)^{-3}\,(\mrs_1 + 2)^{-1}\,(\mrs_3 + 1)^{-1}\, 
          ( - \mrz)^{\mrs_1}\,\mrz^{\mrs_2}\,\mrz^{\mrs_3}
\nl\nl          
\mrH_2 &=& -\,\mrz^2\,\Bigl[ \prod_{i=1}^{2}\,\int_{\mrL_i}\,\frac{\mrd \mrs_i}{2\,i\,\pi} \Bigr]
\,\frac{\eG{ - \mrs_1}\,\eGs{ - \mrs_2}\,\eG{1 + \mrs_1}\,\eGc{1 + \mrs_2}\,\eGs{2 + \mrs_2 + \mrs_1}}
             {\eG{2 + 2\,\mrs_2 + \mrs_1}\,\eG{3 + \mrs_2 + \mrs_1}}
\nl
{}&\times&
          (\mrs_1 + 1)^{-2}\,(\mrs_2 + 1)^{-1}
         ( - \mrz)^{\mrs_1}\,( - \mrz)^{\mrs_2}
\nl\nl          
\mrH_3 &=& \mrz^2\,\Bigl[ \prod_{i=1}^{2}\,\int_{\mrL_i}\,\frac{\mrd \mrs_i}{2\,i\,\pi} \Bigr]\
\,\frac{\eG{ - \mrs_1}\,\eGs{ - \mrs_2}\,\eG{1 + \mrs_1}\,\eGc{1 + \mrs_2}\,\eGs{3 + \mrs_2 + \mrs_1}}
            {\eG{3 + 2\,\mrs_2 + \mrs_1}\,\eG{4 + \mrs_2 + \mrs_1}}
\nl
{}&\times&
          (\mrs_1 + 1)^{-2}\,(\mrs_2 + 1)^{-1}
          ( - \mrz)^{\mrs_1}\,( - \mrz)^{\mrs_2} \spp
\eqa
It is understood that for $\mrz \in \Rf$ we use $\mrz - i\,\delta$; therefore, for $\mrz < 0$ we have
$\marg(\mrz) = - \pi + \delta$ and for $\mrz > 0$ we have $\marg( - \mrz) = \pi - \delta$.

For completeness we also give an example of $2$dHpls~\cite{Gehrmann:2001jv}:
\bqa
\mrG\lpar 1\,;\, - \mrz\,;\,\mry \rpar &=& - \frac{\mry^3}{\mrz}\,
\Bigl[ \prod_{i=1}^{2}\,\int_{\mrL_i}\,\frac{\mrd \mrs_i}{2\,i\,\pi} \Bigr]
\,\frac{\eG{ - \mrs_1}\,\eG{ - \mrs_2}\,\eG{1 + \mrs_1}\,\eG{1 + \mrs_2}\,\eG{3 + \mrs_1 + \mrs_2}}
     {\eG{4 + \mrs_1 + \mrs_2}}
\nl
{}&\times&
(\mrs_1 + 1)^{-1}\,(\mrs_2 + 1)^{-1}\,\lpar - \mry \rpar^{\mrs_1}\,\lpar\frac{\mry}{\mrz}\rpar^{\mrs_2} \spp
\eqa

\section{One{-}loop Feynman integrals with arbitrary space{-}time dimension \label{OLLP}}
In this Section, we consider integrals associated to one{-}loop Feynman diagrams with arbitrary
space{-} time dimension.  We will not consider Feynman integrals as linear combinations of their canonical series, 
\ie as solutions of a certain Gel’fand-Kapranov-Zelevinsky (GKZ) 
system~\cite{delaCruz:2019skx,Klausen:2019hrg} 
but proceed with direct integration once the Lee{-}Pomeransky~\cite{Lee:2013hzt} representation or the
Feynman representation have been introduced.
\subsection{Lee{-}Pomeransky representation: example  \label{OLLPc}}
Consider the following integral:
\bq
\mrI = \int_0^{\infty}\,\Bigl[ \prod_{i=1,3}\,\mrd \mrx_i\,\mrx_i^{\alpha_i-1} \Bigr]\,
\Bigl( \mrM_{\mra} + \mrM_{\mrb}\,\mrx_3 \Bigr)^{ - \beta} \spc
\eq
\bq
\mrM_{\mra} = \mrc_1\,\mrx_1 + \mrc_2\,\mrx_2 + \mrc_4\,\mrx_1\,\mrx_2 \spc \qquad
\mrM_{\mrb} = \mrc_3 + \mrc_5\,\mrx_1 + \mrc_6\,\mrx_2 \spp
\eq
After the $\mrx_3$ integration we obtain
\bq
\mrI = \eB{\alpha_3}{\be - \alpha_3}\,
\mrc_1^{\alpha_3 - \beta}\,\mrc_6^{ - \alpha_2}\,
\int_0^{\infty} \mrd \mrx_1\,\mrd \mrx_2\,
\mrx_1^{\alpha_1 + \alpha_3 - \beta - 1}\,
\mrx_2^{\alpha_2 - 1}\,\lpar 1 + \mrx_2 \rpar^{-\alpha_3}\,
\xi_1^{\alpha_2 - \alpha_3}\,
\lpar 1 + \frac{\xi_1\,\xi_2}{\mrc_1\,\mrc_6}\,\frac{\mrx_2}{\mrx_1} \rpar^{\alpha_3 - \beta} \spc
\eq
where $\mrB$ is the Euler beta function and we have defined
\bq
\xi_1 = \mrc_3 + \mrc_5\,\mrx_1 \spc \qquad
\xi_2 = \mrc_2 + \mrc_4\,\mrx_1 \spp
\eq
The integration over $\mrx_2$ gives
\bq
\eB{\alpha_2}{\beta - \alpha_2}\,
\hyp{\beta - \alpha_3}{\alpha_2}{\beta}{1 - \frac{\xi_1\,\xi_2}{\mrc_1\,\mrc_6\,\mrx_1}} \spc
\eq
which requires $\Re \beta > \Re \alpha_2$. We also assume that $\alpha_3 - \alpha_2$ is not an integer
(logarithmic case); under this assumption we perform a transformation ($\mrz \to 1 - \mrz$) in the
hypergeometric function and obtain
\bqa
\mrI = \int_0^{\infty} \mrd \mrx\,\frac{1}{\eG{\beta}}\,&\Bigl[&
\eG{\alpha_2}\,\eG{\be - \alpha_3}\,\eG{\alpha_3 - \alpha_2}\,
\mrc_1^{\alpha_3 - \beta}\,\mrc_6^{ - \alpha_2}\,
\mrx^{\alpha_1 + \alpha_3 - \beta - 1}\,\xi_1^{\alpha_2 - \alpha_3}\,\mrJ_1
\nl
{}&+&
\eG{\alpha_3}\,\eG{\be - \alpha_2}\,\eG{\alpha_2 - \alpha_3}\,
\mrc_1^{\alpha_2 - \beta}\,\mrc_6^{ - \alpha_3}\,
\mrx^{\alpha_1 + \alpha_2 - \beta - 1}\,\xi_2^{\alpha_3 - \alpha_2}\,\mrJ_2
\Bigl]
\label{numI}
\eqa
\bqa
\mrJ_1 &=& \hyp{\beta - \alpha_3}{\alpha_2}{1 + \alpha_2 - \alpha_2}{\frac{\xi_1\,\xi_2}{\mrc_1\,\mrc_6\,\mrx}} \spc
\nl
\mrJ_2 &=& \hyp{\alpha_3}{\beta - \alpha_2}{1 + \alpha_3 - \alpha_2}{\frac{\xi_1\,\xi_2}{\mrc_1\,\mrc_6\,\mrx}} \spp
\eqa
The result in \eqn{numI} can be used to perform a numerical integration; we define 
$\mry = \xi_1\,\xi_2/(\mrc_1\,\mrc_6\,\mrx)$ and perform a $\mry \to 1/\mry$ transformation in the two $\shyp{2}{1}$ if
$ \mry > 1$. This condition can be better illustrated if we take into account that $\mrI$ corresponds to a three{-}point
function with momenta $\mrp_1$ and $\mrp_2$. Let
\bq
\mrs = - \lpar \mrp_1 + \mrp_2 \rpar^2 \spc \qquad \mrp_i^2 = - \mu_i^2\,\mrs \spc
\eq
and, for the sake of simplicity, let us assume $\mu_1 = \mu_2 = \mu > 0$. We can easily derive that for $\mrx > 0$
and $\mu < 1$, $\mry$ is positive with a minimum at $(1 + \mu)^2/\mu^2 > 1$.  

However, it is interesting to observe that we can perform also the last integration by using \eqn{F21MB}.
The last integration can be performed giving rise to
$\shyp{2}{1}$ functions which are subsequently rewritten in terms of MB integrals, again using \eqn{F21MB}. The
result is written in terms of
\bq
\eta_1 = \frac{\mrc_2\,\mrc_5}{\mrc_3\,\mrc_4} \spc \qquad
\eta_2 = - \frac{\mrc_3\,\mrc_4}{\mrc_1\,\mrc_6} \spp
\eq
\bq
\mrI = \frac{1}{\eG{\beta}}\,\Bigl[ \prod_{i=1}^{2}\,\int_{\mrL_i}\,
         \frac{\mrd \mrs_i}{2\,i\,\pi} \Bigr]\,\Bigl[
\mrK_1\,\Uppsi_1\lpar \mrs_1\,,\,\mrs_2 \rpar\,\theta_{11}\lpar \mrs_1 \rpar\,\theta_{12}\lpar \mrs_2 \rpar +
\mrK_2\,\Uppsi_2\lpar \mrs_1\,,\,\mrs_2 \rpar\,\theta_{21}\lpar \mrs_1 \rpar\,\theta_{22}\lpar \mrs_2 \rpar
\Bigr]\,\eta_2^{ - \mrs_1}\,\lpar \eta_1 - 1 \rpar^{\mrs_2} \spp
\label{H2var}
\eq
\bqa
\mrK_1 &=& \eG{\alpha_2 - \alpha_3}\,\eG{1 - \alpha_2 + \alpha_3}\,
\lpar \frac{\mrc_2}{\mrc_4} \rpar^{\alpha_1}\,
\lpar \frac{\mrc_1}{\mrc_4} \rpar^{\alpha_2}\,
\lpar \frac{\mrc_2}{\mrc_6} \rpar^{\alpha_3}\,
\lpar \frac{\mrc_4}{\mrc_1\,\mrc_2} \rpar^{\beta} \spc
\nl
\mrK_2 &=& \eG{\alpha_3 - \alpha_2}\,\eG{1 + \alpha_2 - \alpha_3}\,
\lpar \frac{\mrc_2}{\mrc_4} \rpar^{\alpha_1}\,
\lpar \frac{\mrc_3}{\mrc_6} \rpar^{\alpha_2}\,
\lpar \frac{\mrc_1\,\mrc_2}{\mrc_3\,\mrc_4} \rpar^{\alpha_3}\,
\lpar \frac{\mrc_4}{\mrc_1\,\mrc_2} \rpar^{\beta} \spc
\eqa
\bqa
\Uppsi_1 &=& \frac{\eG{\mrs_2 + \mrs_1}\,\eG{\alpha_1 + \alpha_2 - \beta + \mrs_2 + \mrs_1}}
                {\eG{\alpha_2 - \alpha_3 + \mrs_2 + 2\,\mrs_1}} \spc
\nl
\Uppsi_2 &= & \frac{\eG{\alpha_3 - \alpha_2 + \mrs_2 + \mrs_1}\,\eG{\alpha_1 + \alpha_3 - \beta + \mrs_2 + \mrs_1}}
                 {\eG{\alpha_3 - \alpha_2 + \mrs_2 + 2\,\mrs_1}} \spc
\label{Hpsi}
\eqa
\bqa
\theta_{11} &=& \frac{\eG{\alpha_3 - \mrs_1}\,\eG{\beta - \alpha_2 - \mrs_1}\,\eG{\beta - \alpha_1 - \alpha_2 + \mrs_1}}
                        {\eG{1 - \alpha_2 + \alpha_3 - \mrs_1}} \spc
\nl
\theta_{12} &=& \theta_{22} = \eG{ - \mrs_2} \spc
\nl
\theta_{21} &=& \frac{\eG{\mrs_1}\,\eG{\alpha_2 - \mrs_1}\,\eG{\beta - \alpha_3 - \mrs_1}\,
                      \eG{\beta - \alpha_1 - \alpha_2 + \mrs_1}}
                     {\eG{\alpha_3 - \alpha_2 + \mrs_1}\,\eG{1 + \alpha_2 - \alpha_3 - \mrs_1}} \spp
\label{Hthet}
\eqa
The result of \eqn{H2var} can be expressed in terms of two bivariate Fox $\mrH$ functions:
\bq
\mrH_1\lpar \eta_1\,,\,\eta_2 \rpar =
\mrH^{0\,,\,2\,;\,2\,,\,4\,;\,1\,,\,0}_{2\,,\,1\,;\,5\,,\,5\,;\,0\,,\,1} \spc
\qquad
\mrH_2\lpar \eta_1\,,\,\eta_2 \rpar =
\mrH^{0\,,\,2\,;\,2\,,\,2\,;\,1\,,\,0}_{2\,,\,1\,;\,3\,,\,3\,;\,0\,,\,1} \spc
\eq
The convergence of the two series can be discussed following \Bref{compH}. With $\Omega_1$ and $\Omega_2$ defined
in \eqn{Hconv} we have that for $\mrH_1$ the result is $\Omega_1 = 3$ and $\Omega_2 = 2$; for $\mrH_2$ we 
find $\Omega_{1,2} = 2$. Therefore, we have $\mid \marg(\eta_{1,2}) \mid < \pi$.

Following the theorem of \Bref{HS} we obtain
\bq
\mid \phi_i \mid < \pi \spc \qquad
\mid \phi_1 - \phi_2 \mid < \pi \spc \quad
\mid \phi_1 - 2\,\phi_2 \mid < 3\,\pi \spp
\eq
\subsection{Feynman representation: general aspects \label{FRga}}
We start with a simple example which, however, is essential in understanding the analytical
structure of the corresponding Fox function, \ie the emergence (if any) of branch cuts. 
We introduce $\mrd = 4 + \ep$, where $\mrd$ is the space{-}time dimension and consider the simplest 
scalar $\mrC_0$ function~\cite{Bardin:1999ak},
\bq
\mrJ_{\mrd} = \mrC_0\lpar \,0\,,\,0\,,\,\mrs\,;\,0\,,\,\mrM\,,\,0 \rpar
 = \pi^{\ep/2}\,\eG{1 - \frac{\ep}{2}}\,\int_0^1 \mrd \mrx_1\,\int_0^{\mrx_1} \mrd \mrx_2\,\Bigl[
\mrs\,\mrx_1\,\mrx_2 + \mrM^2\,\mrx_1 - ( \mrs + \mrM^2)\,\mrx_2 \Bigr]^{ - 1 + \ep/2} \spp
\eq
After performing the $\mrx_2$ integral we obtain
\bq
\mrJ_{\mrd} = \pi^{\ep/2}\,\eG{1 - \frac{\ep}{2}}\,\int_0^1 \mrd \mrx \; \mrx\,\mrb^{\ep/2 - 1}\,
\hyp{1 - \frac{\ep}{2}}{1}{2}{ - \frac{\mra}{\mrb}\,\mrx} \spc
\eq
where we have introduced
\bq
\mra = - \mrs\,( 1 - \mrx ) - \mrM^2 \spc \qquad
\mrb = \mrM^2\,\mrx \spp
\eq
Next we use the MB representation for the Gauss hypergeometric function, see \eqn{F21MB}, and perform the 
$\mrx$ integration by using
\bq
\int_0^1 \mrd \mrx\,\mrx^{\ep/2}\,\Bigl(1 - \frac{\mrs}{\mrs + \mrM^2}\,\mrx \Bigr)^{\mrs_1} =
\frac{\eG{1 + \ep/2}}{\eG{2 + \ep/2}}\,
\hyp{ - \mrs_1}{1 + \frac{\ep}{2}}{2 + \frac{\ep}{2}}{\frac{\mrs}{\mrs + \mrM^2}} \spp
\eq
Using again a MB representation we obtain
\bqa
\mrJ_{\mrd} &=& \pi^{\ep/2}\,\mrM^{\ep - 2}\,\Bigl[ \prod_{i=1}^{2}\,\int_{\mrL_i}\,\frac{\mrd \mrs_i}{2\,i\,\pi} \Bigr]\,
\frac{\mrN}{\mrD}\,
\lpar - \frac{\mrs + \mrM^2}{\mrM^2} \rpar^{\mrs_1}\,
\lpar - \frac{\mrs}{\mrs + \mrM^2} \rpar^{\mrs_2} \spc
\nl
\mrN &=&
\eG{ - \mrs_2}\,\eG{1 + \mrs_1}\,\eG{\mrs_2 - \mrs_1}\,
\eG{1 - \frac{\ep}{2} + \mrs_1}\,\eG{1 + \frac{\ep}{2} + \mrs_2} \spc
\nl
\mrD &=& \eG{2 + \mrs_1}\,\eG{2 + \frac{\ep}{2} + \mrs_2} \spp
\eqa
Therefore, $\mrJ$ is a generalized Fox function of \sect{GMFF} with parameters $\mrr = 2, \mrm = 5$ and $\mrn = 2$. 
We can study the beahvior of $\mrJ$ as a function of $\ep$.

\ovalbox{$\mrd = 4\; (\ep = 0)$}. In this case we have
\bqa
\mrJ_4 &=& \frac{1}{\mrM^2}\,\Bigl[ \prod_{i=1}^{2}\,\int_{\mrL_i}\,\frac{\mrd \mrs_i}{2\,i\,\pi} \Bigr]\,
\frac{\mrN_4}{\mrD_4}\,
\lpar - \frac{\mrs + \mrM^2}{\mrM^2} \rpar^{\mrs_1}\,
\lpar - \frac{\mrs}{\mrs + \mrM^2} \rpar^{\mrs_2} \spc
\nl
\mrN_4 &=& \eG{ - \mrs_2}\,\eGs{1 + \mrs_1}\,\eG{1 + \mrs_2}\,\eG{\mrs_2 - \mrs_1} \spc
\nl
\mrD_4 &=& \eG{2 + \mrs_1}\,\eG{2 + \mrs_2} \spp
\label{labJ4}
\eqa
The integral is defined by ($\mrs_{\mrj} = \mrt_{\mrj} + i\,\gamma_{\mrj}$),
\bq
- 1 < \mrt_1 < 0 \spc \qquad
\mrt_1 < \mrt_2 < 0 \spp
\eq
It is also possible to change variable $\mrs_1 \to - \mrs_1$ and use 
\bq
 \frac{\eGs{1 - \mrs_1}}{\eG{2 - \mrs_1}} = \frac{\mrs_1}{\mrs_1 - 1}\,\eG{ - \mrs_1} \spc
\eq
bringing the integral in \eqn{labJ4} in a form suitable for deriving a Horn system of partial differential equations.

We select the poles at $\mrs_2 = \mrn \in \Zf^*$ to obtain
\bq
\mrJ_4 = \frac{1}{\mrM^2}\,\int_{\mrL_1}\,\frac{\mrd \mrs_1}{2\,i\,\pi}\,
\sum_{n=0}^{\infty}\,( - 1)^{\mrn}\, 
\lpar - \frac{\mrs + \mrM^2}{\mrM^2} \rpar^{\mrs_1}\,
\lpar - \frac{\mrs}{\mrs + \mrM^2} \rpar^{\mrn}\,
\frac{\eG{1 + \mrs_1}\,\eG{\mrn - \mrs_1}}{(\mrs_1 + 1)\,\eG{\mrn + 2}} \spp
\eq
We have a double pole at $\mrs_1 = - 1$. Let us consider the corresponding contribution to $\mrJ_4$. Using
\bqa
\eG{ 1 + \mrs_1} &=& \frac{1}{\mrs_1 + 1} + \uppsi(1) + \ord{(\mrs_1 + 1)} \spc
\nl
\eG{\mrn - \mrs_1} &=& \eG{\mrn + 1}\,\Bigl[ 1 + (\mrs_1 + 1)\,\uppsi(\mrn + 1) + \ord{(\mrs_1 + 1)^2} \Bigr]
\spc
\nl
\mrz^{\mrs_1} &=& \frac{1}{\mrz}\,\Bigl[ 1 + (\mrs_1 + 1)\,\ln(\mrz) + \ord{(\mrs_1 + 1)^2} \Bigr] \spc
\eqa
and also using the Feynman prescription
\bq
\ln \lpar - 1 - \frac{\mrs}{\mrM^2} \rpar = \ln \lpar 1 + \frac{\mrs}{\mrM^2} \rpar - i\,\pi \spc
\eq
we end up with
\bq
\sum_{\mrn=0}^{\infty}\,\lpar 1 + \frac{\mrM^2}{\mrs} \rpar^{- \mrn}\,\frac{1}{1 + \mrn} =
\lpar 1 + \frac{\mrM^2}{\mrs} \rpar\,\ln\lpar 1 + \frac{\mrs}{\mrM^2} \rpar \spc
\eq
which gives an imaginary part to $\mrJ_4$,
\bq
\Im\,\mrJ_4 = \frac{\pi}{\mrs}\,\ln\lpar 1 + \frac{\mrs}{\mrM^2} \rpar \spc
\eq
which is the correct result~\cite{Bardin:1999ak} since we know that
\bq
\mrJ_4 = - \frac{1}{\mrs}\,\Bigl[ \zeta(2) - \li{2}{1 + \frac{\mrs}{\mrM^2}} \Bigr] \spp
\eq

\ovalbox{$\mrd = 6\; (\ep = 2)$}. In this case we introduce $\rho = \ep/2 - 1$ and expand around $\rho = 0$,
obtaining a UV{-}divergent part,
\bq
\pi^{-1}\,\mrJ^{\mathrm{UV}}_6 = \frac{1}{\rho} + \frac{1}{2}\,(1 - \gamma) - \ln(\pi) - \ln(\mrM^2) \spc
\eq
where $\gamma$ is the Euler{-}Mascheroni constant, as well a a finite part,
\bqa
\pi^{-1}\,\mrJ^{\mrf}_6 &=& 2\,\Bigl[ \prod_{i=1}^{2}\,\int_{\mrL_i}\,\frac{\mrd \mrs_i}{2\,i\,\pi} \Bigr]\,
\lpar - 1 - \frac{\mrs}{\mrM^2} \rpar^{\mrs_1}\,
\lpar - 1 - \frac{\mrm^2}{\mrs} \rpar^{- \mrs_2}\,\frac{1}{\mrs_1 + 2}
\nl
{}&\times& \Bigl[
\lpar 1 + \frac{\mrs}{\mrM^2} \rpar\,
\frac{
\eG{ - \mrs_2}\,
\eGs{1 + \mrs_1}\,
\eG{2 + \mrs_2}\,
\eG{\mrs_2 - \mrs_1}}
{\eG{2 + \mrs_1}\,\eG{3 + \mrs_2}}
\nl
{}&-& \frac{\mrs}{\mrM^2}\,
\frac{
\eG{ - \mrs_2}\,
\eGs{1 + \mrs_1}\,
\eG{3 + \mrs_2}\,
\eG{\mrs_2 - \mrs_1}}
{\eG{2 + \mrs_1}\,\eG{4 + \mrs_2}} \Bigr] \spp
\eqa
After selecting the poles at $\mrs_2 = \mrn$ we observe the presence of double poles at $\mrs_1 = - 1$ and $\mrs_1 = - 2$.
Both give a contribution to the imaginary part of $\mrJ_6$. From $\mrs_1 = - 1$ we obtain
\bq
i\,\pi\,\Bigl[ \hyp{1}{2}{3}{\frac{\mrs}{\mrs + \mrM^2}} -
\frac{2}{3}\,\frac{\mrs}{\mrs + \mrM^2}\,\hyp{1}{3}{4}{\frac{\mrs}{\mrs + \mrM^2}} \Bigr] \spp
\eq
From $\mrs_1 = - 2$ we obtain
\bq
i\,\pi\,\Bigl[
\frac{2}{3}\,\frac{\mrs \mrM^2}{(\mrs + \mrM^2)^2}\,\hyp{2}{3}{4}{\frac{\mrs}{\mrs + \mrM^2}} -
\frac{\mrM^2}{\mrs + \mrM^2}\,\hyp{2}{2}{3}{\frac{\mrs}{\mrs + \mrM^2}} \Bigr] \spp
\eq
\subsection{Feynman representation: example \texorpdfstring{$1$}{1} \label{OLFb}}
We introduce $\mrd = 4 + \ep$ where $\mrd$ is the space{-}time dimension.
Consider the following scalar three{-}point function:
\bq
\mrJ = \mrC_0\lpar 0\,,\,0\,,\,\mrv\,;\,\mrM\,,\,0\,,\,\mrm, \rpar \spc
\eq
where $\mrv$ is one of the Mandelstam invariants, $\mrs > 0$ or $\mrt, \mru < 0$. 
The Feynman prescription $\mrm^2 \to \mrm^2 - i\,\delta$ with $\delta \to 0_{+}$ is equivalent to
$\mrv \to \mrv + i\,\delta$.
We obtain
\bq
\mrJ = \pi^{\ep/2}\,\eG{1 - \frac{\ep}{2}}\,
\int_0^1 \mrd \mrx_1\,\int_0^{\mrx_1} \mrd \mrx_2\,
\Bigl[ \mrm^2 - \mrm^2\,\mrx_1 + (\mrM^2 - \mrv)\,\mrx_2 + \mrv\,\mrx_1\,\mrx_2 \Bigr]^{\ep/2 - 1} \spp
\eq
After the integration over $\mrx_2$ we obtain
\bq
\mrJ = \pi^{\ep/2}\,\eG{1 - \frac{\ep}{2}}\,\mrm^{\ep - 2}\,\int_0^1 \mrd \mrx\,
\mrx^{\ep/2 - 1}\,\lpar 1 - \mrx \rpar\,\hyp{1 - \frac{\ep}{2}}{1}{2}{(1 - \mrx)\,(\mrz_1 - \frac{\mrz_2}{\mrx})} \spc
\eq
where $\mrv = \mrz_1\,\mrm^2$ and $\mrM^2 = \mrz_2\,\mrm^2$. Next we use the MB representation for $\shyp{2}{1}$,
(\eqn{F21MB}) perform the last integration, use again the MB representation for the resulting $\shyp{2}{1}$ function and
obtain
\bqa
\mrJ &=& \pi^{\ep/2}\,\mrm^{\ep - 2}\,\Bigl[ \prod_{i=1}^{2}\,\int_{\mrL_i}\,\frac{\mrd \mrs_i}{2\,i\,\pi} \Bigr]\,
\Uppsi(\mrs_1\,,\,\mrs_2)\,
\lpar \frac{\mrm^2}{\mrM^2} \rpar^{ - \mrs_1}\,\lpar - \frac{\mrM^2}{\mrv} \rpar^{ - \mrs_2} \spc
\nl
\Uppsi &=& \eG{\mrs_2 - \mrs_1}\,\eG{\frac{\ep}{2} + \mrs_2 - \mrs_1}\,
         \eG{1 + \mrs_1}\,\eG{1 - \frac{\ep}{2} + \mrs_1}\,
         \frac{\eG{ - \mrs_2}}{\eG{2 + \frac{\ep}{2} + \mrs_2}} \spp
\label{bFF}
\eqa
The contours $\mrL_i$ are defined ($\mrs_{\mrj} = \mrt_{\mrj} + i\,\gamma_{\mrj}$) by
\bq
- 1 + \frac{\ep}{2} < \mrt_1 < 0 \quad (\ep < 2) \spc \qquad
\mrt_1 < \mrt_2 < 0 \spp
\eq
The final result in \eqn{bFF} is a bivariate Fox $\mrH$ function.
If $\phi_1$ is the argument of $\mrm^2/\mrM^2$ and $\phi_2$ is the argument of $- \mrM^2/\mrv + i\,\delta$
the region of convergence is defined by
\bq
\mid \phi_i \mid < \pi \spc \qquad
\mid \phi_1 + \phi_2 \mid < \pi \spp
\eq
We can also change $\mrs_1 \to - \mrs_1$ in \eqn{bFF} and write $\eG{1 - \mrs_1} = - \mrs_1\,\eG{ - \mrs_1}$ which
brings the integral in a form suitable for deriving a Horn system of partial differential equations.

If $\mrm = \mrM$ we introduce $\lambda = \mrM^2/\mrv$ and use
\bqa
{}&{}& \int_{\mrL_1}\,\frac{\mrd \mrs_1}{2\,i\,\pi}\,
\eG{1 + \mrs_1}\,\eG{1 - \frac{\ep}{2} + \mrs_1}\,\eG{\mrs_2 - \mrs_1}\,\eG{\frac{\ep}{2} + \mrs_2 - \mrs_1} =
\nl
{}&{}& \frac{1}{\eG{2 + 2\,\mrs_2}}\,\eGs{1 + \mrs_2}\,\eG{1 - \frac{\ep}{2} + \mrs_2}\,
\eG{1 + \frac{\ep}{2} + \mrs_2} \spp
\eqa
The resulting expression for $\mrJ$ becomes
\bqa
\mrJ &=& \frac{1}{2}\,\pi^{\ep/2 + 1/2}\,\eG{1 - \frac{\ep}{2}}\,\mrM^{\ep - 2}\,\mrK \spc \quad
\mrK = \int_{\mrL}\,\frac{\mrd \mrs}{2\,i\,\pi}\,\frac{\mrN}{\mrD}\,\lpar - \frac{1}{4\,\lambda} \rpar^{\mrs} \spc
\nl
\mrN &=& \eG{ - \mrs}\,\eG{1 + \mrs}\,\eG{1 - \frac{\ep}{2} + \mrs}\,\eG{1 + \frac{\ep}{2} + \mrs} \spc
\nl
\mrD &=& \eG{\frac{3}{2} + \mrs}\,\eG{2 + \frac{\ep}{2} + \mrs} \spp
\eqa
The last integral gives a Meijer function, $\mrG^{1\,,\,3}_{3\,,\,3}$ of argument $ - 1/(4\,\lambda)$ and parameters
\bq
0\,,\,\ep/2\,,\,- \ep/2 \spc \qquad 0\,,\,- 1/2\,,\,- 1 - \ep/2 \spp
\eq
 Next we use the following relation:
\bq
\mrK = \frac{\eG{1 - \ep/2}\,\eG{1 + \ep/2}}{\eG{3/2}\,\eG{2 + \ep/2}}\,
\hhyp{1}{1 - \ep/2}{1 + \ep/2}{3/2}{2 + \ep/2}{\frac{1}{4}\,\lambda^{-1}} \spp
\label{itK}
\eq
The advantage of using \eqn{itK} is that $\uppsi_3 = 1/2 + \ep/2$ (see \eqn{defpsi}); therefore, we see from \eqn{GHFato}
that for $\mrd = 3$ we have a logarithmic singularity at $\mrv = 4\,\mrm^2$ and a simple pole at $\mrd = 2$. 
Obviously $\uppsi > 0$ for $\ep$ positive; as a check we shall let $\ep = 0$ and obtain
\bq
\mrJ = \frac{4}{\mrv}\,\Bigl[ \mathrm{arcsin}\lpar \frac{1}{2}\,\lambda^{ - 1/2} \rpar \Bigr]^2 \spc
\eq
which is the correct result, see \Bref{Bardin:1999ak}.
\subsection{Feynman representation: example \texorpdfstring{$2$}{2} \label{OLFc}}
Consider the four{-}point function with massless external legs and internal legs; we obtain
\bq
\mrJ = \pi^{\ep/2}\,\eG{2 - \frac{\ep}{2}}\,
\int_0^1 \mrd \mrx_1 \,\int_0^{\mrx_1} \mrd \mrx_2 \, \int_0^{\mrx_2} \mrd \mrx_3\,
\Bigl[ - \mrs\,(1 - \mrx_1)\,\mrx_2 + \lpar \mrs + \mru\,\mrx_1 + \mrt\,\mrx_2 \rpar\,\mrx_3 \Bigr]^{\ep/2-2} \spc
\eq
where $\mrs, \mrt$ and $\mru = - \mrs - \mrt$ are the Mandelstam invariants. The Feynman prescription is such that
$\mrs$ must be understood as $\mrs + i\,\delta$ while we have $\mrt(\mru) - i\,\delta$. After integrating over $\mrx_3$ we
obtain
\bq
\mrJ = \pi^{\ep/2}\,\eG{2 - \frac{\ep}{2}}\,\lpar - \mrs \rpar^{\ep/2 - 2}\,
\int_0^1 \mrd \mrx_1 \,\int_0^{\mrx_1} \mrd \mrx_2\,
\lpar 1 - \mrx_1 \rpar^{\ep/2 - 2}\,\mrx_2^{\ep/2 - 1}\,\hyp{2 - \frac{\ep}{2}}{1}{2}{\xi} \spc
\quad
\xi = \frac{\mrs + \mru\,\mrx_1 + \mrt\,\mrx_2}{\mrs\,(1 - \mrx_1)} \spp
\eq
If our goal is to compute the expansion of $\mrJ$ around $\mrd = 4 (\ep = 0)$ then we can use
\bq
\hyp{2 - \frac{\ep}{2}}{1}{2}{\xi} = \frac{1}{1 - \xi} + \frac{\ep}{2}\,\frac{1}{\xi\,(1 - \xi)}\,
\Bigl[ \ln(1 - \xi) + 1 \Bigr] + \ord{\ep^2} \spp
\eq
The remaining integrals will produce logarithms and dilogarithms according to a standard procedure.
If we keep $\mrd$ arbitrary the procedure consists in using a MB representation for the $\shyp{2}{1}$
function, see \eqn{F21MB}. We will have
\bqa
\eG{ - \mrr} &\qquad& \hbox{with poles at} \quad \mrr = \mrn_1 \spc
\nl
\eG{2 - \frac{\ep}{2} + \mrr} &\qquad& \hbox{with poles at} \quad \mrr = \frac{\ep}{2} - 2 - \mrn_2 \spc
\nl
\eG{1 + \mrr} &\qquad& \hbox{with poles at} \quad \mrr = - 1 - \mrn_3 \spp
\eqa
When $\ep = 4 + 2\,(\mrn_1 + \mrn_2)$ the poles cannot be separated. However we can proceed with direct
integration: for instance, for $\mrd = 8$ we have $\hyp{0}{1}{2}{\xi} = 1$ \etc
The calculation of $\mrJ$ is then performed by repeating the same step: an $\shyp{2}{1}$ function is
written in terms of a MB representation, the next integral is performed, giving a new $\shyp{2}{1}$
function. Finally, we obtain 
\bqa
\mrJ &=& \pi^{\ep/2}\,( - \mrs )^{\ep/2 - 2}\,
\Bigl[ \prod_{i=1,3}\,\frac{1}{2\,\pi\,i}\,\int_{\mrL_i} \mrd \mrr_i \Bigr]\,
\frac{\mrN}{\mrD}\,
\exp\{ - \mrr_1\,(i\,\pi - i\,\delta)\}
\lpar \frac{\mrs}{\mrt} \rpar^{ - \mrr_2}\,
\lpar \frac{\mrs}{\mru} \rpar^{ - \mrr_3} \spc
\nl
\mrN &=&
\eG{ - \mrr_2}\,
\eG{ - \mrr_3}\,
\eG{1 + \mrr_1}\,
\eG{\frac{1}{2}\,\ep + \mrr_2}\,
\eG{ - 1 + \frac{1}{2}\,\ep - \mrr_1}\,
\eG{2 - \frac{1}{2}\,\ep + \mrr_1}
\nl
{}&\times&
\eG{\mrr_3 + \mrr_2 - \mrr_1}\,
\eG{1 + \frac{1}{2}\,\ep + \mrr_3 + \mrr_2} \spc
\nl
\mrD &=& 
\eG{2 + \mrr_1}\,
\eG{1 + \frac{1}{2}\,\ep + \mrr_2}\,
\eG{\ep + \mrr_3 + \mrr_2 - \mrr_1}\,
 \label{FRe3}
\eqa
In \eqn{FRe3} we recognize a generalized Fox $\mrH$ function, described in \sect{GMFF}, with paramters
$\mrr = 3$, $\mrm = 8$ and $\mrn = 3$. The contours $\mrL_{\mrj}$ are defined by the following conditions:
let $\tau_{\mrj} = \Re\,\mrr_{\mrj}$,
\bq
 - 1 + \frac{\ep}{2} <  \tau_1  < 0 \spc \quad
\mathrm{max}(\tau_1\,,\, - \frac{\ep}{2}) <  \tau_2  < 0 \spc \quad
\mathrm{max}(\tau_1 - \tau_2\,,\,- 1 - \frac{\ep}{2} - \tau_2) <  \tau_3  < 0 \spp 
\eq
Following the analysis of \sect{GMFF} and discarding the sequences
$(\mrj_1\,,\,\mrj_2)$ when they violate the condition in \eqn{Hseq} we obtain
\bq
\mid \phi_i \mid < \pi \spc \quad
\mid \phi_1 + \phi_2 \mid < 2\,\pi \spc \quad
\mid \phi_1 + \phi_3 \mid < 2\,\pi \spc \quad
\mid \phi_2 - \phi_3 \mid < \pi \spp
 \eq
It is worth noting that in this case we have $55$ sequences but only $45$ have rank $2$ and only $6$ sequences are 
different.

A comment is in order: the result for a one{-}loop box diagrams can be written as the sum of $192$ 
dilogarithms~\cite{tHooft:1978jhc} while, with the present method, we have one Fox function; the series
corresponding to arbitrary one{-}loop diagrams can be of high dimension and we need analytic continuation.
A solution has been developed in \Bref{Kershaw:1973km} where, using the results of Sect.~5.9 of \Bref{HTF}, a simple method 
for finding the system of linear differential equations satisfied by any (one{-}loop) Feynman integral is presented.
For a modern approach see \Brefs{Remiddi_1997,Argeri:2007up,Henn:2013pwa,Henn_2015,Lee_2018}.
Furthermore, \Bref{Kershaw:1973km} gives explicit examples for the continuation of power series to region of 
physical interest.
\subsection{Feynman representation: example \texorpdfstring{$3$}{3} \label{OLFd}}
An interesting case is represented by the function
\bq
\mrJ_{\off} = \mrC_0\lpar \mrM\,,\,\mrM\,,\,\mrs\,;\,\mrm\,,0\,,\,\mrm \rpar \spc
\eq
the scalar QED vertex with off{-}shell legs. we obtain
\bqa
\mrJ_{\off} &=& \pi^{\ep/2}\,\lpar \mrm^2 - \mrM^2 \rpar^{ - 1 + \ep/2}\,
\Bigl[ \prod_{i=1}^{3}\,\int_{\mrL_i}\,\frac{\mrd \mrs_i}{2\,i\,\pi} \Bigr]\,\frac{\mrN}{\mrD}\, 
\mrx_0^{\mrs_1}\,( - \xm )^{ - \mrs_2}\,( - \xp )^{ - \mrs_3} \spc
\nl
\mrN &=&
\eG{ - \mrs_2}\,
\eG{ - \mrs_3}\,
\eG{\mrs_2 - \mrs_1}\,
\eG{\mrs_3 - \mrs_1}\,
\eG{1 + \mrs_3 + \mrs_2}\,
\eG{1 + \ep + \mrs_1}\,
\eG{1 - \frac{1}{2}\,\ep + \mrs_1} \spc
\nl
\mrD &=&
\eG{ - \mrs_1}\,
\eG{2 + \mrs_3 + \mrs_2}\,
\eG{2 + \ep + \mrs_1} \spc
\label{offs}
\eqa
which is a generalized Fox function of \sect{GMFF} with parameters $\mrr = 3, \mrm = 7$ and $\mrn = 3$ and also
\bq
\mrx_0 = \mrm^2/(\mrm^2 - \mrM^2) \spc \qquad
\mrx_{\pm} = \frac{1}{2}\,\lpar 1 \pm\, \beta_{\mrM} \rpar \spc \quad
\beta^2_{\mrM} = 1 - 4\,\frac{\mrM^2}{\mrs} \spp
\eq
The advantage of \eqn{offs} is that we can study the interplay of three different limits,
the on{-}shell limit $\mrM \to \mrm$, the collinear limit $\mrM = \mrm \to 0$ and the infrared
limit $\ep \to 0$. There are interesting consequences for the Sudakov form factor when computed for
off{-}shell fermions~\cite{Forte:2020fbc,Jackiw:1968zz,Collins:2011zzd}. 

The on{-}shell limit gives
\bqa
\mrJ_{\on} &=& \frac{\pi^{\ep/2}}{\ep}\,\mrm^{\ep - 2}\,
\Bigl[ \prod_{i=1}^{2}\,\int_{\mrL_i}\,\frac{\mrd \mrs_i}{2\,i\,\pi} \Bigr]\,\frac{\mrN}{\mrD}\, 
( - \xm )^{ - \mrs_1}\,( - \xp )^{ - \mrs_2} \spc
\nl
\mrN &= &
\eG{ - \mrs_1}\,
\eG{ - \mrs_2}\,
\eG{1 + \mrs_2 + \mrs_1}\,
\eG{1 - \frac{1}{2}\,\ep + \mrs_1}\,
\eG{1 - \frac{1}{2}\,\ep + \mrs_2} \spc
\nl
\mrD &=&
\eG{1 - \frac{1}{2}\,\ep}\,
\eG{2 + \mrs_2 + \mrs_1} \spp
\eqa
\label{ons}
Using \eqn{ons} we can study the infrared limit $\ep \to 0$,
\bqa
\mrJ_{\on} &=& \mrm^{-2}\,\Bigl(\mrJ^{\mrd}_{\on} + \mrJ^{\mrf}_{\on}\Bigr) \spc
\nl\nl
\mrJ^{\mrd}_{\on} &=& \frac{1}{2}\,\Bigl[ \frac{2}{\ep} - \uppsi(1) + \ln \pi + \ln \mrm^2 \Bigr]\,
\sum_{\mrk=0}^{\infty}\,\frac{\xm^{ - \mrk}}{\mrk + 1}\,\sum_{\mrn=0}^{\mrk}\,
\lpar \frac{\xm}{\xp} \rpar^{\mrn} \spc
\nl
\mrJ^{\mrf}{\on} &=& \frac{1}{2}\,\int_0^1\,\frac{\mrd \mrx}{1 - \mrx}\,
\sum_{\mrk=0}^{\infty}\,\frac{1}{\mrk + 1}\,\sum_{\mrn=0}^{\mrk}\,\Bigl[
\lpar \frac{\mrx}{\xm} \rpar^{\mrk}\,\lpar \frac{\xm}{\xp\,\mrx} \rpar^{\mrn} +
\lpar \frac{1}{\xm} \rpar^{\mrk}\,\lpar \frac{\xm\,\mrx}{\xp} \rpar^{\mrn} -
2\,\lpar \frac{1}{\xm} \rpar^{\mrk}\,\lpar \frac{\xm}{\xp} \rpar^{\mrn} \spc
\eqa
where in the expansion we have used
\bq
\uppsi(\mrn + 1) = \uppsi(1) + \int_0^1\,\frac{\mrd \mrx}{1 - \mrx}\,\lpar 1 - \mrx^{\mrn} \rpar \spp
\eq
After summing the series and performing the $\mrx\,${-} integral we arrive at
\bq
\mrJ_{\on}= \frac{1}{\ep}\,\frac{1}{\beta\,\mrs}\,\Bigl[
\ln\lpar 1 - \xp^{-1}\rpar - \ln\lpar 1 - \xm^{-1} \rpar \Bigr] + \dots \spc
\eq
which is the expected result~\cite{Bardin:1999ak}. Other combinations of limits can be obtained with the same method.
\subsection{Comparison with alternative approaches \label{CAA}}
%
Our approach is based on MB expansions and we would like to compare to multinomial expansion. We illustrate
the comparison following the evaluation of one{-}loop integrals given in \Bref{Kershaw:1973km}. The $\mrN\,${-}point
one{-}loop function can be written as follows~\cite{Eden:1966dnq}
\bq
\mrI_{\mrN} = \int_0^1\,\Bigl[ \prod_{\mrj+1}^{\mrN}\,\mrd \alpha_{\mrj} \Bigr]\,
\delta\lpar 1 - \sum_{\mrj}\,\alpha_{\mrj} \rpar\, \Bigl(
\sum_{\mrj}\,\mrm_{\mrj}^2\,\alpha_{\mrj} - \sum_{\mri < \mrj}\,\mrz_{\mri \mrj}\,\alpha_{\mri}\,\alpha_{\mrj}
\Bigr)^{2 - \mrN} \spc
\eq
where the variables $\mrz_{\mri \mrj}$ are external masses (squared) or Mandelstam invariants.
The case $\mrN = 4$ proves that the multinomial expansion gives a very compact and elegant result. 
The procedure given in \Bref{Kershaw:1973km} is based on the following steps:
\begin{enumerate}

\item use $\mrz_{\mri \mrj} = \mry_{\mri \mrj} +(\mrm_\mri - \mrm_\mrj)^2$, \ie go to the pseudo{-}threshold,
obtaining an integrand of the following form:
\bq
\lpar \sum_{\mrj}\,\mrm_{\mrj}\,\alpha_{\mrj} \rpar^2 -
\sum_{\mri < \mrj}\,\mry_{\mri \mrj}\,\alpha_{\mri}\,\alpha_{\mrj} \spc
\eq
\item expand using the multinomial theorem, with and integral over Feynman parameters given by
\bq
\int_0^1\,\Bigl[ \prod_{\mrj+1}^{\mrN}\,\mrd \alpha_{\mrj} \Bigr]\,
\prod_{\mrj+1}^{\mrN}\,\alpha_{\mrj}^{\mrn_{\mrj}}\,
\lpar \sum_{\mrj}\,\mrm_{\mrj}\,\alpha_{\mrj} \rpar^{2\,(\mrN - \mrn - 2)} \spc
\eq
where $\sum\,\mrn_{\mrj} = 2\,\mrn$. When $\mrN \ge 4$ the result becomes
\bq
\Bigl[ \prod_{\mrj=1}^{\mrN}\,\frac{\partial^{\mrn_\mrj}}{\partial\,\mrm_{\mrj}^{\mrn_\mrj}} \Bigr]\,
\Bigl[ \sum_{\mrj=1}^{\mrN}\,\frac{\partial}{\partial\,\mrm_{\mrj}} \Bigr]\,
\Bigl[ \prod_{\mrj=1}^{\mrN}\,\frac{1}{\mrm_{\mrj}} \Bigr] \spp
\eq
In particular for $\mrN = 4$ the result is an hypergeometric power series in six variables. 

\end{enumerate}
This procedure cannot be used in the following cases: $\mrN = 3$, one or more massles internal lines and 
arbitrary space{-}time dimensions. Actually the massless cases can be worked out but the corresponding
result is by far more complicated. We provide a simple example where $\mrN = 4$, only $\mrz_{12} \not= 0$
and where $\mrm_4 = 0$ but $\mrm_{1,2,3} = \mrM$. Furthermore, having in mind the BST approach~\cite{Ferroglia:2002mz}, 
we will consider
\bq
\mrI_{\mrl} =
\int_0^1\,\Bigl[ \prod_{\mrj+1}^{\mrN}\,\mrd \alpha_{\mrj} \Bigr]\,
\delta\lpar 1 - \sum_{\mrj}\,\alpha_{\mrj} \rpar\,
\Bigl[\ln\,\mrM^2 + \ln(1 - \alpha_4) +
\ln\lpar 1 - \frac{\alpha_1\,\alpha_2}{1 - \alpha_4}\,\frac{\mrz_{12}}{\mrM^2} \rpar \Bigr]\spp
\eq
\bq
\mrI_{\mrl} = \frac{1}{6}\,\ln\,\mrM^2 - \frac{5}{9} + \Delta\mrI_{\mrl} \spp
\eq
After expanding, and integrating (only the non{-}trivial part) we obtain
\bq
\Delta\mrI_{\mrl} = - \sum_{\mrn=0}^{\infty}\,
\frac{\eG{\mrn + 1}\,\eG{\mrn + 2}}{\eG{2\,\mrn + 6}}\,
\ghyp{4}{3}\lpar
4 + \mrn,2\,,\,1\,,\,1 + \mrn\,;\,6 + 2\,\mrn\,,\,4 + \mrn\,,\,2\,;\,1 \rpar \,
\lpar \frac{\mrz_{12}}{\mrM^2} \rpar^{\mrn + 1} \spp
\eq
evaluating $\ghyp{4}{3}$ at $1$ we finally obtain
\bq
\Delta\mrI_{\mrl} = - \frac{1}{96\,\sqrt{2}}\,\frac{\mrz{_12}}{\mrM^2}\,
\ghyp{4}{3}\lpar 1\,,\,1\,,\,2\,,\,4\,;\,3\,,\,5\,,\,\frac{5}{2}\,;\,\frac{\mrz_{12}}{\mrM^2} \rpar \spc
\label{F43res}
\eq
where we have assumed $\mrz_{12} < \mrM^2$. Therefore the region outside the unit disk requires analytical 
continuation of the $\ghyp{4}{3}$ hypergeometric function. In this simple case we could use the results given
in \Bref{F43AC} but in more realistic examples the analytic continuation is not available. The
result when we use the MB expansion is
\bq
\Delta\mrI_{\mrl} = - \frac{\sqrt{\pi}}{32}\,\frac{\mrz_{12}}{\mrM^2}\,\int_{\mrL} \frac{\mrd \mrs}{2\,i\,\pi}\,
\frac{\eG{ - \mrs}\,\eGs{1 + \mrs}}
     {(\mrs + 2)\,(\mrs + 4)\,\eG{\frac{5}{2} + \mrs}}\lpar - \frac{1}{4}\,\frac{\mrz_{12}}{\mrM^2} \rpar^{\mrs} \spp
\eq
The integral is defined by $ - 1 < \Re(\mrs) < 0$ and for $\mrz_{12} > \mrM^2$ we can evaluate the residues of the poles
at the left of the integration contour. The dominant contribution is
\bq
\Delta\mrI_{\mrL} \sim -\,\frac{1}{6\,\sqrt{2}}\,\Bigl[
\frac{4}{3} + \gamma + \uppsi(\frac{3}{2}) - \ln\lpar -\,\frac{1}{4}\,\frac{\mrz_{12}}{\mrM^2}\rpar \Bigr] \spp
\eq
The standard procedure will produce a result with $27$ terms containing dilogarithms.
The most general extension of \eqn{F43res} can be found in \Bref{HTF}:
\bq
\ghyp{\mrp}{\mrq}\lpar {\mathbf a}\,;\,{\mathbf b}\,;\,\mrz \rpar =
\frac{\prod_{\mrj}\,\mrb_{\mrj}}{\prod_{\mrj}\,\mra_{\mrj}}\,
\mrG^{1\,,\,\mrp}_{\mrp\,,\,\mrq+1}\lpar {\mathbf \mrA}\,;\,0\,,\,{\mathbf \mrB}\,;\,\mrz \rpar = 
\frac{\prod_{\mrj}\,\mrb_{\mrj}}{\prod_{\mrj}\,\mra_{\mrj}}\,
\mrG^{\mrp\,,\,1}_{\mrq+1\,,\,\mrp}\lpar 1\,,\,{\mathbf b}\,;\,{\mathbf a}\,;\, - \frac{1}{\mrz} \rpar \spc
\eq
where $\mrA_{\mrj}= 1 - \mra_{\mrj}$ and $\mrB_{\mrj}= 1 - \mrb_{\mrj}$.
\paragraph{A more symmetric representation} \hspace{0pt} \\
Finally, if we are not looking for the minimal number of MB integrals, we can use a multiple MB splitting from
the very beginning~\cite{Freitas:2010nx}; for $\mrn$ variables the result is
\bq
\lpar \sum_{\mrj=1}^{\mrn}\,\mrx_{\mrj} \rpar^{-\alpha} =
\Bigl[ \prod_{\mrj=1}^{\mrn-1}\,\int_{\mrL_{\mrj}}\,\frac{\mrd \mrs_{\mrj}}{2\,i\,\pi} \Bigr]\,
\frac{\eG{\beta}}{\eG{\alpha}}\,\mrx_{\mrn}^{ - \beta}\,
\prod_{\mrj=1}^{\mrn - 1}\,\eG{\mrs_{\mrj}}\,\mrx_{\mrj}^{\mrs_{\mrj}} \spc \quad
\beta = \alpha - \sum_{\mrj=1}^{\mrn - 1}\,\mrs_{\mrj} \spp
\label{mMBs}
\eq
For instance, the most general one{-}loop four{-}point function can be writted as
\bqa
\mrD_0 &=& \pi^{\ep/2}\,\mra_{33}^{ - 2 + \ep/2}\,
 \Bigl[ \prod_{\mrj=1}^{9}\,\int_{\mrL_{\mrj}}\,\frac{\mrd \mrs_{\mrj}}{2\,i\,\pi} \Bigr]\, 
\Bigl[ \prod_{\mrj=1}^{9}\,\eG{\mrs_{\mrj}} \Bigr]\,
\frac{\mrN}{\mrD}\,\prod_{\mrj=1}^{9}\,\mrz_{\mrj}^{ - \mrs_{\mrj}} \spc
\nl\nl
\mrN &=& 
\eG{ - 1 + \ep + 2\,\mrs_1 + \sigma-2}\,
\eG{ - 2 + \ep + 2\,\sigma_1 + \sigma_4}\,
\eG{ - 3 + \ep + 2\,\sigma_5 + \sigma_3} \spc
\nl
\mrD &=&
\eG{\ep + 2\,\mrs_1 + \sigma_2}\,
\eG{ - 1 + \ep + 2\,\sigma_1 + \sigma_4}\,
\eG{ - 2 + \ep + 2\,\sigma_5 + \sigma_3} \spc
\label{mores}
\eqa
Where we have introduced
\bq
\sigma_1 = \mrs_{123} \spc \quad
\sigma_2 = \mrs_{269} \spc \quad
\sigma_3 = \mrs_{589} \spc \quad
\sigma_4 = \mrs_{4569} \spc \quad
\sigma_5 = \mrs_{123467} \spc 
\eq
with $\mrs_{\mri \mrj \dots \mrk} = \mrs_{\mri} + \mrs_{\mrj} +\,\dots\,+ \mrs_{\mrk}$.
Furthermore we have
\bqa
\mrz_1 &=& \frac{\mra_0}{\mra_{33}} \spc \quad
\mrz_2 = \frac{\mra_1}{\mra_{33}} \spc \quad
\mrz_3 = \frac{\mra_{11}}{\mra_{33}} \spc \quad
\mrz_4 = \frac{\mra_{12}}{\mra_{33}} \spc
\nl
\mrz_5 &=& \frac{\mra_{13}}{\mra_{33}} \spc \quad
\mrz_6 = \frac{\mra_2}{\mra_{33}} \spc \quad
\mrz_7 = \frac{\mra_{22}}{\mra_{33}} \spc \quad
\mrz_8 = \frac{\mra_{23}}{\mra_{33}} \spc \quad
\mrz_9 = \frac{\mra_3}{\mra_{33}} \spc \quad
\eqa
\bqa
\mra_0 &=& \mrm_1^2 \spc \quad
\mra_1 = \mrm_2^2 - \mrM_1^2 - \mrm_1^2 \spc \quad
\mra_2 = \mrm_2^2 - \mrM_1^2 - \mrm_3^2 + \mrs \spc \quad
\mra_3 = - (\mrm_3^2 + \mrM_4^2 - \mrm_4^2 - \mrs) \spc
\nl
\mra_{11} &=& \mrM_1^2 \spc \quad
\mra_{22}= \mrM_2^2 \spc \quad
\mra_{33} = \mrM_3^2 \spc
\nl
\mra_{12} &=& - (\mrM_2^2 + \mrM_1^2 - \mrs) \spc \quad
\mra_{13} = - (\mrM_1^2 + \mrM_3^2 - \mru) \spc \quad
\mra_{23} = - (\mrM_2^2 + \mrM_3^2 - \mrt) \spp
\eqa
The integration contours are defined by
\bq
\mrt_i > 0 \spc \qquad
\mrt_1 < 2 - \frac{\ep}{2} \spc \qquad
\mrt_{\mrj} < 2 - \frac{\ep}{2} - \sum_{\mrk=1}^{\mrj - 1}\,\mrt_{\mrk} \spc
\eq
where $\mrt_{\mrj} = \Re\, \mrs_{\mrj}$. The result in \eqn{mores} corresponds to a Fox function of \sect{GMFF} 
with parameters $\mrr = 9, \mrm = 12$ and $\mrn = 3$.

Although the symmetric representation has its own advantages it makes difficult to reconstruct the analytic
structure of the integral. For instance, we would like to have a better representation around the normal
threshold of the box diagram; the procedure is well defined since we are looking for the leading Landau
singularity of the bubble sub{-}diagram of the box. The original integral is
\bq
\mrD_0 = \int_0^1 \mrd \mrx_1\,\int_0^{\mrx_1} \mrd \mrx_2\,\int_0^{\mrx_2} \mrd \mrx_3\,
\lpar \mra_0 + \sum_{\mrj=1}^{3}\,\mra_{\mrj}\,\mrx_{\mrj} + 
               \sum_{\mri \le \mrj}\,\mra_{\mri \mrj}\,\mrx_{\mri}\,\mrx_{\mrj} \rpar^{ - 2 + \ep/2} \spc
\eq
where $\mra_0 = \mrm_1^2$. We now change variables
\bq
\mrx_2 = \mrx_1 - \mrx_2^{\prime} \spc \quad
\mrx_1 = \mrx_1^{\prime} + \mrX \spc \qquad
\mrX = - \frac{1}{2\,\mrs}\,( \mrm_3^2 - \mrm_1^2 - \mrs) \spp
\eq
The integrand will have transformed coefficients $\mra^{\prime}$ where 
\bqa
\mra_{11}^{\prime} &=& \mrs \spc\quad
\mra_{22}^{\prime} = \mrM_2^2 \spc\quad
\mra_{33}^{\prime} = \mrM_3^2 \spc
\nl
\mra_{12}^{\prime} &=& \mrM_1^2 - \mrM_2^2 - \mrs \spc\quad
\mra_{13}^{\prime} = \mrM_4^2 - \mrM_3^2 - \mrs \spc\quad
\mra_{23}^{\prime} = \mrM_2^2 + \mrM_3^2 - \mrt \spc
\nl
\mra_{2}^{\prime} &=&
           \frac{1}{2}\,\Bigl[ s - \mrM_1^2 - \mrM_2^2 + 2\,\mrm_2^2 - \mrm_3^2 - \mrm_1^2 + 
            (\mrm_3^2 - \mrm_1^2)\,\frac{\mrM_2^2 - \mrM_1^2}{\mrs} \Bigr] \spc
\nl
\mra_{3}^{\prime} &=&
           \frac{1}{2}\,\Bigl[ s - \mrM_3^2 - \mrM_4^2 - \mrm_3^2 + 2\,\mrm_4^2 - \mrm_1^2 + 
            (\mrm_3^2 - \mrm_1^2)\,\frac{\mrM_3^2 - \mrM_4^2}{\mrs} \Bigr] \spc
\nl
\mra_0^{\prime} &=&  - \frac{1}{4\,\mrs}\,\lambda(s\,,\,\mrm_1^2\,,\,\mrm_3^2) \spc
\eqa
where $\mrs, \mrt$ are the Mandelstam invariants and $\lambda$ is the K\"allen function, the relevant quantity to
study the behavior of $\mrD_0$ around the normal threshold when using \eqn{mMBs}.
\section{Two{-}loop Feynman integrals : partial quadratization of Symanzik polynomials\label{TLFI}}
A two loop diagram with $\mrL$ internal lines is described by the two Symanzik polynomials~\cite{Weinzierl:2013yn}
in $\mrL$ variables $\alpha_1\,,\dots\,,\,\alpha_{\ssL}$. The diagram will have
$l_1$ lines with momentum $\mrq_1$, $l_2$ with momentum $\mrq_2$ and $l_{12}$ with momentum
$\mrq_1 - \mrq_2$. 

Partial quadratization is a change of variables~\cite{Yuasa:2011ff} defined as follows:
to the $l_1$ lines we assign parameters $\alpha_1,\,\dots\,,\alpha_{l_1}$; to the $l_{12}$ lines parameters
$\alpha_{l_1+1},\,\dots\,,\alpha_{l_1+l_{12}}$; to the $l_2$ lines parameters
$\alpha_{l_1+l_{12}+1},\,\dots\,,\alpha_{l_1+l_{12}+l_2}$.

Next we perform the following change of variables:
\bq
\alpha_1 = \rho_1\,\mrx_1\;,\;\dots\;,\;
\alpha_{l_1-1} = \rho_1\,\mrx_{l_1-1}\;,\;
\alpha_{l_1} = \rho_1\,\lpar 1 - \sum_{\mrj=1}^{l_1-1}\,\mrx_{\mrj} \rpar \spp
\eq
For $l_{12} = 1$ we introduce
\bq
\alpha_{l_1+1} = \rho_3\;,\;
\alpha_{l_1+2} = \rho_2\,\mrx_{l_1}\;,\;\dots\;,\;
\alpha_{l_1+l_2} = \rho_2\,\mrx_{l_1+l_2-2}\;,\;
\alpha_{l_1+l_2+1} = \rho_2\,\lpar 1 - \sum_{\mrj=l_1}^{l_1+l_2-2}\,\mrx_{\mrj} \rpar \spp
\eq
For $l_{12} = 2$ we introduce
\bqa
\alpha_{l_1+1} &=& \rho_3\,\mrx_{l_1+l_2-1}\;,\;\alpha_{l_1+2} = \rho_3\,\lpar 1 - \mrx_{l_1+l_2-1} \rpar \spc
\nl
\alpha_{l_1 + 3} &=& \rho_2\,\mrx_{l_1}\;,\;\dots\;,\;
\alpha_{l_1+l_2+1} = \rho_2\,\mrx_{l_1+l_2-2}\;,\;
\alpha_{l_1+l_2+2} = \rho_2\,\lpar 1 - \sum_{\mrj=l_1}^{l_1+l_2-2}\,\mrx_{\mrj} \rpar \spc
\eqa
\etc As a result of the transformation we will have $\sum_i\,\rho_i = 1$; $\mrS_1$ is a funtion of
the $\rho$ variables but not of the $\mrx$ variables and $\mrS_2$ is a quadratic form
in the $\mrx$ variables with coefficients that are $\rho$ dependent. 
We also intoduce complementary variables:
\bq
\rho_i + \sigma_i = 1 \spc \quad i= 1,3 \spp
\eq
We describe the transformation with the following examples:

\begin{example}[Example: the sunrise]

The two-loop sunrise diagram~\cite{Tancredi:2016buj,Broedel:2017siw} is shown in Fig.~\ref{fd_sunrise}; 
to line $[i]$ we assign a Feynman parameter $\alpha_i$ and obtain:
\bq
\mrS_1 = \prod_{i < j}\,\alpha_i\,\alpha_{\mrj} \spc \qquad
\mrS_2 = \mrs\,\alpha_1\,\alpha_2\,\alpha_3 + \mrS_1\,\sum_i\,\mrm^2_i\,\alpha_i \spc
\eq
where $\mrp^2= -\,\mrs$. After a projective transformation~\cite{Davydychev:1995mq,Berends:1997vk} we obtain
\bq
\mrS_2= s\,\alpha_1\,\alpha_2\,\alpha_3 - 
 \mrm^2_3\,\alpha_1\,\alpha_2 - \mrm^2_2\,\alpha_1\,\alpha_3 - \mrm^2_3\,\alpha_2\,\alpha_3 \spc
\eq
Introducing
\bq
\alpha_1 = 1 - \rho \spc \qquad
\alpha_2 = \rho\,\mrx \spc \qquad
\alpha_3 = \rho\,(1 - \mrx) \spc
\label{quadsun}
\eq
we obtain quadratization in $\mrx$, \ie
\bq
\mrS_2 = \rho\,\lpar \mra\,\mrx^2 + \mrb\,\mrx + \mrc \rpar\,\mrs \spc
\eq
\bq
\mra = \rho\,\lpar \rho + \mrl^2_1 - 1 \rpar \spc \quad
\mrb = \mrl^2_2 - \mrl^2_3 + \lpar 1 - \mrl^2_1 - \mrl^2_2 + \mrl^2_3 \rpar\,\rho - \rho^2 \spc \quad
\mrc = \mrl^2_1\,\lpar \rho - 1 \rpar \spc
\eq
where $\mrm^2_i= \mrl^2_i\,s$.
\begin{figure}[t]
   \centering
   \includegraphics[width=0.7\textwidth, trim = 30 250 50 80, clip=true]{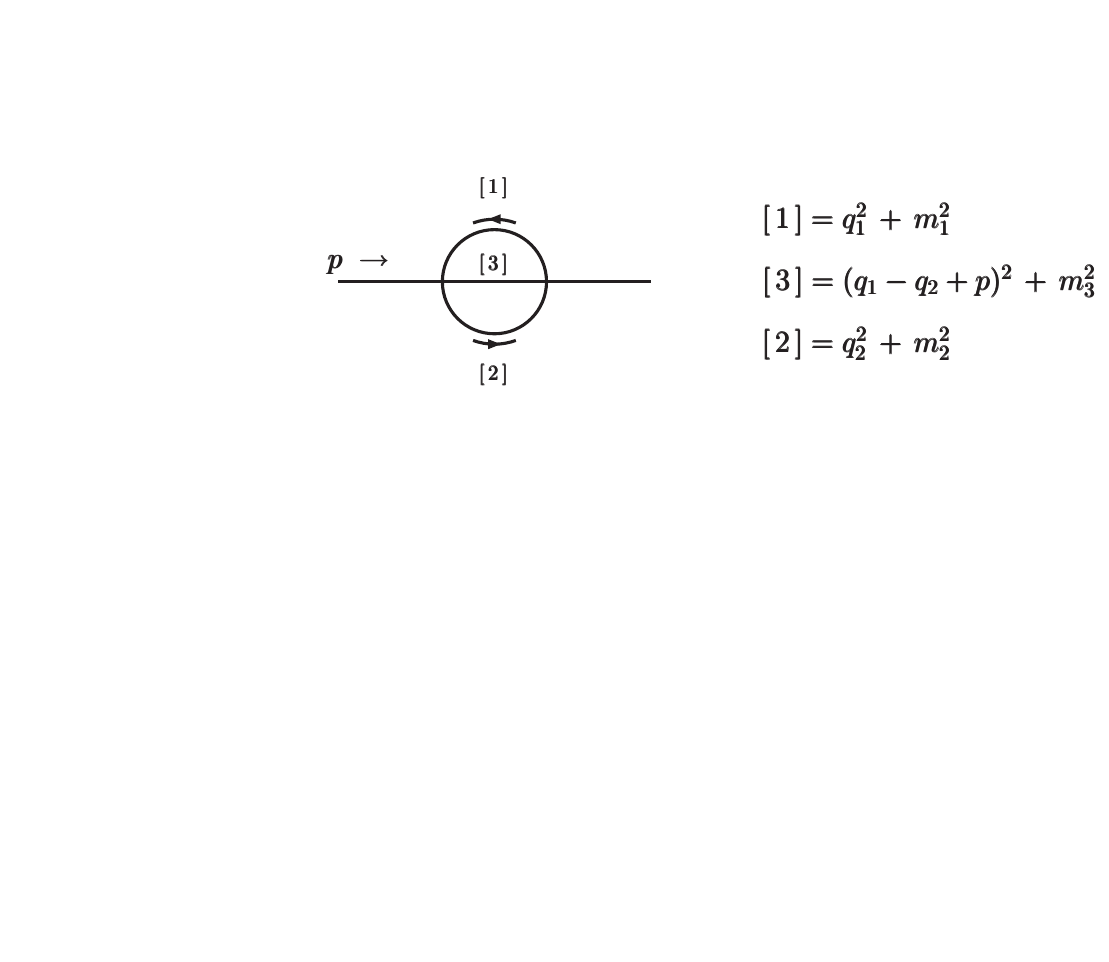}
\vspace{-6.cm}
\caption[]{The two-loop sunrise diagram.}
\label{fd_sunrise}
\end{figure}
\end{example}

\begin{example}[Example: the kite]
The two-loop kite diagram~\cite{Remiddi:2016gno,Bogner:2017vim,Bezuglov:2020ywm,Giroux:2024yxu} is shown in 
Fig.~\ref{fd_kite}; to line $[i]$ we assign a Feynman parameter $\alpha_i$ and obtain:
\bqa
\mrS_1 &=& (\alpha_4 + \alpha_5)\,(\alpha_1 + \alpha_2) + 
           (\alpha_1 + \alpha_2 + \alpha_4 + \alpha_5)\,\alpha_3  \spc
\nl
\mrS_2 &=& \Bigl[ (\alpha_4 + \alpha_5)\,(\alpha_1 + \alpha_2) + 
                  (\alpha_1 + \alpha_2 + \alpha_4 + \alpha_5)\,\alpha_3 \Bigr]\,
           \sum_i\,\mrm^2_i\,\alpha_i
\nl
{}&-&      \Bigl\{ \alpha_1\,\alpha_4\,\alpha_5 + 
           \Bigl[ \alpha_4\,\alpha_5 + \alpha_1\,(\alpha_4 + \alpha_5) \Bigr]\,\alpha_2 + 
           (\alpha_1 + \alpha_4)\,(\alpha_2 + \alpha_5)\,\alpha_3 \Bigr\}\,\mrs \spp
\eqa
\begin{figure}[t]
   \centering
   \includegraphics[width=0.7\textwidth, trim = 30 250 50 80, clip=true]{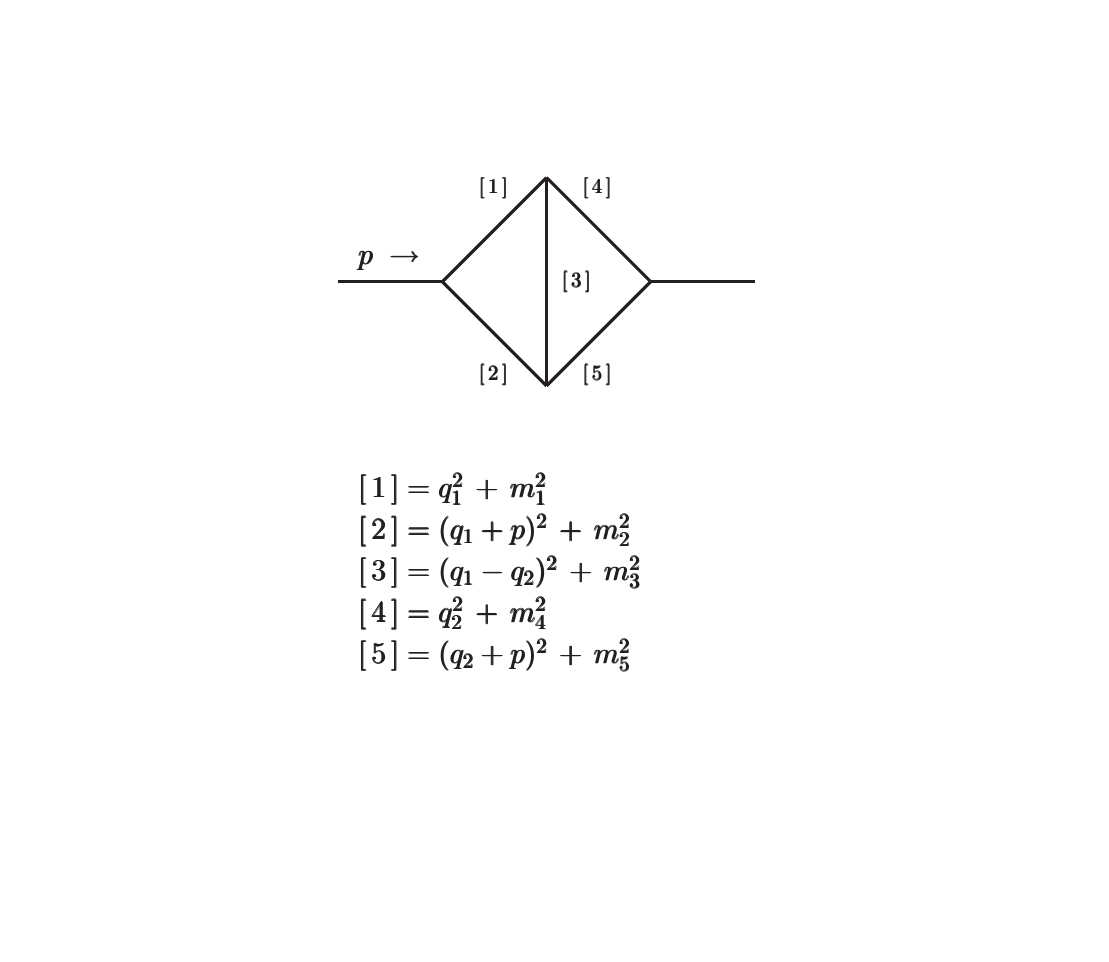}
\vspace{-3.cm}
\caption[]{The two-loop kite diagram.}
\label{fd_kite}
\end{figure}
Next we perform the change of variables described in \tabn{Quadkite}
\begin{table}
\begin{center}
\caption[]{Quadratization of Symanzik polynomials for the kite diagram. \label{Quadkite}}
\vspace{0.2cm}
\begin{tabular}{lllll}
\hline
&&&& \\
$\alpha_1 = \rho_1\,\mrx_1$ & $\alpha_2 = \rho_1\,(1 - \mrx_1)$ &
$\alpha_3 = \rho_3$ & $\alpha_4 = \rho_2\,\mrx_2$ & $\alpha_5 = \rho_2\,(1 - \mrx_2)$ \\
$\rho_3 = 1 - \rho_1 - \rho_2$ &&&& \\ 
&&&& \\
\hline
&&&& \\
$0 \le \mrx_1 \le 1$ & $0 \le \mrx_2 \le 1$ & $0 \le \rho_1 \le 1$ & $0 \le \rho_2 \le 1 - \rho_1$ & \\
&&&& \\
\hline
\end{tabular}
\end{center}
\end{table}
obtaining the following form for the Symanzik polynomials.
\bq
\mrS_1 = \beta = \rho_3\,\rho_1 + \sigma_2\,\rho_2 \spc
\qquad
\mrS_2 = \lpar \sum_{i \le j}\,\mra_{ij}\,\mrx_i\,\mrx_{\mrj} + \sum_i\,\mrb_i\,\mrx_i + \mrc)\rpar\,\mrs \spc
\eq
\bqa
\mra_{11} &=& \sigma_1\,\rho_1^2 \spc
\quad
\mra_{22} = \sigma_2\,\rho_2^2 \spc
\quad
a_{12} = 2\,\rho_3\,\rho_1\,\rho_2 \spc
\nl
\mrb_1 &=& - (1 + \mrl_2^2 - \mrl_1^2)\,
      (\rho_3\,\rho_1 + \sigma_2\,\rho_2)\,\rho_1)  \spc
\quad
\mrb_2 = - (1 + \mrl_5^2 - \mrl_4^2)\,
      (\rho_3\,\rho_1 + \sigma_2\,\rho_2)\,\rho_2)  \spc
\nl
\mrc &=& (\mrl_5^2 - 3\,\mrl_3^2 + \mrl_2^2)\,\rho_1\,\rho_2
       + (\mrl_5^2 - 2\,\mrl_3^2)\,\rho_2^2
       - (\mrl_5^2 - 2\,\mrl_3^2 + \mrl_2^2)\,(\rho_2 + \rho_1)\,\rho_1\,\rho_2  \spc
\nl
{}&-& (\mrl_5^2 - \mrl_3^2)\,\rho_2^3
          + (\mrl_3^2 - \mrl_2^2)\,\rho_1^3
          - (2\,\mrl_3^2 - \mrl_2^2)\,\rho_1^2
          + (\rho_2 + \rho_1)\,\mrl_3^2 \spc
\eqa
where $\mrm^2_i = \mrl_i^2\,\mrs$.
\end{example}
\begin{example}[Example: the delta{-}kites]
The direct, two-loop, delta{-}kite diagram is shown in Fig.~\ref{fd_ddkite}; to line $[i]$ we assign
a Feynman parameter $\alpha_i$ and obtain:
\bqa
\mrS_1 &=& \alpha_3\,\alpha_6 + (\alpha_4 + \alpha_5)\,\alpha_3 + 
      (\alpha_1 + \alpha_2)\,(\alpha_4 + \alpha_3) + (\alpha_1 + \alpha_2)\,(\alpha_5 + \alpha_6)
\nl
\mrS_2 &=&
       - \mrM^2_1\,\Bigl[ (\alpha_4 + \alpha_1)\,\alpha_3 + 
                      (\alpha_1 + \alpha_2)\,\alpha_4 \Bigr]\,\alpha_5
\nl
{}&-& \mrM^2_2\,\Bigl[ (\alpha_6 + \alpha_2)\,\alpha_3 + 
                      (\alpha_1 + \alpha_2)\,\alpha_6 \Bigr]\,\alpha_5
\nl
{}&-& s\,
 \Bigl[ \alpha_3\,\alpha_6\,\alpha_4 + (\alpha_5 + \alpha_3)\,\alpha_2\,\alpha_1 + 
        (\alpha_1 + \alpha_2)\,\alpha_6\,\alpha_4 + (\alpha_1 + \alpha_3)\,\alpha_2\,\alpha_4 + 
        (\alpha_2 + \alpha_3)\,\alpha_1\,\alpha_6) \Bigr]
\nl
{}&+& \sum_{i=1,6}\,\alpha_i\,\mrm^2_i\,
 \Bigl[ \alpha_3\,\alpha_6 + (\alpha_4 + \alpha_5)\,\alpha_3 + 
        (\alpha_1 + \alpha_2)\,(\alpha_3 + \alpha_4 + \alpha_5 + \alpha_6) \Bigr] \spc
\eqa
where $s = -\,(p_1 + p_2)^2$ and $\mrp^2_i = - \mrM^2_i$.
\begin{figure}[t]
   \centering
   \includegraphics[width=0.7\textwidth, trim = 30 250 50 80, clip=true]{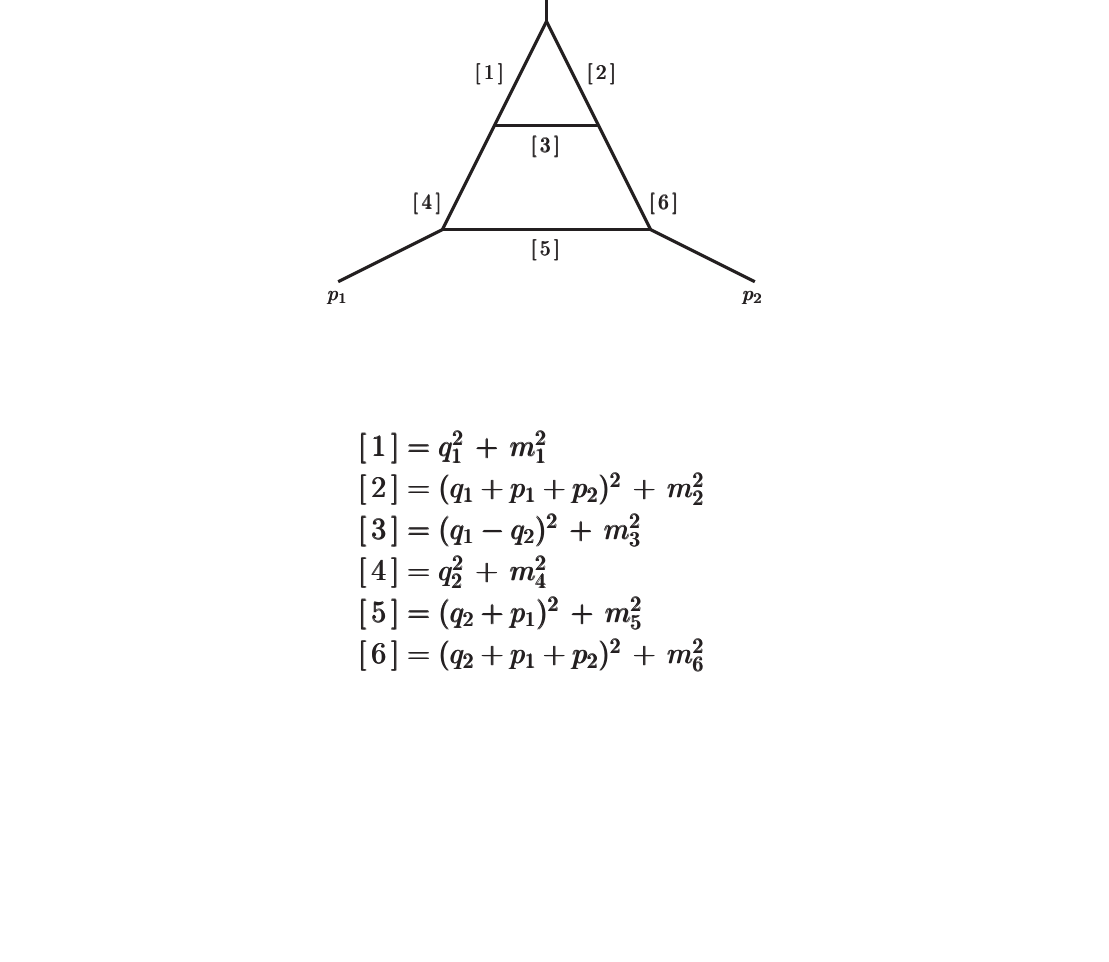}
\vspace{-3.cm}
\caption[]{The two-loop (direct) delta{-}kite diagram. Momenta are flowing inwards.}
\label{fd_ddkite}
\end{figure}
Next we perform the change of variables described in \tabn{Quadddkite}
\begin{table}
\begin{center}
\caption[]{Quadratization of Symanzik polynomials for the direct delta{-}kite diagram. 
\label{Quadddkite}}
\vspace{0.2cm}
\begin{tabular}{llllll}
\hline
&&&&& \\
$\alpha_1 = \rho_1\,x_1$ &
$\alpha_2 = \rho_1\,(1 - x_1)$ &
$\alpha_3 = \rho_3$ &
$\alpha_4 = \rho_2\,x_2$ &
$\alpha_5 = \rho_2\,x_3$ &
$\alpha_6 = \rho_2\,(1 - x_2 - x_3)$ \\ 
$\rho_3 = 1 - \rho_1 - \rho_2$ &&&&& \\ 
&&&&& \\
\hline
&&&&& \\
$0 \le x_1 \le 1$ & $0 \le x_2 \le 1$ & $0 \le x_3 \le 1 - x_2$ &
$0 \le \rho_1 \le 1$ & $0 \le \rho_2 \le 1 - \rho_1$ & \\
&&&&& \\
\hline
\end{tabular}
\end{center}
\end{table}
obtaining the following form for the Symanzik polynomials,
\bqa
\mrS_1 &=& \beta = \rho_3\,\rho_1 + \sigma_2\,\rho_2 \spc
\qquad
\mrS_2 = \Bigl( \sum_{i \le j}\,\mra_{ij}\,\mrx_i\,\mrx_{\mrj} + \sum_i\,\mrb_i\,\mrx_i + \mrc \Bigr)\,\mrs \spc
\nl
\mra_{11} &=& \sigma_1\,\rho_1^2 \spc
\quad
\mra_{22} = \sigma_2\,\rho_2^2 \spc
\quad
\mra_{33} = \sigma_2\,\rho_2^2\,\mrL_2^2 \spc
\nl
\mra_{12} &=& 2\,\rho_1\,\rho_2\,\rho_3 \spc
\quad
\mra_{13} = (1 + \mrL_2^2 - \mrL_1^2)\,\rho_1\,\rho_2\,\rho_3 \spc
\quad
\mra_{23} = (1 + \mrL_2^2 - \mrL_1^2)\,\sigma_2\,\rho_2^2 \spc
\nl
\mrb_1 &=& (1 + \mrl_2^2 - \mrl_1^2)\,\beta\,\rho_1 \spc
\quad
\mrb_2 = - (1 + \mrl_6^2 - \mrl_4^2)\,\beta\,\rho_2 \spc
\quad
\mrb_3 = - (\mrL_2^2 + \mrl_6^2 - \mrl_5^2)\,\beta\,\rho_2 \spc
\nl
\mrc &=&   (\mrl_6^2 - 3\,\mrl_3^2 + \mrl_2^2)\,\rho_1\,\rho_2
         + (\mrl_6^2 - 2\,\mrl_3^2)\,\rho_2^2
         - (\mrl_6^2 - 2\,\mrl_3^2 + \mrl_2^2)\,(\rho_1 + \rho_2)\,\rho_1\,\rho_2
\nl
{}&-&       (\mrl_6^2 - \mrl_3^2)\,\rho_2^3
          + (\mrl_3^2 - \mrl_2^2)\,\rho_1^3
          - (2\,\mrl_3^2 - \mrl_2^2)\,\rho_1^2
          + \mrl_3^2\,(\rho_2 + \rho_1) \spc
\eqa
where $\mrm^2_i = \mrl^2_i\,s$ and $\mrp^2_i = - \mrL^2_i\,s$.
The crossed, two-loop, delta{-}kite diagram is shown in Fig.~\ref{fd_cdkite}; to line $[i]$ we assign
a Feynman parameter $\alpha_i$ and obtain:
\bqa
\mrS_1 &=& (\alpha_3 + \alpha_4)\,(\alpha_5 + \alpha_6) + 
           ( \alpha_1 + \alpha_2)\,(\alpha_5 + \alpha_6) + 
           ( \alpha_1 + \alpha_2)\,(\alpha_3 + \alpha_4) \spc
\nl
\mrS_2 &=& \mrS_1\,\sum_i\,\alpha_i\,\mrm_i^2 
\nl
{}&-& \mrM_1^2\,(\alpha_4\,\alpha_5\,\alpha_6 + \alpha_3\,\alpha_5\,\alpha_6 + 
                 \alpha_2\,\alpha_5\,\alpha_6 + \alpha_2\,\alpha_4\,\alpha_6 + 
                 \alpha_1\,\alpha_5\,\alpha_6 + \alpha_1\,\alpha_3\,\alpha_5)
\nl
{}&-& \mrM_2^2\,(\alpha_2\,\alpha_3\,\alpha_6 + \alpha_1\,\alpha_4\,\alpha_5 + 
                 \alpha_1\,\alpha_2\,\alpha_6 + \alpha_1\,\alpha_2\,\alpha_5 + 
                 \alpha_1\,\alpha_2\,\alpha_4 + \alpha_1\,\alpha_2\,\alpha_3)
\nl
{}&-&     \mrs\,(\alpha_3\,\alpha_4\,\alpha_6 + \alpha_3\,\alpha_4\,\alpha_5 + 
                 \alpha_2\,\alpha_3\,\alpha_5 + \alpha_2\,\alpha_3\,\alpha_4 + 
                 \alpha_1\,\alpha_4\,\alpha_6 + \alpha_1\,\alpha_3\,\alpha_4)
\eqa
Next we perform the change of variables described in \tabn{Quadcdkite}
\begin{table}
\begin{center}
\caption[]{Quadratization of Symanzik polynomials for the crossed delta{-}kite diagram. 
\label{Quadcdkite}}
\vspace{0.2cm}
\begin{tabular}{llllll}
\hline
&&&&& \\
$\alpha_1 = \rho_1\,\mrx_1$ &
$\alpha_2 = \rho_1\,(1 - \mrx_1)$ &
$\alpha_3 = \rho_3\,\mrx_3$ &
$\alpha_4 = \rho_3\,(1 - \mrx_3)$ &
$\alpha_5 = \rho_2\,\mrx_2$ &
$\alpha_6 = \rho_2\,(1 - \mrx_2)$ \\ 
$\rho_3 = 1 - \rho_1 - \rho_2$ &&&&& \\ 
&&&&& \\
\hline
&&&&& \\
$0 \le \mrx_1 \le 1$ & $0 \le \mrx_2 \le 1$ & $0 \le \mrx_3 \le 1$ &
$0 \le \rho_1 \le 1$ & $0 \le \rho_2 \le 1 - \rho_1$ & \\
&&&&& \\
\hline
\end{tabular}
\end{center}
\end{table}
obtaining the following form for the Symanzik polynomials,
\bqa
\mrS_1 &=& \beta = \rho_3\,\rho_1 + \sigma_2\,\rho_2 \spc
\qquad
\mrS_2 = \Bigl( \sum_{i \le j}\,\mra_{ij}\,\mrx_i\,\mrx_{\mrj} + \sum_i\,\mrb_i\,\mrx_i + \mrc \Bigr)\,\mrs \spc
\nl
\mra_{11} &=& \sigma_1\,\rho_1^2\,\mrL_2^2 \spc \quad
       \mra_{22} = \sigma_2\,\rho_2^2\,\mrL_1^2 \spc \quad
       \mra_{33} = \sigma_3\,\rho_3^2 \spc
\nl
\mra_{12} &=& \rho_1\,\rho_2\,\rho_3\,(1 - \mrL_1^2 - \mrL_2^2) \spc \quad
       \mra_{13} = \rho_1\,\rho_2\,\rho_3\,(1 - \mrL_1^2 + \mrL_2^2) \spc \quad
       \mra_{23}= \rho_1\,\rho_2\,\rho_3\,(\mrL_2^2 - \mrL_1^2 - 1) \spc
\nl
\mrb_1 &=& \rho_1\,\rho_2\,\rho_3\,(\mrL_1^2 - 1)
          + \beta\,\rho_1\,(\mrl_1^2 - \mrl_2^2)
          - \sigma_1\,\rho_1^2\,\mrL_2^2 \spc \quad
       \mrb_2 = \beta\,\rho_2\,(\mrl_5^2 - \mrl_6^2)
          - (2\,\rho_1^2 + \beta - 2\,\sigma_2\,\rho_1)\,\rho_2\,\mrL_1^2 \spc
\nl
\mrb_3 &=& \rho_1\,\rho_2\,\rho_3\,(\mrL_1^2 - \mrL_2^2)
          + \beta\,\rho_3\,(\mrl_3^2 - \mrl_4^2)
          - \sigma_3\,\rho_3^2 \spc
\nl
\mrc &=& - \rho_1\,\rho_2\,\rho_3\,\mrL_1^2
       + \beta\,(\rho_1\,\mrl_2^2 + \rho_2\,\mrl_6^2 + \rho_3\,\mrl_4^2) \spp
\eqa

\begin{figure}[t]
   \centering
   \includegraphics[width=0.7\textwidth, trim = 30 250 50 80, clip=true]{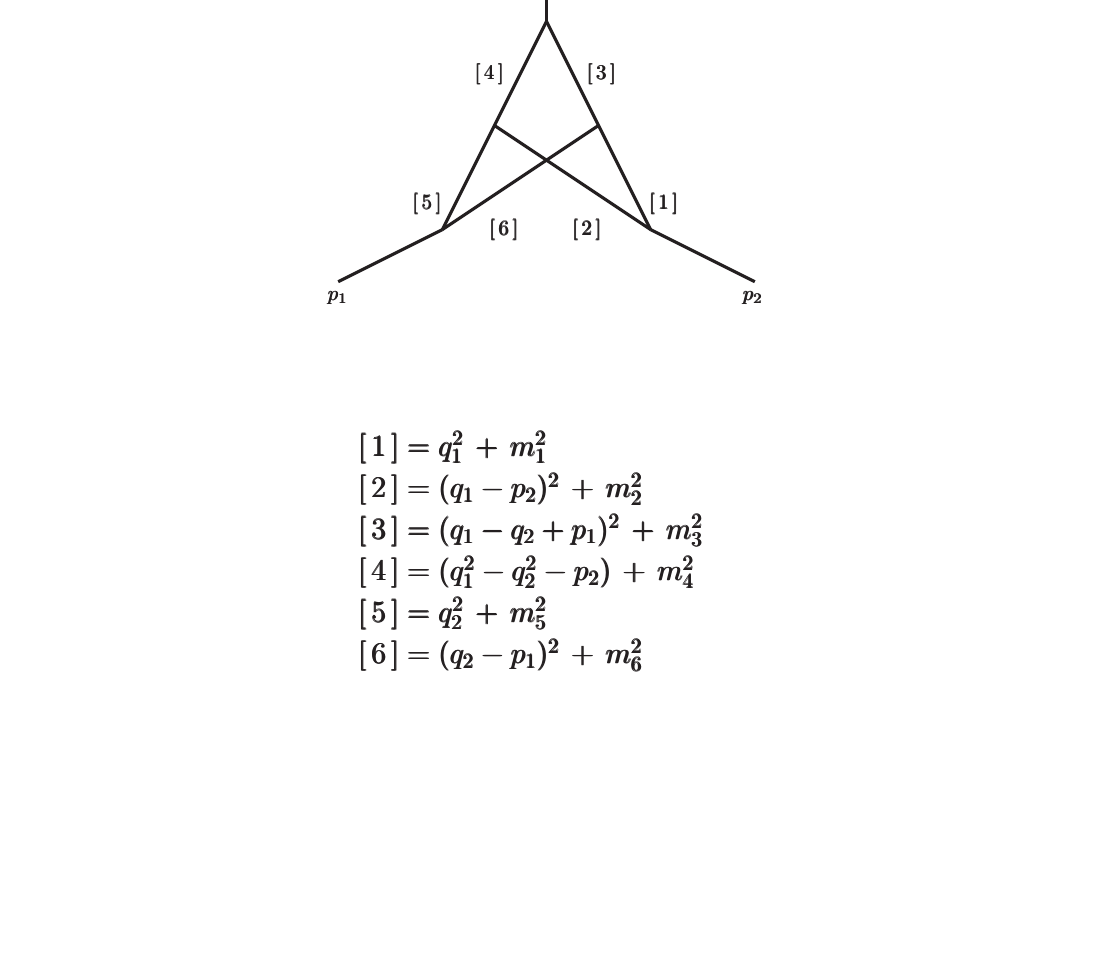}
\vspace{-3.cm}
\caption[]{The two-loop (crossed) delta{-}kite diagram. Momenta are flowing inwards.}
\label{fd_cdkite}
\end{figure}
\end{example}
\begin{example}[A three{-}loop example: the double{-}kite]
The procedure of partial quadratization can be extended beyond two{-}loop. Consider the three{-}loop diagram
described by
\bq
\prod_{i=1}^{8}\,[\,i\,] \spc \qquad [\,i\,] = \spro{\mrk_i}{\mrk_i} + \mrm_i^2 \spc
\eq
\bqa
\mrk_1 &=& \mrq_1 \spc \quad
\mrk_2 = \mrq_2 + \mrp \spc \quad
\mrk_3 = \mrq_2 - \mrq_1 \spc \quad
\mrk_4 = \mrq_2 \spc
\nl
\mrk_5 &=& \mrq_2 + \mrp \spc \quad
\mrk_6 = \mrq_3 \spc \quad
\mrk_7 = \mrq_3 - \mrq_2 \spc \quad
\mrk_8 = \mrq_3 + \mrp \spc
\eqa
where $\mrp$ is the external momentum. To each $[\,i\,]$ we assign a variable $\alpha_i$, compute the Symanzik
polynomials and perform a change of variable,
\bqa
\alpha_1 &=& \rho_1\,\mrx_1 \spc \quad
\alpha_2 = \rho_1\,(1 - \mrx_1) \spc \quad
\alpha_6 = \rho_3\,\mrx_3 \spc \quad
\alpha_8 = \rho_3\,(1 - \mrx_3) \spc 
\nl
\alpha_4 &=& \rho_2\,\mrx_2 \spc \quad
\alpha_5 = \rho_2\,(1 - \mrx_2) \spc \quad
\alpha_3 = \rho_4 \spc \quad
\alpha_7 = \rho_5 \spc 
\eqa
with $\sum_i\,\rho_i = 1$. Assume $\mrm_3 = \mrm_7 = 0$ with equal remaining masses: the first Symanzik polynomial is
\bq
\mrS_1 = \rho_4\,\rho_5\,\rho_{123} + \rho_3\,\rho_4\,\rho_{12} + \rho_1\,\rho_5\,\rho_{23} +
\rho_1\,\rho_2\,\rho_3 \spc
\eq
with a second Symanzik polynomial being a quadratic form
\bq
\mrs^{-1}\,\mrS_2 =
\mbx^{\mrt}\,\mbH\,\mbx + \mbK^{\mrt}\,\mbx + \mrM \spc
\eq
where $\mrs = - \spro{\mrp}{\mrp}$ and 
$\rho_{\mri \mrj\,\dots\,\mrl} = \rho_{\mri} + \rho_{\mrj} + \,\dots\, + \rho_{\mrl}$. The $\rho\,${-}dependent
coefficients are
\bqa
\mrH_{11} &=& (\rho_5\,\rho_{23} + \rho_4\,\rho_{35} + \rho_2\,\rho_3)\,\rho_1^2
\spc \quad          
\mrH_{22} = \rho_{35}\,\rho_{14}\,\rho_2^2
\spc \quad          
\mrH_{33} = (\rho_5\,\rho_{14} + \rho_4\,\rho_{12} + \rho_1\,\rho_2)\,\rho_3^2
\spc \nl       
\mrH_{12} &=& \rho_1\,\rho_2\,\rho_4\,\rho_{35}
\spc \quad          
\mrH_{13} = \rho_1\,\rho_3\,\rho_4\,\rho_5
\spc \quad
\mrH_{23} = \rho_2\,\rho_3\,\rho_5\,\rho_{14}
\eqa
\bq
\mrK_{i} = - \mrS_1\,\rho_i \spc \qquad i = 1,2,3 \;\spc
\eq          
\bq
\mrM =
\Bigl\{
\rho_1\,\rho_{123}\,(\rho_5\,\rho_{23} + \rho_2\,\rho_3)  + 
\Bigl[ \rho_{12}\,(\rho_{12}\,\rho_{35} + 2\,\rho_3\,\rho_5 + \rho_3^2) + \rho_3^2\,\rho_5 \Bigr]\,\rho_4
\Bigr\}\,\lambda_{\mrm} \spc
\eq
with $\mrm^2 = \lambda_{\mrm}\,\mrs$.
The Symanzik polynomials can be written ia more symmetric form,
\bq
\mrS_1 = 
\rho_{123}\,\rho_4\,\rho_5 +
\rho_{12}\,\rho_3\,\rho_4 +
\rho_{23}\,\rho_1\,\rho_5 +
\rho_1\,\rho_2\,\rho_3 \spc
\eq
\bq
\mrs^{-1}\,\mrS_2 =
\mby^{\mrt}\,\mbH\,\mby + \mrC + \mrM \spc
\eq
where 
\bq
\mry_i = \mrx_i - \frac{1}{2} \spc \qquad \mrC = - \frac{1}{4}\,\mrS_1\,\rho_{123} \spc \qquad
\mrH_{ij} = \mrh_{ij}\,\rho_i\,\rho_j \spc
\eq
\bqa
\mrh_{11} &=& \rho_{24}\,\rho_3 + \rho_{234}\,\rho_5 \spc \quad
\mrh_{22} = \rho_{14}\,\rho_{35} \spc \quad
\mrh_{33} = \rho_{245}\,\rho_1 + \rho_{25}\,\rho_4  \spc
\nl
\mrh_{12} &=& \rho_{35}\,\rho_4 \spc \quad
\mrh_{13} = \rho_4\,\rho_5 \spc \quad
\mrh_{23} = \rho_{14}\,\rho_5 \spp
\eqa  
Note that the quadratic form $\mby^{\mrt}\,\mbH\,\mby$ is positive{-}semidefinite.

\end{example}
\begin{example}[A three{-}loop example: the banana]
The propagators for the three{-}loop banana diagram are $[\,i\,] = \spro{\mrk_i}{\mrk_i} + \mrm_i^2$, with
\bq
\mrk_1 = \mrq_3 \spc \quad
\mrk_2 = \mrq_1 \spc \quad
\mrk_3 = \mrq_1 - \mrq_2 + \mrq_3 + \mrp \spc \quad
\mrk_4 = \mrq_2 \spc
\eq
where $\mrp$ is the external momentun. The integral will be given in terms of non{-}canonical powers, \ie
\bq
\prod_{i=1}^{4}\,[\,i\,]^{- \nu_i} \spp
\eq
We work in dimension $\mrd = 4 + \ep$ and consider two cases:
\bqa
\mra) \quad & \quad \nu_1 = \nu_3 = 2 \spc \quad \nu_2 = \nu_4 = 1 \spc
\nl
\mrb) \quad & \quad \nu_1 = \nu_2 = \nu_4  = 2 \spc \quad \nu_3 = 1 \spc
\eqa
furthermore we assume equal masses, \ie $\mrm_i^2 = \lambda_{\mrm}\,\mrs$ where $\spro{\mrp}{\mrp} = - \mrs$.
The Symanzik polynomials are written in terms of $\alpha_i$; subsequently we perform a change of variables:
\bq
\alpha_1 = \rho_1\,\mrx \spc \quad
\alpha_2 = \rho_1\,(1 - \mrx) \spc \quad
\alpha_3 = \rho_2 \spc \quad
\alpha_4 = \rho_3 \spc
\eq
with $\sum_i\,\rho_i = 1$. In this case, due to a non{-}separability of loops, both Symanzik polynomials are
quadratic forms in $\mrx$,
\bqa
\mrS_1 &=& \rho_1\,\Bigl[ \rho_2\,\rho_3 + \rho_1\,\sigma_1\,\bigl( \frac{1}{4} - \mry^2 \bigr) \Bigr] \spc
\nl
\mrS_2 &=& \rho_1\,\Bigl[ \lambda_{\mrm}\,\rho_2\,\rho_3 + \frac{1}{4}\,\rho_1\,\bigl(
\lambda_{\mrm}\,\sigma_1 - \rho_2\,\rho_3 - 4\,\lambda_{\mrm}\,\sigma_1\,\mry^2 \bigr) \Bigr]\,\mrs \spc
\eqa
where $\mry = \mrx - 1/2$ and $\sigma_1 = 1 - \rho_1$. The Jacobian of the transformation is $\rho_1$. For
the two configurations we obtain
\bqa
\mrB_{\mra} &=& \eG{ - \frac{3}{2}\,\ep}\,\int_0^1 \mrd \mrx\,\int_0^1 \mrd \rho_1\,\int_0^{1 - \rho_1} \mrd \rho_2\,
\rho_1^2\,\rho_2\,\mrx\,\mrS_1^{ - 2 - 2\,\ep}\,\mrS_2^{3/2\,\ep} \spc
\nl
\mrB_{\mrb} &=& \eG{1 - \frac{3}{2}\,\ep}\,
\rho_1^3\,\rho_3\,\mrx\,(1 - \mrx)\,\mrS_1^{ - 1 - 2\,\ep}\,\mrS_2^{3/2\,\ep - 1} \spp
\eqa
The rest of the procedure is standard: we perform the integration over $\mrx$ giving rise to a $\mrF^{(2)}_{\sPD}$
function, use the corresponding MB representation (\eqn{FDMB}) and integrate over $\rho_2, \rho_1$ obtaining a Fox function.
It is interesting to consider the massless case, $\lambda_{\mrm} = 0$ where $\mrS_2 = - 1/4\,\rho_1^2\,\rho_2\,\rho_3$.
After expanding the Fox function around $\ep = 0$ both configurations can be written as follows:
\bq
\mrB_i = \mrB^{\pole}_i\,\frac{1}{\ep} + \mrB^{\fin}_i + \ord{\ep} \spc
\eq
where the $\mrB$ coefficients are given by multiple MB integrals, eventually containing $\uppsi$ functions. Their
explicit expessions will not be reported here.

\end{example}


It is worth noting that analytical results for all master integrals for massless three{-}point
functions, with one off{-}shell leg, at four loops have been presented in \Bref{Lee:2023dtc}.
\subsection{Two{-}loop:  the strategy \label{str}}
In the following we will make use of the relations,
\bqa
(\mrx\,\mry)^{\alpha} &=& \mrx^{\alpha}\,\mry^{\alpha} \quad \mathrm{for} \quad
- \pi < \marg(\mrx) + \marg(\mry) \le \pi \spc
\nl
( - \mrx \pm\, i\,\delta )^{\alpha} &=& \exp\{\pm\,i\,\alpha\,(\pi - \delta)\}\,\mrx^{\alpha} \qquad 
\mathrm{for} \quad \mrx > 0 \spc \quad \delta \to 0_{+} \spp
\eqa
Therefore, the symbol $(\mrx \mry)^{\alpha}$, appearing as argument of a Fox function, must always be understood as
\bq
( \mrx \mry )^{\alpha}\,\exp\{ - \alpha\,\eta(\mrx\,,\,\mry) \} \spc
\eq
where $\eta$ is the Veltman eta function~\cite{tHooft:1978jhc},
\bq
\eta(\mrx\,,\,\mry) = 2\,i\,\pi\,\Bigl[
\theta( - \Im \mrx)\,\theta( - \Im \mry)\,\theta(   \Im \mrx \mry) -
\theta(   \Im \mrx)\,\theta(   \Im \mry)\,\theta( - \Im \mrx \mry) \Bigr] \spp
\label{veeta}
\eq
For a given Feynman integral we start by performing the integration over the $\mrx\,${-}parameters where the integrand
is a function of a quadratic form in the $\mrx\,${-}variables. There are different options:
\bei

\item[$1)$] the $\alpha$ trick of \Bref{tHooft:1978jhc} which, however, is not very useful when $\alpha$ is
complex, due to a cumbersome analytic continuation, see \sect{numi} for a discussion.

\item[$2)$] Rationalization of the square roots. It is well{-}known that the computation of Feynman integrals often 
involves square roots and one needs to rationalize these square roots by a suitable change of
variables~\cite{Besier:2019kco}. For instance, given
\bqa
\mrQ &=& \mra\,\mrx_1^2 + \mrb\,\mrx_2^2 + \mrc\,\mrx_1\,\mrx_2 + \mrd\,\mrx_1 + \mre\,\mrx_2 + \mrf =
       \mrb\,\lpar \mrx_2 - \mrx_{2-} \rpar\,\lpar \mrx_2 - \mrx_{2+} \rpar \spc
\nl
\mrx_{2\pm} &=& \mrL(\mrx_1) \pm \mrq^{1/2}(\mrx_1) \spc
\eqa
where $\mrL$ is linear and $\mrq$ is quadratic and positive; we need a change variable
\bq
\mrt = \frac{\mrx_1 - \alpha}{\mrq^{1/2} - \beta} \spc \qquad
\beta = \mrq^{1/2}(\alpha) \spp
\eq

\item[$3)$] In general the best solution is as follows:

\item[$3 \mra)$] do the innermost integral and write the solution in terms of (generalized) hypergeometric
functions, in general $\mrF^{(\mrN)}_{\sPD}$ Lauricella functions defined in \eqn{FDEM} and keep trace of the
corresponding convergence conditions.

\item[$3 \mrb)$] Rewrite the HTFs in terms of their MB representations, \eg using \eqn{FDMB}.

\item[$3 \mrc)$] It is also useful to define a MB splitting:
\bq
( \mrx + \mry)^{-\alpha} = \frac{1}{2\,i\,\pi}\,\int_{\mrL}\,\mrd \mrs\,\eB{\mrs}{\alpha - \mrs}\,
 \mrx^{ - \mrs}\,\mry^{\mrs - \alpha} \spc
\label{MBsplit}
\eq
valid for $ 0 < \Re \mrs < \alpha$ and $\mid \marg(\mrx) - \marg(\mry) \mid < \pi$.
We can also apply \eqn{MBsplit} with $\alpha$ negative which requires a deformation of the contour. For instance,
for $\alpha = - 1/2$ the poles ar $\mrs =  - \mrn$ must be at the left of $\mrL$ but $\mrL$ keeps the $\mrs = - 1/2$ pole
to the right. Of course we can avoid deformation: for $\alpha > 0$ let $\mrm$ be the integer such that
$\mrm - 1 \le \alpha < \mrm$ (the existence of $\mrm$ follows from the Dedekind completeness property); it follows
that we can write
\bq
( \mrx + \mry )^{\alpha} = ( \mrx + \mry )^{\mrm}\,
\frac{1}{2\,i\,\pi}\,\int_{\mrL}\,\mrd \mrs\,\eB{\mrs}{\mrm - \alpha - \mrs}\,
 \mrx^{ - \mrs}\,\mry^{\mrs + \alpha - \mrm} \spp
\label{MBsplitp}
\eq
To give an example of the correct treatment of the integrals involved we consider
\bq
\mrI = \int_0^1\,\mrd \mrx\,\mrx^{\alpha - 1}\,\lpar \mra + \mrb\,\mrx \rpar^{ - \beta} \spp
\eq
A careless derivation, valid only for $\mra$ and $\mrb$ positive, gives
\bq
\mrI = \frac{\eG{\alpha}}{\eG{\alpha + 1}}\,\mra^{ - \beta}\,\hyp{\beta}{\alpha}{\alpha + 1}{ - \frac{\mrb}{\mra}} =
\frac{1}{\eG{\beta}}\,\int_{\mrL}\,\frac{\mrd \mrs}{2\,i\,\pi}\,
\frac{\eG{ - \mrs}\,\eG{\alpha + \mrs}\,\eG{\beta + \mrs}}{\eG{\alpha + 1 + \mrs}}\,
\mra^{ - \beta - \mrs}\,\mrb^{\mrs} \spp
\eq
Using the Feynman prescription in the general case give
\bqa
\mrI &=& \int_{\mrL}\,\frac{\mrd \mrs}{2\,i\,\pi}\,\eB{ - \mrs}{\beta + \mrs}\,
\lpar \mra - i\,\delta \rpar^{ - \beta - \mrs}\,\int_0^1\,\mrd \mrx\,
\mrx^{\alpha - 1}\,\lpar \mrb\,\mrx - i\,\delta \rpar^{\mrs}
\nl
{}&=&
\frac{1}{\eG{\beta}}\,\int_{\mrL}\,\frac{\mrd \mrs}{2\,i\,\pi}\,
\frac{\eG{ - \mrs}\,\eG{\alpha + \mrs}\,\eG{\beta + \mrs}}{\eG{\alpha + 1 + \mrs}}\,
\lpar \mra - i\,\delta \rpar^{ - \beta - \mrs}\,\lpar \mrb - i\,\delta \rpar^{\mrs} \spc
\eqa
with $\delta \to 0_{+}$.

When square roots are present we split according to \eqn{MBsplit}, \ie
\bq
\Bigl[ \mrl(\mrx) + \mrq^{1/2}(\mrx) \Bigr]^{ - \alpha} =
\frac{1}{2\,i\,\pi}\,\int_{\mrL}\,\mrd \mrs\,\eB{\mrs}{\alpha - \mrs}\,\mrl^{ - \mrs}\,\mrq^{1/2\,(\mrs - \alpha)} \spp
\eq

\item[$3 \mrd)$] Perform the next $\mrx\,${-}integral and express the result in terms of (generalized) HTFs; use again
the corresponding MB representations.

\item[$3 \mre)$] Repeat the procedure for the remaining $\mrx\,${-}integrals and continue with the $\rho\,${-} integrals.

\item[$3 \mrf)$] Rewrite the result in terms of generalized, multivariate, Fox functions. 

\eei
For a review on MB integrals and their applications see \Bref{Dubovyk:2022obc}.
\paragraph{Contours of integration} \hspace{0pt} \\

Definition of the contours of integration follows from the following conditions: 
\begin{enumerate}

\item a EM integral as given in \eqn{FDEM} produces $\eB{\mra}{\mrc - \mra}$ where $\mrB$ is 
the Euler beta function; therefore, we require $\Re\,\mra > 0$ and $\Re(\mrc - \mra) > 0$.

\item When we use the MB representation for a $\mrF^{(\mrN)}_{\sPD}$ function, see \eqn{FDMB}, the poles
of the integrand come from
\bq
\eG{\mra + \sum_{\mrj}\,\mrs_{\mrj}}\,\prod_{\mrj=1}^{\mrN}\,\eB{\mrb_{\mrj} + \mrs_{\mrj}}{ - \mrs_{\mrj}} \spp
\eq
$\mrL_{\mrj}$ is a standard contour in the $\mrs_{\mrj}\,$-plane which is a deformed imaginary axis,
that is, it connects $- i\,\infty$ and $+ i\,\infty$ but is curved so that among all the poles of the
integrand only the poles of $\eG{ - \mrs_{\mrj}}$ lie to the right of $\mrL_{\mrj}$. Therefore, we require
\bq
\Re\,\mrs_{\mrj} < 0 \spc \quad
\Re(\mra + \sum_{\mrj}\,\mrs_{\mrj}) > 0 \spc \quad
\Re(\mrb_{\mrj} + \mrs_{\mrj}) > 0 \;\;\forall j \spp
\eq
Inconsistency of these condition requires deformation(s). A simple example is as follows:
\bq
\mrI_{\mra} = \int_{\mrL}\,\frac{\mrd \mrs}{2\,i\,\pi}\,\eG{1 + \mrs}\,\eG{ - \mrs}\,\mrz^{ - \mrs} \spc \qquad
\mrI_{\mrb} = \int_{\mrL}\,\frac{\mrd \mrs}{2\,i\,\pi}\,\eG{ - \frac{1}{2} + \mrs}\,\eG{ - \mrs}\,
\mrz^{ - \mrs} \spc
\eq
with poles at $\mrs = \mrn_1$ and $\mrs= - \mra - \mrn_2$, $\mra = 1$ or $\mra = - 1/2$. 
With $\mra = 1$ $\mrL$ is the line parallel to the imaginary axis with $- 1 < \Re\,\mrs < 0$.
With $\mra = - 1/2$ the contour must be deformed in order to keep the pole at $\mrs = 0$ to the right and
the pole at $\mrs = 1/2$ to the left of the contour.

There is another way to handle square roots: consider the following integral
\bq
\mrJ = \int_0^1 \mrd \mrx\,\lpar \mrx + \sqrt{\mrx^2 + \mra\,\mrx} \rpar^{\mrs} \spc \quad
\Re \mrs > - 2 \spc 
\eq
we can write
\[
\mrJ = - \frac{\mrs\,\mrx^{\mrs}}{2\,\sqrt{\pi}}\,\int_0^1 \mrd \mrx\,
\mrG^{1,2}_{2,2}\,\left( \frac{\mrx}{\mra}\,,\; \Biggl[
\begin{array}{cc}
1 & 1 - \mrs \\
\frac{1 - \mrs}{2} & - \frac{\mrs}{2} \\
\end{array}
\Biggr ]
\right) 
\]
and derive
\[
\mrJ = - \frac{\mrs}{2\,\sqrt{\pi}}\,
\mrG^{2,2}_{3,3}\,\left( \frac{1}{\mra}\,,\; \Biggl[
\begin{array}{ccc}
- \mrs & 1 & 1 - \mrs \\
\frac{1 - \mrs}{2} & - \frac{\mrs}{2} & - \mrs - 1 \\
\end{array}
\Biggr ]
\right) 
\]
A more general example is the following one:
\bq
\mrJ_2 = \int_0^1 \mrd \mrx \Bigl[ \mrl(\mrx) + \sqrt{\mrq(\mrx)} \Bigr]^{\lambda} \spc \quad
\mrl = \alpha\,\mrx + \beta \spc \quad
\mrq = \mrA\,\mrx^2 + 2\,\mrB\,\mrx + \mrC \ge 0 \spc
\label{roots}
\eq
where all the parameters are real and $\lambda < 0$. We discuss this example in details to illustrate the 
main features of the calculations given in the rest of the paper. We introduce
\bq
\mra = \alpha\,\mrx + \beta \spc \quad
\mrz= (\mrA - \alpha^2)\,\mrx^2 + 2\,(\mrB - \alpha\,\beta)\,\mrx + \mrC - \beta^2 \spc
\eq
and derive
\[
\mrJ_2 = - \frac{\lambda}{2\,\sqrt{\pi}}\,\int_0^1 \mrd \mrx\,\mra^{\lambda}\,
\mrG^{1,2}_{2,2}\,\left( \frac{\mrz}{\mra^2}\,,\; \Biggl[
\begin{array}{cc}
\frac{\lambda + 1}{2} & \frac{\lambda}{2} + 1 \\
0                     & \lambda \\
\end{array}
\Biggr ]
\right) 
\]
If we have $\mrl - \sqrt{\mrq}$ we can use the following relation:
\bq
\lpar 1 + \sqrt{1 + \mrz} \rpar^{\lambda} +
\lpar 1 - \sqrt{1 + \mrz} \rpar^{\lambda} =
2\;\hyp{ - \frac{\lambda}{2}}{\frac{1 - \lambda}{2}}{\frac{1}{2}}{1 + \mrz} \spp
\eq
Furthermore, when $\mrz < 0$ we use
\bq
2^{ - \lambda}\,\lpar 1 + \sqrt{1 - \mrz} \rpar^{\lambda} = 
\hyp{ - \frac{\lambda}{2}}{\frac{1 - \lambda}{2}}{1 - \lambda}{\mrz} \spp 
\eq
When $\mrq < 0$ in \eqn{roots} we introduce $\zeta_{\pm} = 1 \,\pm\, i\,\sqrt{\mrz}$ and use~\cite{msf}
\bqa
\zeta_{-}^{\lambda} + \zeta_{+}^{\lambda} &=& 2\;
\hyp{ - \frac{\lambda}{2}}{\frac{1 - \lambda}{2}}{\frac{1}{2}}{ - \mrz} \spc
\nl
\zeta_{-}^{\lambda} - \zeta_{+}^{\lambda} &=& 2\,i\,(\lambda + 1)\,\sqrt{\mrz}\;
\hyp{ - \frac{\lambda}{2}}{\frac{1 - \lambda}{2}}{\frac{3}{2}}{ - \mrz} \spp
\eqa
If we introduce $\mrx_0 = - \beta/\alpha$ and $\mrx_{\pm}$ are the roots of $\mrz = 0$ (here $\mrz$ is assumed to be
real and positive) then the result follows by using \eqn{Gtheta} and  \eqn{FDEM},
\bqa
\mrJ_2 &=& - \frac{\lambda}{2\,\sqrt{\pi}}\,\int_{\mrL}\,\frac{\mrd \mrs}{2\,i\,\pi}\,
\frac{\eG{ - \mrs}\,\eG{ - \frac{\lambda}{2} + \mrs}\,\eG{\frac{1 - \lambda}{2} + \mrs}}
     {\eG{1 - \lambda + \mrs}}\,
\lpar \gamma\,\xp \xm \rpar^{\mrs}\,\lpar - \alpha\,\mrx_0 \rpar^{\lambda - 2\,\mrs}
\nl
{}&\times&
\mrF^{(3)}_{\sPD}\lpar 1\,;\,2\,\mrs - \lambda\;,\; - \mrs\;,\; - \mrs\,;\, 
- \mrx_0^{-1}\,,\, - \xm^{-1}\,,\, - \xp^{-1} \rpar \spp
\eqa
If needed, a system of analytic continuation formulae for the Lauricella function is given in \Bref{FDMB}.
Using \eqn{FDMB} we finally obtain,
\bqa
\mrJ_2 &=& - \frac{\lambda}{2\,\sqrt{\pi}}\,\Bigl[ \prod_{i=1}^{4}\,\int_{\mrL_i} \frac{\mrd \mrs_i}{2\,i\,\pi} \Bigr]\,
\frac{\mrN}{\mrD}\,
\lpar - \alpha\,\mrx_0 \rpar^{\lambda}\,
\lpar \frac{\xp\,\xm}{\alpha^2\,\mrx_0^2}\,\gamma \rpar^{\mrs_1}\,
\lpar \mrx_0 \rpar^{ - \mrs_2}\,
\lpar \xm \rpar^{ - \mrs_3}\,
\lpar \xp \rpar^{ - \mrs_4} \spc
\nl\nl
\mrN &=&
\eG{\mrs_1 - \frac{1}{2}\,\lambda}\,
\eG{\frac{1}{2} + \mrs_1 - \frac{1}{2}\,\lambda}\,
\eG{ - \mrs_2}\,
\eG{ - \mrs_3}\,
\eG{ - \mrs_4}\,
\nl {}&\times&
\eG{\mrs_3 - \mrs_1}\,
\eG{\mrs_4 - \mrs_1}\,
\eG{\mrs_2 + 2\,\mrs_1 - \lambda}\,
\eG{1 + \mrs_4 + \mrs_3 + \mrs_2}
\nl\nl
\mrD &=&
\eG{ - \mrs_1}\,
\eG{2\,\mrs_1 - \lambda}\,
\eG{1 + \mrs_1 - \lambda}\,
\eG{2 + \mrs_4 + \mrs_3 + \mrs_2} \spc
\eqa
where $\gamma = \mrA - \alpha^2$; the result corresponds to a Fox function of \sect{GMFF} with parameters
$\mrr = 4$, $\mrm = 9$ and $\mrn = 4$. The integration contours are defined by the following conditions
($\mrs_i = \mrr_i + \mrt_i$):
\bq
\frac{\lambda}{2} < \mrr_1 < 0 \spc \quad
\lambda - 2\,\mrr_1 < \mrr_2 < 0 \spc \quad
\mrr_1 < \mrr_{3,4} < 0 \spp
\eq
\end{enumerate}
\ovalbox{Examples}: The general case: no scale in the problem is zero and the integrand is an arbitrary power
of some indefinite multivariate polynomial. 
To illustrate the complexity of the calculation involved in the most general case we consider the following
example: 
\bq
\mrV\lpar \mrx_1\,,\,\mrx_2 \rpar = 
   \mathbf{x}^{\mrt}\,\mbH\,\mbx +
   2\,\mbK^{\mrt}\,\mbx + \mrL - i\,\delta \spc
\eq
with $\delta \to 0_+$. The parameters may depend on $\rho\,${-}variables in the two loop case.
When we have
\bq
\mrI_1 = \int_0^1 \mrd \mrx_1\,\int_0^{1 - \mrx_1} \mrd \mrx_2\,\mrV^{ - \lambda} \spc
\eq
we perform the transformation $\mrx = 1 - \mrx^{\prime}$ and obtain
\bq 
\mrV = 
   \mbx^{\mrt}\,\mathbf{h}\,\mbx +
   2\,\mathbf{k}^{\mrt}\,\mbx + \mrL^{\prime} - i\,\delta \spc
\eq
with parameters defined by
\bqa
{}&{}& \mrh_{11} = \mrH_{11} \spc \quad
       \mrh_{22} = \mrH_{22} \spc \quad
       \mrh_{12} = - \mrH_{12} \spc
\nl
{}&{}& \mrk_1 = - \mrH_{11} - \mrK_1 \spc \quad
       \mrk_2 = \mrH_{12} + \mrK_2 \spc \quad
       \mrL^{\prime} = \mrL + \mrH_{11} + 2\,\mrK_1 \spp
\eqa
Therefore, we only have to consider the following integral:
\bq
\mrI_2 = \int_0^1 \mrd \mrx_1\,\int_0^{\mrx_1} \mrd \mrx_1\,\mrV^{ - \lambda} \spp
\eq
Using a sector decomposition~\cite{Binoth:2003ak} we obtain
\bq
\mrI_2 = \int_0^1\,\mrd \mrx\,\mrd \mry\,\mrx\,
       \Bigl[ \mrx\,\Bigl(
2\,\mrK_2\,y + 2\,\mrK_1 + \mrH_{22}\,\mrx\,\mry^2 + \mrH_{11}\,\mrx + 2\,\mrH_{12}\,\mrx\,\mry - i\,\delta \Bigr) +
\mrL - i\,\delta \Bigr]^{ - \lambda} \spp
\eq
Next we perform a MB splitting, \eqn{MBsplit},
\bq
\Bigl[ \mrx\,\mrU(\mrx\,,\,\mry) + \mrL - i\,\delta]^{ - \lambda} =
\int_{\mrL_1}\,\frac{\mrd \mrs_1}{2\,i\,\pi}\,\eB{\mrs_1}{\lambda - \mrs_1}\,
\lpar \mrL - i\,\delta \rpar^{\mrs_1 - \lambda}\,\lpar \mrx\,\mrU \rpar^{ - \mrs_1} \spp
\eq
In order to proceed with the $\mrx$ integration we must use the following relation:
\bq
\lpar \mra\,\mrb \rpar^{\alpha} = \mra^{\alpha}\,\mrb^{\alpha}\,\exp\{\alpha\,\eta(\mra\,,\,\mrb) \} \spc
\label{psplit}
\eq
where $\eta$ is the Veltman eta function defined in \eqn{veeta}.
In our case the eta functions are zero and 
the $\mrx$ integration produces an hypergeometric function for wich we use a MB representation. For the $\mry$ integration 
we will present the result in the case $\mrG = {\mathrm{det}}\,\mrH > 0$. We introduce
\bq
\mry_0 = \frac{\mrK_2}{\mrK_1 - i\,\delta} \spc \qquad
\mry_{\pm} = \frac{1}{\mrH_{22}}\,\lpar - \mrH_{12} \,\pm\, i\,\sqrt{\mrG} \rpar \spc
\eq
and obtain a $\mrF^{(3)}_{\sPD}$ Lauricella function (\eqn{FDEM}) for which we use again a MB representation (\eqn{FDMB}). 
The final result is
\bqa
\mrI_2 &=& \frac{1}{\eG{\lambda}}\,\Bigl[ \prod_{i=1}^{5}\,\int_{\mrL_i}\,\frac{\mrd \mrs_i}{2\,i\,\pi} \Bigr]\,
           \frac{\mrN}{\mrD}\,\lpar \mrL - i\,\delta \rpar^{- \lambda}\,\mrJ \spc
\nl\nl
\mrN &=& \eG{ - \mrs_1 + \lambda}\,
\eG{2 + \mrs_2 - \mrs_1}\,
\eG{ - \mrs_3}\,
\eG{\mrs_3 + \mrs_2 + \mrs_1}\,
\eG{ - \mrs_4}\,
\eG{\mrs_4 - \mrs_2}\,
\eG{ - \mrs_5}\,
\eG{\mrs_5 - \mrs_2}\,
\eG{1 + \mrs_5 + \mrs_4 + \mrs_3} \spc
\nl
\mrD &=& \eG{ - \mrs_2}\,
\eG{3 + \mrs_2 - \mrs_1}\,
\eG{2 + \mrs_5 + \mrs_4 + \mrs_3} \spc
\nl
\mrJ &=&
\mrq\lpar - \mrs_1\,,\,\mrL - i\,\delta \rpar\,
\mrq\lpar - \mrs_1\,,\,2\,\mrK_1 - i\,\delta \rpar\,
\mrq\lpar - \mrs_2\,,\,\mrH_{22} - i\,\delta \rpar\,
\mrq\lpar - \mrs_2\,,\,\mry_{-}^{-1}\,\mry_{+}^{-1} \rpar\,
\nl
{}&\times&
\mrq\lpar - \mrs_2\,,\,2\,\mrK_1 - i\,\delta \rpar\,
\mrq\lpar - \mrs_3\,,\,\mry_0^{-1} \rpar\,
\mrq\lpar - \mrs_4\,,\, - \mry_{-} \rpar\,
\mrq\lpar - \mrs_5\,,\, - \mry_{+} \rpar \spc
\eqa
where $\mrq(\mrs\,,\,\mrz)= \mrz^{\mrs}$. The result
corresponds to a Fox function of \sect{GMFF} with parameters $\mrr = 5, \mrm = 9$ and $\mrn = 3$.
It is possible to combine powers in $\mrJ$ but only using \eqn{psplit}.

If $\mrG  = {\mathrm{det}}\,\mrH \approx 0$ we can use the BST 
algorithm~\cite{JB,MS,Tkachov:1996wh,Ferroglia:2002mz} and compute
\bq
\mrI_{\mrl} = \int_0^1 \mrd \mrx_1\,\int_0^{\mrx_1} \mrd \mrx_2\,\ln \lpar \mrV \rpar \spp
\eq
Writing $\mrV = \mrQ(\mrx_1\,,\,\mrx_2) + \mrL - i\,\delta$ we obtain
\bq
\ln\,\mrV = \ln\lpar \mrL - i\,\delta \rpar + \ln\lpar 1 + \mrz \rpar =
\ln\lpar \mrL - i\,\delta \rpar + \mrz\;\hyp{1}{1}{2}{ - \mrz} \spc \quad
\mrz = \frac{\mrQ}{\mrL - i\,\delta} \spc
\eq
After using the MB representation for the hypergeometric function (\eqn{F21MB}) we can perform the integrations as done 
before obtaining the following result ($\mrG > 0$):
\bqa
\mrI_{\mrl} &=& \frac{1}{2}\,\ln\lpar \mrL - i\,\delta \rpar +
\Bigl[ \prod_{i=1}^{5}\,\int_{\mrL_i}\,\frac{\mrd \mrs_i}{2\,i\,\pi} \Bigr]\,\frac{\mrN}{\mrD}\,
\mrq( - 1 - \mrs_1\,,\,\mrL - i\,\delta )\,\mrJ_{\mrl} \spc
\nl\nl
\mrN &= &
\eG{ - \mrs_1}\,
\eGs{1 + \mrs_1}\,
\eG{ - \mrs_3}\,
\eG{3 + \mrs_2 + \mrs_1}\,
\nl
{}&\times&
\eG{ - 1 + \mrs_3 + \mrs_2 - \mrs_1}\,
\eG{ - \mrs_4}\,
\eG{\mrs_4 - \mrs_2}\,
\eG{ - \mrs_5}\,
\eG{\mrs_5 - \mrs_2}\,
\eG{1 + \mrs_5 + \mrs_4 + \mrs_3} \spc
\nl
\mrD &=& 
\eG{ - 1 - \mrs_1}\,
\eG{2 + \mrs_1}\,
\eG{ - \mrs_2}\,
\eG{4 + \mrs_2 + \mrs_1}\,
\eG{2 + \mrs_5 + \mrs_4 + \mrs_3} \spc
\nl
\mrJ_{\mrl} &=&
\mrq\lpar \mrs_2\,,\,\mrH_{22} - i\,\delta \rpar\,
\mrq\lpar 1 - \mrs_2 + \mrs_1\,,\,2\,\mrK_1 - i\,\delta \rpar\,
\mrq\lpar \mrs_3\,,\,\mry_0 \rpar\,
\nl
{}&\times&
\mrq\lpar  - \mrs_4 + \mrs_2\,,\, - \mry_{-} \rpar\,
\mrq\lpar  - \mrs_5 + \mrs_2\,,\, - \mry_{+} \rpar \spp
\eqa
Finally we consider the following integral:
\bq
\mrI_3 = \int_0^1\,\mrd \mrx_1 \mrd \mrx_2\,\mrV^{ - \lambda}\lpar \mrx_1\,,\,\mrx_2 \rpar \spp
\eq
We can always write
\bq
\int_0^1\,\mrd \mrx_1 \mrd \mrx_2 = 
\int_0^1\,\mrd \mrx_1\,\int_0^{\mrx_1}\,\mrd \mrx_2 +
\int_0^1\,\mrd \mrx_2\,\int_0^{\mrx_2}\,\mrd \mrx_1 \spc
\eq
and use the previous results. However, we want to present an alternative derivation of $\mrI_3$. Let us 
introduce~\cite{Ferroglia:2002mz}
\bq
\mbbx = - \mbH^{-1}\,\mbK \spc \qquad
\mrC = \mrL + \mbK^{\mrt}\,\mbbx \spp
\eq
It follows
\bqa
\mrI_3 &=& \int_{ - \mrX_1}^{1 - \mrX_1}\,\mrd \mrx_1\,\int_{ - \mrX_2}^{1 - \mrX_2}\,\mrd \mrx_2\,
\Bigl(
\mbx^{\mrt}\,\mbH\,\mbx + \mrC \Bigr)^{ - \lambda} =
\mrJ\lpar 1 - \mrX_1 \rpar - \mrJ \lpar - \mrX_1 \rpar \spc
\nl
\mrJ\lpar \mrX \rpar &=&
\int_0^{\mrX}\,\mrd \mrx_1\,\int_{ - \mrX_2}^{1 - \mrX_2}\,\mrd \mrx_2\, 
\Bigl( \mbx^{\mrt}\,\mbH\,\mbx + \mrC \Bigr)^{ - \lambda} \spp
\eqa
We will consider the case $\mrX > 0$; for $\mrX < 0$ we change variable $\mrx_1 = - \mrx_1^{\prime}$ and compute
$\mrJ( - \mrX)$ with the replacement $\mrH_{12}\to - \mrH_{12}$. Next we use a MB splitting, \eqn{MBsplit}:
\bq
\Bigl( \mbx^{\mrt}\,\mbH\,\mbx + \mrC - i\,\delta\Bigr)^{ - \lambda} =
\int_{\mrL_1}\,\frac{\mrd \mrs_1}{2\,i\,\pi}\,
\eB{\mrs_1}{\lambda - \mrs_1}\,
\lpar \mrC - i\,\delta \rpar^{\mrs_1 - \lambda}\,
\lpar \mbx^{\mrt}\,\mbH\,\mbx - i\,\delta \rpar^{ - \mrs_1} \spp
\eq
Two cases follow:
\begin{enumerate}

\item $\mrG = {\mathrm{det}}\,\mrH > 0$. Here we introduce
\bq
\mrx_{2\,\pm} = \mrz_{\pm}\,\mrx_1 \spc \qquad
\mrz_{\pm} = \frac{1}{\mrH_{22}}\,\lpar - \mrH_{12} \pm i\,\sqrt{\mrG} \rpar \spp
\eq
Before performing the $\mrx_2$ integration let us consider 
\bq
\mrA = - \mrb \spc \quad
\mrB= 1 - \frac{\mra}{\mrb}\,\mry \spc \quad
\mrA\,\mrB = \mra\,\mry - \mrb \spc
\eq
with $\mry > 0$ and $\mra \in \Rf$. Since
\bq
\Im (\mrA\,\mrB ) = - \Im (\mrb) \spc \quad
\Im ( \mrA ) = - \Im (\mrb) \spc \quad
\Im ( \mrB) = \frac{\mra}{\mid \mrb \mid^2}\,\mry\,\Im ( \mrb) \spc
\eq
we can use $( \mrA\,\mrB )^{\alpha} = \mrA^{\alpha}\,\mrB^{\alpha}$, which means
\bqa
\Bigl[ \lpar 1 - \mrX_2 \rpar\,\mry - \mrx_{2\,\pm} \Bigr]^{\alpha} &=&
\lpar - \mrx_{2\,\pm} \rpar^{\alpha}\,\Bigl[ 1 - \frac{1 - \mrX_2}{\mrx_{2\,\pm}}\,\mry \Bigr]^{\alpha} \spc
\nl
\Bigl[ \lpar - \mrX_2 \rpar\,\mry - \mrx_{2\,\pm} \Bigr]^{\alpha} &=&
\lpar - \mrx_{2\,\pm} \rpar^{\alpha}\,\Bigl[ 1 + \frac{\mrX_2}{\mrx_{2\,\pm}}\,\mry \Bigr]^{\alpha} \spp
\eqa
The $\mrx_2$ integration can be performed giving rise to a combination of $\mrF^{(2)}_{\sPD}$ Lauricella
functions, \eqn{FDEM}. After introducing the corresponding MB representation (\eqn{FDMB}) we proceed with 
the $\mrx_1$ integration.
The final result is as follows:
\bqa
\mrJ(\mrX) &=& \frac{1}{\eG{\lambda}}\,\mrX\,\Bigl[ \prod_{i=1}^{3}\,\int_{\mrL_i}\,\frac{\mrd \mrs_i}{2\,i\,\pi} \Bigr]\,
\frac{\mrN}{\mrD}\,\mrJ_0\Bigl[ \lpar 1 - \mrX_2 \rpar\,\mrJ_{1} + \mrX_2\,\mrJ_{2} \Bigr] \spc
\nl\nl
\mrN &=&
\eG{ - \mrs_1 + \lambda}\,
\eG{ - \mrs_2}\,
\eG{\mrs_2 + \mrs_1}\,
\eG{ - \mrs_3}\,
\nl
{}&\times&
\eG{\mrs_3 + \mrs_1}\,
\eG{1 + \mrs_3 + \mrs_2}\,
\eG{1 - \mrs_3 - \mrs_2 - 2\,\mrs_1} \spc
\nl
\mrD &=& 
\eG{\mrs_1}\,
\eG{2 + \mrs_3 + \mrs_2}\,
\eG{2 - \mrs_3 - \mrs_2 - 2\,\mrs_1} \spc
\nl
\mrJ_0 &=&  \mrq\lpar - \mrs_1\,,\,\zm\,\zp \rpar\,
            \mrq\lpar - \mrs_1\,,\,\mrH_{22} - i\,\delta \rpar\,
            \mrq\lpar\mrs_1 - \lambda\,,\,\mrC - i\,\delta \rpar\,
            \mrq\lpar - \mrs_3 - \mrs_2 - 2\,\mrs_1\,,\,\mrX \rpar \spc
\nl
\mrJ_1 &=&  \mrq\lpar \mrs_2\,,\, - \frac{1 - \mrX_2}{\zm} \rpar\,
            \mrq\lpar \mrs_3\,,\, - \frac{1 - \mrX_2}{\zp} \rpar \spc
\nl
\mrJ_2 &=&  \mrq\lpar \mrs_2\,,\,\frac{\mrX_2}{\zm} \rpar\,
            \mrq\lpar \mrs_3\,,\,\frac{\mrX_2}{\mrz_+} \rpar \spp
\eqa
\item $\mrG = {\mathrm{det}}\,\mrH < 0$. In this case we introduce
\bq
\mrx_{2\,\pm} = \mrz_{\pm}\,\mrx_1  \mp i\,\delta \spc \qquad
\mrz_{\pm} = \frac{1}{\mrH_{22}}\,\lpar - \mrH_{12} \pm \sqrt{ - \mrG} \rpar \spc
\eq
and use the following relation ($\mry > 0$ and $\delta \to 0_{+}$):
\bq
\lpar \mra\,\mry \mp i\,\mrb\,\delta \rpar^{\alpha} =
\theta(\mra)\,\mra^{\alpha}\,\mrb^{\alpha} +
\theta( - \mra)\,\Bigl[ \theta(\mrb)\,\mrE( - \alpha) + \theta( - \mrb)\,\mrE(\alpha) \Bigr]\,
( - \mra)^{\alpha}\,\mry^{\alpha} \spc
\eq
where we have introduced $\mrE(\alpha) = \exp\{i\,\alpha(\pi - \delta)\}$. 
The final result is as follows:
\bqa
\mrJ(\mrX) &=& \frac{\mrX}{\eG{\lambda}}\,
\Bigl[ \prod_{i=1}^{3}\,\int_{\mrL_i}\,\frac{\mrd \mrs_i}{2\,i\,\pi} \Bigr]\,
\frac{\mrN}{\mrD}\,
\mrq\lpar - \mrs_1\,,\,\mrH_{22} - i\,\delta \rpar\,\mrq\lpar \mrs_1 - \lambda\,,\, \mrC - i\,\delta \rpar\,
\mrq\lpar - 2\,\mrs_1 - \mrs_2 - \mrs_3\,,\,\mrX \rpar\,
\lpar \mrJ_{-} + \mrJ_{+} \rpar \spc 
\nl\nl  
\mrN &=&
\eG{ - \mrs_1 + \lambda}\,
\eG{ - \mrs_2}\,
\eG{\mrs_2 + \mrs_1}\,
\eG{ - \mrs_3}\,
\eG{\mrs_3 + \mrs_1}\,
\eG{1 + \mrs_3 + \mrs_2}\,
\eG{1 - \mrs_3 - \mrs_2 - 2\,\mrs_1} \spc
\nl
\mrD &=&
\eG{\mrs_1}\,
\eG{2 + \mrs_3 + \mrs_2}\,
\eG{2 - \mrs_3 - \mrs_2 - 2\,\mrs_1} \spc
\eqa
\bq
\mrJ_{\pm} = \mrq\lpar - \mrs_1\,,\, \pm \;\zp\,\zm \rpar\,\mrT_{\pm} \spc
\eq
\bqa
\mrT_{-} &=& 
       \mrq( - \mrs_2\,,\, - \frac{\zm}{\mrX_2})\,\mrq( - \mrs_3\,,\,\frac{\zp}{\mrX_2})\,\mrX_2 \, 
\nl 
{}&\times& \Bigl[
          \mrE( - \mrs_2 + \mrs_1)\,\theta(\zm)\,\theta( - \mrX_2)\,\theta( - \zp)
          + \mrE(\mrs_2 - \mrs_1)\,\theta(\mrX_2)\,\theta(\zp)\,\theta( - \zm)
          \Bigr]
\nl
 {}&+& \mrq( - \mrs_2\,,\,\frac{\zm}{\mrX_2})\,\mrq( - \mrs_3\,,\, - \frac{\zp}{\mrX_2})\,\mrX_2 \, 
\nl
{}&\times& \Bigl[
          \mrE( - \mrs_3 + \mrs_1)\,\theta(\mrX_2)\,\theta(\zm)\,\theta( - \zp)
          + \mrE(\mrs_3 - \mrs_1)\,\theta(\zp)\,\theta( - \mrX_2)\,\theta( - \zm)
          \Bigr]
\nl
  {}&+& \mrq( - \mrs_2\,,\, - \frac{\zm}{1 - \mrX_2})\,\mrq( - \mrs_3\,,\,\frac{\zp}{1 - \mrX_2})\,( 1 - \mrX_2) \, 
\nl
{}&\times& \Bigl[
          \mrE( - \mrs_3 - \mrs_1)\,\theta(1 - \mrX_2)\,\theta(\zp)\,\theta( - \zm)
          + \mrE(\mrs_3 + \mrs_1)\,\theta(\zm)\,\theta(\mrX_2 - 1)\,\theta( - \zp)
          \Bigr]
\nl
 {}&+& \mrq( - \mrs_2\,,\,\frac{\zm}{1 - \mrX_2})\,\mrq( - \mrs_3\,,\, - \frac{\zp}{1 - \mrX_2})\,( 1 - \mrX_2) \, 
\nl
{}&\times& \Bigl[
          \mrE( - \mrs_2 - \mrs_1)\,\theta(\zp)\,\theta(\mrX_2 - 1)\,\theta( - \zm)
          + \mrE(\mrs_2 + \mrs_1)\,\theta(1 - \mrX_2)\,\theta(\zm)\,\theta( - \zp)
          \Bigr] \spc
\nl\nl
\mrT_{+} &=& 
       \mrq( - \mrs_2, - \frac{\zm}{\mrX_2})\,\mrq( - \mrs_3, - \frac{\zp}{\mrX_2})\,\mrX_2 \,
\nl
{}&\times& \Bigl[
          \mrE( - \mrs_3 + \mrs_2)\,\theta(\mrX_2)\,\theta( - \zm)\,\theta( - \zp)
          + \mrE(\mrs_3 - \mrs_2)\,\theta(\zm)\,\theta(\zp)\,\theta( - \mrX_2)
          \Bigr]
\nl
      {}&+& \mrq( - \mrs_2,\frac{\zm}{\mrX_2})\,\mrq( - \mrs_3,\frac{\zp}{\mrX_2})\,\mrX_2 \,
\nl
{}&\times& \Bigl[
          \theta( - \mrX_2)\,\theta( - \zm)\,\theta( - \zp)
          + \theta(\mrX_2)\,\theta(\zm)\,\theta(\zp)
          \Bigr]
\nl
       {}&+& \mrq( - \mrs_2, - \frac{\zm}{1 - \mrX_2})\,\mrq( - \mrs_3, - \frac{\zp}{1 - \mrX_2})\,(1 - \mrX_2) 
\,
\nl
{}&\times& \Bigl[
          \theta(1 - \mrX_2)\,\theta( - \zm)\,\theta( - \zp)
          + \theta(\zm)\,\theta(\zp)\,\theta(\mrX_2 - 1)
          \Bigr]
\nl
       {}&+& \mrq( - \mrs_2,\frac{\zm}{1 - \mrX_2})\,\mrq( - \mrs_3,\frac{\zp}{1 - \mrX_2})\,(1 - \mrX_2) \,
\nl
{}&\times& \Bigl[
          \mrE( - \mrs_3 + \mrs_2)\,\theta(1 - \mrX_2)\,\theta(\zm)\,\theta(\zp)
          + \mrE(\mrs_3 - \mrs_2)\,\theta(\mrX_2 - 1)\,\theta( - \zm)\,\theta( - \zp)
          \Bigr] \spp
\eqa
\end{enumerate}
\subsection{The sunrise (sunset) diagram \label{srexa}}
Sunrise integrals with non{-}canonical powers are connected by integration{-}by{-}parts (IBP) 
identities~\cite{Chetyrkin:1981qh} which can also be used to obtain a quasi-finite basis for multi-loop 
Feynman integrals~\cite{vonManteuffel:2014qoa}. 
We will study in details the following integral:
\bq
\mrS_{\mrD} = 
\int\,\mrd^{\mrd} \mrq_1\,\mrd^{\mrd} \mrq_2\,
(\mrq^2_1 + \mrm^2_1)^{-2}\,
((\mrq_1 - \mrq_2 + \mrp)^2 + \mrm^2_2)^{-2}\,
(\mrq^2_2 + \mrm^2_3)^{-1} \spp
\label{SD}
\eq
With equal masses we have the following expression:
\bq
\mrS_{\mrD} = \int\,\prod_i\,d\alpha_i\,\theta(\alpha_i)\,\delta \lpar \sum_i\,\alpha_i - 1\rpar\,
\alpha_3\,\Bigl[ \mrs\,\alpha_1\,\alpha_2\,\alpha_3 - 
\mrm^2\,\prod_{\mrj > i}\,\alpha_i\,\alpha_{\mrj} \Bigr]^{-1} \spc
\eq
corresponding to $a = b = c = 1$, in \eqn{Gdiag}. Partial quadratization of the integrand 
is obtained by introducing new variables, as shown in \eqn{quadsun};
the result is,
\bq
\mrS_{\mrD} = \int_0^1\,\mrd \rho\,\rho^2\,\mrd \mrx\,\rho\,(1 - \mrx)\,\mrQ^{-1}(\rho\,,\,\mrx) \spc
\qquad
\mrQ = \mra\,\mrx^2 + \mrb\,\mrx + \mrc \spc
\label{SDem}
\eq
where the quadratic form $\mrQ$ is defined by
\bq
\mra = - \mrb = \rho\,\mrm^2 - \rho\,\sigma\,\mrs \spc
\qquad
\mrc = - \mrm^2\,\sigma \spc
\eq
with $\sigma = 1 - \rho$.
Using the properties of Mellin transforms,
\bq
\mrS^{(2)}_{\mrD} = \int_0^1\,\mrd \rho\;\rho^2\,\mrI(\rho) \spc \quad
\int_0^1\,\mrd \rho\;\rho^{\mrt - 1}\,\mrI(\rho) = \mrg(\mrt) \spc \quad
\int_0^1\,\mrd \rho\;\rho^{\mrt}\,\mrI(\rho) = \mrg(\mrt + 1) \spc 
\eq
we will give explicit results for $\mrt= 2$. We introduce $\mrs = \lambda^2\,\mrm^2$ and
\bq
\mrB = - \frac{1}{4}\,\lambda^2\,\rho^2 + \frac{1}{4}\,(\lambda^2 + 3)\,\rho - 1 \spc
\eq
obtaining 
\bq
\mrS^{(1)}_{\mrD} = \frac{1}{\mrm^2}\,\int_0^1 \mrd \rho\,\mrd \mrx\,\rho\,(1 - \mrx)\,
\Bigl[ \mra\,\lpar \mrx - \frac{1}{2} \rpar^2 + \mrB \Bigr]^{-1} \spc
\eq
where now 
\bqa
\mra &=& \rho\,\lambda^2\,\lpar \rho - \rhz \rpar \spc
\qquad
\rhz = 1 - \frac{1}{\lambda^2} \spc
\nl
\mrB &=& - \frac{1}{4}\,\lambda^2\,(\rho - \rho_-)\,(\rho - \rho_+) \spc \qquad
\rho_{\pm} = \frac{1}{2\,\lambda^2}\,\Bigl[ \lambda^2 + 3 \pm
(\lambda^2 - 1)^{1/2}\,(\lambda^2 - 9)^{1/2} \Bigr] \spp
\eqa
Note that $\lambda = 1$ corresponds to the pseudo-threshold while $\lambda = 3$ to
the normal threshold. We have four different scenarios: 

\bei

\item[a)] If $\mrm^2 < \mrs < 9\,\mrm^2$ then the two roots are complex, 

\item[b)] for $\lambda^2 < 0$ we have $\rho_- > 1$ and $\rho_+ < 0$,

\item[c)] for $0 < \lambda < 1$  we have $\rho_{\pm} > 1$ and

\item[d)] for $\lambda > 3$ we have $0 < \rho_- < \rho_+ < 1$.

\eei
\paragraph{The case $\lambda < 3$} \hspace{0pt} \\
For a), b) and c) $\mrB$ is never zero for $0 \le \rho \le 1$ and we can write:
\bq
\int_0^1 \mrd \mrx\,\lpar 1 - \mrx \rpar\,\Bigl[ \mra\,\lpar \mrx - \frac{1}{2} \rpar^2 + \mrB \Bigr]^{-1} =
\frac{1}{2}\,\mrB^{-1}\;\hyp{1}{\frac{1}{2}}{\frac{3}{2}}{ - \frac{1}{4}\,\frac{\mra}{\mrB}} \spp 
\eq
As a consequence we obtain
\bq
\mrS_{\mrD} = - \frac{1}{\mrm^4}\,\int_0^1 \mrd \rho\,\int_{\mrL_0}\,\frac{\mrd \mrs_0}{2\,\pi\,i}\,
 \frac{\eG{ - \mrs_0}\,\eG{1 + \mrs_0}\,\eG{\frac{1}{2} + \mrs_0}}
      {\eG{\frac{3}{2} + \mrs_0}}\,
      \rho^{\mrs_0 + 1}\,\lpar \rho - \rhz \rpar^{\mrs_0}\,
      \lpar \rho - \rhm \rpar^{ - \mrs_0 - 1}\,
      \lpar \rho - \rhp \rpar^{ - \mrs_0 - 1} \spp
\eq
After performing the $\rho$ integration we obtain
\bqa
\mrS_{\mrD} &=& - \frac{1}{4\,\mrm^2}\,\int_{\mrL_0}\,\frac{\mrd \mrs_0}{2\,\pi\,i}\,
 \frac{\eG{ - \mrs_0}\,\eG{1 + \mrs_0}\,\eG{\frac{1}{2} + \mrs_0}}
      {\eG{\frac{3}{2} + \mrs_0}}\,
 \lpar \frac{4\,\rhz}{\lambda^2} \rpar^{\mrs_0}
\nl
{}&\times&
\mrF^{(3)}_{\sPD} \lpar 2 + \mrs_0\,;\, - \mrs_0\,,\,1 + \mrs_0\,,\,1 + \mrs_0\,;\,3 + \mrs_0\,;\,
\rhz^{-1}\,,\,\rhm^{-1}\,,\,\rhp^{-1} \rpar \spp
\eqa
For $\mrF_{\sPD}$ we use the Mellin{-}Barnes representation of \eqn{FDMB}; the final result corresponds to a Fox
function with parameters $\mrr = 4$, $\mrm = 8$ and $\mrn = 3$
\bq
\mrS^{(1)}_{\mrD} =
- \frac{1}{4}\,\Bigl[ \prod_{i=1}^{4}\,\int_{\mrL_i}\,\frac{\mrd \mrs_i}{2\,i\,\pi} \Bigr]\,
\frac{\mrN}{\mrD}\,
\lpar - \frac{4}{1 - \lambda^2} \rpar^{ - \mrs_1}\;
\lpar - \rho_0 \rpar ^{ - \mrs_2}\,
\lpar - \rho_{-} \rpar ^{ - \mrs_3}\,
\lpar - \rho_{+} \rpar ^{ - \mrs_4} \spc
\eq
where
\bqa
\mrN &=&
\eG{(\frac{1}{2} + \mrs_1}\,
\eG{( - \mrs_2}\,
\eG{( - \mrs_3}\,
\eG{( - \mrs_4}\,
\eG{(\mrs_2 - \mrs_1}
\nl
{}&\times&
\eG{(1 + \mrs_3 + \mrs_1}\,
\eG{(1 + \mrs_4 + \mrs_1}\,
\eG{(2 + \mrs_4 + \mrs_3 + \mrs_2 + \mrs_1} \spc
\nl
\mrD &=&
\eG{(1 + \mrs_1}\,
\eG{(\frac{3}{2} + \mrs_1}\,
\eG{(3 + \mrs_4 + \mrs_3 + \mrs_2 + \mrs_1}\spp
\eqa
The contours $\mrL_i$ are defined by
\bqa
- \frac{1}{2} < \mrr_1 < 0 & \qquad & \mrr_1 < \mrr_2 < 0 \spc
\nl
- 1 - \mrr_1 < \mrr_3 < 0 & \qquad &
\mathrm{max}(- 1 - \mrr_1\,,\,- 2 + \mrr_1 + \mrr_2 + \mrr_3) < \mrr_4 < 0 \spc
\eqa
where $\mrr_i = \Re \mrs_i$. The Feynman prescription gives $\lambda^2 + i\,\delta$ with $\delta \to 0_{+}$;
therefore, $\marg( - \rho_0) = - \pi + \delta$. With $1 < \lambda < 3$ we have 
$ - \pi/6 < \marg(\rho_{-}) < 0$ and $0 < \marg(\rho_{+}) < \pi/6$.
Convergence conditions are given by $\mid \phi_i \mid < \pi$ plus a set of additional strips; after discarding the 
irrelevant ones we obtain
\bqa
{}&{}& 
\mid \phi_1 + \phi_2 \mid < \pi \spc \quad
\mid \phi_1 - \phi_3 \mid < \pi \spc \quad
\mid \phi_1 - \phi_4 \mid < \pi \spc 
\nl
{}&{}&
\mid \phi_1 + \phi_2 - \phi_3 \mid < \pi \spc \quad
\mid \phi_1 - \phi_3 - \phi_4 \mid < \pi \spc 
\nl
{}&{}&
\mid \phi_1 + \phi_2 - 2\,\phi_3 \mid < 2\,\pi \spc \quad
\mid \phi_1 + \phi_2 - 2\,\phi_4 \mid < 2\,\pi \spc \quad
\mid \phi_1 + \phi_2 - \phi_3 - \phi_4\mid < \pi \spp
\eqa
\paragraph{The case $\lambda > 3$} \hspace{0pt} \\
In the following we will consider explicitly case c), $\mrs > 9\,\mrm^2$, where $\rho_{\pm} \in [0\,,\,1]$.
For $\lambda > 3$ we have $0 \le \rhm \le \rhp \le \rhz \le 1$. 
We have
\bq
\mrm^2\,\mrS^{(1)}_{\mrD} = \int_0^1 \mrd \rho\,\rho\,\mrI(\rho)
\eq
We introduce an additional variable:
\bq
\rhb= \frac{1}{2}\,(\rhp + \rhz) \spc 
\eq
and split the $[0\,,\,1]$ interval as follows
\bq
[0\,,\,1] = [0\,,\,\rhb]\;\cup\;[\rhb\,,\,1] \spp
\eq
Furthermore, we introduce
\bqa
\mrL_{+}(\mrz) &=& {}_2\mrF_{1}\lpar 1\,,\,\frac{1}{2}\,;\,\frac{3}{2}\,;\,\mrz^2 \rpar =
\frac{1}{2\,\mrz}\,\ln\frac{1 + \mrz}{1 - \mrz} \spc
\nl
\mrL_{-}(\mrz) &=& {}_2\mrF_{1}\lpar 1\,,\,\frac{1}{2}\,;\,\frac{3}{2}\,;\,-\,\mrz^2 \rpar =
\frac{1}{\mrz}\,\arctan(\mrz) \spp
\eqa
In $\mrL_{+}$, when $\mrz > 1$, we replace $\mrz^2 \to 
\mrz^2 + \mathrm{sign}(a - \mrB)\,i\,\delta$, with $\delta \to 0_{+}$. 
Consider the inner integral,
\bqa
\mrI(\rho) &=& \int_0^1 \mrd \mrx\,\,\lpar 1 - \mrx \rpar\,
\Bigl[ \mra\,\lpar \mrx - \frac{1}{2} \rpar^2 + \mrB \Bigr]^{-1} =
\int_{-\,1/2}^{+\,1/2} \mrd \mrx\,\lpar \frac{1}{2} - \mrx \rpar\,\lpar \mra\,\mrx^2 + \mrB \rpar^{-1}
\nl
{}&=& 
\frac{1}{2}\,\int_{-\,1/2}^{+\,1/2} \mrd \mrx\,\lpar \mra\,\mrx^2 + \mrB \rpar^{-1}
= \frac{1}{2}\,\int_0^1 \mrd \mrx \lpar \frac{1}{4}\,\mra\,\mrx^2 + \mrB \rpar^{-1} =
\frac{1}{2}\,\lpar \mrI_{\mrs} - \mrI_{\mrr} \rpar
\spp
\label{Irho}
\eqa
\bq
\mrI_{\mrr}(\rho)= \frac{4}{\mra}\;\hyp{1}{\frac{1}{2}}{\frac{3}{2}}{ - 4\,\frac{\mrB}{\mra}} \spc
\eq
\bq
\mrI_{\mrs}(\rho) = \int_0^{\infty} \mrd \mrx\,\lpar \frac{1}{4}\,\mra\,\mrx^2 + \mrB \rpar^{-1} =
\pi\,\lpar \mra\,\mrB \rpar^{-1/2} =
\frac{2\,\pi}{\lambda^2}\,\rho^{-1/2}\,\lpar \rhz - \rho \rpar^{-1/2}\,
\lpar \alpha\,\rho^2 + \beta\,\rho + \gamma \rpar^{-1/2}
\eq
There are different cases depending on the value of $\rho$. For a numerical evaluation we can use Gauss{-}Jacobi quadrature:
\bei

\item[1)] For $0 < \rho < \rhm$ we have
\bq
\mrJ_{1\mrs} = \int_0^{\rhm} \mrd \rho\,\rho\,\mrI_{\mrs}(\rho) =
               \int_0^{\rhm} \mrd \rho\,\rho^{1/2}\,\lpar \rhm - \rho \rpar^{-1/2}\,
               \Bigl[ \lpar \rhz - \rho \rpar\, \lpar \rhp - \rho \rpar \Bigr]^{-1/2} \spc
\eq
interval $[0\,,\,\rhm]$, weights $(1/2\,,\, - 1/2)$.
\item[2)] For $\rhm < \rho < \rhp$ we have
\bq
\mrJ_{2\mrs} = \int_{\rhm}^{\rhp} \mrd \rho\,\rho\,\mrI_{\mrs}(\rho) =
               i\,\int_{\rhm}^{\rhp} \mrd \rho\,
               \lpar \rho - \rhm \rpar^{-1/2}\,\lpar \rhp - \rho \rpar^{-1/2}\,
               \Bigl[ \rho^{1/2}\, \lpar\rhz - \rho\rpar^{-1/2} \Bigr] \spc
\eq
interval $[\rhm\,,\,\rhp]$, weights $( - 1/2\,,\, - 1/2)$.
\item[3)] For $\rhp < \rho < \rhb$ we have
\bq
\mrJ_{3\mrs} = \int_{\rhp}^{\rhb} \mrd \rho\,\rho\,\mrI_{\mrs}(\rho) =
               \int_{\rhp}^{\rhb} \mrd \rho\,
               \lpar \rho - \rhp \rpar^{-1/2}\,
               \Bigl[ \rho^{1/2}\,\lpar \rhz - \rho \rpar^{-1/2}\,\lpar \rho - \rhm \rpar^{-1/2} \Bigr] \spc
\eq
interval $[\rhp\,,\,\rhb]$, weights $( - 1/2\,,\,0)$.
\eei
We can also express the integrals through $\mrF_{\mrD}$ functions:
\bei

\item[1)] For $0 < \rho < \rhm$ we have
\bq
\mrJ_{1\mrs} = \frac{\pi}{2}\,\rhm\,(\rhz\,\rhp)^{-1/2}\,
\mrF^{(2)}_{\sPD}\lpar \frac{3}{2}\,;\,\frac{1}{2}\,,\,\frac{1}{2}\,;\,2\,;\,\frac{\rhm}{\rhz}\,,\,\frac{\rhm}{\rhp} \rpar \spp
\eq
\item[2)] For $\rhm < \rho < \rhp$ we have
\bqa
\mrJ_{2\mrs} &=& \frac{\pi}{2}\,\Bigl[
\rhp\,( - \rhz\,\rhm)^{-1/2}\,
\mrF^{(2)}_{\sPD}\lpar \frac{3}{2}\,;\,\frac{1}{2}\,,\,\frac{1}{2}\,;\,2\,;\,\frac{\rhp}{\rhz}\,,\,\frac{\rhp}{\rhm} \rpar 
\nl
{}&-&
\rhm\,( - \rhz\,\rhp)^{-1/2}\,
\mrF^{(2)}_{\sPD}\lpar \frac{3}{2}\,;\,\frac{1}{2}\,,\,\frac{1}{2}\,;\,2\,;\,\frac{\rhm}{\rhz}\,,\,\frac{\rhm}{\rhp} 
 \rpar\Bigr] \spp
\eqa
\item[3)] For $\rhp < \rho < \rhb$ we have
\bqa
\mrJ_{3\mrs} &=& \frac{2}{3}\,\Bigl[
\rhb^{3/2}\,(\rhz\,\rhm\,\rhp)^{-1/2}\,
\mrF^{(3)}_{\sPD}\lpar \frac{3}{2}\,;\,\frac{1}{2}\,,\,\frac{1}{2}\,,\,\frac{1}{2}\,;\,\frac{5}{2}\,;\,
\frac{\rhb}{\rhz}\,,\,\frac{\rhb}{\rhm}\,,\,\frac{\rhb}{\rhp} \rpar
\nl
{}&-&
\rhp\,(\rhz\,\rhm)^{-1/2}\,
\mrF^{(2)}_{\sPD}\lpar \frac{3}{2}\,;\,\frac{1}{2}\,,\,\frac{1}{2}\,;\,2\,;\,\frac{\rhp}{\rhz}\,,\,\frac{\rhp}{\rhm} \rpar \Bigr]
\spp
\eqa
The final integration over $\rho$ can be performed by using the MB representation for the Lauricella 
functions (\eqn{FDMB}) with a result expressed again in terms of Lauricella functions.
\eei
There is also a third way to compute the integrals: we can proceed by defining
\bq
\mrQ_{1/2}(\mrz) =
\int_0^1 \mrd \mry\,\Bigl( \alpha\,\mrz^2\,\mry^2 + \beta\,\mrz\,\mry + \gamma \Bigr)^{-1/2}
= \int_0^1 \mrd \mry\,\Bigl[ \alpha\,\lpar \mry - \mrY \rpar^2 + \Sigma \Bigr]^{-1/2} \spp
\eq
\bei

\item[1)] $0 < \rho < \rhm$. Here we have
\bq
\alpha = 1 \spc \quad \mrz\,\mrY= - \frac{\lambda^2 + 3}{2\,\lambda^2} = - \rhc < 0 \spc \quad
\Sigma = - \frac{1}{\lambda^4}\,\lpar 1 - \lambda^2 \rpar\,\lpar 9 - \lambda^2 \rpar \spp
\label{Sdef}
\eq
Therefore, the result is
\bq
\mrQ_{1/2}(\mrz) = \Sigma^{-1/2}\,\sum_{i=1}^{2}\,( - 1 )^i\,\mrY_i\;
\hyp{\frac{1}{2}}{\frac{1}{2}}{\frac{3}{2}}{\xi_i} \spc
\eq
where $\mrY_1 = - \mrY$, $\mrY_2 = 1 - \mrY$ and $\xi_i = - \mrY_i^2/\Sigma$. Depending on $\xi_i$ we will write
\bqa
\hyp{\frac{1}{2}}{\frac{1}{2}}{\frac{3}{2}}{\xi^2} &=& 
\frac{\pi}{2} - \xi^{-2}\,\lpar 1 - \xi^2 \rpar^{1/2}\;
\hyp{\frac{1}{2}}{\frac{1}{2}}{\frac{3}{2}}{1 - \xi^{-2}} \spc \quad \xi > 1 \spc
\nl
{}&=& \frac{1}{\xi}\,\mathrm{arcsin}(\xi) \spc \quad 0 < \xi < 1 \spc
\nl
{}&=& \frac{1}{\eta}\,\mathrm{arctn}(\eta) \spc \quad \eta^2 = - \xi^2 > 0 \spp
\eqa
In this case we have $\Sigma < 0$. We will present explicit results only for the first interval (the remaining
intervals can be treated with the same procedure) where
\bqa
\mrJ_{1\mrs} &=&
\rhm^{3/2}\,\lpar \rhz - \rhm \rpar^{-1/2}\,\mrQ_{1/2}(\rhm) +
\frac{1}{2}\,\mrJ^{\mrR}_{1\mrs}
\nl
\mrJ^{\mrR}_{1\mrs} &=&
\,\int_0^{\rhm} \mrd \rho\,\Bigl[
\rho^{3/2}\,\lpar \rhz - \rho \rpar^{-3/2} +
\rho^{1/2}\,\lpar \rhz - \rho \rpar^{-1/2} \Bigr]\,\mrQ_{1/2}(\rho) \spp
\eqa
The final result has the following form:
\bq
\mrJ^{\mrR}_{1\mrs} = \rho_{\mrc}\,\sum_{i=1}^{4}\,\mrH_i \spc
\label{thirdway}
\eq
\bqa
\mrH_1 &=& - \mrC_1\,\Bigl[ \prod_{i=1}^{2}\,\int_{\mrL_i}\,\frac{\mrd \mrs_i}{2\,i\,\pi} \Bigr]\,
\nl
{}&\times&
\frac{
\eG{ - \mrs_1}\,
\eG{ - \mrs_2}\,
\eGs{\frac{1}{2} + \mrs_1}\,
\eG{\frac{3}{2} + \mrs_2}
\eG{\frac{3}{2} + \mrs_2 - 2\,\mrs_1}\,
}{
\eG{\frac{3}{2} + \mrs_1}\,
\eG{\frac{5}{2} + \mrs_2 - 2\,\mrs_1}\,
}\,
\nl
{}&\times&
\lpar\frac{\rho_{\mrc}^2}{\rho^2_{-}\,\Sigma} \rpar^{\mrs_1}\,\lpar - \frac{\rho_{-}}{\rho_0} \rpar^{\mrs_2}
\nl\nl
\mrH_2 &=& - \frac{1}{2}\mrC_2\,\,\Bigl[ \prod_{i=1}^{2}\,\int_{\mrL_i}\,\frac{\mrd \mrs_i}{2\,i\,\pi} \Bigr]\,
\nl
{}&\times&
\frac{
\eG{ - \mrs_1}\,
\eG{ - \mrs_2}\,
\eGs{\frac{1}{2} + \mrs_1}\,
\eG{\frac{1}{2} + \mrs_2}\,
\eG{\frac{1}{2} + \mrs_2 - 2\,\mrs_1}\,
}{
\eG{\frac{3}{2} + \mrs_1}\,
\eG{\frac{3}{2} + \mrs_2 - 2\,\mrs_1}\,
}\,
\nl
{}&\times&
\lpar \frac{\rho_{\mrc}^2}{\rho^2_{-}\,\Sigma}\rpar^{\mrs_1}\,\lpar - \frac{\rho_{-}}{\rho_0} \rpar^{\mrs_2}
\nl\nl
\mrH_3 &=& - \mrC_1\,\,\Bigl[ \prod_{i=1}^{3}\,\int_{\mrL_i}\,\frac{\mrd \mrs_i}{2\,i\,\pi} \Bigr]\,
\nl
{}&\times&
\frac{
\eG{ - \mrs_1}\,
\eG{ - \mrs_2}\,
\eG{ - \mrs_3}\,
\eGs{\frac{1}{2} + \mrs_1}\,
\eG{\frac{3}{2} + \mrs_2}\,
\eG{ - 1 + \mrs_3 - 2\,\mrs_1}\,
\eG{\frac{3}{2} + \mrs_3 + \mrs_2 - 2\,\mrs_1}\,
}{
\eG{ - 1 - 2\,\mrs_1}\,
\eG{\frac{3}{2} + \mrs_1}\,
\eG{\frac{5}{2} + \mrs_3 + \mrs_2 - 2\,\mrs_1}\,
}\,   
\nl
{}&\times&     
\lpar \frac{\rho_{\mrc}^2}{\rho^2_{-}\,\Sigma} \rpar^{\mrs_1}\,\lpar - \frac{\rho_{-}}{\rho_0}\rpar^{\mrs_2}\,
\lpar - \frac{\rho_{-}}{\rho_{\mrc}} \rpar^{\mrs_3}
\nl\nl
\mrH_4 &=& - \frac{1}{2}\,\mrC_2\,\,\Bigl[ \prod_{i=1}^{3}\,\int_{\mrL_i}\,\frac{\mrd \mrs_i}{2\,i\,\pi} \Bigr]\,
\nl
{}&\times&
\frac{
\eG{ - \mrs_1}\,
\eG{ - \mrs_2}\,
\eG{ - \mrs_3}\,
\eGs{\frac{1}{2} + \mrs_1}\,
\eG{\frac{1}{2} + \mrs_2}\,
\eG{ - 1 + \mrs_3 - 2\,\mrs_1}\,
\eG{\frac{1}{2} + \mrs_3 + \mrs_2 - 2\,\mrs_1}\,
}{
\eG{ - 1 - 2\,\mrs_1}\,
\eG{\frac{3}{2} + \mrs_1}\,
\eG{\frac{3}{2} + \mrs_3 + \mrs_2 - 2\,\mrs_1}
}\,
\nl
{}&\times&
\lpar \frac{\rho_{\mrc}^2}{\rho^2_{-}\,\Sigma} \rpar^{\mrs_1}\,\lpar - \frac{\rho_{-}}{\rho_0} \rpar^{\mrs_2}\,
\lpar - \frac{\rho_{-}}{\rho_{\mrc}} \rpar^{\mrs_3}
\eqa
where we have introduced
\bq 
\mrC_1 = \pi^{3/2}\,\lpar \frac{\rho_{-}}{\rho_0} \rpar^{3/2}\,\Sigma^{ - 1/2} \spc
\qquad
\mrC_2 = \pi^{3/2}\,\lpar \frac{\rho_{-}}{\rho_0} \rpar^{1/2}\,\Sigma^{ - 1/2} \spp
\eq
Here we use the Feynman presciption:
\bqa
{}&{}& \Sigma < 0 \spc \quad
\frac{\rho_{-}}{\rho_0} > 0 \spc \quad
\frac{\rho_{-}}{\rho_{\mrc}} > 0 \spc
\nl
{}&{}& \Sigma \to \Sigma + i\,\delta \spc \quad
\frac{\rho_{-}}{\rho_0} \to \frac{\rho_{-}}{\rho_0} + i\,\delta  \spc \quad
\frac{\rho_{-}}{\rho_{\mrc}} \to \frac{\rho_{-}}{\rho_{\mrc}} + i\,\delta  \spc \quad
\frac{\rho^2_{-}}{\rho^2_{\mrc}}\,\Sigma \to 
\frac{\rho^2_{-}}{\rho^2_{\mrc}}\,\Sigma - i\,\delta \spp
\eqa
The result based on \eqn{thirdway} has the advantage of having an explicit dependence on $\Sigma$ introduced
in \eqn{Sdef}, which allows us to study the behavior of the sunset at the normal threshold using \eqn{zbeh}.
If we define
\bq
\mrz_1 = \frac{\rho^2_{-}\,\Sigma}{\rho^2_{\mrc}} \spc \quad
\mrz_2 = - \frac{\rho_{0}}{\rho_{-}} \spc \quad
\mrz_3 = - \frac{\rho_{\mrc}}{\rho_{-}} \spc
\eq
and define $\phi_i = \marg( \mrz_i)$, the region of convergence for $\mrH_{1\,,\,2}$ is $\mid \phi_i \mid < \pi$, while
for $\mrH_{3\,,\,4}$ is $\mid \phi_i \mid < \pi$ plus an additional condition, $\mid \phi_1 + 2\,\phi_3 \mid < \pi$.

From a computational point of view it could be better to stop before the last step; for instance, we will have
\bqa
\mrH_4 &=& - \frac{1}{2}\,\frac{\mrC_2}{\pi}\,\,\int_{\mrL}\,\frac{\mrd \mrs}{2\,i\,\pi}\,
\frac{\eG{ - \mrs}\,\eGs{\frac{1}{2} + \mrs}\,\eG{\frac{1}{2} - 2\,\mrs}}
     {\eG{\frac{3}{2} + \mrs}\,\eG{\frac{3}{2} - 2\,\mrs}}\,
\lpar \frac{\rho_{\mrc}^2}{\rho^2_{-}\,\Sigma} \rpar^{\mrs}
\nl {}&\times&
\mrF^{(2)}_{\sPD}\lpar \frac{1}{2} - 2\,\mrs\,;\,\frac{1}{2}\,,\, - 1 - 2\,\mrs\,;\,
                       \frac{3}{2} - 2\,\mrs\,;\,\frac{\rho_{-}}{\rho_0}\,,\,
                                                 \frac{\rho_{-}}{\rho_{\mrc}} \rpar \spp
\label{FDexa}
\eqa
\eei
\subsection{The kite diagram \label{kited}}
The kite diagram has been discussed in terms of elliptic polylogaritms in \Bref{Adams:2016xah}; for a numerical
treatment see \Bref{Passarino:2001jd}.

The kite integral, although non{-}trivial, is simple enough to illustrate some of the basic aspects of
our procedure. Let us define the following function
\bq
\mrQ = \rho_1^2\,\sigma_1\,\lpar \mrx_1 - \frac{1}{2} \rpar^2 + 
       \rho_2^2\,\sigma_2\,\lpar \mrx_2 - \frac{1}{2} \rpar^2 +
       2\,\rho_1\,\rho_2\,\rho_3\,\lpar \mrx_1 - \frac{1}{2} \rpar\,\lpar \mrx_2 - \frac{1}{2} \rpar \spc
\eq
and also a reduced Symanzik polynomial $\mrS_2 = {\overline{\mrS}}_2\,\mrs$. We consider $3$ different
configurations:
\bq
\mrm_i = \mrm \spc \quad
{\overline{\mrS}}_{2\,\mra} = \mrQ + \beta\,\lpar \lambda^2 - \frac{1}{4}\,\sigma_3 \rpar \spc
\eq
with $\lambda^2 = \mrm^2/\mrs$.
\bq
\mrm_3 = 0 \spc\; \mrm_i = \mrm\,(i \not= 3) \spc \quad
{\overline{\mrS}}_{2\,\mrb} = \mrQ + \beta\,\sigma_3\,\lpar \lambda^2 - \frac{1}{4} \rpar \spp
\eq
\bq
\mrm_{1,2,3} = \mrm \spc \; \mrm_{4,5} = \mrM \spc \quad
{\overline{\mrS}}_{2\,\mrc} = \mrQ + \beta\,\lpar \rho_2\,\lambda^2_{\mrM} + \sigma_2\,\lambda^2_{\mrm} - 
\frac{1}{4}\,\sigma_3 \rpar \spp
\eq
It is easy to see that the following $3$ quantities
\bq
\mrC_{\mra} = \beta\,\lpar \lambda^2 - \frac{1}{4}\,\sigma_3 \rpar \spc \quad
\mrC_{\mrb} = \beta\,\sigma_3\,\lpar \lambda^2 - \frac{1}{4} \rpar \spc \quad
\mrC_{\mrc} = \beta\,\lpar \rho_2\,\lambda^2_{\mrM} + \sigma_2\,\lambda^2_{\mrm} - 
              \frac{1}{4}\,\sigma_3 \rpar \spc
\eq
are the ratios Cayley{/}Gram determinants~\cite{Melrose:1965kb} of the Symanzik polynomials and we can write
\bq
{\overline{\mrS}}_{2\,\mrj} = \mrQ + \mrC_{\mrj} \spp
\label{pinch}
\eq
The C{-}parameters are the quantities to be used in the MB splitting (\eqn{MBsplit}),
\bq
\mrI_{\mrj} = \int_{\mrL}\,\frac{\mrd \mrs_1}{2\,i\,\pi}\,\eB{\mrs_1}{1 - \mrs_1}\,\mrC_{\mrj}^{ - \mrs_1}\,
\int_0^1 \mrd \mrx_1\,\mrd \mrx_2\,\mrQ^{\mrs_1 - 1} \spc
\eq
after which we can perform the $\mrx_{1,2}$ integration obtaining
\bq
\int_0^1 \mrd \mrx_1\,\mrd \mrx_2\,\mrQ^{\mrs_1 - 1} = \mrJ\lpar \rho_1\,,\,\rho_2 \rpar \spc
\eq
\bqa
\mrJ &=& - \frac{1}{\beta}\,\Bigl[ \prod_{i=1}^{5}\,\int_{\mrL_i}\,\frac{\mrd \mrs_i}{2\,i\,\pi} \Bigr]\,
\frac{\mrN}{\mrD}\,\lpar \frac{1}{\rho_1}\,\mrJ_1 + \frac{1}{\rho_2}\,\mrJ_2 \rpar \spc
\nl
\mrN &=&
\eG{\mrs_1}\,
\eG{ - \mrs_3}\,
\eG{ - \mrs_4}\,
\eG{ - \mrs_5}\,
\eG{\mrs_4 - \mrs_2}\,
\eG{\mrs_5 - \mrs_2}\,
\eG{1 + \mrs_2 + \mrs_1}\,
\eG{1 + \mrs_3 + \mrs_2 - \mrs_1}\,
\eG{1 + \mrs_5 + \mrs_4 + \mrs_3} \spc
\nl
\mrD &=&
\eG{ - \mrs_2}\,
\eG{2 + \mrs_2 + \mrs_1}\,
\eG{2 + \mrs_5 + \mrs_4 + \mrs_3} \spc
\nl
\mrJ_1 &=&
\mrq\lpar - \mrs_1, - \frac{1}{\beta\,\rho_1} \rpar\,
\mrq\lpar - \mrs_2, - \frac{\beta}{\rho_2\,\sigma_2\,\mrv_{+}\,\mrv_{-}} \rpar\,
\mrq\lpar - \mrs_3,\frac{\rho_1}{\rho_2} \rpar\,
\mrq\lpar - \mrs_4, - \mru_{-} \rpar\,
\mrq\lpar - \mrs_5, - \mru_{+}\rpar \spc
\nl
\mrJ_2 &=& 
\mrq\lpar - \mrs_1, - \frac{1}{\beta\,\rho_2} \rpar\,
\mrq\lpar - \mrs_2, - \frac{\beta\,\rho_2}{\rho_1^2\,\sigma_1\,\mru_{+}\,\mru_{-}} \rpar\,
\mrq\lpar - \mrs_3,\frac{\rho_2}{\rho_1} \rpar\,
\mrq\lpar - \mrs_4, - \mru_{-} \rpar\,
\mrq\lpar - \mrs_5, - \mru_{+} \rpar \spc
\eqa
where we have introduced $\mrq( \mrs\,,\,\mrz )= \mrz^{\mrs}$ and
\bq
\mru_{\pm} = \frac{\rho_2}{\sigma_1\,\rho_1}\,\lpar - \rho_3 \pm i\,\sqrt{\beta} \rpar \spc \quad
\mrv_{\pm} = \frac{\rho_1}{\sigma_2\,\rho_2}\,\lpar - \rho_3 \pm i\,\sqrt{\beta} \rpar \spp
\eq
The $\rho_1, \rho_2$ integration depends on the configuration of the kite and can be performed using again the strategy 
outlined in \sect{str}. The (configuration independent) $\mrJ$ function has been reduced to generalized Fox functions 
with parameters $\mrr = 5$, $\mrm = 9$ and $\mrn = 3$.

Consider again \eqn{pinch}, it is immediately seen that $\mrx_1 = \mrx_2 = 1/2$ is a pinch singularity when
$\mrC_{\mrj} \to 0$, \ie the leading Landau singularity~\cite{Passarino:2001jd,Passarino:2018wix} 
for the $\mrx_1, \mrx_2$ integral of $\mrS_2^{-1}$. Let us consider configuration $\mrb$, the corresponding BST
factor is proportional to $\lambda^2 - 1/4$. As a check we write the Landau equations for the 
kite~\cite{Passarino:2001jd}, keeping $\mrm_3 = \mrM$:
\bqa
{}&{}& - \mrm^2\,\alpha_1 + \frac{1}{2}\,(\mrs - 2\,\mrm^2)\,\alpha_2 - \frac{1}{2}\,\mrM^2\,\alpha_3 = 0 \spc 
\nl
{}&{}& (\mrM^2 - 2\,\mrm^2)\,\alpha_1 + (\mrs + \mrM^2 - 2\,\mrm^2)\,\alpha_2 + \mrM^2\,\alpha_3 = 0 \spc 
\nl
{}&{}& \alpha_2 = \alpha_1 \spc \quad \mrs \not= 0 \spc
\nl
{}&{}& \mrM^2\,\alpha_3 + (\mrM^2 - 2\,\mrm^2)\,\alpha_4 + (\mrs + \mrM^2 - 2\,\mrm^2)\,\alpha_5 = 0 \spc 
\nl
{}&{}& \mrM^2\,\alpha_3 + 2\,\mrm^2\,\alpha_4 + (2\,\mrm^2 - \mrs)\,\alpha_5 = 0 \spc 
\nl
{}&{}& \alpha_5 = \alpha_4 \spc \quad \mrs \not= 0 \spp
\eqa
For $\mrM = 0$ we have a non trivial solution ($\alpha_i \not= 0, \forall i$) if $\mrs = 4\,\mrm^2$, \ie
$\lambda = 1/2$.
If we are interested in the behavior around $\lambda = 1/2$ we can use the result of \Bref{Ferroglia:2002mz}
which gives
\bq
\int_0^1 \mrd \mrx_1\, \mrd \mrx_2\,{\overline{\mrS}}_{2\,\mrb}^{ - 1} \sim
- \frac{i}{2}\,\frac{1}{\rho_1\,\rho_2\,\sqrt{\beta}}\,
\Bigl( \ln\frac{\mrA_{+}}{\mrA_{-}} + \ln\frac{\mrB_{+}}{\mrB_{-}} \Bigr)\,\ln \mrC_{\mrb} \spc
\eq
where $\lambda \to 1/2$ and
\bqa
\mrA_{\pm} &=& \sqrt{\beta}\,\Bigl[ \lpar \rho_1 + \rho_2 \rpar\,\lambda^2 - \frac{1}{2}\,\rho_2 \Bigr] \pm
\frac{i}{2}\,\rho_1\,\rho_2 \spc 
\nl
\mrB_{\pm} &=& \sqrt{\beta}\,\Bigl[ \lpar \rho_1 + \rho_2 \rpar\,\lambda^2 - \frac{1}{2}\,\rho_1 \Bigr] \pm
\frac{i}{2}\,\rho_1\,\rho_2 \spp
\eqa
The presence of $\sqrt{\beta}$ is common in our approach, therefore, we will use this example to show how to
proceed. Given $\beta = \rho_1\,\rho_3 + \sigma_2\,\rho_2$ we use the following change of variable:
\bq
\rho_2 = \frac{\sigma_1}{\mrt_2^2 + 1}\,\lpar \mrt_2^2 - 2\,\omega\,\mrt_2 \rpar \spc
\qquad
\sqrt{\beta} = \frac{\sqrt{\sigma_1\,\rho_1}}{\mrt_2^2 + 1}\,\lpar t^2 + \frac{1}{\omega}\,\mrt_2 - 1 \rpar \spc
\eq
where $\omega^2 = \rho_1/\sigma_1$. We obtain $2\,\omega \le \mrt_2 \le \infty$ with a Jacobian
\bq
\mrJ_2 = 2\,\frac{\sqrt{\sigma_1\,\rho_1}}{(\mrt_2^2 + 1)^2}\,\lpar \mrt_2^2 + \frac{1}{\omega}\,\mrt_2 - 1 \rpar \spp
\eq
We are left with $\sqrt{\sigma_1\,\rho_1}$ for which we introduce
\bq
\rho_1 = \frac{\mrt_1^2}{\mrt_1^2 + 1} \spc \qquad
\sqrt{\sigma_1\,\rho_1} = \frac{\mrt_1}{\mrt_1^2 + 1} \spc
\eq
with $0 \le \mrt_1 \le \infty$ amd $\omega= \mrt_1$. The Jacobian is
\bq
\mrJ_1 = 2\,\frac{\mrt_1}{(\mrt_1^2 + 1)^2} \spp
\eq 
Clearly the leading singularity should not be confused with normal(pseudo){-}thresholds. Consider the
configuration $\mra)$ (equal masses): the three{-}particle cut can be studied by observing that the
equal masses sunrise has the same cut and the sunrise is a contraction of the kite corresponding to
$\mrx_1 = 1$ and $\mrx_2 = 0$. This fact suggests a change of variables:
\bq
\mrx_1 \to 1 - \mrx_1 = \mru_1 \spc \qquad 
\mrx_2 = \mru_2 \spc \qquad
\rho_{1,2} \to \rho_{1,2} - \frac{1}{3} = \mru_{3,4} \spc
\eq
transforming the Symanzik polynomial into the following form:
\bqa
{}&{}&- \mrr_1\,\mru_1^2
- \mrr_2\,\mru_2^2
- 2\,\lpar \mrr_3 + \frac{1}{27} \rpar\,\mru_1\,\mru_2
+ \lpar \mrr_1 + \mrr_3 + \frac{1}{27} \rpar\,\mru_1
+ \lpar \mrr_2 + \mrr_3 + \frac{1}{27} \rpar\,\mru_2
\nl
{}&{}& - \lambda^2\,\mrr_4 - \mrr_3
       + \frac{1}{3}\,\lpar \lambda^2 - \frac{1}{9}\rpar \spc
\eqa
\bqa
\mrr_1 &=& \mru_3^3 - \frac{1}{3}\,\mru_3 - \frac{2}{27} \spc
\qquad
\mrr_2 = \mru_4^3 - \frac{1}{3}\,\mru_4 - \frac{2}{27} \spc
\nl
\mrr_3 &=& - \mru_3\,\mru_4^2 - \mru_3^2\,\mru_4 - \frac{1}{3}\,\mru_4 \spc
\qquad
\mrr_4 = \mru_3^2 + \mru_4^2 + \mru_3\,\mru_4 \spp
\eqa
Therefore, the point $\mru_i = 0$ is a zero of the Symanzik polynomial for $\lambda = 1/3$. 
Of course our function is not anymore a quadratic in 
two variables but this way of transforming the Symanzik polynomial suggests to use $\lambda^2 - 1/9$ as the
relevant object when performing the MB splitting of \eqn{MBsplit}. After integarting over $\mru_1$ and $\mru_2$ we obtain
\bqa
\mrJ &=& \frac{1}{4\,\beta}\,\Bigl[ \prod_{i=1}^{6}\,\int_{\mrL_i}\,\frac{\mrd \mrs_i}{2\,i\,\pi} \Bigr]\,
\frac{\mrN}{\mrD}\,\mrq( - \mrs_1\,,\,\frac{1}{3}\,\mrC)\,
\mrq( - \mrs_2\,,\, - \mrr_3 - \lambda^2\,\mrr_4)\,\sum_{i=1}^{8}\,\mrJ_i \spc
\nl\nl
\mrN &=&
\eG{\mrs_1}\,
\eG{\mrs_2}\,
\eG{ - \mrs_4}\,
\eG{ - \mrs_5}\,
\eG{ - \mrs_6}\,
\eG{\mrs_5 - \mrs_3}\,
\eG{\mrs_6 - \mrs_3}
\nl
{}&\times& \eGs{2 + \mrs_2 + \mrs_1}\,
\eG{1 + \mrs_3 + \mrs_2 + \mrs_1}\,
\eG{1 + \mrs_6 + \mrs_5 + \mrs_4}\,
\eG{1 + \mrs_4 + \mrs_3 - \mrs_2 - \mrs_1} \spc
\nl\nl
\mrD &=&
\eG{ - \mrs_3}\,
\eGs{1 + \mrs_2 + \mrs_1}\,
\eG{2 + \mrs_3 + \mrs_2 + \mrs_1}\,
\eG{2 + \mrs_6 + \mrs_5 + \mrs_4} \spc
\eqa
\bqa
\mrJ_1 &=& \mrq(\mrs_3, - \frac{1}{4}\,\mrr_2)\,\mrq(\mrs_4, - \mrx_0)\,
\mrq(\mrs_5, - \mrx_1^{-1})\,\mrq(\mrs_6, - \mrx_2^{-1})\,
\mrq( - 1 - \mrs_3 + \mrs_2 + \mrs_1, - \frac{1}{54}\,\mrc_2)\,\mrq(1 + \mrs_3 - \mrs_2 - \mrs_1, - \mrx_0)
\spc \nl
\mrJ_2 &=& \mrq(\mrs_3, - \frac{1}{4}\,\mrr_2)\,\mrq(\mrs_4, - \mrx_0)\,\mrq(\mrs_5, - \mrx_1^{-1})\,
\mrq(\mrs_6, - \mrx_2^{-1})\,
\mrq( - 1 - \mrs_3 + \mrs_2 + \mrs_1,\frac{1}{54}\,\mrc_2)\,\mrq(1 + \mrs_3 - \mrs_2 - \mrs_1, - \mrx_0)
\spc \nl
\mrJ_3 &=& \mrq(\mrs_3, - \frac{1}{4}\,\mrr_2)\,\mrq(\mrs_4,\mrx_0)\,\mrq(\mrs_5, - \mrx_5^{-1})\,
\mrq(\mrs_6, - \mrx_6^{-1})\,
\mrq( - 1 - \mrs_3 + \mrs_2 + \mrs_1, - \frac{1}{54}\,\mrc_2)\,\mrq(1 + \mrs_3 - \mrs_2 - \mrs_1,\mrx_0)
\spc \nl
\mrJ_4 &=& \mrq(\mrs_3, - \frac{1}{4}\,\mrr_2)\,\mrq(\mrs_4,\mrx_0)\,\mrq(\mrs_5, - \mrx_5^{-1})\,
\mrq(\mrs_6, - \mrx_6^{-1})\,\mrq( - 1 - \mrs_3 + \mrs_2 + \mrs_1,\frac{1}{54}\,\mrc_2)\,
\mrq(1 + \mrs_3 - \mrs_2 - \mrs_1,\mrx_0)
\spc \nl
\mrJ_5 &=& \mrq(\mrs_3, - \frac{1}{4}\,\mrr_1)\,\mrq(\mrs_4, - \mrx_0^{-1})\,
\mrq(\mrs_5, - \mrx_3^{-1})\,\mrq(\mrs_6, - \mrx_4^{-1})\,
\mrq( - 1 - \mrs_3 + \mrs_2 + \mrs_1, - \frac{1}{54}\,\mrc_1\,\mrx_0)
\spc \nl
\mrJ_6 &=& \mrq(\mrs_3, - \frac{1}{4}\,\mrr_1)\,\mrq(\mrs_4, - \mrx_0^{-1})\,\mrq(\mrs_5, - \mrx_3^{-1})\,
\mrq(\mrs_6, - \mrx_4^{-1})\,
\mrq( - 1 - \mrs_3 + \mrs_2 + \mrs_1,\frac{1}{54}\,\mrc_1\,\mrx_0)
\spc \nl
\mrJ_7 &=& \mrq(\mrs_3, - \frac{1}{4}\,\mrr_1)\,\mrq(\mrs_4,\mrx_0^{-1})\,
\mrq(\mrs_5, - \mrx_1^{-1})\,\mrq(\mrs_6, - \mrx_2^{-1})\,
\mrq( - 1 - \mrs_3 + \mrs_2 + \mrs_1, - \frac{1}{54}\,\mrc_1\,\mrx_0)
\spc \nl
\mrJ_8 &=& \mrq(\mrs_3, - \frac{1}{4}\,\mrr_1)\,\mrq(\mrs_4,\mrx_0^{-1})\,
\mrq(\mrs_5, - \mrx_1^{-1})\,\mrq(\mrs_6, - \mrx_2^{-1})\,
\mrq( - 1 - \mrs_3 + \mrs_2 + \mrs_1,\frac{1}{54}\,\mrc_1\,\mrx_0) \spp
\eqa       
We have introduced $\mrC = \lambda^2 - 1/9$, $\mrq(\mrs\,,\,\mrz) = \mrz^{\mrs}$ and also
\bq
\mrc_1 = 1 + 27\,(\mrr_2 + \mrr_3) \spc \quad
\mrc_2 = 1 + 27\,(\mrr_1 + \mrr_3) \spc \quad
\mrc_3 = 1 + 27\,\mrr_3 \spc
\eq
$\mrx_0 = - \mrc_2/\mrc_1$ while the remaing $\mrx$ variables are
\bqa
\mrx_{1,2} \quad &\hbox{roots of}& \quad \mrr_2\,\mrx^2 - \frac{2}{27}\,\mrc_3\,\mrx + \mrr_1 \spc
\nl 
\mrx_{3,4} \quad &\hbox{roots of}& \quad \mrr_2\,\mrx^2 + \frac{2}{27}\,\mrc_3\,\mrx + \mrr_1 \spc
\nl
\mrx_{5,6} \quad &\hbox{roots of}& \quad \mrr_1\,\mrx^2 - \frac{2}{27}\,\mrc_3\,\mrx + \mrr_2 \spc
\nl
\mrx_{7,8} \quad &\hbox{roots of}& \quad \mrr_1\,\mrx^2 + \frac{2}{27}\,\mrc_3\,\mrx + \mrr_2 \spp
\eqa
The behavior at threshold is controlled by the pole at $\mrs_1 = 0$. The result
corresponds to a Fox function of \sect{GMFF} with parameters $\mrr = 6, \mrm = 12$ and $\mrn = 5$.

Finally, let us consider a configuration with $\mrm_3 = \mrM$ and $\mrm_i = \mrm$ for $i \not= 3$. If we are interested in
the $\mrs = 4\,\mrm^2$ threshold we perform the transformation
\bq
\mru_{1,2} = \mrx_{1,2} - \frac{1}{2} \spc \qquad \mru_{3,4} = \rho_{1,2} - \frac{1}{2} \spc
\eq
giving the following form for the reduced Symanzik polynomial
\bq
{\overline{\mrS}}_2 = \frac{1}{8}\,\Bigl[
(1 + \mrr_1)\,\mru_1^2 + (1 + \mrr_2)\,\mru_2^2 +
(\mrr_4 - \mrr_1 - \mrr_2)\,\mru_1\,\mru_2 +
\mrr_3\lambda^2_{\mrM} + (2 + \mrr_4)\,(\lambda^2_{\mrm} - \frac{1}{4}) \Bigr] \spc
\eq
where we have defined
\bqa
\mrr_1 &=& 2\,\mru_3\,(1 - 2\,\mru_3 - 4\,\mru_3^2) \spc
\nl
\mrr_2 &=& 2\,\mru_4\,(1 - 2\,\mru_4 - 4\,\mru_4^2) \spc
\nl
\mrr_3 &=& 8\,\mru_3\,\mru_4\,(1 + 16\,\mru_3 + 16\,\mru_4) - \mrr_2 - \mrr_1 \spc
\nl
\mrr_4 &=& \mrr_1 + \mrr_2 + \mrr_3 + 4\,(\mru_3 + \mru_4) + 8\,(\mru_3^2 + \mru_3\,\mru_4 + \mru_4^2) \spp
\eqa
The point $\mru_i = 0$ is a zero of the Symanzik polynomial for $\lambda_{\mrm} = 1/2$; therefore, $\lambda^2_{\mrm} - 1/4$ 
is the releant object to be used in the MB splitting of \eqn{MBsplit}.
\subsection{The crossed delta{-}kite diagram \label{cdkiteexa}}
For the diagram of Fig.~\ref{fd_cdkite} in the configuration where the external masses are zero while
the internal masses are equal we have
\bq
\frac{\mrS_2}{\mrs} =
 \sigma_3\,\rho_3^2\,\mrx_3^2
+ \rho_1\,\rho_2\,\rho_3\,\lpar \mrx_1\,\mrx_2 + \mrx_1\,\mrx_3 - \mrx_2\,\mrx_3 - \mrx_1 \rpar
- \sigma_3\,\rho_3^2\,\mrx_3 + \lambda^2\,\beta \spc
\eq
where $\lambda^2 = \mrm^2/\mrs$.
The integral over the $\mrx$ variables gives
\bq
\mrI_{123} = \int_0^1\,\Bigl[ \prod_{i=1}^{3}\,\mrd \mrx_i \Bigr]\,\mrS_2^{-2} \spp
\eq
The integral over the $\mrx_1$ and $\mrx_2$  gives the following result:
\bqa
\mrI_{123} &=& \frac{1}{(\rho_1\,\rho_2\,\rho_3\,\mrs)^2}\,\int_0^1 \mrd \mrx\,\mrJ(\mrx) \spc
\nl
\mrJ(\mrx) &= &
\Bigl[\mrx^2\,(\mrx - \mrx_{1})\Bigr]^{-1}\;\hyp{1}{1}{2}{\mrx^{-1}} + 
\Bigl[\mrx\,(1 - \mrx)\,(\mrx - \mrx_{2}\Bigr]^{-1}\;\hyp{1}{1}{2}{\mrx^{-1}} 
\nl
{}&-&
\Bigl[ \mrx\,\mrx_{1}\,(\mrx - \mrx_{1})\Bigr]^{-1}\;\hyp{1}{1}{2}{\mrx_{1}^{-1}} -
\Bigl[ (1 - \mrx)\,\mrx_{2}\,(\mrx - \mrx_{2})\Bigr]^{-1}\;\hyp{1}{1}{2}{\mrx_{2}^{-1}} \spc
\eqa
where we have introduced the roots
\bq
\mrx_{1} = \frac{\rho_3\,\beta\,\mrx - \sigma_3\,\rho_3^2\,\mrx^2 - \lambda^2\,\beta}
                {\rho_1\,\rho_2\,\rho_3\,\mrx} \spc
\qquad
\mrx_{2} = \frac{ - \sigma_3\,\rho_3^2\,\mrx\,(1 - \mrx) + \lambda^2\,\beta}
                {\rho_1\,\rho_2\,\rho_3\,(1 - \mrx)} \spp
\eq
The next step consists in using \eqn{F21MB} obtaining
\bq
\mrI_{123}= \int_{\mrL}\,\frac{\mrd \mrs_1}{2\,i\,\pi}\,
\frac{\eG{ - \mrs_1}\,\eGs{1 + \mrs_1}}{\eG{2 + \mrs}}\,
\lpar \rho_1\,\rho_2\,\rho_3 \rpar^{\mrs_1}\,\int_0^1\,\mrd \mrx\,\mrQ_1^{-1}\,
\Bigl[ (1 - \mrx)^{1 + \mrs_1}\,\mrQ_3^{- 1 - \mrs_1} - 
       \mrx^{1 + \mrs_1}\,\mrQ_2^{- 1 - \mrs_1} \Bigr] \spc
\label{cdkx}
\eq
\bq
\mrQ_1 =  \rho_3\,\beta\,\mrx\,(1 - \mrx) - \lambda^2\,\beta \spc \quad
\mrQ_2 = \sigma_3\,\rho_3^2\,\mrx^2 - \rho_3\,\beta\,\mrx + \lambda^2\,\beta \spc \quad
\mrQ_3 = \sigma_3\,\rho_3^2\,\mrx\,(1 - \mrx) - \lambda^2\,\beta \spp
\label{Qquad}
\eq
\subsubsection{Method I}
We perform a MB splitting (\eqn{MBsplit}) in the quadratic forms of \eqn{Qquad}, and perform the $\mrx\,${-}integral
in \eqn{cdkx} producing Gauss hypergeometric functions. After using \eqn{F21MB} we derive
\bq
\mrI_{123} = \mrI^{(4)}_{123} + \mrI^{(5)}_{123} \spc 
\qquad
\mrI^{\mrn}_{123} = \Bigl[ \prod_{i=1}^{\mrn}\,\frac{\mrd \mrs_i}{2\,i\,\pi} \Bigr]\,
\frac{\mrN_{\mrn}}{\mrD_{\mrn}}\,(\lambda^2)^{ - 2 + \mrs_3 + \mrs_2 - \mrs_1}\,\mrK_{\mrn} \spc
\eq
\bqa
\mrN_4 &=&
\eG{ - \mrs_1}\,\eG{\mrs_2}\,\eG{\mrs_3}\,\eG{\mrs_4}\,
\eG{1 + \mrs_1}\,\eG{1 - \mrs_2}\,\eG{1 - \mrs_3 + \mrs_1}\,\eG{1 - \mrs_3 - \mrs_2}\,
\nl
{}&\times&
\eG{2 - \mrs_3 - \mrs_2 + \mrs_1}\,\eG{2 - \mrs_4 - \mrs_3 + \mrs_1} \spc
\nl
\mrD_4 &=&
\eG{2 + \mrs_1}\,\eG{2 - \mrs_3 + \mrs_1}\,\eG{3 - 2\,\mrs_3 - 2\,\mrs_2 + \mrs_1} \spc
\nl
\mrN_5 &=& 
\eG{ - \mrs_1}\,\eG{\mrs_2}\,\eG{ - \mrs_4}\,\eG{\mrs_5}\,
\eG{1 + \mrs_1}\,\eG{1 - \mrs_2}^2\,\eG{\mrs_4 + \mrs_3}\,\eG{1 - \mrs_3 + \mrs_1}\,
\nl
{}&\times&
\eG{2 - \mrs_5 + \mrs_4 + \mrs_1}\,\eG{2 + \mrs_4 - \mrs_3 - \mrs_2 + \mrs_1} \spc
\nl
\mrD_5 &=&
\eG{2 + \mrs_1}\,\eG{2 + \mrs_4 + \mrs_1}\,\eG{3 + \mrs_4 - \mrs_3 - 2\,\mrs_2 + \mrs_1} \spc
\eqa
\bqa
\mrK_4 &=&
\mrE_{-}(1 - \mrs_3 + \mrs_1)\,
(\rho_1)^{ - \mrs_4 + \mrs_1}\,
(\rho_2)^{ - 2 + \mrs_4 - \mrs_3 + 2\,\mrs_1}\,
(\rho_3)^{ - \mrs_4 - 2\,\mrs_3 - \mrs_2 + \mrs_1}\,
(\sigma_2)^{ - 2 + \mrs_4 - \mrs_3 + \mrs_1}\,
(\sigma_3)^{ - \mrs_3} \spc
\nl
\mrK_5 &=&
\mrE_{-}(\mrs_3 + \mrs_4)\,
(\rho_1)^{ - \mrs_5 + \mrs_1}\,
(\rho_2)^{ - 2 + \mrs_5 - \mrs_4}\,
(\rho_3)^{ - \mrs_5 + \mrs_4 - \mrs_3 - \mrs_2 + \mrs_1}\,
(\sigma_2)^{ - 2 + \mrs_5 - \mrs_4 - \mrs_1}\,
(\sigma_3)^{\mrs_4} \spc
\eqa
where $\mrE_{\pm}$ is defined by
\bq
\mrE_{\pm}(\mrs) = \exp\{ \pm i\,\frac{\mrs}{2}\,\pi\} \spp 
\eq
The croseed delta{-}kite integral is
\bq
\mrD\mrK_{\mrc} = \int_0^1\,\mrd \rho_1\,\int_0^{1 - \rho_1}\,\mrd \rho_2\,
\rho_1\,\rho_2\,\rho_3\,\mrI_{123} \spp
\eq
First we perform the $\rho_2$ integral giving rise to $\mrF^{(2)}_{\sPD}$ functions; we use \eqn{FDMB} and
perform the $\rho_1$ integral obtaining the following result:
\bq
\mrD\mrK_{\mrc}= \mrD\mrK^{(6)}_{\mrc} + \mrD\mrK^{(7)}_{\mrc} \spc
\quad
\mrD\mrK^{(\mrn)}_{\mrc} = \Bigl[ \prod_{i=1}^{\mrn}\,\frac{\mrd \mrs_i}{2\,i\,\pi} \Bigr]\,
\lpar \lambda^2 \rpar^{\mrs_3 + \mrs_2 - \mrs_1 - 2}\,\frac{\mrN^{(\mrn)}}{\mrD^{(\mrn)}}\,\mrE^{(\mrn)} \spc
\eq
\bqa
\mrN^{(6)} &=&
\eG{ - \mrs_1}\,
\eG{1 + \mrs_1}\,
\eG{\mrs_2}\,
\eG{1 - \mrs_2}\,
\eG{\mrs_4}\,
\eG{ - \mrs_5}\,
\eG{ - \mrs_6}\,
\eG{\mrs_6 + \mrs_3}\,
\eG{1 - \mrs_3 + \mrs_1}\,
\eG{1 - \mrs_3 - \mrs_2}
\nl
{}&\times&
\eG{2 - \mrs_3 - \mrs_2 + \mrs_1}\,
\eG{\mrs_6 + \mrs_5 + \mrs_4 + \mrs_3}\,
\eG{2 - \mrs_4 - 2\,\mrs_3 - \mrs_2 + \mrs_1}\,
\eG{2 + \mrs_5 - \mrs_4 - \mrs_3 + \mrs_1}
\nl
{}&\times&
\eG{2 - \mrs_6 - \mrs_4 - \mrs_3 + \mrs_1} \spc
\nl
\mrD^{(6)} &=&
\eG{2 + \mrs_1}\,
\eG{2 - \mrs_3 + \mrs_1}\,
\eG{3 - 2\,\mrs_3 - 2\,\mrs_2 + \mrs_1}\,
\eG{4 + \mrs_5 - \mrs_4 - 2\,\mrs_3 - \mrs_2 + 2\,\mrs_1} \spc
\nl
\mrN^{(7)} &=&
\eG{ - \mrs_1}\,
\eG{1 + \mrs_1}\,
\eG{\mrs_2}\,
\eGs{1 - \mrs_2}\,
\eG{\mrs_5}\,
\eG{ - \mrs_6}\,
\eG{ - \mrs_7}\,
\eG{\mrs_4 + \mrs_3}\,
\eG{\mrs_7 - \mrs_4}\,
\eG{1 - \mrs_3 + \mrs_1}
\nl
{}&\times&
\eG{\mrs_7 + \mrs_6 + \mrs_5 - \mrs_4}\,
\eG{2 + \mrs_4 - \mrs_3 - \mrs_2 + \mrs_1}\,
\eG{2 + \mrs_6 - \mrs_5 + \mrs_4 + \mrs_1}\,
\eG{2 - \mrs_7 - \mrs_5 + \mrs_4 + \mrs_1}
\nl
{}&\times&
\eG{2 - \mrs_5 + \mrs_4 - \mrs_3 - \mrs_2 + \mrs_1} \spc
\nl
\mrD^{(7)} &=&
\eG{2 + \mrs_1}\,
\eG{2 + \mrs_4 + \mrs_1}\,
\eG{3 + \mrs_4 - \mrs_3 - 2\,\mrs_2 + \mrs_1}\,
\eG{4 + \mrs_6 - \mrs_5 + \mrs_4 - \mrs_3 - \mrs_2 + 2\,\mrs_1} \spc
\eqa
\bq 
\mrE^{(6)} = \exp\{i\,(\mrs_5 + \mrs_3 - \mrs_1)\,(\pi - \delta)\} \spc \quad
\mrE^{(7)} = \exp\{i\,(\mrs_6 - \mrs_4 - \mrs_3)\,(\pi - \delta)\} \spc \qquad
\delta \to 0_{+} \spp
\eq
Following the theorem of \Bref{HS} we have $6438$ sequences out of $11628$ with $200$ unequal sequences for
$\mrD\mrK^{(6)}_{\mrc}$. Given the conditions $\mid \phi_i \mid < \pi$ we are left with $103$ relevant strips
containing from two to six variables. The explicit expression for the relevant strips will not be shown but
we have verified that the $103$ convergence conditions are satisfied by the actual arguments of the Fox
function.

The integration contour for $\mrD\mrK^{(6)}_{\mrc}$ is defined by the following relations ($\mrs_i = \mrr_i + i\,\mrt_i$):
\bqa
{}&{}&
 - 1 < \mrr_1 < 0 \spc \quad
   0 < \mrr_2 < 1 \spc \quad
   0 < \mrr_3 < \mathrm{min}(1 + \mrr_1\,,\,1 - \mrr_2) \spc \quad
   9 < \mrr_4 < 1 + \mrr_1 - \mrr_3 \spc
\nl
{}&{}&
 - 2 - \mrr_1 + \mrr_3 + \mrr_4 < \mrr_5 < 0 \spc \quad
 \mathrm{max}( - \mrr_3\,,\, - \mrr_3 - \mrr_4 - \mrr_5\,,\, - 2 - \mrr_1 + \mrr_2 + \mrr_3 - \mrr_5) < \mrr_6 < 0 \spp
\eqa
Both $\mrz_4$ and $\mrz_6$ are equal to $1$, which rises the question of computing Meijer or Fox fumctions with
one or more arguments equal to $1$. We give a simple example: consider the integral over $\mrs_6$ in 
$\mrD\mrK^{(6)}_{\mrc}$:
\bq
\mrJ_6 = \int_{\mrL_6}\,\frac{\mrd \mrs_6}{2\,i\,\pi}\,
\eG{ - \mrs_6}\,\eG{2 + \mrs_1 - \mrs_3 - \mrs_4 - \mrs_6}\,
\eG{\mrs_3 + \mrs_6}\,\eG{\mrs_3 + \mrs_4 + \mrs_5 + \mrs_6} \spp
\eq
According to the definitions given in 
https://functions.wolfram.com/PDF/MeijerG.pdf
$\quad \mrJ_6$ corresponds to a Meijer $\mrG^{2,2}_{2,2}$ with parameters
\bq
\mra_1 = 1 \spc \quad
\mra_2 = - 1 - \mrs_1 + \mrs_3 + \mrs_4\spc \quad
\mrb_1 = \mrs_3 \spc \quad \mrb_2 = \mrs_3 + \mrs_4 + \mrs_5 \spp
\eq
We can use the result reported at p. $91$ of https://functions.wolfram.com/PDF/MeijerG.pdf
\[
\mrG^{2,2}_{2,2}\,\left( \mrz\,,\; \Biggl[
\begin{array}{cc}
\mra_1 & \mra_2 \\
\mrb_1 & \mrb_2 \\
\end{array}
\Biggr ]
\right)
=
\Uppsi\,\mrz^{\mrb_1}\,
\hyp{1 - \mra_1 + \mrb_1}{1 - \mra_2 + \mrb_1}{2 - \mra_1 - \mra_2 + \mrb_1 + \mrb_2}{1 - \mrz} \spc
\]
\bq
\Uppsi= \eG{1 - \mra_1 + \mrb_1}\,
      \eG{1 - \mra_2 + \mrb_1}\,
      \eG{1 - \mra_2 + \mrb_2}\,
      \eG{1 - \mra_1 + \mrb_2}\,\frac{1}{\eG{2 - \mra_1 - \mra_2 + \mrb_1 + \mrb_2}} \spc
\eq
giving the following result for $\mrJ_6$:
\bq
\mrJ_6 = \eG{\mrs_3}\,\eG{2 + \mrs_1 - \mrs_4}\,\eG{2 + \mrs_1 + \mrs_5}\,
\frac{\eG{\mrs_3 + \mrs_4 + \mrs_5}}
     {\eG{2 + \mrs_1 + \mrs_3 + \mrs_5}} \spp
\eq
Actually this result can be obtained directly by using Barnes's lemma.

The integration contour for $\mrD\mrK^{(7)}_{\mrc}$ is defined by the following conditions:
\bqa
{}&{}&
 - 1 < \mrr_1 < 0 \spc \quad
   0 < \mrr_2 < 1 \spc \quad
   0 < \mrr_3 < 1 + \mrr_1 \spc
\nl
{}&{}&
 - 2 - \mrr_1 + \mrr_2 + \mrr_3 < \mrr_4 < 0 \spc \quad
   0 < \mrr_5 < 2 + \mrr_1 + \mrr_4 \spc \quad
 - 2 - \mrr_1 - \mrr_4 + \mrr_5 < \mrr_6 < 0 \spc
\nl
{}&{}&
\mathrm{max}(\mrr_4\,,\,\mrr_4 - \mrr_5 - \mrr_6\,,\, - 2 - \mrr_1 + \mrr_2 + \mrr_3 - \mrr_6) < \mrr_7 <
\mathrm{min}(0\,,\,2 + \mrr_1 + \mrr_4 - \mrr_5) \spp
\eqa
There are $211$ relevant strips for $\mrD\mrK^{(7)}_{\mrc}$, all the convergence conditions are satisfied by the arguments 
of the corresponding Fox function.

As a simpler example let us consider the integral over $\mrs_1 , \mrs_2 , \mrs_3$ in $\mrD\mrK^{(6)}_{\mrc}$.
Here we have
\bq
\mrz_1 = \lambda^{-2}\,\exp\{ - i\,\pi\} \spc \qquad
\mrz_2 = \lambda^2 \spc \qquad
\mrz_3 = \lambda^2\,\exp\{i\,\pi\} \spp
\eq
We recognize a generalized Fox $\mrH$ function, described in \sect{GMFF}, with paramters
$\mrr = 3$, $\mrm = 12$ and $\mrn = 4$. Following the analysis of \sect{GMFF} and discarding the sequences
$(\mrj_1\,,\,\mrj_2)$ when they violate the condition in \eqn{Hseq} we obtain $20$ unequal sequences out of $112$:
we have $\mid \phi_i \mid < \pi$ (where $\phi_i$ is the argument of $\mrz_i$) plus other $17$ irrelevant strips involving 
two or three arguments. 
\subsubsection{Method II}
Here we do not split the quadratic forms of \eqn{Qquad} and start by studying their sign. Any $\mrQ_i$ is
generically witten as $\mra\,\mrx^2 + \mrb\,\mrx + \mrc$, where the coefficients are functions of $\rho_1, \rho_3$
and $0 \le \rho_1 \le 1$, $0 \le \rho_3 \le 1 - \rho_1$. Furthermore, $\mrx_{i \pm}$ are the roots of $\mrQ_i$.
\bei

\item[\ovalbox{$\mrQ_1$}] In this case $\mra < 0$ and
\bq
\mrx_{1 \pm} = \frac{1}{2}\,\Bigl(1 \mp \sqrt{\frac{\Delta_1}{\rho_3}} \Bigr) \spc \qquad
\Delta_1 = \rho_3 - 4\,\lambda^2 \spp
\label{Q1roots}
\eq 
\begin{enumerate}

\item $\rho_3 < 4\,\lambda^2$; in this case $\Delta_1 < 0$, $\mrx_{1 \pm} \in \Cf$ and $\mrQ_1 < 0$. This corresponds to
\bqa
0 \le \rho_1 \le 1 - 4\,\lambda^2 \spc \quad
0 \le \rho_3 \le 4\,\lambda^2 \spc &\qquad&
\lambda < \frac{1}{2} \spc
\nl
1 - 4\,\lambda^2 \le \rho_1 \spc \quad
0 \le \rho_3 \le 1 - \rho_1 \spc &\qquad&
\lambda < \frac{1}{2} \spc
\nl
0 \le \rho_1 \le 1 \spc \quad
0 \le \rho_3 \le 1 - \rho_1 \spc &\qquad&
\lambda > \frac{1}{2} \spp
\eqa

\item $\rho_3 > 4\,\lambda^2$; in this case $\Delta_1 > 0$, $\mrx_{1 \pm} \in \Rf$ and the sign of $\mrQ_1$ depends
on $\mrx_{1 \pm}$. This corresponds to
\bq
0 \le \rho_1 \le 1 - 4\,\lambda^2 \spc \quad
4\,\lambda^2 \le \rho_3 \le 1 - \rho_1 \qquad
\lambda < \frac{1}{2} \spp
\eq
From \eqn{Q1roots} we see that $\mrx_{1 -} < 0$ and $\mrx_{1 +} > 1$ so that $\mrQ_1 > 0$

\end{enumerate}

\item[\ovalbox{$\mrQ_2$}] In this case $\mra > 0$ and
\bq
\mrx_{2 \pm} = \frac{\sqrt{\beta}}{2\,\sigma_3\,\rho_3}\,\Bigl(
\sqrt{\beta} \pm \sqrt{\Delta_2} \Bigr) \spc \qquad
\Delta_2 = - \rho_1^2 + \sigma_3\,\rho_1 + \sigma_3\,\Delta_1 \spp
\label{Q2roots}
\eq
Furthermore the roots of $\Delta_2$ are
\bq
\mra_{\pm} = \frac{\sqrt{\sigma_3}}{2}\,\Bigl( \sqrt{\sigma_3} \mp \sqrt{\delta_2} \Bigr) \spc
\qquad
\delta_2 = 3\,\rho_3 + 1 - 16\,\lambda^2 \spp
\eq
\begin{enumerate}

\item If $\mrx_{2 \pm} \in \Cf$ then $\mrQ_2 > 0$; this requires $\delta_2 < 0$ or $\rho_3 < \mrr$ with
$\mrr = (16\,\lambda^2 - 1) /3$. This corresponds to
\bqa
0 \le \rho_1 \le\frac{4}{3}\,(1 - 4\,\lambda^2) \spc \quad
0 \le \rho_3 \le \frac{1}{3}\,(16\,\lambda^2 - 1) \spc &\qquad&
\frac{1}{4} < \lambda < \frac{1}{2} \spc
\nl
\frac{4}{3}\,(1 - 4\,\lambda^2) \le \rho_1 \le 1 \spc \quad
0 \le \rho_3 \le 1 - \rho_1 \spc &\qquad&
\frac{1}{4} < \lambda < \frac{1}{2} \spc
\nl
0 \le \rho_1 \le 1 \spc \quad
0 \le \rho_3 \le 1 - \rho_1 \spc &\qquad&
\lambda > \frac{1}{2} \spp
\eqa

\item If $\mrx_{2 \pm} \in \Rf$ then the sign of $\mrQ_2$ depends on $\mrx_{2 \pm}$; this requires $\rho_3 > \mrr$ and
corresponds to
\bqa
0 \le \rho_1 \le 1 \spc \quad
0 \le \rho_3 \le 1 - \rho_1 \spc &\qquad&
\lambda < \frac{1}{4} \spc
\nl
0 \le \rho_1 \le \frac{4}{3}\,(1 - 4\,\lambda^2) \spc \quad
\frac{1}{3}\,(16\,\lambda^2 - 1) \le \rho_3 \le 1 - \rho_1 \spc &\qquad&
\frac{1}{4} < \lambda < \frac{1}{2} \spp
\eqa

\end{enumerate}

\item[\ovalbox{$\mrQ_3$}] In this case $\mra < 0$ and
\bq
\mrx_{3 \pm} = \frac{1}{2}\,\Bigl( 1 \mp \frac{1}{\rho_3}\,\sqrt{\frac{\Delta_3}{\sigma_3}} \Bigr) \spc \quad
\Delta_3 = 4\,\lambda^2\,\rho_1\,( \rho_1 - \sigma_3) + \sigma_3\,\rho_3\,\Delta_1 \spp
\eq
Furthermore the roots of $\Delta_3$ are
\bq
\mrb_{\pm} = \frac{\sigma_3}{2} \pm \frac{1}{2\,\lambda}\,\sqrt{\sigma_3\,\delta_3} \spc
\qquad
\delta_3 = - \rho_3^2 + 3\,\lambda^2\,\rho_3 + \lambda^2 = - (\rho_3 - \mrc_{-})\,(\rho_3 - \mrc_{+}) \spp
\eq
The roots $\mrc_{\pm}$ are real with $\mrc_{-} < 0$ and
\bqa
0 < \mrc_{+} < 1 \qquad &\hbox{for}& \qquad \lambda < \frac{1}{2} \spc
\nl 
\mrc_{+} > 1 \qquad &\hbox{for}& \qquad \lambda > \frac{1}{2} \spp
\eqa
We have the following cases:

\begin{enumerate}

\item $\lambda > \frac{1}{2}$, giving $\delta_3 > 0$ and $\mrb_{\pm} \in \Rf$. Since $\mrb_{-} < 0$ and
$0 < \mrb_{+} < 1$ but $\mrb_{+} \ge 1 - \rho_3$ we have
\bq
0 \le \rho_3 \le 1 \qquad \spc 0 \le \rho_1 \le 1 - \rho_3 \spc \qquad
\Delta_3 < 0 \spc \;\; \mrQ_3 < 0 \spp
\eq

\item $\lambda < \frac{1}{2}$ We have

\bq
0 \le \rho_1 \le 1 - \mrc_{+} < \mrb_{+} \spc \quad
0 \le \rho_3 \le 4\,\lambda^2 \spc \qquad \Delta_3 < 0 \spc
\eq
\bq
4\,\lambda^2 \le \rho_3 \le \mrc_{+} \spc
\eq
\bqa
0 \le \rho_1 \le \mrb_{-} \spc &\qquad& \Delta_3 > 0 \spc
\nl
\mrb_{-} \le \rho_1 \le \mrb_{+} \spc &\qquad& \Delta_3 < 0 \spc
\nl
\mrb_{+} \le \rho_1 \le 1 \spc &\qquad& \Delta_3 > 0 \spp
\eqa
Other regions are
\bq
0 \le \rho_1 \le 1 - \mrc_{+} \spc \quad
\mrc_{+} \le \rho_3 \le 1 - \rho_1 \spc \qquad
\Delta_3 > 0 \spc
\eq
\bqa
0 \le \rho_3 \le 1 - \rho_1 \spc \quad
1 - \mrc_{+} \le \rho_1 \le \mrb_{+} \spc &\qquad&
\Delta_3 < 0 \spc
\nl
0 \le \rho_3 \le 1 - \rho_1 \spc \quad
\mrb_{+} \le \rho_1 \le 1 \spc &\qquad&
\Delta_3 > 0 \spp
\eqa
When $\Delta_3 < 0$ we have $\mrQ_3 < 0$. For $\Delta_3 > 0$ we must examine the roots $\mrx_{3 \pm}$. 

\end{enumerate}
\eei
In the following we will present results for the case $\lambda > 1/2$.
Going back to \eqn{cdkx} we have
\bqa
\mrQ_1^{\alpha} &=& \lpar - \beta\,\rho_3 + i\,\delta \rpar^{\alpha}\,\prod_{\omega=\pm}\,
                    \lpar \mrx - \mrx_{1 \omega} \rpar^{\alpha} \spc
\nl
\mrQ_2^{\alpha} &=& \lpar \sigma_3\,\rho_3 \rpar^{\alpha}\,\prod_{\omega=\pm}\,
                    \lpar \mrx - \mrx_{2 \omega} \rpar^{\alpha} \spc
\nl
\mrQ_3^{\alpha} &=& \lpar - \sigma_3\,\rho_3 + i\,\delta \rpar^{\alpha}\,\prod_{\omega=\pm}\,
                    \lpar \mrx - \mrx_{3 \omega} \rpar^{\alpha} \spp
\eqa
The integral over $\mrx$ gives rise a pair of $\mrF^{(3)}_{\sPD}$ functions 
\bqa
{}&{}&
\lfdt{3}
    {2 + \mrs_1}
    {1}
    {1 + \mrs_1}
    {1 + \mrs_1}
    {3 + \mrs_1}
    {\mrx_{1 \pm}^{-1}}
    {\mrx_{2{-}}^{-1}}
    {\mrx_{2{+}}^{-1}} \spc
\nl
{}&{}&
\lfdt{3}
    {1}
    {1}
    {1 + \mrs_1}
    {1 + \mrs_1}
    {3 + \mrs_1}
    {\mrx_{1 \pm}^{-1}}
    {\mrx_{2{-}}^{-1}}
    {\mrx_{2{+}}^{-1}} \spc
\eqa
for which we use the MB representation of \eqn{FDMB} producing a four{-}fold MB integral.
In order to proceed with the $\rho_1, \rho_3$ integrations we first perform the relevant set of MB splitting
reaching a seven{-}fold MB integral.
The integral over $\rho_1$ gives again $\mrF^{(3}_{\sPD}$ functions; given the parameters $\mra_{\pm}$ and $\mrb_{\pm}$
\bqa
\lpar - \mra_{\pm} \rpar^{\alpha} &=& 2^{- \alpha}\,\sigma_3^{\alpha/2}\,
                                    \lpar - \sqrt{\sigma_3} \pm i\,\sqrt{\Delta_2} \pm i\,\delta \rpar^{\alpha} \spc 
\nl
\lpar - \mrb_{\pm} \rpar^{\alpha} &=& 2^{ - \alpha}\,
                                      \lpar \sigma_3 \pm \frac{\sqrt{\sigma_3\,\Delta_3}}{\lambda}\rpar^{\alpha} \spc
\eqa
the Lauricella functions are
\bqa
{}&{}&
\lfdt{3}
   {2 - \mrs_8 + \mrs_1}
   {1 - \mrs_8}
   {\mrs_7 + \mrs_6}
   {\mrs_7 + \mrs_6}
   {3 + 2\,\mrs_1}
   { - \frac{\sigma_3}{\rho_3}}
   {\frac{\sigma_3}{\mrb_{+}}}
   {\frac{\sigma_3}{\mrb_{-}}} \spc
\nl
{}&{}& 
\lfdt{3}
    {2 - \mrs_8 + \mrs_1}
    {3 - \mrs_8- \frac{1}{2}\,(\mrs_7 + \mrs_6) + \mrs_4 + \mrs_3 + 2\,\mrs_1}
    {\mrs_7 + \mrs_6}
    {\mrs_7 + \mrs_6}
    {1 + \frac{1}{2}\,(\mrs_7 + \mrs_6) - \mrs_4 - \mrs_3}
    { - \frac{\sigma_3}{\rho_3}}
    {\frac{\sigma_3}{\mra_{+}}}
    {\frac{\sigma_3}{\mra_{-}}} \spp
\eqa
They are transformed into their MB representations and we will also use a MB splitting, \eg
\bq
\lpar \sigma_3 \pm \frac{\sqrt{\sigma_3\,\Delta_3}}{\lambda}\rpar^{\alpha} =
\int_{\mrL} \frac{\mrd \mrs}{2\,i\,\pi}\,\eB{\mrs}{ - \alpha - \mrs}\,
\sigma_3^{\alpha + \mrs}\,\lpar \pm\,\frac{\sqrt{\sigma_2\,\Delta_3}}{\lambda} \rpar^{ - \mrs} \spp
\eq
Finally, the integral over $\rho_3$ gives Lauricella functions with two or three variables (\eqn{FDEM}), \ie
\bqa
{}&{}&
\lfdd{2}
    { - 2 + \mrs_9 + \frac{1}{2}\,(\mrs_7 + \mrs_6 - \mrs_2) + \mrs_5}
    {1 + \mrs_5}
    {\frac{1}{2}\,(\mrs_{13} + \mrs_{12})}
    {\frac{1}{2}\,(\mrs_{13} + \mrs_{12} - \mrs_2) + 2\,\mrs_9 - \mrs_7 - \mrs_6 + \mrs_5 + \mrs_1}
    {\frac{1}{4\,\lambda^2}}
    {\mrc_0} \spc
\nl
{}&{}&
\lfdt{3}
    { - 2 + \mrs_9 + \mrs_7 + \mrs_6 + \mrs_5 + \frac{1}{2}\,(\mrs_4 + \mrs_3 - \mrs_2)}
    {1 + \mrs_5}
    {\frac{1}{2}\,(\mrs_{13} + \mrs_{12})}
    {\frac{1}{2}\,(\mrs_{13} + \mrs_{12})}
    { - 1 + \frac{1}{2}\,(\mrs_{13} + \mrs_{12} - \mrs_7 - \mrs_6 - \mrs_2) + 2\,\mrs_9 + \mrs_5}
       {\frac{1}{4\,\lambda^2}}
       {\mrc_{+}^{-1}}
       {\mrc_{-}^{-1}} \spc
\eqa
where we have defined 
\bq
\mrc_0 = \frac{3}{16\,\lambda^2 - 1} \spc \qquad
\mrc_{\pm} = \frac{1}{2}\,\lpar 3\,\lambda^2 \pm \sqrt{9\,\lambda^4 + 4\,\lambda^2} \rpar \spp
\eq
The final result contains two Fox functions, see \sect{GMFF}: one with $\mrr = 15$, $\mrm= 28, \mrn = 9$ and another 
with $\mrr= 16$, $\mrm= 30, \mrn= 8$. Let us define
\bq
\Gamma_{\mrr} = \frac{\mrN_{\mrr}}{\mrD_{\mrr}} \spc
\eq
where we have defined
\bqa
\mrN_{1} &=&
\Bigl[ \prod_{i=5,6,7,8,12,13}\,\eG{\mrs_i} \Bigr]\,
\Bigl[ \prod_{i=1,2,3,4,9,10,11,14,15}\,\eG{ - \mrs_i} \Bigr]\,
\eG{1 - \mrs_5 + \mrs_2}\,
\eG{1 + \mrs_{14} + \mrs_5}\,
\eG{\mrs_{15} + \frac{1}{2}\,(\mrs_{13} + \mrs_{12})}\,
\nl
{}&\times&
\eG{1 - \mrs_6 + \mrs_3 + \mrs_1}\,
\eG{1 - \mrs_7 + \mrs_4 + \mrs_1}\,
\eG{ - \mrs_{12} + \mrs_{10} + \mrs_7 + \mrs_6}\,
\eG{ - \mrs_{13} + \mrs_{11} + \mrs_7 + \mrs_6}\,
\nl
{}&\times&
\eG{2 + \mrs_4 + \mrs_3 + \mrs_2 + \mrs_1}\,
\eG{2 + \mrs_{11} + \mrs_{10} + \mrs_9 - \mrs_8 + \mrs_1}\,
\eG{ - 1 + \mrs_8 + \frac{1}{2}\,(\mrs_7 + \mrs_6) - \mrs_4 - \mrs_3 - \mrs_1}\,
\nl
{}&\times&
\eG{2 + \frac{1}{2}\,(\mrs_{13} + \mrs_{12}) + \mrs_9 - \frac{3}{2}\,(\mrs_7 + \mrs_6) + \mrs_1}\,
\eG{ - 2 + \mrs_{15} + \mrs_{14} + \mrs_9 + \frac{1}{2}\,(\mrs_7 + \mrs_6 - \mrs_2) + \mrs_5}\,
\nl
{}&\times&
\eG{3 + \mrs_9 - \mrs_8 - \frac{1}{2}\,(\mrs_7 + \mrs_6) + \mrs_4 + \mrs_3 + 2\,\mrs_1} \spc
\nl
\nl
\mrD_{1} &=&
\eG{1 + \mrs_5}\,
\eG{2 + \mrs_1}\,
\eGs{\mrs_7 + \mrs_6}\,
\eG{\frac{1}{2}\,(\mrs_{13} + \mrs_{12})}\,
\eG{3 + \mrs_4 + \mrs_3 + \mrs_2 + \mrs_1}\,
\nl
{}&\times&
\eG{3 - \frac{1}{2}\,(\mrs_7 + \mrs_6) + \mrs_4 + \mrs_3 + 2\,\mrs_1}\,
\eG{1 + \mrs_{11} + \mrs_{10} + \mrs_9 + \frac{1}{2}\,(\mrs_7 + \mrs_6) - \mrs_4 - \mrs_3}\,
\nl
{}&\times&
\eG{\mrs_{15} + \mrs_{14} + \frac{1}{2}\,(\mrs_{13} + \mrs_{12} - \mrs_2) + 2\,\mrs_9 - \mrs_7 - \mrs_6 + \mrs_5 + \mrs_1}\,
\nl\nl
\mrN_{2} &=&
\Bigl[ \prod_{i=5,6,7,8,12,13}\,\eG{\mrs_i} \Bigr]\,
\Bigl[ \prod_{i=1,2,3,4,9,10,11,14,15,16}\,\eG{ - \mrs_i} \Bigr]\,
\eG{1 - \mrs_5 + \mrs_2}\,
\eG{1 + \mrs_8 + \mrs_1}\,
\eG{1 + \mrs_9 - \mrs_8}
\nl
{}&\times&
\eG{1 + \mrs_{14} + \mrs_5}\,
\eG{\mrs_{15} + \frac{1}{2}\,(\mrs_{13} + \mrs_{12})}\,
\eG{\mrs_{16} + \frac{1}{2}\,(\mrs_{13} + \mrs_{12})}\,
\eG{1 + \mrs_4 + \mrs_3 + \mrs_2}\,
\eG{1 - \mrs_6 + \mrs_3 + \mrs_1}\,
\nl
{}&\times&
\eG{1 - \mrs_7 + \mrs_4 + \mrs_1}\,
\eG{ - \mrs_{12} + \mrs_{10} + \mrs_7 + \mrs_6}\,
\eG{ - \mrs_{13} + \mrs_{11} + \mrs_7 + \mrs_6}\,
\eG{2 + \mrs_{11} + \mrs_{10} + \mrs_9 - \mrs_8 + \mrs_1}\,
\nl
{}&\times&
\eG{1 + \frac{1}{2}\,(\mrs_{13} + \mrs_{12} - \mrs_4 - \mrs_3) + \mrs_9 - \frac{3}{2}\,(\mrs_7 + \mrs_6)}
\nl
{}&\times&
\eG{ - 2 + \mrs_{16} + \mrs_{15} + \mrs_{14} + \mrs_9 + \mrs_7 + \mrs_6 + \mrs_5 + \frac{1}{2}\,(\mrs_4 + \mrs_3 - \mrs_2)}\,
\nl
\nl
\mrD_{2} &=&
\eG{1 + \mrs_5}\,
\eGs{\mrs_7 + \mrs_6}\,
\eGs{\frac{1}{2}\,(\mrs_{13} + \mrs_{12})}\,
\eG{3 + \mrs_4 + \mrs_3 + \mrs_2 + \mrs_1}\,
\eG{3 + \mrs_{11} + \mrs_{10} + \mrs_9 + 2\,\mrs_1}\,
\nl
{}&\times&
\eG{ - 1 + \mrs_{16} + \mrs_{15} + \mrs_{14} + \frac{1}{2}\,(\mrs_{13} + \mrs_{12} - \mrs_7 - \mrs_6 - \mrs_2) + 2\,\mrs_9 + \mrs_5} \spp 
\eqa
Collecting the various conditions we introduce $s_{\mrj} = \mrr_{\mrj} + i\,\mrt_{\mrj}$ and derive
\bqa
{}&{}& - 1 < \mrr_1 < 0 \spc \quad - 1 < \mrr_2 < 0 \spc \quad - 1 - \mrr_1 < \mrr_3 < 0 \spc
\nl
{}&{}&
\mathrm{max}( - 1 - \mrr_1\,,\,- 2 - \mrr_1 - \mrr_2 - \mrr_3) < \mrr_4 < 0 \spc
\nl
{}&{}&
0 < \mrr_5 < 1 + \mrr_2 \spc \quad 0 < \mrr_6 < 1 + \mrr_1 + \mrr_3 \spc \quad
0 < \mrr_7 <1 + \mrr_1 + \mrr_4 \spc
\nl
{}&{}&
\mathrm{max}(0\,,\,1 + \mrr_1 + \mrr_3 + \mrr_4 - \frac{1}{2}\,(\mrr_6 + \mrr_7)) < \mrr_8 
< \mathrm{min}(2 + \mrr_1\,,\,3 + 2\,\mrr_1 + \mrr_3 + \mrr_4 - \frac{1}{2}\,(\mrr_6 + \mrr_7)) \spc
\nl
{}&{}&
- 3 - 2\,\mrr_1 - \mrr_3 - \mrr_4 + \frac{1}{2}\,(\mrr_6 + \mrr_7) + \mrr_8 < \mrr_9 
< \mathrm{min}(0\,,\,- \frac{3}{2} + \frac{1}{2}\,(\mrr_5 + \mrr_6 + \mrr_7)) \spc
\nl
{}&{}&
- \mrr_6 - \mrr_7 < \mrr_{10} < 0 \spc \quad
\mathrm{max}( - \mrr_6 - \mrr_7\,,\,- 2 - \mrr_1 + \mrr_8 - \mrr_9 - \mrr_{10}) < \mrr_{11} < 0 \spc
\nl
{}&{}&
0 < \mrr_{12} < \mrr_6 + \mrr_7 + \mrr_{10} \spc 
\nl
{}&{}&
\mathrm{max}(0\,,\,- 4 - 2\,\mrr_1 + 3\,(\mrr_6 + \mrr_7) - 2\,\mrr_9 - \mrr_{12}) < \mrr_{13} 
< \mrr_6 + \mrr_7 + \mrr_{11} \spc \quad
- 1 - \mrr_5 < \mrr_{14} < 0
\nl
{}&{}&
\mathrm{max}( - \frac{1}{2}\,(\mrr_{12} + \mrr_{13})\,,\,\frac{3}{2} 
- \frac{1}{2}\,(\mrr_5 + \mrr_6 + \mrr_7) + \mrr_9 - \mrr_{14}) < \mrr_{15} < 0 \spc
\eqa
for $\Gamma_{1}$ and
\bqa
{}&{}&
- 1 < \mrr_1 < 0 \spc \quad
- 1 < \mrr_2 < 0 \spc \quad
- 1 - \mrr_1 < \mrr_3 < 0 \spc
\nl
{}&{}& 
\mathrm{max}(- 1 - \mrr_1\,,\,- 1 - \mrr_2 - \mrr_2) < \mrr_4 < 0 \spc
\nl
{}&{}&
0 < \mrr_5 < 1 + \mrr_2 \spc\quad
0 < \mrr_6 < 1 + \mrr_1 + \mrr_3 \spc \quad
0 < \mrr_7 < 1 + \mrr_1 + \mrr_4 \spc
\nl
{}&{}&
\mathrm{max}(0\,,\,- 1 - \mrr_1) < \mrr_8 < \mathrm{min}(1\,,\,2 + \mrr_1) \spc
\nl
{}&{}& - 1 + \mrr_8 < \mrr_9 < \mathrm{min}(0\,,\,- \frac{5}{2} - \mrr_1 + \frac{1}{2}\,\mrr_5 + \mrr_6 + \mrr_7) \spc
\nl
{}&{}& - \mrr_6 - \mrr_6 < \mrr_{10} < 0 \spc
\nl
{}&{}&
\mathrm{max}(- \mrr_6 - \mrr_7\,,\,- 2 - \mrr_1 + \mrr_8 - \mrr_9 - \mrr_{10}) < \mrr_{11} < 0 \spc \quad
0 < \mrr_{12} < \mrr_6 + \mrr_7 + \mrr_{10} \spc 
\nl
{}&{}&
\mathrm{max}(0\,,\,- 4 - 2\,\mrr_1 + 3\,(\mrr_6 + \mrr_7) - 2\,\mrr_9 - \mrr_{12}) < \mrr_{13} 
< \mrr_6 + \mrr_7 + \mrr_{11} \spc
\nl
{}&{}&
- 1 - \mrr_5 < \mrr_{14} < 0 \spc \quad
- \frac{1}{2}\,(\mrr_{12} + \mrr_{13}) < \mrr_{15} < 0 \spc 
\nl
{}&{}&
\mathrm{max}(- \frac{1}{2}\,(\mrr_{12} + \mrr_{13})\,,\,\frac{5}{2} + \mrr_1 - \frac{1}{2}\,\mrr_5 - \mrr_6 - \mrr_7 + \mrr_9 - \mrr_{14} - \mrr_{15}) < \mrr_{16} < 0 \spc
\eqa
for $\Gamma_{2}$.
It is understood that violation of some of these inequalities, \ie $\mrr_{i\,\mathrm{min}} > \mrr_{i\,\mathrm{max}}$, requires
deformation of the corresponding contour as explained in \sect{str}.

The final result is as follows:
\bq
\mrI= \int_0^1 \mrd \rho_3\,\int_0^{1 - \rho_3} \mrd \rho_1\,\rho_1\,\rho_2\,\rho_3\,\mrI_{123} =
2\,\sum_{i=1}^{2}\,\Gamma_i\,\lpar \mrC_{i1} + \mrC_{i2} \rpar \spc
\eq
\bqa
\mrC_{11} &=& - i\,
         \mrC(\mrs_1\,,\,4\,,\,0)\,
         \mrC(\mrs_2\,,\,2\,,\,0)\,
         \mrC(\mrs_3\,,\,2\,,\, - \pi + \delta)\,
         \mrC(\mrs_4\,,\,2\,,\,\pi - \delta)\,
\nl {}&\times&
         \mrC(\mrs_5\,,\,\frac{1}{4}\,\lambda^{-2}\,,\,\frac{\pi}{2})\,
         \mrC(\mrs_6\,,\,4\,,\,\frac{\pi}{2})\,
         \mrC(\mrs_7\,,\,4\,,\, - \frac{\pi}{2})\,
         \mrC(\mrs_{10}\,,\,2\,,\, - \pi + \delta)
\nl {}&\times&
         \mrC(\mrs_{11}\,,\,2\,,\,\pi - \delta)\,
         \mrC(\mrs_{12}\,,\,\sqrt{\frac{\mrc_0}{3}}\,,\,\frac{\pi}{2})\,
         \mrC(\mrs_{13}\,,\,\sqrt{\frac{\mrc_0}{3}}\,,\, - \frac{\pi}{2})\,
         \mrC(\mrs_{14}\,,\,\frac{1}{4}\,\lambda^{-2}\,,\, - \pi + \delta)\,
         \mrC(\mrs_{15}\,,\,\mrc_0\,,\,\pi - \delta) \spc
\nl\nl
\mrC_{12} &=& - i\,
         \mrC(\mrs_1\,,\,4\,,\,0)\,
         \mrC(\mrs_2\,,\,2\,,\, - \pi + \delta)\,
         \mrC(\mrs_3\,,\,2\,,\, - \pi + \delta)\,
         \mrC(\mrs_4\,,\,2\,,\,\pi - \delta)\,
\nl {}&\times&
         \mrC(\mrs_5\,,\,\frac{1}{4}\,\lambda^{-2}\,,\,\frac{\pi}{2})\,
         \mrC(\mrs_6\,,\,4\,,\,\frac{\pi}{2})\,
         \mrC(\mrs_7\,,\,4\,,\, - \frac{\pi}{2})\,
         \mrC(\mrs_{10}\,,\,2\,,\, - \pi + \delta)\,
\nl {}&\times&
         \mrC(\mrs_{11}\,,\,2\,,\,\pi - \delta)\,
         \mrC(\mrs_{12}\,,\,\sqrt{\frac{\mrc_0}{3}}\,,\,\frac{\pi}{2})\,
         \mrC(\mrs_{13}\,,\,\sqrt{\frac{\mrc_0}{3}}\,,\, - \frac{\pi}{2})\,
         \mrC(\mrs_{14}\,,\,\frac{1}{4}\,,\,\lambda^{-2}\,,\, - \pi + \delta)\,
         \mrC(\mrs_{15}\,,\,\mrc_0\,,\,\pi - \delta) \spc
\nl\nl          
\mrC_{21} &=& - i\,
         \mrC(\mrs_1\,,\,4\,,\, - \pi + \delta)\,
         \mrC(\mrs_2\,,\,2\,,\, - \pi + \delta)\,
         \mrC(\mrs_3\,,\,2\,,\,\pi - \delta)\,
         \mrC(\mrs_4\,,\,2\,,\, - \pi + \delta)\,
\nl {}&\times&
         \mrC(\mrs_5\,,\,\frac{1}{4}\,\lambda^{-2}\,,\,\frac{\pi}{2})\,
         \mrC(\mrs_6\,,\,\lambda^{-2}\,,\,\frac{\pi}{2})\,
         \mrC(\mrs_7\,,\,\lambda^{-2}\,,\, - \frac{\pi}{2})\,
         \mrC(\mrs_{10}\,,\,2\,,\,0)\,
\nl {}&\times&
         \mrC(\mrs_{11}\,,\,2\,,\,0)\,
         \mrC(\mrs_{13}\,,\,1\,,\, - \frac{\pi}{2})\,
         \mrC(\mrs_{14}\,,\,\frac{1}{4}\,\lambda^{-2}\,,\, - \pi + \delta)\,
         \mrC(\mrs_{15}\,,\,\mrc_{+}^{-1}\,,\, - \pi + \delta)\,
         \mrC(\mrs_{16}\,,\, - \mrc_{-}^{-1}\,,\,0) \spc
\nl\nl
\mrC_{22} &=& - i\,
         \mrC(\mrs_1\,,\,4\,,\, - \pi + \delta)\,
         \mrC(\mrs_2\,,\,2\,,\,0)\,
         \mrC(\mrs_3\,,\,2\,,\,\pi - \delta)\,
         \mrC(\mrs_4\,,\,2\,,\, - \pi + \delta)\,
\nl {}&\times&
         \mrC(\mrs_5\,,\,\frac{1}{4}\,,\,\lambda^{-2}\,,\,\frac{\pi}{2})\,
         \mrC(\mrs_6\,,\,\lambda^{-2}\,,\,\frac{\pi}{2})\,
         \mrC(\mrs_7\,,\,\lambda^{-2}\,,\, - \frac{\pi}{2})\,
         \mrC(\mrs_{10}\,,\,2\,,\,0)\,
\nl {}&\times&
         \mrC(\mrs_{11}\,,\,2\,,\,0)\,
         \mrC(\mrs_{13}\,,\,1\,,\, - \frac{\pi}{2})\,
         \mrC(\mrs_{14}\,,\,\frac{1}{4}\,,\,\lambda^{-2}\,,\, - \pi + \delta)\,
         \mrC(\mrs_{15}\,,\,\mrc_{+}^{-1}\,,\, - \pi + \delta)\,
         \mrC(\mrs_{16}\,,\, - \mrc_{-}^{-1}\,,\,0) \spc
\eqa
where we have defined
\bq
\mrC\lpar \mrs\,,\,\zeta\,,\,\phi\rpar = \lpar \zeta\,\exp\{i\,\phi\} \rpar^{\mrs} \spp
\eq         
\subsection{The planar double{-}box diagram \label{pdb}}
Our next example is the planar double{-}box integral with $\mrp_i^2 = 0$, $\mrm_4 = 0$ and the remaining masses equal
to $\mrm$.
This is one of the scalar contributions to $\PGg \PGg \to \PGg \PGg$ and $\Pg \Pg \to \Pg\Pg$.
As usual we change variables in the Symanzik polynomials and write the integral in terms of
\bq
0 \le \mrx_{1,3} \le 1 \spc \quad 0 \le \mrx_{2,4} \le 1 - \mrx_{1,3} \spc \quad 
0 \le \rho_1 \le 1 \spc \quad 0 \le \rho_2 \le 1 - \rho_1 \spp
\eq
The integrand becomes
\bqa
\mrQ &=& \rho_1\,\rho_2\,\rho_3\,\lu\,\mrx_2\,\mrx_3
     + \rho_1\,\rho_2\,\rho_3\,\lt\,(\mrx_2 - \mrx_4)
     + \rho_1\,\rho_2\,\rho_3\,(\mrx_1\,\mrx_3 - \mrx_1\,\mrx_4 - \mrx_3)
\nl
{}&-& \rho_2^2\,\sigma_2\,\mrx_3\,\mrx_4
     + \rho_1^2\,\sigma_1\,(\mrx_1 + \mrx_2 - 1)\,\mrx_1
     + \lm\,\beta\,\sigma_3 \spc
\eqa
\bq
\beta= \rho_1\,\rho_3 + \rho_2\,\sigma_2 \spc \quad \lm = \frac{\mrm^2}{\mrs} \spc
\quad \lt = - \frac{\mrt}{\mrs} \spc 
\quad \lu = - \frac{\mru}{\mrs} \spc
\eq
where $\rho_3= 1 - \rho_1 - \rho_2$ and $\sigma_{\mrj} = 1 - \rho_{\mrj}$.
After perfoming the integration over the $\mrx$ variables we are left with the following result:
\bq
\mrI = \Bigl[ \prod_{\mri=1}^{8}\,\int_{\mrL_{\mri}}\,\frac{\mrd \mrs_{\mri}}{2\,i\,\pi}\,
\prod_{\mrj =\{2,4,5,7,8\}}\,\eG{ - \mrs_{\mrj}}\,\frac{\mrN}{\mrD}
\,
\lm^{\alpha_1}\,\lt^{\alpha_2}\,\lu^{\alpha_3}\,\exp\{i\,\alpha_4\,(\pi - \delta)\}\,
\mrF\lpar \mra_1\,,\,\dots\,,\,\mra_6\,;\,\rho_1\,,\,\rho_2 \rpar \spc
\eq 
where we have introduced
\bqa
\alpha_1 &=& - 3 + \ep - \mrs_7 - \mrs_6 - \mrs_3 - \mrs_1 \spc \quad
\alpha_2 =  - \mrs_8 + \mrs_6 + \mrs_5 \spc 
\nl
\alpha_3 &=& \mrs_4 \spc \quad
\alpha_4 = - \mrs_7 + \mrs_4 - \mrs_3 - \mrs_1 \spc 
\eqa
\bqa
\mra_1 &=& \mrs_8 + 2\,\mrs_7 + \mrs_6 + \mrs_5 + \mrs_3 \spc \quad
\mra_2 = \mrs_8 + \mrs_7 \spc \quad
\mra_3 = - \mrs_8 + \mrs_6 + \mrs_5 + \mrs_3 + 2\,\mrs_2 \spc
\nl
\mra_4 &=& \mrs_2 \spc \quad
\mra_5 = - \mrs_8 + \mrs_6 + \mrs_5 + \mrs_3 \spc \quad 
\mra_6 = - 3 + \ep - \mrs_7 - \mrs_6 - \mrs_3 - \mrs_1 \spp
\eqa
Furthermore,
\bq
\mrF\lpar \mra_1\,,\,\dots\,,\,\mra_6\,;\,\rho_1\,,\,\rho_2 \rpar =
\rho_1^{\mra_1}\,\sigma_1^{\mra_2}\,
\rho_2^{\mra_3}\,\sigma_2^{\mra_4}\,
\rho_3^{\mra_5}\,\sigma_3^{\mra_6}\,\beta^{\mra_6} \spc
\eq
\bqa
\mrN &=&
\eG{1 + \mrs_1}\,
\eG{1 + \mrs_3 + \mrs_2}\,
\eG{\mrs_4 - \mrs_3}
\nl
{}&\times&
\eG{\mrs_5 + \mrs_2 - \mrs_1}\,
\eG{1 + \mrs_6 + \mrs_5 + \mrs_4}\,
\eG{2 + \mrs_7 + \mrs_6 + \mrs_5 + \mrs_3}
\nl
{}&\times&
\eG{3 - \ep + \mrs_7 + \mrs_6 + \mrs_3 + \mrs_1}\,
\eG{\mrs_8 - \mrs_6}\,
\eG{1 + \mrs_8 + \mrs_7 - \mrs_5 - \mrs_2 + \mrs_1} \spc
\nl
\mrD &=&
\eG{3 + \mrs_3 + \mrs_2 + \mrs_1}\,
\eG{2 + \mrs_6 + \mrs_5 + \mrs_4}\,
\eG{3 + \mrs_8 + 2\,\mrs_7 + \mrs_6 + \mrs_3 - \mrs_2 + \mrs_1} \spp
\eqa
The final result is obtained by using
\bqa
\mrP_{\rho} &=& \int_0^1 \mrd \rho_1\,\int_0^{1 - \rho_1} \mrd \rho_2\; \rho_1^2\,\rho_2^2\,\mrF \,= 
\Bigl[ \prod_{\mrj=1}^{3}\,\int_{\mrL_{\mrj}}\,\frac{\mrd \mrs_{\mrj}}{2\,i\,\pi} \Bigr]\,
\Bigl[ \prod_{\mri=1}^{3}\,\eG{ - \mrs_{\mri}} \Bigr]\,\exp\{ - i\,\mrs_2\,(\pi - \delta)\}\,\frac{\mrX}{\mrY}
\nl\nl
\mrX &=&
\eG{ - \mra_6 + \mrs_1}\,
\eG{ - \mra_6 + \mrs_3}\,
\eG{1 + \mra_5 + \mrs_1}\,
\eG{ - \mra_6 - \mra_4 + \mrs_2 + \mrs_1}
\nl
{}&\times&
\eG{3 + \mra_6 + \mra_1 - \mrs_3 + \mrs_1}\,
\eG{3 + \mra_6 + \mra_3 + \mrs_3 + \mrs_2 - \mrs_1}\,
\eG{4 + \mra_6 + \mra_5 + \mra_3 + \mra_2 + \mrs_3 + \mrs_2} \spc
\nl
\mrY &=&
\eGs{ - \mra_6}\,
\eG{ - \mra_6 - \mra_4 + \mrs_1}\,
\eG{4 + \mra_6 + \mra_5 + \mra_3 + \mrs_3 + \mrs_2}
\nl
{}&\times&
\eG{7 + 2\,\mra_6 + \mra_5 + \mra_3 + \mra_2 + \mra_1 + \mrs_2 + \mrs_1}
\eqa
\subsection{The pentagons  \label{penta}}
For processes like $\mrq_1 + \mrq_2 \to \sum_{\mrj=1}^{\mrn}\,\mrp_{\mrj}$ the number of linearly independent
Mandelstam invariants is $3\,(\mrn + 2) - 10$; as a result, for pentagons we have $5$ linearly independent 
invariants~\cite{Kumar:1970cr}, \ie
\bqa
{}&{} \mrs = - (\mrq_1 + \mrq_2)^2 \spc \qquad \mrs_1 = - (\mrq_1 + \mrq_2 - \mrp_1)^2 \spc
\nl
{}&{} \mrt_0 = - (\mrq_1 - \mrp_1)^2 \spc \qquad \mrt_1 = - (\mrq_1 - \mrp_2)^2 \spc \qquad
\mru_1 = - (\mrq_1 + \mrq_2 - \mrp_2)^2 \spp
\eqa
\paragraph{one loop} \hspace{0pt} \\
Our example is given by the following propagators:
\bq
[\,1\,] = \mrQ^2 \spc \quad
[\,2\,] = (\mrQ + \mrq_1)^2 \spc \quad
[\,3\,] = (\mrQ + \mrq_1 + \mrq_2)^2 \spc \quad
[\,4\,] = (\mrQ + \mrq_1 + \mrq_2 - \mrp_1)^2 \spc 
\eq
\bq
[\,5\,] = (\mrQ + \mrq_1 + \mrq_2 - \mrp_1 - \mrp_2)^2 + \mrm^2 \spc 
\eq
with $\mrq_1^2 = \mrq_2^2 = \mrp_1^2 = 0$ and $\mrp_2^2 = \mrp_3^2 = - \mrm^2$. The loop momentum is $\mrQ$.
The pentagon is given by the following integral:
\bq
\mrP_1 = \int_0^1 \mrd \mrx_1\,\int_0^{\mrx_1} \mrd \mrx_2\,\int_0^{\mrx_2} \mrd \mrx_3\,\int_0^{\mrx_3} \mrd \mrx_4\,
\mrs^{ - 3 - \ep}\,\mrV^{ - 3 - \ep}(\mrx_1\,,\,\mrx_2\,,\,\mrx_3\,,\,\mrx_4) \spc
\eq
\bq
\mrV = \mrz_{4 4}\,\mrx_4^2 + \sum_{\mri < \mrj}\,\mrz_{\mri \mrj}\,\mrx_{\mri}\,\mrx_{\mrj} +
       \sum_{\mrj=2}^{4}\,\mrz_{0 \mrj}\,\mrx_{\mrj} \spc
\eq
where we have introduced
\bqa
{}&{}& \mrz_{4 4}\,\mrs = \mrm^2 \spc \quad
\mrz_{1 2} = \mrs \spc \quad
\mrz_{1 3}\,\mrs = \mrt_0 \spc \quad 
\mrz_{1 4}\,\mrs = \mrt_1 - \mrm^2 \spc \quad
\mrz_{2 3}\,\mrs = \mrs_1 - \mrt_0 - \mrs \spc
\nl
{}&{}& \mrz_{2 4}\,\mrs = \mru_1 - \mrt_1 - \mrs \spc \quad
\mrz_{3 4}\,\mrs = \mrm^2 - \mru_1 - \mrs_1 + \mrs \spc \quad
\mrz_{0 2} = \mrs \spc \quad
\mrz_{0 3}\,\mrs = \mrs_1 - \mrs \spc \quad
\mrz_{0 4}\,\mrs = \mrs_1 \spp
\eqa
The final result is as follows:
\bqa
\mrP_1 &=& -\,\frac{\mrs^{ - 3 - \ep}}{\eG{3 + \ep}}\,
\Bigl[ \prod_{\mrj=1}^{9}\,\int_{\mrL_{\mrj}}\,\frac{\mrd \mrs_{\mrj}}{2\,i\,\pi} \Bigr]\,
\prod_{\mrj=2,4,5,6,7,8,9}\,\eG{ - \mrs_{\mrj}}
\nl
{}&\times&
\Bigl[ \frac{\mrN_1}{\mrD_1}\, - \frac{\mrN_2}{\mrD_2}\, - \frac{\mrN_3}{\mrD_3}\, + \frac{\mrN_4}{\mrD_4} \Bigr]\,
\zeta_3^{3 + \ep}\,\prod_{\mrj=1}^{9}\,\zeta_{\mrj}^{\mrs_{\mrj}} \spc
\eqa
\bqa
\zeta_1 &=& - \frac{\mrz_{03}}{\mrz_{04}} \spc \quad
\zeta_2 = \frac{\mrz_{23}}{\mrz_{03}} \spc \quad
\zeta_3 = - \frac{1}{\mrz_{04}} \spc
\nl
\zeta_4 &=& \mrz_{24}  \spc \quad
\zeta_5 = \frac{\mrz_{34}}{\mrz_{03}} \spc \quad
\zeta_6 = - \frac{\mrz_{44}}{\mrz_{04}} \spc
\nl
\zeta_7 &=& 1  \spc \quad
\zeta_8 = \frac{\mrz_{13}}{\mrz_{03}}  \spc \quad
\zeta_9 = - \frac{\mrz_{14}}{\mrz_{04}} \spc
\eqa
\bqa
\mrN_1 &=&
\eG{1 + \mrs_1}\,
\eG{1 + \mrs_3 + \mrs_2}\,
\eG{\mrs_7 + \mrs_4 - \mrs_3}\,
\eG{\mrs_8 + \mrs_5 + \mrs_2 - \mrs_1}\,
\eG{ - 1 - \ep + \mrs_6 + \mrs_5 + \mrs_4 - \mrs_3}
\nl
{}&\times&
\eG{3 + \ep + \mrs_9 + \mrs_6 + \mrs_3 + \mrs_1}\,
\eG{1 - \ep + \mrs_9 + \mrs_8 + \mrs_7 + \mrs_6 + \mrs_5 + \mrs_4 + \mrs_2} \spc
\nl
\mrD_1 &=&
\eG{2 + \mrs_1}\,
\eG{2 + \mrs_3 + \mrs_2}\,
\eG{ - \ep + \mrs_6 + \mrs_5 + \mrs_4 - \mrs_3}\,
\eG{2 - \ep + \mrs_9 + \mrs_8 + \mrs_7 + \mrs_6 + \mrs_5 + \mrs_4 + \mrs_2} \spc
\nl\nl
\mrN_2 &=&
\eG{1 + \mrs_1}\,
\eG{1 + \mrs_3 + \mrs_2}\,
\eG{\mrs_7 + \mrs_4 - \mrs_3}\,
\eG{\mrs_8 + \mrs_5 + \mrs_2 - \mrs_1}\,
\eG{ - \ep + \mrs_6 + \mrs_5 + \mrs_4 + \mrs_2}
\nl
{}&\times&
\eG{3 + \ep + \mrs_9 + \mrs_6 + \mrs_3 + \mrs_1}\,
\eG{1 - \ep + \mrs_9 + \mrs_8 + \mrs_7 + \mrs_6 + \mrs_5 + \mrs_4 + \mrs_2} \spc
\nl
\mrD_2 &=&
\eG{2 + \mrs_1}\,
\eG{2 + \mrs_3 + \mrs_2}\,
\eG{1 - \ep + \mrs_6 + \mrs_5 + \mrs_4 + \mrs_2}\,
\eG{2 - \ep + \mrs_9 + \mrs_8 + \mrs_7 + \mrs_6 + \mrs_5 + \mrs_4 + \mrs_2} \spc
\nl\nl
\mrN_3 &=&
\eG{1 + \mrs_1}\,
\eG{\mrs_7 + \mrs_4 - \mrs_3}\,
\eG{2 + \mrs_3 + \mrs_2 + \mrs_1}\,
\eG{\mrs_8 + \mrs_5 + \mrs_2 - \mrs_1}\,
\eG{3 + \ep + \mrs_9 + \mrs_6 + \mrs_3 + \mrs_1}
\nl
{}&\times&
\eG{ - 2 - \ep + \mrs_6 + \mrs_5 + \mrs_4 - \mrs_3 - \mrs_1}\,
\eG{1 - \ep + \mrs_9 + \mrs_8 + \mrs_7 + \mrs_6 + \mrs_5 + \mrs_4 + \mrs_2} \spc
\nl
\mrD_3 &=&
\eG{2 + \mrs_1}\,
\eG{3 + \mrs_3 + \mrs_2 + \mrs_1}\,
\eG{ - 1 - \ep + \mrs_6 + \mrs_5 + \mrs_4 - \mrs_3 - \mrs_1}\,
\eG{2 - \ep + \mrs_9 + \mrs_8 + \mrs_7 + \mrs_6 + \mrs_5 + \mrs_4 + \mrs_2} \spc
\nl\nl
\mrN_4 &=&
\eG{1 + \mrs_1}\,
\eG{\mrs_7 + \mrs_4 - \mrs_3}\,
\eG{2 + \mrs_3 + \mrs_2 + \mrs_1}\,
\eG{\mrs_8 + \mrs_5 + \mrs_2 - \mrs_1}\,
\eG{ - \ep + \mrs_6 + \mrs_5 + \mrs_4 + \mrs_2}
\nl
{}&\times&
\eG{3 + \ep + \mrs_9 + \mrs_6 + \mrs_3 + \mrs_1}\,
\eG{1 - \ep + \mrs_9 + \mrs_8 + \mrs_7 + \mrs_6 + \mrs_5 + \mrs_4 + \mrs_2} \spc
\nl\nl
\mrD_4 &=&
\eG{2 + \mrs_1}\,
\eG{3 + \mrs_3 + \mrs_2 + \mrs_1}\,
\eG{1 - \ep + \mrs_6 + \mrs_5 + \mrs_4 + \mrs_2}\,
\eG{2 - \ep + \mrs_9 + \mrs_8 + \mrs_7 + \mrs_6 + \mrs_5 + \mrs_4 + \mrs_2} \spp
\eqa
Analytic continuation, when needed, can be performed by using standard procedures. For instance, suppose that we need 
to continue in $\zeta_1$ since $0 \le \mrz_1 \le \mrs/(4\,\mrm^2)$; the $\mrs_1\,${-}integral gives rise to a 
Meijer $\mrG^{1,2}_{2,2}$ which can be rewritten in terms of a $\ghyp{2}{1}$.
\paragraph{two loops, the pentabox} \hspace{0pt} \\
The configuration of the pentabox~\cite{Badger:2024fgb} considered here is defined by the following propagators:
\bqa
[\,1\,] &=& \mrQ_1^2 \spc \quad
[\,2\,] = (\mrQ_1 + \mrq_1)^2 \spc \quad
[\,3\,] = (\mrQ_1 + \mrq_1 + \mrq_2)^2 \spc \quad
[\,4\,] = (\mrQ_2 - \mrQ_1)^2 \spc
\eqa
\bqa
[\,5\,] &=& (\mrQ_2 + \mrq_1 + \mrq_2)^2 \spc \quad
[\,6\,] = (\mrQ_2 + \mrq_1 + \mrq_2 - \mrp_1)^2 \spc \quad
[\,7\,] = (\mrQ_2 + \mrq_1 + \mrq_2 - \mrp_1 - \mrp_2 + \mrm^2)^2 \spc \quad
[\,8\,] = \mrQ_2^2 \spc
\nl
\eqa
where $\mrQ_{1,2}$ are the loop momenta. 
The two{-}loop pentabox is given by an integral over $\alpha_1,\dots,\alpha_8$ depending on the two Symanzik polynomials.
We introduce the following change of variables:
\bqa
\alpha_1 &=& \rho_1\,\mrx_1 \spc \quad
\alpha_2 = \rho_1\,\mrx_2 \spc \quad
\alpha_3 = \rho_1\,(1 - \mrx_1 - \mrx_2) \spc
\nl
\alpha_4 &=& \rho_3 \spc
\nl
\alpha_5 &=& \rho_2\,(1 - \mrx_3 - \mrx_4 - \mrx_5) \spc \quad
\alpha_6 = \rho_2\,\mrx_5 \spc \quad
\alpha_7 = \rho_2\,\mrx_4 \spc \quad
\alpha_8 = \rho_2\,\mrx_3 \spc
\eqa
corresponding to $\rho_3 = 1 - \rho_1 - \rho_2$ with $0 \le \rho_1 \le 1$ and $0 \le \rho_2 \le 1 - \rho_1$;
furthermore we have
\bqa
{}&{}& 0 \le \mrx_1 \le 1 \spc \qquad 0 \le \mrx_2 \le 1 - \mrx_1 \spc
\nl
{}&{}&
0 \le \mrx_3 \le 1 \spc \quad
0 \le \mrx_4 \le 1 - \mrx_3 \spc \quad
0 \le \mrx_5 \le 1 - \mrx_3 - \mrx_4 \spp
\label{xvarpb}
\eqa
It is more convenient to perform another change of variables:
\bqa
{}&{}& \mrx_1 = 1 - \mrx_1^{\prime} \spc \qquad \mrx_2 = \mrx_2^{\prime} \spc
\nl
{}&{}& \mrx_3 = 1 - \mrx_3^{\prime} \spc \qquad
\mrx_4 = \mrx_3^{\prime} - \mrx_4^{\prime} \spc \qquad
\mrx_5 = \mrx_5^{\prime} \spp
\eqa
It is also convenient to define the following combinations of Mandelstam invariants:
\bqa
{}&{}& \mrm^2 = \mrs\,\mrZ_0 \spc \quad
\mrs_1 - \mrs = \mrs\,\mrZ_1 \spc \quad
\mrm^2 - \mru_1 = \mrs\,\mrZ_2 \spc \quad
\mrm^2 - \mru_1 - \mrs_1 = \mrs\,\mrZ_3 \spc 
\nl
{}&{}&
\mrt_0 - \mrs_1 + \mrs = \mrs\,\mrZ_4 \spc \quad
3\,\mrm^2 - \mru_1 - \mrs_1 = \mrs\,\mrZ_5 \spc 
\nl
{}&{}& \mrm^2 - \mru_1 - \mrs_1 + \mrs = \mrs\,\mrZ_6 \spc \quad
\mrm^2 - \mru_1 - \mrs_1 + 2\,\mrs = \mrs\,\mrZ_7 \spc \quad
2\,\mrm^2 - \mru_1 - \mrs_1 + \mrs = \mrs\,\mrZ_8 \spc 
\nl
{}&{}&
\mru_1 - \mrt_1 - \mrt_0 + \mrs_1 - 2\,\mrs = \mrs\,\mrZ_9 \spc \quad
\mru_1 - \mrt_1 - \mrt_0 + \mrs_1 - \mrs = \mrs\,\mrZ_{10} \spp
\eqa
Feynman prescription assigns $-\,i\,\delta$ to $\mrZ_{0,2,4,7,8,9}$ and $+\,i\,\delta$ to
$\mrZ_{3,5,6,10}$.
In order to have compact result we have introduced the following abbreviation:
\bq
[\,\mrx\,;\,\mra_1,\dots,\mra_{\mrp}\,;\,\mrb_1,\dots,\mrb_{\mrq}\,] =
\mrx + \sum_{\mrj=1}^{\mrp}\,\mra_{\mrj} - \sum_{\mrj=1}^{\mrq}\,\mrb_{\mrj} \spp
\eq
The result, after integrating over the $\mrx\,${-} variables is
\bqa
\mrP_2 &=&
\frac{\mrq( - 4 + \ep\,,\,\mrs\,\mrZ_8)}{\eG{4 - \ep}}\,
\Bigl[ \prod_{\mrj=1}^{16}\,\int_{\mrL_{\mrj}}\,\frac{\mrd \mrs_{\mrj}}{2\,i\,\pi} \Bigr]\,
\Bigl[ \prod_{\mrj=2,5,7,9,10,11,12,13,14,15,16}\,\eG{ - \mrs_{\mrj}} \Bigr]\,
\frac{\mrN}{\mrD}
\nl
{}&\times&
\Bigl[ \prod_{\mrj=1}^{13}\,\mrq(\mrs_{\mrj}\,,\,\zeta_{\mrj})\,
\mrF({\mathbf s},\rho_1,\rho_2)\,
\mrq([\,0\,;\,\mrs_{16},\mrs_{14},\mrs_{13},\mrs_{12},\mrs_{10},\mrs_9,\mrs_7,\mrs_2\,;\,\mrs_{15},\mrs_8,\mrs_6,\mrs_1
\,]\,,\exp\{i\,(\pi - \delta)\})
\spc
\nl\nl
{}&{}& \zeta_1 = \frac{\mrZ_1}{\mrZ_8} \spc\quad
\zeta_2 = \frac{\mrZ_4}{\mrZ_1} \spc\quad
\zeta_3 = \mrZ_8^{-1} \spc\quad
\zeta_4 = \frac{\mrZ_0}{\mrZ_8}) \spc\quad
\zeta_5 = \mrZ_9 \spc
\nl
{}&{}&
\zeta_6 = \mrZ_0^{-1} \spc \quad
\zeta_7 = - \frac{\mrZ_6}{\mrZ_1} \spc \quad
\zeta_8 = \mrZ_0^{-1} \spc \quad
\zeta_9 = - \mrZ_5 \spc
\nl
{}&{}&
\zeta_{10} = - \mrZ_3 \spc \quad
\zeta_{11} = - \mrZ_{10} \spc \quad
\zeta_{12} = \frac{\mrZ_2}{\mrZ_1} \spc \quad
\zeta_{13} = \mrZ_7 \spc
\nl\nl
\mrN &=&
\eG{[\,-6+2\,\ep\,;\,\mrs_7,\mrs_6,\mrs_5\,;\,2\,\mrs_4,2\,\mrs_3,\mrs_1\,]}\,
\eG{[\,-5+2\,\ep\,;\,\mrs_{12},\mrs_{11},\mrs_{10},\mrs_9,\mrs_7,\mrs_5\,;\,2\,\mrs_8,\mrs_6,2\,\mrs_3,\mrs_1\,]}\,
\nl
{}&\times&
\eG{[\,1\,;\,\mrs_1\,]}\,
\eG{[\,1\,;\,\mrs_3,\mrs_2\,]}\,
\eG{[\,2\,;\,\mrs_{16},\mrs_{15},\mrs_{14},\mrs_{13},\mrs_8,\mrs_3,\mrs_2\,]}\,
\eG{[\,4 - \ep\,;\,\mrs_4,\mrs_3,\mrs_1\,]}\,
\eG{[\,0\,;\,\mrs_8,\mrs_6\,;\,\mrs_4\,]} 
\nl
{}&\times&
\eG{[\,0\,;\,\mrs_{13},\mrs_9\,;\,\mrs_6\,]}\,
\eG{[\,0\,;\,\mrs_{14},\mrs_{10}\,;\,\mrs_8\,]}\,
\eG{[\,0\,;\,\mrs_{15},\mrs_{11},\mrs_5\,;\,\mrs_3\,]}\,
\eG{[\,0\,;\,\mrs_{16},\mrs_{12},\mrs_7,\mrs_2\,;\,\mrs_1\,]} \spc
\nl\nl
\mrD &=&
\eG{[\,-5+2\,\ep\,;\,\mrs_7,\mrs_6,\mrs_5\,;\,2\,\mrs_4,2\,\mrs_3,\mrs_1\,]}\,
\eG{[\,-4+2\,\ep\,;\,\mrs_{12},\mrs_{11},\mrs_{10},\mrs_9,\mrs_7,\mrs_5\,;\,2\,\mrs_8,\mrs_6,2\,\mrs_3,\mrs_1\,]}
\nl
{}&\times&
\eG{[\,2\,;\,\mrs_1\,]}\,
\eG{[\,2\,;\,\mrs_3,\mrs_2\,]}\,
\eG{[\,3\,;\,\mrs_{16},\mrs_{15},\mrs_{14},\mrs_{13},\mrs_8,\mrs_3,\mrs_2\,]} \spc
\nl\nl
\mrF &=&
\mrq([\,0\,;\,\mrs_{16},\mrs_{15},\mrs_{14},\mrs_{13},\mrs_8,\mrs_3,\mrs_2\,],\rho_1)\,
\mrq([\,0\,;\,\mrs_{15},\mrs_{14}\,],\sigma_1)\,
\mrq([\,-8+2\,\ep\,;\,\mrs_{12},\mrs_{11},\mrs_{10},\mrs_9,\mrs_7,\mrs_5\,;\,2\,\mrs_8,\mrs_6,2\,\mrs_3,\mrs_1\,],\rho_2)
\nl
{}&\times&
\mrq([\,-4+\ep\,;\,\mrs_{12},\mrs_9,\mrs_7\,;\,\mrs_8,\mrs_6,\mrs_3,\mrs_1\,],\sigma_2)\,
\mrq([\,0\,;\,\mrs_{16},\mrs_{13},\mrs_{11},\mrs_{10},\mrs_5,\mrs_2\,],\rho_3)
\nl
{}&\times&
\mrq([\,0\,;\,\mrs_8,\mrs_6,\mrs_3,\mrs_1\,;\,\mrs_{16},\mrs_{15},\mrs_{14},\mrs_{13},\mrs_{12},\mrs_{11},\mrs_{10},\mrs_9,\mrs_7,\mrs_5,\mrs_2\,],\beta) \spc
\eqa
where $\mrq(\mrs\,,\,\zeta) = \zeta^{\mrs}$.
\paragraph{A more symmetric representation} \hspace{0pt} \\
When more scales are present it is better to use the symmetric representation. Consider the configuration where we
have $\mrm_4 = \mz$ and the other internal masses equal to $\mt$; furthermore $\mrq_{1,2}^2 = 0$ and
$\mrp^2_{1,2,3} = - \mh^2$. 

The result, after the $\mrx\,${-} integrations  is
\bq
\mrP_{2\,\mrH} = \frac{\mra_{55}^{ - 4 + \ep}}{\eG{4 - \ep}}\,
\Bigl[ \prod_{\mrj=1}^{20}\,\int_{\mrL_{\mrj}}\,\frac{\mrd \mrs_{\mrj}}{2\,i\,\pi} \Bigr]\,
\Bigl[ \prod_{\mrj=1}^{20}\,\eG{\mrs_{\mrj}} \Bigr]\,
\frac{ \prod_{\mrj=1}^{6}\,\eG{\mrr_{\mrj}}}{\eG{1 + \mrr_2 + \mrr_3 + \mrr_4}\,\eG{1 + \mrr_5 + \mrr_6}}\,
\prod_{\mrj=1}^{20}\,\mrq(\zeta_{\mrj}\,,\, - \mrs_{\mrj} ) \spc
\eq
where we have introduced (in the general case)
\bqa
\mrr_1 &=& 4 - \ep - \sum_{\mrj=1}^{20}\,\mrs_{\mrj} \spc\quad
\mrr_2 = 7 - 2\,\ep - \sum_{\mrj=7,12,16,19,20}\,\mrs_{\mrj} - 2\,\sum_{\mrj\not=7,12,16,19,20}\,\mrs_{\mrj} \spc\quad
\nl
\mrr_3 &=& 13 - 4\,\ep - \mrs_{19} - 2\,\sum_{\mrj=7,12,16,18,20}\,\mrs_{\mrj}
         - 3\,\sum_{\mrj=6,11,15,17}\,\mrs_{\mrj}
         - 4\,\sum_{\mrj=1,2,3,4,5,,8,9,10,13,14}\,\mrs_{\mrj} \spc\quad
\nl
\mrr_4 &=& 1 - \sum_{\mrj=5,10,13,15,16}\,\mrs_{\mrj} - 2\,\mrs_{14} \spc\quad
\mrr_5 = 1 - \sum_{\mrj=4,8,10,11,12}\mrs_{\mrj} - 2\,\mrs_9 \spc\quad
\mrr_6 = 1 - \sum_{\mrj=2,4,5,6,7}\,\mrs_{\mrj} - 2\,\mrs_3 \spc
\eqa
\bqa
\zeta_1 &=& \frac{\mra_0}{\mra_55} \spc\quad
\zeta_2 = \frac{\mra_1}{\mra_55} \spc\quad
\zeta_3 = \frac{\mra_{11}}{\mra_55} \spc\quad
\zeta_4 = \frac{\mra_{12}}{\mra_55} \spc
\nl
\zeta_5 &=& \frac{\mra_{13}}{\mra_55} \spc\quad
\zeta_6 = \frac{\mra_{14}}{\mra_55} \spc\quad
\zeta_7 = \frac{\mra_{15}}{\mra_55} \spc\quad
\zeta_8 = \frac{\mra_2}{\mra_55} \spc
\nl
\zeta_9 &=& \frac{\mra_{22}}{\mra_55} \spc\quad
\zeta_{10} = \frac{\mra_{23}}{\mra_55} \spc\quad
\zeta_{11} = \frac{\mra_{24}}{\mra_55} \spc\quad
\zeta_{12} = \frac{\mra_{25}}{\mra_55} \spc
\nl
\zeta_{13} &=& \frac{\mra_3}{\mra_55} \spc\quad
\zeta_{14} = \frac{\mra_{33}}{\mra_55} \spc\quad
\zeta_{15} = \frac{\mra_{34}}{\mra_55} \spc\quad
\zeta_{16} = \frac{\mra_{35}}{\mra_55} \spc
\nl
\zeta_{17} &=& \frac{\mra_4}{\mra_55} \spc\quad
\zeta_{18} = \frac{\mra_{44}}{\mra_55} \spc\quad
\zeta_{19} = \frac{\mra_{45}}{\mra_55} \spc\quad
\zeta_{20} = \frac{\mra_5}{\mra_55} \spp
\eqa
For the present configuration we have
\bqa
\mra_{11} &=& \mrs\,\rho_1^2\,\sigma_1 \spc\quad
\mra_{33} = \mrs\,\rho_2^2\,\sigma_2 \spc\quad
\mra_{44} = (3\,\mhs - \mru_1 - \mrs_1 + \mrs)\,\rho_2^2\,\sigma_2 \spc\quad
\mra_{55} = \rho_2^2\,\sigma_2\,\mhs \spc
\nl          
\mra_{12} &=& \mrs\,\rho_1^2\,\sigma_1 \spc\quad
\mra_{13} = 2\,\mrs\,\rho_1\,\rho_2\,\rho_3 \spc\quad
\mra_{14} = (2\,\mhs - \mru_1 - \mrs_1 + 2\,\mrs)\,\rho_1\,\rho_2\,\rho_3 \spc\quad
\mra_{15} = (\mhs - \mrs_1 + \mrs)\,\rho_1\,\rho_2\,\rho_3 \spc
\nl          
\mra_{23} &=& \mrs\,\rho_1\,\rho_2\,\rho_3 \spc\quad
\mra_{24} = - (\mru_1 - \mrt_1 - \mrt_0 + \mrs_1 - 2\,\mrs)\,\rho_1\,\rho_2\,\rho_3 \spc\quad
\mra_{25} = (\mrt_0 - \mrs_1 + \mrs)\,\rho_1\,\rho_2\,\rho_3 \spc
\nl
\mra_{34} &=& (2\,\mhs - \mru_1 - \mrs_1 + 2\,\mrs)\,\rho_2^2\,\sigma_2 \spc
\nl          
\mra_{35} &=& (\mhs - \mrs_1 + \mrs)\,\rho_2^2\,\sigma_2 \spc\quad
\mra_{45} = (3\,\mhs - \mru_1 - \mrs_1 + \mrs)\,\rho_2^2\,\sigma_2 \spc\quad
\mra_1 = - \mrs\,\beta\,\rho_1 \spc\quad
\mra_2 = - \mrs\,\beta\,\rho_2 \spc
\nl          
\mra_4 &=& - (3\,\mhs - \mru_1 - \mrs_1 + \mrs)\,\beta\,\rho_2 \spc\quad
\mra_5 = - \beta\,\rho_2\,\mhs \spc\quad
\mra_0 = (\sigma_3\,\mts + \rho_3\,\mzs)\,\beta \spc
\eqa
Where $\mrs, \mrs_1, \mru_1, \mrt_0, \mrt_1$ are Mandelstam invariants.
The integration over $\rho_1, \rho_2$ proceeds with the same strategy; indeed
\bq
\prod_{\mrj=1}^{20}\,\mrq(\zeta_{\mrj}\,,\, - \mrs_{\mrj} ) =
\mrH({\mathbf s}\,;\,\{ \mrM \})\,
\rho_1^{\lambda_1}\,\sigma_1^{\lambda_2}\,
\rho_2^{\lambda_3}\,\sigma_2^{\lambda_4}\,
\beta^{\lambda_5}\,
(\rho_3\,\mzs + \sigma_3\,\mts)^{ - \mrs_1} \spc
\eq
where $\{ \mrM \}$ is the set of Mandelstam invariants and the $\lambda_{\mrj}$ are linear combination
of the $\mrs_{\mrk}$. Furthermore $\sigma_{\mrj} = 1 - \rho_{\mrj}$, $\beta= \rho_1\,\rho_3 + \rho_2\,\sigma_2$
and $\rho_3 = 1 - \rho_1 - \rho_2$.

As already discussed for the one{-}loop box the symmetric representation can be better formulated in order to 
study the analytic properties of the integral. A a simple example, consider the case where $\mh = \mz = 0$;
strating from \eqn{xvarpb} with an integrand containing
\bq 
\mra_0 + \sum_{\mrj=1}^{5}\,\mra_{\mrj}\,\mrx_{\mrj} + \sum_{\mri \le \mrj}\,\mra_{\mri \mrj}\,\mrx_{\mri}\,\mrx_{\mrj}
\spc
\eq
we perform the transformation $\mrx_{1,3}= \mrx_{1,3}^{\prime} + 1/2$ and
derive $\mra_0^{\prime} = \sigma_3\,(\mts - \mrs/4)$, a convenient parameter to study the behavior at the
normal threshold $\mrs = 4\,\mts$ when using \eqn{mMBs}. 
\subsection{The tria{-}box \label{TB}}
For processes like $\mrQ \to \sum_{\mrj=1}^{4}\,\mrp_{\mrj}$ we have $5$ linearly independent Mandelstam 
invariants~\cite{Kumar:1970cr}:
\bqa
{}&{}& 
\mrs_1 = - (\mrQ - \mrp_1)^2 \spc \qquad
\mrs_2 = - (\mrQ - \mrp_1 - \mrp_2)^2 \spc \qquad
\mrt_2 = - (\mrQ - \mrp_2 - \mrp_3)^2 \spc
\nl
{}&{}&
\mru_1 = (\mrQ - \mrp_2)^2 \spc \qquad
\mru_2 = (\mrQ - \mrp_3)^2 \spp
\eqa
The tria{-}box is defined by the following propagators:
\bqa
[\,1\,] &=& \mrq_1 \spc\qquad
[\,2\,] = \mrq_1 + \mrQ \spc\qquad
[\,3\,] = \mrq_2 - \mrq_1 - \mrp_4 \spc
\eqa
\bqa
[\,4\,] &=& \mrq_2 + \mrQ - \mrp_1 - \mrp_4 \spc\qquad
[\,5\,] = \mrq_2 + \mrQ - \mrp_1 - \mrp_2 - \mrp_4 \spc\qquad
[\,6\,] = \mrq_2 \spc
\eqa
where $\mrq_{1,2}$ are the loop momenta. We will study the configuration where
\bqa
\mrQ^2 &=& - \mrs \spc \quad
\mrp_1^2 = \mrp_4^2 = 0 \spc \quad
\mrp_2^2 = \mrp_3^2 = - \mrm^2 \spc
\nl
\mrm_{3,5} &=& 0 \spc \quad \mrm_{1,2,4,6}= \mrM \spp
\eqa
After introducing the Feynman parapetrization we change variables according to
\bqa
\alpha_1 &=& \rho_1\,\mrx_1 \spc \quad \alpha_2 = \rho_1\,(1 - \mrx_1) \spc \quad \alpha_3 = \rho_3 \spc
\nl
\alpha_4 &=& \rho_2\,\mrx_2 \spc \quad
\alpha_5 = \rho_2\,\mrx_3 \spc \quad
\alpha_6 = \rho_2\,(1 - \mrx_2 - \mrx_3 \spc 
\eqa
whith $\rho_3 = 1 - \rho_1 - \rho_2$. The first Symanzik polynomial is $\mrS_1 = \beta = \rho_1\,\rho_3 +
\rho_2\,\sigma_2$ while for the second polynomial we introduce the following quantities:
\bqa
{}&{}&
\mrm^2 + \mrM^2 = \mrs\,\mrF_1 \spc \quad
\mrm^2 - \mru_2 + \mrs = \mrs\,\mrF_2 \spc \quad
2\,\mrm^2 - \mru_2 - \mru_1 + 2\,\mrs = \mrs\,\mrF_3 \spc \quad
\mrm^2 - \mrM^2 - \mru_2 - \mrs_2 + \mrs = \mrs\,\mrF_4 \spc 
\nl
{}&{}&
2\,\mrm^2 - \mru_2 - \mru_1 + \mrt_2 + \mrs = \mrs\,\mrF_5 \spc \quad
2\,\mrm^2 - \mru_2 - \mru_1 - \mrs_1 + 2\,\mrs = \mrs\,\mrF_6 \spc 
\nl
{}&{}&
2\,\mrm^2 - 2\,\mrM^2 - \mru_2 - \mru_1 - \mrs_1 + 2\,\mrs = \mrs\,\mrF_7 \spc 
\eqa
\bqa
\mrz_{11} &=& \rho_1^2\,\sigma_1 \spc \quad
\mrz_{22} = \mrF_5\,\rho_2^2\,\sigma_2 \spc \quad
\mrz_{33} = \lambda_{\mrm}\,\rho_2^2\,\sigma_2 \spc 
\nl
\mrz_{12} &=& \mrF_3\,\rho_1\,\rho_2\,\rho_3 \spc \quad
\mrz_{13} = \mrF_2\,\rho_1\,\rho_2\,\rho_3 \spc \quad
\mrz_1 = \mrz_{22} - \mrz_{11} + \mrz_2 \spc 
\nl
\mrz_2 &=& \mrF_6\,\rho_1\,\rho_2\,\rho_3 - \mrz_{22} \spc \quad
\mrz_3= \rho_2\,(\mrF_4\,\rho_1\,\rho_3 - \mrF_1\,\rho_2\,\sigma_2) \spc \quad
\mrz_0 = \lambda_{\mrM}\,(\rho_2^2\,\sigma_2 - \rho_1^2\,\sigma_1) + \mrF_7\,\rho_1\,\rho_2\,\rho_3 \spc
\eqa
where $\lambda_{\mrm} = \mrm^2/\mrs$ and $\lambda_{\mrM} = \mrM^2/\mrs$.
The integral over the $\mrx\,${-}variables becomes
\bqa
\mrI_{\mrt\mrb} &=& \mrs^{ - 2 + \ep}\,\int_0^1 \mrd \mrx_1\,\int_0^1 \mrd \mrx_2\,\int_0^{1 - \mrx_2} \mrd \mrx_3\,
\mrV^{ - 2 + \ep}(\mrx_1\,,\,\mrx_2\,,\,\mrx_3) \spc
\nl
\mrV &=& \sum_{\mrj}\,\mrz_{\mrj \mrj}\,\mrx_{\mrj}^2 +
\sum_{\mri < \mrj}\,\mrz_{\mri \mrj}\,\mrx_{\mri}\,\mrx_{\mrj} + 
\sum_{\mrj}\,\mrz_{\mrj}\,\mrx_{\mrj} + \mrz_0 \spc
\eqa
with $\mrz_{23}= \mrz_{22}$. The result is as follows
\bqa
\mrI_{\mrt\mrb} &=& 
\frac{(\mrz_{33}\,\mrs)^{ - 2 + \ep}}{\eG{2 - \ep}}\,
 \Bigl[ \prod_{\mrj=1}^{8}\,\int_{\mrL_{\mrj}}\,\frac{\mrd \mrs_{\mrj}}{2\,i\,\pi} \Bigr]
 \frac{\mrN}{\mrD}\,
\prod_{\mrj=1}^{8}\,\mrq(\mrs_{\mrj}\,,\,\zeta_{\mrj}) \spc
\nl\nl
\zeta_1 &=& \frac{\mrz_{22}}{\mrz_{33}} \spc\quad
\zeta_2 = 1 \spc\quad
\zeta_3 = \frac{\mrz_{11}}{\mrz_3} \spc\quad 
\zeta_4 = \frac{\mrz_{2}}{\mrz_1} \spc\quad
\zeta_5 = \frac{\mrz_0}{\mrz_{11}} \spc
\nl
\zeta_6 &=& - \frac{\mrz_{12}}{\mrz_2} \spc\quad
\zeta_7 = - \frac{\mrz_{13}}{\mrz_3} \spc\quad
\zeta_8 = \frac{\mrz_1}{\mrz_0} \spc
\nl\nl
\mrN &=&
\eG{ - \mrs_6}\,
\eG{ - \mrs_7}\,
\eG{ - \mrs_8}\,
\eG{\mrs_2 - \mrs_1}\,
\eG{\mrs_3 - \mrs_1}\,
\eG{\mrs_3 - \mrs_2}\,
\nl
{}&\times&
\eG{\mrs_4 - \mrs_3}\,
\eG{\mrs_5 - \mrs_3}\,
\eG{\mrs_6 - \mrs_4}\,
\eG{\mrs_7 + \mrs_3}\,
\eG{\mrs_8 - \mrs_5}\,
\eG{2 - \ep + \mrs_1}\,
\nl
{}&\times&
\eG{1 + \mrs_4 - \mrs_3 + 2\,\mrs_1}\,
\eG{ - 3 + 2\,\ep + \mrs_2 - 2\,\mrs_1}\,
\eG{1 + \mrs_8 + \mrs_7 + \mrs_6 - 2\,\mrs_5 + 2\,\mrs_3} \spc
\nl\nl
\mrD &= &
\eG{ - \mrs_1}\,
\eG{\mrs_3}\,
\eG{ - 1 + 2\,\ep + \mrs_4 - \mrs_3 + \mrs_2}\,
\eG{2 + \mrs_8 + \mrs_7 + \mrs_6 - 2\,\mrs_5 + 2\,\mrs_3} \spc
\eqa
where $\mrq(\mrs\,,\,\zeta) = \zeta^{\mrs}$.
\section{Expansion parameters and Landau equations \label{LEQS}}
We repeat part of the results stated in \Bref{Passarino:2018wix};
for a given Feynman integral there exists a discriminant
\bq 
\mrD\lpar \spro{\mrp_\mri}{\mrp_\mrj}\,,\,\mrm^2_\mri\,,\,\alpha_\mri \rpar \spc
\eq
which is an homogeneous polynomial in the $\alpha_{\mrj}$ and whose coefficients 
are linear in the $\spro{\mrp_{\mri}}{\mrp_{\mrj}}$
and $\mrm^2_{\mrj}$, such that the equations
$\partial \mrD/\partial \alpha_{\mrj} = 0$ are equivalent to the usual Landau conditions
for the existence of the singularity, as described in \Brefs{Landau:1959fi,Nakanishi:1968hu,doi:10.1143/PTP.22.128}. 
As it is well known, given any $\mrm$ 
homogeneous polynomials in $\mrm$ unknowns there exists a unique minimal homogeneous polynomial in the coefficients
($\mrR$ the resultant) such that $\mrR = 0$ is a necessary and sufficient condition for the existence of a
solution to the system of equations (Landau-Nakanishi equations), distinct from the trivial solution
$\alpha_1 = \,\dots\,\alpha_{\mrm} = 0$
Note that $\alpha_{\mrj} \ge 0$ is required for the process to be physical (the so-called $+\,\alpha$ Landau surfaces,
as opposite to ``mixed-$\,\alpha$'' solutions).

The leading Landau singularity requires all of the$\alpha_{\mrj}$'s to be non zero; the case where some of the 
parameters vanish can be interpreted as the leading singularity of a diagram obtained from the original one
contracting the lines associated to the vanishing $\alpha$'s. 
Furthermore, for a given set of values $( \mrp_1,\,\dots\,,p_{\ssN} )$ which lie on the given physical{-}region Landau 
singularity there exists only one unique set of values for the internal momenta which satisfy the Landau
equations.

In the theory of (generalized) hypergeometric functions the power series are obtained by expanding about points
which are singular points of the function. We want to define a scheme  wher $\mrR$ is the expansion parameter in
the MB splitting. Note that we usually not interested in the case where $
\mrR = 0$ represents the anomalous threshold of the Feynman integral; therefore, $\mrR= 0$ refers to one of the 
multi{-}particle cuts of the integral, \ie the leading Landau dingularity of one of its sub{-}integrals.

For simplicity we start with the one{-}loop vertex with propagators
\bq
[\,1\,] = \mrq^2 + \mrm_1^2 \spc \quad
[\,2\,] = (\mrq + \mrp_1)^2 + \mrm_2^2 \spc \quad
[\,3\,] = (\mrq + \mrp_1 + \mrp_2)^2 + \mrm_3^2 \spc 
\eq
where we are interested in the two{-}particle threshold, $\mrs = (\mrm_1 + \mrm_3)^2$, where $s = (\mrp_1 + \mrp_2)^2$.
The integral contains an integrand of the following form:
\bq
\sum_{\mrj}\,\mrm_{\mrj}^2\,\alpha_{\mrj} + \mrp_1^2\,\alpha_1\,\alpha_2 + \mrp_2^2\,\alpha_2\,\alpha_3 +
\mrP^2\,\alpha_1\,\alpha_2 \spc \quad \mrP= \mrp_1 + \mrp_2 \spp
\eq
Writing the Landau equations we have
\bq
\mrP^2 = - \mrs \spc \qquad
\mrq^2 = - \mrm_1^2 \spc \qquad
\spro{\mrq}{\mrP} = \frac{1}{2}\,(\mrs + \mrm_1^2 - \mrm_3^2) \spp
\eq
At this point we are not integrested in the behavior around the anomalous threshold but only in the behavior
around the normal threshold. Therefore, we set $\alpha_2 = 0$ and consider the equation
\bq
(\alpha_1 + \alpha_3)\,\mrq_{\mu} + \alpha_3\,\mrP_{\mu} = 0 \spp
\label{C0nt}
\eq
We multiply \eqn{C0nt} by $\mrq_{\mu}$ and by $\mrP_{\mu}$ obtaining a system which has non{-}trivial solutions
only if $\lambda(\mrs\,,\,\mrm_1^2\,,\,\mrm_3^2) = 0$, where $\lambda$ is the K\"allen function. In this case the 
solutions are
\bq
\alpha_1 = \mrA_1 = \frac{1}{2\,\mrs}\,(\mrs - \mrm_1^2 + \mrm_3^2) \spc \qquad
\alpha_3 = \mrA_3 = \frac{1}{2\,\mrs}\,(\mrs + \mrm_1^2 - \mrm_3^2) \spp
\eq
Next we return to the original integral and make the following change of variables
\bq
\alpha_1 = \alpha_1^{\prime} + \mrA_1 \spc \quad
\alpha_2 = \alpha_2^{\prime} \spc \quad
\alpha_3 = \alpha_3^{\prime} + \mrA_3 \spc
\eq
\bq
\sum_{\mrj}\,\alpha_{\mrj} = 1 \qquad \mapsto \qquad
\sum_{\mrj}\,\alpha_{\mrj}^{\prime} = 0 \spp
\eq
The integral, after the transformation is
\bq
\mrC_0 = - \mrs^{-1}\,\int_{\Gamma}\,\mrd \alpha_1^{\prime}\,\mrd \alpha_2^{\prime}\,
\Bigl[ \mrQ(\alpha_1^{\prime}\,,\,\alpha_2^{\prime}) + \frac{1}{4\,\mrs^2}\,\lambda(\mrs\,,\,\mrm_1^2\,,\,\mrm_3^2) 
\Bigr]^{-1} \spc
\eq
where the quadratic form $\mrQ$ has coefficients
\bqa
\mra_2 &=& - \nu_2^2 - \frac{1}{2}\,\Bigl[
1 - \mu_1^2 - \mu_2^2 + (\nu_3^2 - \nu_1^2)\,(\mu_2^2 - \mu_1^2) - \nu_1^2 - \nu_3^2 \Bigr] \spc
\nl
\mra_{11} &=& - 1 \spc \quad
\mra_{22} = - \mu_2^2 \spc \quad
\mra_{12} = - 1 + \mu_1^2 - \mu_2^2 \spc
\eqa
where we have defined $\mrm_i^2 = \nu_i^2\,\mrs$ and $\mrp_i^2 = - \mu_i^2\,\mrs$.
The integration domain $\Gamma$ is defined by
\bq
- \mrA_1 \;\le\; \alpha_1^{\prime}\;\le \mrA_3 \spc \qquad
0\;\le \alpha_2^{\prime}\;\le\,\mrA_3 - \alpha_1^{\prime} \spc
\eq
and we can MB expand with the parameter given by the K\"allen function (pseudo{-}threshold or normal threshold).
The multi{-}MB expansion is not always the best choice and a mixed strategy can give simpler results. If we
define
\bqa
\mrQ_{\mrj} &=& \mrq_{\mrj 2}\,\mrx^2 + \mrq_{\mrj 1}\,\mrx + \mrq_{\mrj 0} 
= \eta_{\mrj}^2\,\mra_{22}\,\mrx^2 - \eta_{\mrj}\,(\mra_2 + \mrA_3\,\mra_{12})\,\mrx + \mrA_3^2\,\mra_{11} \spc
\nl\nl
\mrL_{\mrj} &=& \mrl_{\mrj 1}\,\mrx + \mrl_{\mrj 0}
= \eta_{\mrj}\,(\mra_2 + \mrA_3\,\mra_{12}) - 2\,\mrA_3\,\mra_{11} +
\eta_{\mrj}\,(\mra_{12} - 2\,\eta_{\mrj}\,\mra_{22})\,\mrx \spc
\eqa
with $\eta_1 = 1 - \eta$, $\eta_2 = \eta$ and
\bq
\eta = \frac{1}{2\,\mu_2^2}\,\Bigl[ \mu_1^2 - \mu_2^2 - 1 + \lambda^{1/2}(1\,,\,\mu_1^2\,,\,\mu_2^2) \Bigr] \spc
\eq
the result can be written as follows:
\bq
\mrC_0 = \Bigl[ \prod_{\mrj}\,\int_{\mrL_{\mrj}}\,\frac{\mrd \mrs_{\mrj}}{2\,i\,\pi} \Bigr]\,
\frac{\mra_0^{ - \mrs_1}}{1 + \mrs_2}\,\eG{\mrs_1}\,\eG{ - \mrs_2}\,\eG{1 + \mrs_2 - \mrs_1}\,
\sum_{\mrj=1}^{2}\,\mrK_{\mrj}
\,\lpar \mrF_{\mrj 1} - \frac{1}{2 + \mrs_2}\,\mrF_{2 j} \rpar \spc
\eq
\bq
\mrK_{\mrj} = \eta_{\mrj}\,\mrl_{\mrj 0}^{\mrs_2}\,\mrq_{\mrj 0}^{\mrs_1 - \mrs_2 - 1} \spc
\eq
\bqa
\mrF_{j 1} &=& \mrF^{(3)}_{\sPD}\lpar 1\,;\, - \mrs_2\,,\,1 + \mrs_2 - \mrs_1\,;\,2\,;\,
- \frac{\mrl_{\mrj 1}}{\mrl_{\mrj 0}}\,,\,\mrx_{\mrj -}^{-1}\,,\,\mrx_{\mrj +}^{-1} \rpar \spc
\nl
\mrF_{j 2} &=& \mrF^{(3)}_{\sPD}\lpar 2 + \mrs_2\,;\, - \mrs_2\,,\,1 + \mrs_2 - \mrs_1\,;\,3 + \mrs_2\,;\,
- \frac{\mrl_{\mrj 1}}{\mrl_{\mrj 0}}\,,\,\mrx_{\mrj -}^{-1}\,,\,\mrx_{\mrj +}^{-1} \rpar \spc
\eqa
where $\mrx_{\mrj \pm}$ are the roots of $\mrQ_{\mrj}$ and
\bq
\mra_0 = \frac{1}{4\,\mrs^2}\,\lambda(1\,,\,\nu_1^2\,,\,\nu_2^2) \spp
\eq
The general strategy is clear:

\begin{enumerate}

\item identify the sub{-}diagram where the leading Landau singularity is the expansion parameter,

\item set the releavant $\alpha_{\mrj}$ to xero,

\item find the solution of the Landau equations for the remaining $\alpha_{\mrk}$ and shift them.

\end{enumerate}

As an example we consider the kite integral with $\mrm_{\mrj} = \mrm$, where we are interested in the behavior
aroud the threshold $s = 9\,\mrm^2$. Given the propagators
\bq
[\,1\,] = \mrq_1^2 + \mrm^2 \spc \quad
[\,2\,] = (\mrq_1 + \mrp)^2 + \mrm^2 \spc \quad
[\,3\,] = (\mrq_1 - \mrq_2 + \mrp)^2 + \mrm^2 \spc 
\eq
\bq
[\,4\,] = (\mrq_2 - \mrp)^2 + \mrm^2 \spc \quad
[\,5\,] = \mrq_2^2 + \mrm^2 \spc 
\eq
The three{-}particle cut of the kite integral is the leading Landau singularity of the sub{-}integral corresponding
to $\alpha_2 = \alpha_4 = 0$. After writing
\bq
\alpha_1\,\mrq_{1\,\mu} + \alpha_3\,(\mrq_1 - \mrq_2 + \mrp)_{\mu} = 0 \spc \qquad
\alpha_3\,(\mrq_2 - \mrq_1 - \mrp)_{\mu} + \alpha_5\,\mrq_{2\,\mu} = 0 \spc
\eq
we multiply both equations by $\mrq_{1\,\mu}, \mrq_{2\,\mu}$ and $\mrp_{\mu}$ and use
\bq
\mrq_{1,2}^2 = - \mrm^2 \spc \qquad
\spro{\mrq_1}{\mrq_2} = \frac{1}{2}\,(2\,\spro{\mrq_1}{\mrp} - 2\,\spro{\mrq_2}{\mrp} - \mrs - \mrm^2) \spp
\eq
Consistency of the equations requires
\bq
\spro{\mrq_1}{\mrp} = - \spro{\mrq_2}{\mrp} =
\frac{1}{4}\,\frac{1}{\mrs - \mrm^2}\,(\mrs^2 + 2\,\mrm^2\,\mrs - 3\,\mrm^4) \spp
\eq
Having two sets of solutions implies the following conditions:
\bq
\alpha_1 = \mrA_1 \spc \quad
\alpha_3 = \mrA_3 = \frac{1}{2}\,\frac{\mrs + 3\,\mrm^2}{\mrs - 3\,\mrm^2}\,\mrA_1 \spc \quad
\alpha_5 = \mrA_5 = \mrA_1 \spc
\eq
with $\mrs = 9\,\mrm^2$. Setting $\mrA_1 + \mrA_3 + \mrA_5 = 1$ gives
\bq
\mrA_1 = \frac{2}{5}\,\frac{\mrs - 3\,\mrm^2}{\mrs - \frac{9}{5}\,\mrm^2} \spp
\eq
The region of interest is given by $\mrs > 3\,\mrm^2$ where $0 \; \le \; \mrA_{1,3,5} \; \le \; 1$. Given 
$\mrA_{1,3,5}$ we perform the change og variables
\bq
\alpha_{1,3,5} = \beta_{1,3,5} + \mrA_{1,3,5} \spc \qquad
\alpha_{2,4} = \beta_{2,4} \spc
\eq
followed by
\bq
\beta_1 = \rho_1\,\mrx_1 \spc \quad
\beta_2 = \rho_1\,(1 - \mrx_1) \spc \quad
\beta_3 = \rho_3 \spc \quad
\beta_4 = \rho_2\,(1 - \mrx_2) \spc \quad
\beta_5 = \rho_2\,\mrx_2 \spc 
\label{betakite}
\eq
which implies $\rho_3 =  - \rho_1 - \rho_2$. With these transformation the first Symanzik polynomial depends only
on $\rho_1, \rho_2$ and the second Symanzik polynomial is a quadratic form in $\mrx_1, \mrx_2$ with
$\rho\,${-}dependent coefficients. The MB expansion parameter is
\bq
\mrR = \mrP(\mrs\,,\,\mrm^2)\,\mrT(\mrs\,,\,\mrm^2) \spc \quad
\mrP = \frac{\mrs - 3\,\mrm^2}{(\mrs - \frac{9}{5}\,\mrm^2)^3} \spc \quad
\mrT= (\mrs - \mrm^2)\,(\mrs - 9\,\mrm^2) \spc
\eq
showing that $\mrT = 0$ at the pseudo{-}threshold and at the normal threshold; furthermre, for $\mrs > 3\,\mrm^2$,
we have $\mrs^2\,\mrP < 1$. The integration domain $\Gamma$ follows from
\bq
\delta\lpar 1 - \sum_{\mrj=1}^{5}\,\alpha_{\mrj} \rpar\,\prod_{\mrj=1}^{5}\,\theta(\alpha_{\mrj}) \spp
\eq
There are four regions defined by
\bqa
\Gamma_1 \quad &;& \quad
0 \le \rho_1 \le \mrA_3 \spc \quad
0 \le \rho_2 \le \mrA_3 - \rho_1 \spc \quad
- \frac{\mrA_1}{\rho_1} \le \mrx_1 \le 1 \spc \quad
- \frac{\mrA_1}{\rho_2} \le \mrx_2 \le 1 \spc 
\nl
\Gamma_2 \quad &;& \quad
- \mrA_1 \le \rho_1 \le 0 \spc \quad
- \mrA_1 \le \rho_2 \le 0 \spc \quad
1 \le \mrx_1 \le - \frac{\mrA_1}{\rho_1} \spc \quad
1 \le \mrx_2 \le - \frac{\mrA_1}{\rho_2} \spc \quad
\nl
\Gamma_3 \quad &;& \quad
- \mrA_1 \le \rho_1 \le 0 \spc \quad
0 \le \rho_2 \le \mrA_3 - \rho_1 \spc \quad
1 \le \mrx_1 \le - \frac{\mrA_1}{\rho_1} \spc \quad
- \frac{\mrA_1}{\rho_2} \le \mrx_2 \le 1 \spc 
\nl
\Gamma_4 \quad &;& \quad
0 \le \rho_1 \le \mrA_3 - \rho_2 \spc \quad
- \mrA_1 \le \rho_2 \le 0 \spc \quad
- \frac{\mrA_1}{\rho_1} \le \mrx_1 \le 1 \spc \quad
1 \le \mrx_2 \le - \frac{\mrA_1}{\rho_2} \spp
\eqa
If we are interested in the two{-}particle cut of the kite integral we can set $\alpha_{3,4,5} = 0$ in the Landau equations
obtaining the solution $\alpha_{1,2} = 1/2$ with the constraint $\mrs= 4\,\mrm^2$. Consequently, we
define $\beta_{1,2} = \alpha_{1,2} + 1/2$ and $\beta_{3,4,5}= \alpha_{3,4,5}$ with $\sum_{\mrj}\,\beta_{\mrj} = 0$ and
proceed with the transformation of \eqn{betakite}.

Similar results hold for the other two{-}loop diagrams; for instance, the three{-}particle cut corresponding
to lines $2, 3, 4$ of the delta{-}kite of Fig.~\ref{fd_ddkite} is represented by the leading Landau
singularity of the sub{-}diagram where $\alpha_{1,5,6} = 0$.
\section{Numerical approach \label{numi}}
In this Section, we discuss the alternative of a numerical integration. To show an example
we consider configuration $\mrc$) for the kite integral; the $\mrx_1, \mrx_2$ integral is given by
\bq
\mrI_{\mrc} = \int_0^1 \mrd \mrx_1\,\mrd \mrx_2\,\Bigl[
\sigma_1\,\rho_1^2\,(\mrx_1 -\frac{1}{2})^2 +
\sigma_2\,\rho_2^2\,(\mrx_2 -\frac{1}{2})^2 +
2\,\rho_1\,\rho_2\,\rho_3\,(\mrx_1 - \frac{1}{2})\,(\mrx_2 - \frac{1}{2}) +
\mrC \Bigr]^{-1} \spc
\eq
where we have introduced
\bq
\mrC = \frac{1}{4}\,\beta\,\lambda\,(\rhob - \rho_2) \spc \quad
\beta = \rho_1\,\rho_3 + \sigma_2\,\rho_2 \spc \quad
\lambda = 1 - 4\,\lambda_{\mrM}^2 + 4\,\lambda_{\mrm}^2 \spc \quad
\rhob = \lambda^{-1}\,\lpar 4\,\lambda_{\mrm}^2 - \rho_1 \rpar \spp
\eq
There are two scenarios:
\begin{enumerate}

\item $\mrC \not = 0$, where we can apply a BST algorithm~\cite{JB,MS,Tkachov:1996wh,Ferroglia:2002mz},

\item $\mrC \approx 0$, where the behavior of $\mrI_{\mrc}$ must be investigated.

\end{enumerate}
We assume that
\bq 
\lambda_{\mrm} < \frac{1}{2} \spc \qquad
\lambda_{\mrM} < \frac{1}{2} \spc \qquad
\lambda_{\mrm} > \lambda_{\mrM} \spc
\eq
so that $\lambda > 0$. We have $\mrC = 0$ for $\rho_2 = \rhob$, the straight line connecting the points
\bq
\lpar \rho_1 = 0\,,\,\rho_2 = 4\,\frac{\lambda_{\mrm}^2}{\lambda} \rpar \spc
\qquad
\lpar \rho_1 = 4\,\lambda^2_{\mrm}\,,\,\rho_2 = 0 \rpar \spp
\eq
Note that the curve corresponding to $\beta = 0$ is always outside the $\rho$ domain of integration.
When $\mrC$ is large we can use
\bq
\mrI_{\mrc} = \int_0^1\,\mrd \mrx_1\,\mrd \mrx_2\,\mrF(\mrx_1\,,\,\mrx_2)
= \int_0^1\,\mrd \mrx_1\,\int_0^{\mrx_1}\,\mrd \mrx_2\,\mrF(\mrx_1\,,\,\mrx_2) +
      \int_0^1\,\mrd \mrx_2\,\int_0^{\mrx_2}\,\mrd \mrx_1\,\mrF(\mrx_1\,,\,\mrx_2) = \mrI_0 + {\overline{\mrI}}_0 \spc
\eq
where the second integral is obtained from the first one exchanging $\mrx_1$ and $\mrx_2$ in $\mrF$. Therefore,
the result is
\bq
\mrI_0 = \frac{1}{\mrC}\,\Bigl\{ \frac{1}{2} +
\int_0^1\,\mrd \mrx_1\,\int_0^{\mrx_1}\,\mrd \mrx_2\,\ln \mrQ(\mrx_1\,,\,\mrx_2) -
\frac{1}{4}\,\int_0^1\,\mrd \mrx\,\Bigl[ \ln \mrQ_0(\mrx) + \ln \mrQ_2(\mrx) \Bigr] \Bigr\} \spc
\eq
and $\mrQ_0(\mrx)= \mrQ(1\,,\,\mrx)$ while $\mrQ_2(\mrx)= \mrQ(x,0)$. Next we proceed with numerical integration.
In this formulation we can see when the integral develops an imaginary part: if we write
\bq
\mrQ = \mby^{\mrt}\,\mbH\,\mby + \mrC \spc \qquad \mry_i = \mrx_i - \frac{1}{2} \spc
\eq
and introduce $\mrG = \mathrm{det}\,\mrH$ it follows that $\mrG = \rho_1^2\,\rho_2^2\,\beta \ge 0$. Since
$\mrH_{11} \ge 0$ we derive that the quadratic form ${\overline{\mrQ}} = 
\mby^{\mrt}\,\mbH\,\mby$
is positive{-}semidefinite. Therefore, a necessary condition for having an imaginary part in $\ln\,\mrQ$ is
$\mrC < 0$, \eg $\lambda > 0$ and $\rho_2 > \rhb$. 
Consider now $\mrQ_2$ which can be written as
\bq
\mrQ_2 = \sigma_1\,\rho_1^2\,\mrx^2 -
\lpar \sigma_1\,\rho_1 + \rho_2\,\sigma_2 \rpar\,\rho_1\,\mrx +
\beta\,\lpar \rho_2\,\lambda_{\mrM}^2 + \sigma_2\,\lambda_{\mrm^2} \rpar \spp
\eq
If we define
\bq
\mra = \sigma_1\,\rho_1^2 \spc \quad
\mrb_2^2 = \mrb_1^2 - \rho_1\,\rho_2\,\rho_3 \spc \quad
\mrb_1^2 = \beta\,\lpar \sigma_2\,\lambda_{\mrm}^2 + \rho_2\,\lambda_{\mrM}^2 \rpar \spc
\eq
the necesary condition for having an imaginary part is
\bq
\lambda\lpar \mra\,,\,\mrb_1^2\,,\,\mrb_2^2 \rpar > 0 \spc
\eq
where now $\lambda$ is the K\"allen function. A similar result can be obtained for $\mrQ_0$.  
We can apply the same startegy also for more complicated integrals like the delta{-}kite where
the quadratic is
\bqa
\mrQ &=& \mbx^{\mrt}\,\mbH\,\mbx + 2\,\mbK^{\mrt}\,\mbx + \mrL =
         \mby^{\mrt}\,\mbH\,\mby + \mrC \spc 
\nl
\mry_i &=& \mrx_i - \mrX_i \spc \quad i = 1,2,3 \spc
\nl
\mbbx &=& \mbH^{-1}\,\mbK \spc \quad 
\mrC = \mrL + \mbK^{\mrt}\,\mbbx \spp
\eqa
However, in this case, the quadratic $\mby^{\mrt}\,\mbH\,\mby$ is an
isotropic quadratic form and the imaginary part depends on the roots of $\mrQ$. 

At the two{-}loop level the region $\mrC \approx 0$ requires particular care. Consider a simple example:
\bq
\mrI = \int_0^1 \mrd \mrx\,\Bigl[ \mrA\,\lpar \mrx - \frac{1}{2} \rpar^2 + \mrC \Bigr]^{-1} \spp
\eq
The resulting expansion depends on the ratio $\mrA/\mrC$. The full result is
\bq
\mrI= \frac{1}{2}\,\mrC^{-1}\;\hyp{1}{\frac{1}{2}}{\frac{3}{2}}{ - \frac{\mrA}{4\,\mrC}} \spp
\eq
If $\mid \mrA \mid < \mid \mrC \mid$ and $\mrA \to 0$ we obtain
\bq
\mrI = \mrC^{-1}\,\Bigl[ 1 + \ord{\frac{\mrA}{4\,\mrC}} \Bigr] \spp
 \eq
When $\mid \mrC \mid < \mid \mrA \mid\;$, $\mrA/\mrC > 0$ and $\mrC \to 0$  we have
\bq
\mrI = \lpar \mrA\,\mrC \rpar^{ - 1/2}\,\mathrm{arctn}\lpar \frac{1}{2}\,\sqrt{\frac{\mrA}{\mrc}} \rpar \sim
\frac{\pi}{2}\,\lpar \mrA\,\mrC \rpar^{ - 1/2} \spc
\eq
while for $\mid \mrC \mid < \mid \mrA \mid\;$, $\mrA/\mrC < 0$ and $\mrC \to 0$  we have
\bq
\mrI \sim \pm i\,\frac{\pi}{2}\,\lpar \mrA\,\mrC \rpar^{ - 1/2} \spc
\eq
depending on the sign of $\mrA$.

For the kite integral we start with
\bqa
\mrI_{\mrc} &=& \int_{-1/2}^{+1/2}\,\mrd \mrx_1\,\int_{-1/2}^{\mrx_1}\,\mrd \mrx_2\,
\Bigl[
\sigma_1\,\rho_1^2\,\mrx_1^2 +
\sigma_2\,\rho_2^2\,\mrx_2^2 +
2\,\rho_1\,\rho_2\,\rho_3\,\mrx_1\,\mrx_2 +
\mrC
\Bigr]^{-1}
\nl
{}&=&
\int_{-1/2}^{+1/2}\,\mrd \mrx_1\,\int_{-1/2}^{\mrx_1}\,\mrd \mrx_2\,
\Bigl[ \mrQ(\mrx_1\,,\,\mrx_2) +
\mrC
\Bigr]^{-1} \spp
\eqa
We use a MB splitting, \ie
\bq
\lpar \mrQ + \mrC \rpar^{-1} = \int_{\mrL}\,\frac{\mrd \mrs}{2\,i\,\pi}\,\eB{\mrs}{1 - \mrs}\,
\mrC^{ - \mrs}\,\mrQ^{ - 1 + \mrs} \spc
\eq
where $0 < \Re(\mrs) < 1$. After the splitting we perform the $\mrx_1\,,\mrx_2$ integration obtaining
\bqa
\mrI_{\mrc} &=&     \frac{1}{8}\,\int_{\mrL}\,\frac{\mrd \mrs}{2\,i\,\pi}
\,\eG{\mrs}\,\eG{1 - \mrs}\,\mrs^{-1}\,\mrC^{ - \mrs}\,\mrq(1 - \mrs\,,\,4)\,
\Bigl( \mrJ_{\mra \mrb} + \mrJ_{\mrc \mrd} \Bigr) \spc
\nl
\nl
\mrJ_{\mra \mrb} &=& 
           \mrq(- 1 +\mrs,\mrz_1)\,
            \mrq(- 1 +\mrs,\frac{\mrz_2}{\mrz_1})\,
         \mrF^{(2)}_{\sPD}(1\,;\,1 - \mrs\,,\,1 - \mrs\,;\,2\,;\,\mra_-^{-1}\,,\,\mra_+^{-1}) \spc
\nl
{}&+&
           \mrq(- 1 +\mrs,\mrz_1)\,
            \mrq(- 1 +\mrs,\frac{\mrz_2}{\mrz_1})\,
         \mrF^{(2)}_{\sPD}(1\,;\,1 - \mrs\,,\,1 - \mrs\,;\,2\,;\,\mrb_-^{-1}\,,\,\mrb_+^{-1}) \spc
\nl
\mrJ_{\mrc \mrd} &=&
          \mrq(- 1 +\mrs\,,\,\mrz_2)\,
            \mrq(- 1 +\mrs\,,\,\frac{\mrz_1}{\mrz_2})\,
         \mrF^{(2)}_{\sPD}(1\,;\,1 - \mrs\,,\,1 - \mrs\,;\,2\,;\,\mrc_-^{-1}\,,\,\mrc_+^{-1}) \spc
\nl
{}&+&         
           \mrq(- 1 +\mrs,\mrz_2)\,
            \mrq(- 1 +\mrs,\frac{\mrz_1}{\mrz_2})\,
         \mrF^{(2)}_{\sPD}(1\,;\,1 - \mrs\,,\,1 - \mrs\,;\,2\,;\,\mrd_-^{-1}\,,\,\mrd_+^{-1}) \spc
\eqa
where $\mrz_i = \sigma_i\,\rho_i^2$ and $\mrq(\mrs,\zeta)= \zeta^{\mrs}$. 
Furthermore we have defined
\bq
\mra_{\pm} = \frac{\rho_2\,\rhpm}{\sigma_1\,\rho_1} \spc \quad
\mrb_{\pm} = - \frac{\rho_2\,\rhmp}{\sigma_1\,\rho_1} \spc \quad
\mrc_{\pm} = \frac{\rho_1\,\rhpm}{\sigma_2\,\rho_2} \spc \quad
\mrd_{\pm} = - \frac{\rho_1\,\rhmp}{\sigma_2\,\rho_2} \spc \quad
\rhpm= \rho_3 \pm \sqrt{ - \beta} \spp
\eq
It is easily seen that the expansion parameters are
\bq
\zeta_1 = 4\,\frac{\mrC}{\sigma_1\,\rho_1^2} \spc
\qquad
\zeta_2 = 4\,\frac{\mrC}{\sigma_2\,\rho_2^2} \spp
\eq
Therefore, for the ``singular'' configuration, \ie
\bq
\mrC \approx 0 \spc \quad \hbox{\ie} \quad \rho_2 \approx \rhb \quad \hbox{or} \quad
\rho_2 = \rhb \pm \delta_2 \spc
\eq
(where $\delta_2$ is a small parameter to be adjusted in the numerical calculation), we still have to
distinguish and separate different regions in the $\rho_1 {-} \rho_2$ plane. Asuming that we are in a region where
$\zeta_1 < 1$ and $\zeta_2 < 1$ we select the poles at $\mrs = - n$ where we
have a double pole at $\mrs = 0$ and single poles at $\mrs = - 1, - 2,\dots$.        
If one or both conditions are violated we select the poles at $\mrs = \mrn + 1$. 
First we define the following functions:
\bq
\mrf_{\mrn}( \mrx\,,\,\mry) = 
\mrF^{(2)}_{\sPD}\lpar 1\,;\,\mrn + 1\,,\,\mrn + 1\,;\,2\,;\,\mrx\,,\,\mry \rpar \spc
\quad
\mrg( \mrx\,,\,\mry ) = \frac{\partial}{\partial\,\mrs}\,
\mrF^{(2)}_{\sPD}\lpar 1\,;\,1 - \mrs\,,\,1 - \mrs\,;\,2\,;\,\mrx\,,\mry \rpar \,\bmid_{\mrs = 0} \spp
\eq
If $\mid \zeta_i \mid < 1$ the simple poles give the following contribution:
\bqa
\mrI_{\mathrm{sp}} &=& - \frac{1}{2}\,\sum_{\mrn = 1}^{\infty} \frac{1}{\mrn}\,
\Bigl[ \frac{1}{\sigma_1\,\rho_1^2}\,\mrH_1 + \frac{1}{\sigma_2\,\rho_2^2}\,\mrH_2 \Bigr] \spc
\nl\nl
\mrH_1 &=& \lpar - \zeta_1 \rpar^{\mrn}\,\Bigl[
\mrf_{\mrn}\lpar \mra_-^{-1}\,,\,\mra_+^{-1} \rpar +
\mrf_{\mrn}\lpar \mrb_-^{-1}\,,\,\mrb_+^{-1} \rpar \Bigr] \spc
\nl
\mrH_2 &=& \lpar - \zeta_2 \rpar^{\mrn}\,\Bigl[
\mrf_{\mrn}\lpar \mrc_-^{-1}\,,\,\mrc_+^{-1} \rpar +
\mrf_{\mrn}\lpar \mrd_-^{-1}\,,\,\mrd_+^{-1} \rpar \Bigr] \spp
\eqa
The Lauricella functions can be simplified giving
\bq
\mrf_1(\mrx\,,\,\mry) = 
2\,\frac{\mrx\,\mry}{(\mry - \mrx)^3}\,\Bigl[ \ln(1 - \mry) - \ln(1 - \mrx) \Bigr] +
\frac{1}{(\mry - \mrx)^2}\,\Bigl( \frac{\mrx^2}{1 - \mrx} + \frac{\mry^2}{1 - \mry} \Bigr) \spc
\eq
\etc
The contribution from the double pole is
\bqa
\mrI_{\mathrm{dp}} &=&
       \frac{1}{2}\,\frac{1}{\tau}\,\ln(\zeta_1)\, \Bigl[
            \ln(1 + \omega_3)
          - \ln(1 - \omega_3)
          + \ln(1 - \omega_4)
          - \ln(1 + \omega_4)
          \Bigr]
\nl
{}&+& \frac{1}{2}\,\frac{1}{\tau}\,\ln(\zeta_2)\, \Bigl[
            \ln(1 + \omega_1)
          - \ln(1 - \omega_1)
          + \ln(1 - \omega_2)
          - \ln(1 + \omega_2)
          \Bigr]
\nl
{}&-& \frac{1}{2}\,\frac{1}{\sigma_1\,\rho_1^2}\,\Bigl[ \mrg(\omega_3,\omega_4) + \mrg( - \omega_4, - \omega_3) \Bigr]
    - \frac{1}{2}\,\frac{1}{\sigma_2\,\rho_2^2}\,\Bigl[ \mrg(\omega_1,\omega_2) + \mrg( - \omega_2, - \omega_1) \Bigr]
\spc
\eqa
where we have defined
\bq
\omega_1 = \frac{\sigma_1\,\rho_1}{\rhm\,\rho_2} \spc \quad
\omega_2 = \frac{\sigma_1\,\rho_1}{\rhp\,\rho_2} \spc \quad
\omega_3 = \frac{\sigma_2\,\rho_2}{\rhm\,\rho_1} \spc \quad
\omega_4 = \frac{\sigma_2\,\rho_2}{\rhp\,\rho_1} \spc \quad
\tau = (\rhp - \rhm)\,\rho_1\,\rho_2 \spp
\eq
Furthermore we obtain the following result for $\mrg$:
\bqa
\mrg( \mrx\,,\,\mry ) &=& \mrg_1(\mrx\,,\,\mry) + \mrg_1(\mry\,,\,\mrx) \spc
\nl\nl
\mrg_1(\mrx\,,\,\mry) &=& - \frac{1}{2}\,\frac{1}{\mrx - \mry}\,\ln^2(2,1 - \mrx) 
\nl          
{}&+& \frac{1}{\mrx\,(\mrx - \mry)}\,\Bigl[
         \li{2}{\frac{\mrx^2}{\mrx - \mry}} \, 
       - \li{2}{\frac{\mrx^2\,(1 - \mry)}{\mrx - \mry}} \, 
       - \ln(1 - \mry)\,\ln(1 - \frac{\mrx^2\,(1 - \mry)}{\mrx - \mry}) \Bigr] \spp
\eqa
If we select the right poles we obtain
\bqa
\mrI_{\mathrm{rp}} &=& \frac{1}{8}\,\mrC^{-1}\,\sum_{\mrn=0}^{\infty}\,\frac{1}{\mrn + 1}\,
  \Bigl\{
  \lpar - \zeta_2 \rpar^{ - \mrn}\,\Bigl[
    \mrf_{ - \mrn}(\mra_-^{-1}\,,\,\mra_+^{-1} +
    \mrf_{ - \mrn}(\mrb_-^{-1}\,,\,\mrb_+^{-1} \Bigr] 
\nl
{}&+&
  \lpar - \zeta_1 \rpar^{ - \mrn}\,\Bigl[
    \mrf_{ - \mrn}(\mrc_-^{-1}\,,\,\mrc_+^{-1} +
    \mrf_{ - \mrn}(\mrd_-^{-1}\,,\,\mrd_+^{-1} \Bigr] \Bigr\} \spc
\eqa
where $\mrf_0 = 1$ etc
Another interesting approach is based on the partial quadratization of the Symanzik polinomials. Let us consider
the kite integral in the configuration wher $\mrm_3 = 0$, $\mrm_\mrj= m$ for $\mrj \not= 3$. In this case the kite
integrand with external momentum $\mrP$ can be written as a $\rho\,${-}convolution of a one{-}loop, three{-}point
scalar function with parameters
\bqa
{}&{}& \mrp_1^2 = \rho_2^2\,\sigma_2\,\mrP^2 \spc \quad
       \mrp_2^2 = \rho_1^2\,\sigma_1\,\mrP^2 \spc \quad
       \mrQ^2 = (\mrp_1 + \mrp_2)^2 = \beta\,\sigma_3\,\mrP^2 \spc
\nl
{}&{}& \mrm_1^2 = \beta\,\sigma_3\,\mrm^2 \spc \quad
       \mrm_2^2 = \beta\,\sigma_3\,\mrm62 + \rho_1\,\rho_2\,\rho_3\,\mrP^2 \spc \quad
       \mrm_3^2 = \beta\,\sigma_3\,\mrm^2 \spp
\eqa
Following \Bref{tHooft:1978jhc} we apply the so{-}called $\alpha\,${-}trick in order to linearize the integrand in 
the variable $\mrx_1$. This amounts to shifting $\mrx_2 \to \mrx_2 + \alpha\,\mrx_1$ and slect $\alpha$ such that
the $\mrx_1^2$ term cancel in the integrand. The resulting expression is given in Eq.~(5.6) of
\Bref{tHooft:1978jhc}. However this equation is not valid in the region $\mrp_1^2 < 0$, $\mrp_2^2 < 0$ and
$\alpha \in \Cf$ which is what happens when $\mrP^2 < 0$. 
To summarize, two{-}loop
Feynman integrals can be written as $\rho\,${-}convolutions of one{-}loop integrals but, most of the times, they 
correspond to unphysical momenta, \ie momenta that are complex (even when embedded in Minkowsky space).
\section{Loop amplidudes \label{OLA}}
The first, simple, example is given by the amplitude for $\PH \to \PGg \PGg$ in arbitrary space{-}time dimensions. Let us
consider the top quark contribution:
\bqa
\mrA_{\PH \to \PGg \PGg} &=& - \frac{1}{72\,\sqrt{2}}\,\pi^{1/2 + \ep/2}\,\mt^{ - 1 + \ep/2}\,
\sum_{\mrn=0}^{\infty}\,\mra_{\mrn}\,\frac{\lpar \frac{1}{4}\,\lambda_{\mrs} \rpar^{\mrn}}{\mrn\,!} \spc
\nl
\mra_{\mrn} &=& \Bigl( 4 - \frac{3}{\mrn + 1} - \frac{2}{\mrn + 2} \Bigr)\,\Phi_0(\mrn) \spc
\nl
\Phi_0(\mrn) &=& \frac{\eG{\mrn + 1}\,\eG{\mrn + 1 - \ep/2}}
                      {\eG{\mrn + 5/2}} \spc
\eqa
where $\mrs = \lambda_s\,\mts$ and $\mrs$ is the Higgs boson virtuality.
It is also possible to rewrite the sum in terms of generalized hypergeometric function, therefore,
offering an analytic continuation:
\bqa
\mrA_{\PH \to \PGg \PGg} &=& - \frac{1}{18\,\sqrt{2}}\,\pi^{\ep/2}\,\mt^{ - 1 + \ep/2}\,\eG{1 - \frac{\ep}{2}}
\nl
{}&\times& \Bigl[
  4\,\ghyp{2}{1}\lpar 1\,,\,1 - \frac{\ep}{2}\,;\,\frac{5}{2}\,;\,\mrz\rpar 
- 3\,\ghyp{3}{2}\lpar 1\,,\,1\,,\,1 - \frac{\ep}{2}\,;\,2\,,\,\frac{3}{2}\,;\,\mrz \rpar
 - \ghyp{4}{3}\lpar 1\,,\,2\,,\,4\,,\,1 - \frac{\ep}{2}\,;\,3\,,\,3\,,\,\frac{5}{2}\,;\,\mrz \rpar
\Bigr] \spc
\eqa
where $\mrz= \lambda_{\mrs}/4$. The important result is that the generalized hypergeometric functions involved are finite
at $\mrs = 4\,\mts$ and they show a branch cut for $\mrs \in [4\,\mts\,,\,\infty]$.
It is interesting to observe that for $\mrd = 3$ the behavior at $\mrs = 4\,\mts$
shows a logarithmic singularity.         

The next example will be the amplitude for $\PH(\mrp_1) \to \Pg(\mrp_2) + \Pg(\mrp_3) + \Pg(\mrp_4)$ where 
we will use the \textit{color flow} approach of \Bref{Maltoni:2002mq}.
Several preliminar ingredients are needed.
To discuss the color structure it is more convenient to use
\bq
\mrA^{\mu_2\mu_3\mu_4}_{\{i,j\}} \lpar
\PH(\mrp_1) + \Pg(\mrp_2) + \Pg(\mrp_3) + \Pg(\mrp_4) \to 0\rpar =
\sum_{\sigma \in \mrS_3}\,
\delta^{\sigma(i_2)}_{\sigma(j_3)}\,
\delta^{\sigma(i_3)}_{\sigma(j_4)}\,
\delta^{\sigma(i_4)}_{\sigma(j_2)}\,
{\mathcal A}^{\mu_2\mu_3\mu_4}\lpar \sigma(2),\sigma(3),\sigma(4)\rpar,
\label{colorflow}
\eq
and the sum in \eqn{colorflow} is over the permutations of the three 
gluons $\{2,3,4\}$.
The most general Lorentz decomposition for each color-flow amplitude 
in $\PH \to \Pg \Pg \Pg$ will be
\bqa
\mcA^{\mu_2\mu_3\mu_4}(2\,,\,3\,,\,4) &=& 
 \sum_{i=2,4}\,\Bigl[ \mrB_{1i}(\mrs,\mrt,\mru)\,\delta^{\mu_2,\mu_3}\,\mrp_i^{\mu_4}+
                      \mrB_{2i}(\mrs,\mrt,\mru)\,\delta^{\mu_2,\mu_4}\,\mrp_i^{\mu_3}+
                      \mrB_{3i}(\mrs,\mrt,\mru)\,\delta^{\mu_3,\mu_4}\,\mrp_i^{\mu_2}\Bigr]
\nl
{}&+& \sum_{i,j,k=2,4}\,\mrC_{ijk}(\mrs,\mrt,\mru)\,\mrp_i^{\mu_2}\,\mrp_j^{\mu_3}\,\mrp_k^{\mu_4}.
\label{Ldec}
\eqa
Imposing WST identities~\cite{Ward:1950xp,Taylor:1971ff,Slavnov:1972fg}
requires the following relations among the form factors of \eqn{Ldec},
\bqa
\mrB_{13}(\mrs,\mrt,\mru) &=& - \frac{\mru}{\mrs}\,\mrB_{12}(\mrs,\mrt,\mru), \quad
\mrB_{24}(\mrs,\mrt,\mru) = - \frac{\mrt}{\mrs}\,\mrB_{22}(\mrs,\mrt,\mru), \quad
\mrB_{34}(\mrs,\mrt,\mru) = - \frac{\mrt}{\mru}\,\mrB_{33}(\mrs,\mrt,\mru),
\nl
\mrC_{323}(\mrs,\mrt,\mru) &=& - \frac{\mru}{\mrs}\,\mrC_{322}(\mrs,\mrt,\mru), \quad
\mrC_{442}(\mrs,\mrt,\mru) = - \frac{\mrt}{\mrs}\,\mrC_{422}(\mrs,\mrt,\mru), \quad
\mrC_{443}(\mrs,\mrt,\mru) = - \frac{\mrt}{\mrs}\,\mrC_{343}(\mrs,\mrt,\mru),
\nl
\mrC_{423}(\mrs,\mrt,\mru) &=& \frac{2}{\mrs}\,\Bigl[ \mrB_{22}(\mrs,\mrt,\mru) 
- \frac{\mru}{2}\,\mrC_{422}(\mrs,\mrt,\mru)\Bigr], \quad
\mrC_{343}(\mrs,\mrt,\mru) = \frac{2}{\mrs}\,\Bigl[ \mrB_{33}(\mrs,\mrt,\mru) 
- \frac{\mru}{2}\,\mrC_{342}(\mrs,\mrt,\mru)\Bigr], 
\nl
\mrC_{342}(\mrs,\mrt,\mru) &=& \frac{2}{\mrs}\,\Bigl[ \mrB_{12}(\mrs,\mrt,\mru) 
- \frac{\mrt}{2}\,\mrC_{322}(\mrs,\mrt,\mru)\Bigr], 
\nl
\mrC_{422}(\mrs,\mrt,\mru) &=& \frac{2}{\mru}\,\Bigl[ \mrB_{12}(\mrs,\mrt,\mru) 
+ \mrB_{22}(\mrs,\mrt,\mru) - \frac{\mrt}{2}\,\mrC_{322}(\mrs,\mrt,\mru)\Bigr],
\label{gggWI}
\eqa
with Mandelstam invariants,
\bq
\mrs= - (\mrp_1+\mrp_2)^2 = - (\mrp_3+\mrp_4)^2, \quad
\mrt= - (\mrp_1+\mrp_4)^2 = - (\mrp_2+\mrp_3)^2, \quad
\mru= - (\mrp_1+\mrp_3)^2 = - (\mrp_2+\mrp_4)^2,
\eq
satisfying $\mrs+\mrt+\mru= \mhs$. As usual, we have considered WST identities for
the $S$-matrix elements, i.e. when a gluon line is contracted the remaining
two are assumed to be physical.

To write the complete expression for the amplitude 
we introduce the following combinations of form factors:
\bqa
\mrF_1(\mrs,\mrt,\mru) &=&
           \mrB_{12}(\mrs,\mrt,\mru)
          - \frac{\mrs}{\mru}\,\mrB_{22}(\mru,\mrs,\mrt)
          + \mrB_{33}(\mrt,\mru,\mrs) \spc
\nl
\mrF_2(\mrs,\mrt,\mru) &=&
          - \frac{2}{\mrt}\,\mrB_{22}(\mrs,\mrt,\mru)
          + 2\,\frac{\mrs}{t u}\,\mrB_{33}(\mrs,\mrt,\mru)
          + \mrC_{322}(\mrs,\mrt,\mru) \spc
\nl
\mrF_3(\mrs,\mrt,\mru) &=&
           \mrF_2(\mrs,\mrt,\mru)
          + \frac{\mrs \mru}{t^2}\,\mrF_2(\mrt,\mru,\mrs)
          + \frac{s^2}{t u}\,\mrF_2(\mru,\mrs,\mrt) \spp
\eqa
It turns out that, after summing over all permutations of the gluons, only 
two form factors, i.e. $\mrF_1$ and $\mrF_3$, are needed to parametrize the 
amplitude. Neglecting terms which will vanish after contraction with the 
polarization vectors we may write
\bq
\mrA^{\mu_2\mu_3\mu_4}_{\{i,j\}} =
\delta^{i_2}_{j_3}\,\delta^{i_3}_{j_4}\,\delta^{i_4}_{j_2}\,
\mrA^{\mu_2\mu_3\mu_4}_a +
\delta^{i_2}_{j_4}\,\delta^{i_3}_{j_2}\,\delta^{i_4}_{j_3}\,
\mrA^{\mu_2\mu_3\mu_4}_b \spp
\label{tCF}
\eq
The two color flows of \eqn{tCF} have coefficients defined by,
\bqa
\mrA^{\mu_2\mu_3\mu_4}_a &=&
  \mrF_1(\mrs,\mrt,\mru)\,\mrO^{\mu_2\mu_3\mu_4}_1
+ \mrF_1(\mrt,\mru,\mrs)\,\mrO^{\mu_2\mu_3\mu_4}_2
\nl
{}&+& \mrF_1(\mru,\mrs,\mrt)\,\mrO^{\mu_2\mu_3\mu_4}_3
+ \mrF_3(\mrs,\mrt,\mru)\,\mrO^{\mu_2\mu_3\mu_4}_4 \spc
\nl
\mrA^{\mu_2\mu_3\mu_4}_b &=&
- \mrF_1(\mru,\mrt,\mrs)\,\mrO^{\mu_2\mu_3\mu_4}_1
- \mrF_1(s,u,t)\,\mrO^{\mu_2\mu_3\mu_4}_2
\nl
{}&-& \mrF_1(\mrt,\mrs,\mru)\,\mrO^{\mu_2\mu_3\mu_4}_3
- \mrF_3(\mrt,\mrs,\mru)\,\mrO^{\mu_2\mu_3\mu_4}_4 \spc
\eqa
where we have defined the following operators,
\bqa
\mrO_1^{\mu_2 \mu_3 \mu_4} &=& (\mrp_2 - \frac{\mru}{\mrs} \mrp_3)^{\mu_4}\,\Bigl[
   \frac{2}{\mru}\,\mrp_2^{\mu_3}\,\mrp_4^{\mu_2} 
 + \frac{2}{\mrs \mru}\,\mrp_4^{\mu_3}\,(\mru \mrp_3 - \mrt \mrp_4)^{\mu_2}
 + \delta^{\mu_2\mu_3} \,\Bigr] \spc
\nl
\mrO_2^{\mu_2 \mu_3 \mu_4} &=& - (\frac{\mrs}{\mrt} \mrp_2 - \mrp_4)^{\mu_3}\,\Bigl[
  \frac{2}{\mrs}\,\mrp_3^{\mu_4}\,\mrp_4^{\mu_2} 
+ \frac{2}{\mrs \mrt}\,\mrp_3^{\mu_2}\,(\mrs \mrp_2 - \mru \mrp_3)^{\mu_4}
+ \delta^{\mu_2\mu_4}  \,\Bigr] \spc
\nl
\mrO_3^{\mu_2 \mu_3 \mu_4} &=& ( \mrp_3 - \frac{\mrt}{\mru} \mrp_4)^{\mu_2}\,\Bigl[
 \frac{2}{\mru}\,\mrp_2^{\mu_4}\,\mrp_4^{\mu_3}
- \frac{2}{\mrt \mru}\,\mrp_2^{\mu_3}\,(\mrs \mrp_2 - \mru \mrp_3)^{\mu_4}
+ \delta^{\mu_3\mu_4}  \,\Bigr] \spc
\nl
\mrO_4^{\mu_2 \mu_3 \mu_4} &=& 
\frac{1}{\mrs^2 \mru}\,(\mru \mrp_3 - \mrt \mrp_4)^{\mu_2}\,
(\mrs \mrp_2 - \mru \mrp_3)^{\mu_4}\,
(\mrs \mrp_2 - \mrt \mrp_4)^{\mu_3} \spp
\eqa
It is worth noting that the following identities hold:
\bq
\mrp^{\mu_2}_2\,\mrO_i^{\mu_2 \mu_3 \mu_4} =
\mrp^{\mu_3}_3\,\mrO_i^{\mu_2 \mu_3 \mu_4} =
\mrp^{\mu_4}_4\,\mrO_i^{\mu_2 \mu_3 \mu_4} = 0 \spp
\eq
Dirac statistics, confirmed by an explicit calculation, requires the following
relations:
\bqa
\mrF_1(\mrs,\mrt,\mru) &=& \mrF_1(\mru,\mrt,\mrs) \spc \qquad
\mrF_1(\mru,\mrs,\mrt) = \mrF_1(\mrt,\mrs,\mru) \spc
\nl
\mrF_1(\mrt,\mru,\mrs) &=& \mrF_1(s,u,t) \spc \qquad
\mrF_3(\mrs,\mrt,\mru) = \mrF_3(\mrt,\mrs,\mru) \spp
\eqa
As a consequence we arrive at the final expression for the amplitude,
\bq
\mrA^{\mu_2\mu_3\mu_4}_{\{i,j\}} =
\mrA^{\mu_2\mu_3\mu_4}_a\,\Bigl(
\delta^{i_2}_{j_3}\,\delta^{i_3}_{j_4}\,\delta^{i_4}_{j_2} -
\delta^{i_2}_{j_4}\,\delta^{i_3}_{j_2}\,\delta^{i_4}_{j_3} \Bigr) \spp
\label{tCFfin}
\eq
To proceed further, projectors are introduced; first we define
\bq
\mrH_{ij} = -2\,\spro{\mrp_i}{\mrp_j}, \quad (i,j = 2,\dots,4), \qquad
\mrG= {\mathrm det}\mrH = 2\,\mrs \mrt \mru,
\eq
\bq
\mrP^{\mu\nu}= \delta^{\mu\nu} - 2\,\mrp^{\mu}\,\mrH^{-1}\,\mrp^{\nu} \spc
\qquad
\mrR^{\mu}_i = 2\,\mrH^{-1}_{ij}\,\mrp^{\mu}_j \spp
\eq
Next we introduce
\bq
{\mathcal P}^{j_2,j_3,j_4}_{i_3,i_4,i_2} = - \frac{1}{240}\,\Bigl(
\delta^{j_2}_{i_3}\,\delta^{j_3}_{i_4}\,\delta^{j_4}_{i_2}
- 9\,\delta^{j_2}_{i_4}\,\delta^{j_3}_{i_2}\,\delta^{j_4}_{i_3}\Bigr) \spc
\qquad
{\mathcal Q}^{j_2,j_3,j_4}_{i_3,i_4,i_2} =  \frac{1}{240}\,\Bigl(
 9\,\delta^{j_2}_{i_3}\,\delta^{j_3}_{i_4}\,\delta^{j_4}_{i_2}
- \delta^{j_2}_{i_4}\,\delta^{j_3}_{i_2}\,\delta^{j_4}_{i_3} \Bigr) \spp
\eq
and define projections
\bqa
\mrA_{1i} &=& {\mathcal P}^{j_2,j_3,j_4}_{i_3,i_4,i_2}\,
           \mrA^{\mu_2\mu_3\mu_4}_{\{i,j\}}\,\mrP_{\mu_2\mu_3}\,\mrR^i_{\mu_4} \spc 
\quad
\mrA_{2i} = {\mathcal P}^{j_2,j_3,j_4}_{i_3,i_4,i_2}\,
           \mrA^{\mu_2\mu_3\mu_4}_{\{i,j\}}\,\mrP_{\mu_2\mu_4}\,\mrR^i_{\mu_3} \spc
\nl
\mrA_{3i} &=& {\mathcal P}^{j_2,j_3,j_4}_{i_3,i_4,i_2}\,
           \mrA^{\mu_2\mu_3\mu_4}_{\{i,j\}}\,\mrP_{\mu_3\mu_4}\,\mrR^i_{\mu_2}, 
\quad
\mrA_{4i} = {\mathcal Q}^{j_2,j_3,j_4}_{i_3,i_4,i_2}\,
           \mrA^{\mu_2\mu_3\mu_4}_{\{i,j\}}\,\mrP_{\mu_2\mu_3}\,\mrR^i_{\mu_4} \spc 
\nl
\mrA_{5i} &=& {\mathcal Q}^{j_2,j_3,j_4}_{i_3,i_4,i_2}\,
           \mrA^{\mu_2\mu_3\mu_4}_{\{i,j\}}\,\mrP_{\mu_2\mu_4}\,\mrR^i_{\mu_3} \spc
\quad
\mrA_{6i} = {\mathcal Q}^{j_2,j_3,j_4}_{i_3,i_4,i_2}\,
           \mrA^{\mu_2\mu_3\mu_4}_{\{i,j\}}\,\mrP_{\mu_3\mu_4}\,\mrR^i_{\mu_2} \spc 
\nl
\mrA_{p1,kln} &=& {\mathcal P}^{j_2,j_3,j_4}_{i_3,i_4,i_2}\,
        \mrA^{\mu_2\mu_3\mu_4}_{\{i,j\}}\,\mrR^k_{\mu_2}\,\mrR^l_{\mu_3}\,\mrR^n_{\mu_4} \spc
\quad
\mrA_{p2,kln} = {\mathcal Q}^{j_2,j_3,j_4}_{i_3,i_4,i_2}\,
        \mrA^{\mu_2\mu_3\mu_4}_{\{i,j\}}\,\mrR^k_{\mu_2}\,\mrR^l_{\mu_3}\,\mrR^n_{\mu_4} \spc
\label{Aprj}
\eqa
One finds by direct calculation that $\mrF_1(\mrs,\mrt,\mru)$ and $\mrF_3(\mrs,\mrt,\mru)$ are given
by
\bqa
\mrF_1(\mrs,\mrt,\mru) &=&  \frac{1}{2-\mrd}\,\Bigl[
         \mrA_{12}
       + \frac{1}{2}\,\frac{\mrs}{\mrt}\,\mrA_{14}
       + \frac{\mrs}{\mru}\,\mrA_{23}
       - t\,\mrA_{\mrp1,322} \Bigr] \spc
\nl
\mrF_3(\mrs,\mrt,\mru) &=& \frac{1}{2-\mrd}\,\Bigl[
        \frac{3}{\mrt}\,\mrA_{12}
       + \frac{\mrd+1}{2}\,\frac{\mrs}{\mrt^2}\,\mrA_{14}
       + (\mrd+1)\,\frac{\mrs}{\mrt \mru}\,\mrA_{23}
       + (\mrd-5)\,\mrA_{\mrp1,322} \Bigr] \spp
\label{expl}
\eqa

We now return to $\PH \to \Pg \Pg \Pg$ with internal top quark lines; the amplitude splits into a part
containing boxes and a part containing triangles. For the latter the main ingredient is defined by
\bq
\mrT(\mrm_1\,,\,\mrm_2) =
\eG{1 - \frac{\ep}{2}}\,\pi^{\ep/2}\,\mt^{\ep - 2}\,
\chi_{\mrT}\lpar - 1 + \frac{\ep}{2}\,;\,\lambda_{\mrs}\,;\,\mrm_1\,,\,\mrm_2 \rpar \spc
\eq
where we have defined
\bq
\chi_{\mrT}\lpar \mrn\,;\,\lambda_{\mrs}\,;\,\mrm_1\,,\,\mrm_2 \rpar =
\int_0^1 \mrd \mrx_1\,\int_0^{\mrx_1} \mrd \mrx_2\,
\mrx_1^{\mrm_1}\,\mrx_2^{\mrm_2}\,\mrP^{\mrn}(\mrx_1\,,\,\mrx_2) \spc
\eq
\bq
\mrP(\mrx_1\,,\,\mrx_2) = 1 - \lh\,\mrx_2 + \ls\,\mrx_2^2 + (\lh - \ls)\,\mrx_1\,\mrx_2 \spc
\eq
where $\mhs = \lh\,\mts$. The result can be written as a multiple sum,
\bqa
\chi_{\mrT}\lpar - 1 + \frac{\ep}{2}\,;\,\lambda_{\mrs}\,;\,\mrm_1\,,\,\mrm_2 \rpar &=& 
       \sum_{\mrn_1,\mrn_2\,\mrn_3=0}^{\infty}\,
       \frac{\Phi_0(\mathbf n\,,\,\ep)}{\mrn_1\,!\;\mrn_2\,!\,\mrn_3\,!}\,
       \Phi({\mathbf m}\,,\,{\mathbf n}\,,\,\ep)\,
       (\lh - \ls)^{\mrn_1}\,\mrx_{\mrs\,-}^{ - \mrn_2}\,\mrx_{\mrs\,+}^{ - \mrn_3}
\nl
{}&=& \sum_{\mrn_1,\mrn_2\,\mrn_3=0}^{\infty}\,
       \Phi({\mathbf m}\,,\,{\mathbf n}\,,\,\ep)\,
       \Phi_0(\mathbf n\,,\,\ep)\,
       \mrp({\mathbf n}\,,\,\ls) \spc
\eqa
where ${\mathbf n} = \{\mrn_1\,,\,\mrn_2\}$ and where
\bqa
\Phi_0(\mathbf n\,,\,\ep) &=& \frac{\eG{\mrn_1 + \mrn_2 + 1 - \ep/2}\,\eG{\mrn_1 + \mrn_3 + 1 - \ep/2}}
                {\eG{\mrn_1 + 1 - \ep/2}} \spc
\nl
\Phi({\mathbf m}\,,\,{\mathbf n}\,,\,\ep) &=& 
\lpar 1 + \mrm_1 + \mrn_1 \rpar^{-1}\,\lpar 1 + \mrm_2 + \mrn_3 + \mrn_2 + \mrn_1 \rpar^{-1} \spc
\nl
&-& \lpar 1 + \mrm_1 + \mrn_1 \rpar^{-1}\,\lpar 2 + \mrm_2 + \mrm_1 + \mrn_3 + \mrn_2 + 2\,\mrn_1 \rpar^{-1} 
\spc
\eqa
\bq
\mrx_{\mrs\,\pm} = \frac{1}{2}\,\frac{\lh}{\ls}\,\Bigl( 1 \pm \sqrt{1 - 4\,\frac{\ls}{\lh^2}} \Bigr) \spp
\eq
The phase space for the process is defined by
\bq
0 \le \ls \le \lh \spc \qquad
0 \le \lt \le \lh \spc \qquad
0 \le \lu \le \lh \spc \qquad \lh = \frac{\mhs}{\mts} < 1 \spp
\eq
The roots $\mrx_{\mrs\,\pm}$ are real and positive for $\mrs <  \mhq/(4\,\mts)$ or complex 
conjugate for $\mhq/(4\,\mts) < \mrs < \mhs$.

The projections of \eqn{Aprj} can be written in a compact form, \eg
\bqa
\mrA_{14} &=& \pi^{ - \ep/2}\,\eG{1 + \frac{\ep}{2}}\,\mt^{ - \ep}\,
       \sum_{\mrn_1,\mrn_2\,\mrn_3=0}^{\infty}\,
       \Phi_0({\mathbf n}\,,\,\ep)\,
\Bigl[ \mrp({\mathbf n}\,,\,\ls)\,\mrA^{\mrs}_{14}\,\ls^{-1} + 
       \mrp({\mathbf n}\,,\,\lu)\,\mrA^{\mru}_{14}\,\lu^{-1} \Bigr] \spc
\nl
\mrA^{\mrs}_{14} &=&
          - \Phi(0,0\,,\,{\mathbf n}\,,\,\ep)\,(\lu - \lt - \lh - \ls)
          - 2\,\Phi(0,1\,,\,{\mathbf n}\,,\,\ep)\,(2\,\lu - 2\,\lt + \ls)
\nl
{}&+& 2\,\Phi(1,0\,,\,{\mathbf n}\,,\,\ep)\,(2\,\lu - 2\,\lt - \lh - \ls)
          +  2\,\Phi(0,2\,,\,{\mathbf n}\,,\,\ep)\,\ls
\nl
{}&+& 4\,\Phi(1,1\,,\,{\mathbf n}\,,\,\ep)\,(\lu - \lt)
          - 2\,\Phi(2,0\,,\,{\mathbf n}\,,\,\ep)\,(2\,\lu - 2\,\lt - \lh) \spc
\nl
\mrA^{\mru}_{14} &=&
          -  \Phi(0,0\,,\,{\mathbf n}\,,\,\ep)\,(\lu - \lt + \lh + \ls)
          + 2\,\Phi(0,1\,,\,{\mathbf n}\,,\,\ep)\,(\lu + 2\,\lt - 2\,\ls)
\nl
{}&+& 2\,\Phi(1,0\,,\,{\mathbf n}\,,\,\ep)\,(\lu - 2\,\lt + \lh + 2\,\ls)
          - 2\,\Phi(0,\,,\,{\mathbf n}\,,\,\ep2)\,\ls
\nl
{}&-& 2\,\Phi(1,1\,,\,{\mathbf n}\,,\,\ep)\,(\lu + 2\,\lt - 3\,\ls)
          + 2\,\Phi(2,0\,,\,{\mathbf n}\,,\,\ep)\,(2\,\lt - \lh - 2\,\ls) \spp
\eqa
For boxes we define
\bq
\mrB(\mrm_1\,,\,\mrm_2\,,\,\mrm_3) =
\eG{2 - \frac{\ep}{2}}\,\pi^{\ep/2}\,\mt^{\ep - 4}\,
\chi_{\mrB}
\lpar - 2 + \frac{\ep}{2}\,;\,\lambda_{\mrs}\,,\,\lambda_{\mrt}\,,\,\lambda_{\mru}\,,\,
\mrm_1\,,\,\mrm_2\,,\,\mrm_3 \rpar \spc
\eq
where we have defined
\bq
\chi_{\mrB}\lpar \mrn\;\,\lambda_{\mrs}\,,\,\lambda_{\mrt}\,,\,\lambda_{\mru}\,,\,\mrm_1\,,\,\mrm_2 \rpar =
\int_0^1 \mrd \mrx_1\,\int_0^{\mrx_1} \mrd \mrx_2\,\int_0^{\mrx_2} \mrd \mrx_3\,
\mrx_1^{\mrm_1}\,\mrx_2^{\mrm_2}\,\mrx_3^{\mrm_3}\,\mrP^{\mrn} \spc
\eq
\bq
\mrP = 1 - \lh\,\mrx_1 + (\lh - \ls)\,\mrx_2 + \ls\,\mrx_3 +
\lh\,\mrx^2_1 + (\ls - \lh)\,\mrx_1\,\mrx_2 + (\lu - \lh)\,\mrx_1\,\mrx_3 + \lt\,\mrx_2\,\mrx_3 \spc
\eq
where $\mhs = \lh\,\mts$ \etc

\section{Conclusions \label{Conc}}
In this work, we have analyzed Feynman integrals; following \Bref{TR}, it was suggested in \Bref{Kershaw:1973km}
that Feynman integrals possess \textit{simple} power{-}series expansions. We have shown that Fox functions are
the necessary tool, \ie Feyman integrals are multivariate analytic (Fox) functions and the singularities of the functions
are described by Landau equations~\cite{Landau:1959fi}. We have made clear that the crucial step is related to the 
process of analytic continuation.

After discussing general aspects of generalized hypergeometric functions,
in \sect{OLLP} we presented examples for one{-}loop Feynman integrals using the Feynman representation.
In \sect{TLFI} we have discussed a procedure for the partial quadratization of the Symanzik polynomials 
discussing few explicit examples, with emphasis on the representation of 
these integrals through the generalized, multivariate, Fox function. 
Stated differently, we can say that Fox functions provide a general framework and a concise notation for
Feynman integrals.

It should be stressed that general hypergeometric functions can alternatively be defined as solutions of the
hypergeometric A-systems of Gelfand, Kapranov, and Zelevinsky~\cite{GZK}. For a MB representation of
GKZ hypergeometric functions see \Bref{MH}.
To summarize: we have shown the connection between Feynman integrals and Fox functions; next, we have shown that
a large class of Fox functions can be written in terms of general Horn hypergeometric functions 
(terminology introduced in \Bref{Ghs}). The Horn series and the more populat GKZ series are related~\cite{Ghs}.
It is clear that describing a basis for the solutions of the Horn systems, including an analytic continuation of this
basis, is a well{-}known but difficult problem.

We have addressed the problem of obtaining a computable form for the Fox $\mrH$ function: in \sect{fFtM} we have shown 
how to expand the H function in terms of Meijer $\mrG$ functions; the latter are usually reducible to generalized 
hypergeometric functions as described in \sect{MGfun}. The recursive algorithm of \sect{hmff} can be used in several 
cases. The analytical structure of Fox functions has been discussed in \sect{FRga}. 
The alternative approach of a numerical integration (ab initio) of the Feynman integrals has been discussed in\sect{numi}.

\clearpage
\bibliographystyle{elsarticle-num}
\bibliography{FIFF}
\end{document}